\documentclass[aps,final,notitlepage,oneside,twocolumn,nobibnotes,nofootinbib,superscriptaddress,noshowpacs,centertags]{revtex4-1}
\pdfoutput=1 

\usepackage[utf8]{inputenc}
\usepackage[T2A]{fontenc}
\usepackage[english]{babel}
\usepackage{graphicx}
\usepackage{latexsym}
\usepackage{amssymb}
\usepackage{amsmath}
\usepackage{float}
\usepackage[dvipsnames]{xcolor}
\usepackage[colorlinks,urlcolor=Blue,citecolor=Blue,linkcolor=Blue,pdfusetitle,breaklinks]{hyperref}
\hypersetup{breaklinks=true,colorlinks=true}
\usepackage{natbib}
\usepackage[caption=false]{subfig}
\usepackage{bm}
\renewcommand{\thesection}{\arabic{section}}

\makeatletter
\renewcommand{\p@subsection}{}
\makeatother

\newcommand{\eq}{\text{eq}}
\newcommand{\dec}{\text{dec}}
\newcommand{\dm}{\text{\tiny DM}}

\newcommand{\pbh}{\text{\tiny PBH}}

\newcommand{\PK}[1]{{#1}}
\begin{document}
\title{Primordial black holes}
\author{V.I.~Dokuchaev}\thanks{e-mail: dokuchaev@inr.ac.ru}
\affiliation{Institute for Nuclear Research of the Russian Academy of Sciences, prospekt 60-letiya Oktyabrya 7a, 117312 Moscow, Russia}
\author{Yu.N.~Eroshenko}\thanks{e-mail: eroshenko@inr.ac.ru}
\affiliation{Institute for Nuclear Research of the Russian Academy of Sciences, prospekt 60-letiya Oktyabrya 7a, 117312 Moscow, Russia}
\author{V.V.~Nikulin}\thanks{e-mail: vvnikulin@mephi.ru}
\affiliation{National Research Nuclear University MEPhI, 115409 Moscow, Russia}
\author{K.A.~Postnov}\thanks{e-mail: kpostnov@gmail.com}
\affiliation{Sternberg Astronomical Institute, Lomonosov Moscow State University, Universitetskii prospekt 13, 119234 Moscow, Russia}
\affiliation{Institute for Nuclear Research of the Russian Academy of Sciences, prospekt 60-letiya Oktyabrya 7a, 117312 Moscow, Russia}
\author{S.G.~Rubin}\thanks{e-mail: sergeirubin@list.ru}
\affiliation{National Research Nuclear University MEPhI, 115409 Moscow, Russia}
\author{V.D.~Stasenko}\thanks{e-mail: stasenkovd@gmail.com}
\affiliation{National Research Nuclear University MEPhI, 115409 Moscow, Russia}
\begin{abstract}
Primordial black holes, whose formation in the early Universe was first hypothesized by Ya.B.~Zel'dovich and I.D.~Novikov in 1966 (the text of this pioneering article is provided in Appendix II to this review), have attracted increasing attention in recent years. Modern astronomical data across the electromagnetic spectrum, together with observations of gravitational waves and cosmic neutrinos, have revealed a number of unique astrophysical phenomena that may be related to primordial black holes formed in the pre-stellar cosmological epoch. This review describes various models of primordial black hole formation, as well as methods for testing these models with astronomical and astrophysical observations. In particular, we discuss in detail the formation of primordial black holes and their clusters through the collapse of closed domain walls and through effects associated with extra spatial dimensions of the Universe.
\end{abstract}
\maketitle 
\tableofcontents

\section{Introduction}

Recent observations of gravitational waves with the ground-based LIGO/Virgo/KAGRA detectors~\cite{Rei17} and of the shadows (silhouettes) of the supermassive black holes (BHs) in the galaxy M87 and in the Milky Way~\cite{DokNaz20} have greatly strengthened confidence in the existence of BHs. Figuratively speaking, a BH is an object whose gravity is so strong that it traps even light. The first hypothesis about the existence of such invisible objects was formulated within Newtonian theory by John Michell in 1783 and, somewhat later, by Pierre-Simon Laplace in 1799. From the modern point of view, BHs are described by solutions of the equations of General Relativity (GR), and these objects are already observed with a high degree of confidence.

Several possible classes of BHs can be distinguished. Stellar-origin BHs, born in the core collapse of massive stars, are observed in close binary systems through the X-ray emission produced by accretion~\cite{Cherepashchuk:2014}. Mergers of BHs with stellar masses are currently being detected by the ground-based gravitational-wave observatories LIGO/Virgo/KAGRA~\cite{Rei17}. Supermassive BHs reside at the centers of most large galaxies~\cite{Dok91,Dol18}. In particular, they manifest themselves as extremely luminous objects --- quasars. Several formation mechanisms of supermassive BHs have been proposed, but this issue has not yet been finally resolved. There is also evidence for intermediate-mass BHs, which are detected through the dynamics of stars in the centers of dwarf galaxies and of some globular clusters. Possible production of microscopic BHs in collisions of high-energy particles has been discussed theoretically, but this process has not been observed so far. One more possible class of BHs is primordial black holes (PBHs), which are born in the early Universe, before the formation of the first stars and galaxies.

PBHs have not yet been reliably detected, although some known astrophysical objects may be PBHs or may be evolutionarily related to them, so the existence of PBHs is quite plausible (see, e.g., the reviews~\cite{Dol18,2024PhR..1054....1C}). Several well-motivated scenarios of PBH formation in the early Universe have been proposed. In this review, we discuss various mechanisms of PBH formation and their possible cosmological and astrophysical manifestations. The main idea is that the early Universe contained inhomogeneities which could collapse into PBHs long before the formation of stars and galaxies. Moreover, the nature of inhomogeneities leading to PBH formation may be entirely different: they may be, for example, curvature perturbations or closed domain walls.

A prelude to the idea of PBHs was the 1964 paper by I.D.~Novikov~\cite{Nov64}, which considered the possible existence in the Universe of regions whose cosmological expansion is delayed (``delayed cores''). These regions could have a density exceeding the mean density of the Universe. I.D.~Novikov proposed the collision of the surface of such regions with matter falling onto them from outside as a model for the emission of quasars. A similar idea of regions with delayed cosmological expansion was put forward independently by Y.~Ne'eman~\cite{1965ApJ...141.1303N}. The idea of delayed expansion, in turn, goes back to the ideas of V.A.~Ambartsumian about superdense massive objects in the nuclei of active galaxies~\cite{1962PrCmg...8....2A,Nov64}. As we now know, these objects are supermassive BHs. In modern terms, the delayed expansion discussed in~\cite{Nov64} can be associated with the presence of adiabatic density perturbations (curvature perturbations), since in the Lagrangian description the evolution of such perturbations corresponds to a slower expansion of an overdense region compared with the mean expansion rate of the Universe.

The 1966 paper by Ya.B.~Zel'dovich and I.D.~Novikov~\cite{ZelNov66} already explicitly discussed the possibility of gravitational collapse of regions with delayed expansion, which marks the birth of the idea of PBHs. As long as the size of a density perturbation exceeds the size of the cosmological horizon ($2ct$ during the radiation-dominated era), matter cannot collapse into a BH, because different parts of the perturbation are not causally connected. At the radiation-dominated stage (when the density of the early Universe was dominated by a mixture of relativistic particles, conventionally called radiation), the radius of the cosmological horizon is of the order of the gravitational radius $r_g=2GM_{\rm H}/c^2$ of the mass contained within the horizon: the mass $M_{\rm H}$ equals the volume $4\pi(2ct)^3/3$ multiplied by the mean radiation density $\bar\rho=3/(32\pi Gt^2)$, i.e., $M_\textrm{H}=(c^3/G)t$. Therefore, gravitational collapse is possible immediately after horizon crossing, and the relation between the collapse time and the PBH mass is given by~\cite{ZelNov66}
\begin{equation}
t_{\rm H}=\frac{GM_{\rm PBH}}{c^3}=
0.5
\left(\frac{M_{\rm PBH}}{10^5M_{\odot}}\right)\mbox{~s}.
\label{th}
\end{equation}
Thus, the smaller the PBH mass, the earlier it could have formed, and PBHs with masses smaller than typical stellar masses could form in the early Universe. The ``hoop'' criterion for the formation of a BH of mass $M$ from a density fluctuation $\delta \rho/\bar\rho$ yields a simple condition on the fluctuation amplitude: $\delta\rho/\bar\rho\ge(M_{\rm H}/M)^2$, i.e., PBHs with masses of the order of the mass inside the cosmological horizon will preferentially form from perturbations with $\delta \rho/\bar\rho\sim 1$. However, a more detailed analysis, discussed below, somewhat lowers the threshold amplitude of the fluctuation.

By the time~\cite{ZelNov66} was published, the cosmic microwave background (CMB) had already been discovered and the hot model of the Universe had become established. Ref.~\cite{ZelNov66} considered the accretion of radiation onto PBHs and asked whether it could become catastrophic, leading to unlimited mass growth. However, the approximate method used in~\cite{ZelNov66} did not allow this point to be fully clarified, so the question of catastrophic accretion was left open. The paper~\cite{ZelNov66} obtained, for the first time, a constraint on the fraction of collapsing matter (from the requirement that the present PBH density be of the order of or less than the critical density), and noted that this constraint is in fact a constraint on the spectra of density perturbations that lead to PBH formation.

A model of PBH formation was also discussed by S.~Hawking in 1971~\cite{Haw71}. That paper considered the possibility that PBHs carry an electric charge. Charged PBHs could capture electrons or protons and form atom-like bound systems. The paper~\cite{Haw71} also discussed the capture of PBHs into the Sun and stars, as well as accretion of matter onto a PBH that has ended up inside a neutron star after a supernova explosion. Such accretion could eventually consume the entire neutron star. Finally, it was proved for the first time in~\cite{1974MNRAS.168..399C} that on a cosmological background there is no inevitable catastrophic accretion of matter onto a PBH that would lead to unlimited growth of its mass. Thus, Ref.~\cite{1974MNRAS.168..399C} was the first to demonstrate the self-consistency of the PBH hypothesis with respect to accretion.
 
The initial surge of interest in PBHs occurred after S.~Hawking showed theoretically that PBHs should undergo quantum evaporation, emitting photons and other particles~\cite{1974Natur.248...30H}. PBH evaporation would produce an additional gamma-ray background, as well as individual gamma-ray bursts~\cite{Car76} (see the recent reviews~\cite{2021RPPh...84k6902C, 2023PrPNP.13104040A}). This effect offered a way to test Hawking's prediction observationally, but quantum evaporation of PBHs has not been reliably detected to date. Nevertheless, interest in the quantum evaporation of PBHs has never faded, and it remains an active research topic. For example, it has been hypothesized that the ultra-high-energy neutrino ($\sim220$~PeV) detected by the KM3NeT neutrino telescope on 13 February 2023 may be related to the evaporation and explosion of a PBH~\cite{KliKai25}.

The PBH hypothesis became especially popular after the LIGO/Virgo/KAGRA interferometers began detecting gravitational-wave events, since some of these events may be explained by mergers of PBH binaries. The idea of such an origin of gravitational-wave events was put forward in~\cite{1997ApJ...487L.139N}, long before the first detections of gravitational waves. This model was subsequently developed in~\cite{Ioka:1998nz,Sasaki:2016jop} and other works. In this context, PBHs are of interest because some of the LIGO/Virgo/KAGRA events correspond to mergers of BHs with spins and masses that are not expected for BHs of stellar origin (see~\cite{PosKurMit19}).

PBHs are also of interest as a possible explanation for a number of astrophysical processes. We list only some of them here (these processes are discussed in more detail later in the review). PBHs have been proposed as dark matter (DM) candidate of the Universe. However, this possibility is already strongly constrained~\cite{Carr:2021bzv}. Constraints on PBHs as cold DM have been obtained, in particular, from the analysis of gravitational microlensing events~\cite{Gre26}. Gas accretion onto PBHs could affect a number of processes and lead to interesting effects. Sufficiently massive PBHs could serve as seeds for the formation of supermassive BHs in early quasars and galaxies~\cite{Blietal16,Zipetal25}. The motion of PBHs in galaxies could exert gravitational dynamical influence on stellar subsystems, and the absence of such an influence constrains the PBH abundance.

The mechanism of PBH formation from adiabatic density perturbations considered in~\cite{ZelNov66} was historically the first, but several other scenarios were proposed later, including the collapse of density perturbations during early matter-dominated stages~\cite{KhlPol80,PolKhl82,PolKhl85}, the collapse of domain walls~\cite{Satetal81,Maeetal82,BerKuzTka83,RubKhlSak00, RubSakKhl01}, of cosmic string loops, or of regions with baryon charge fluctuations~\cite{DolSil93,Dol18}. In addition to isolated PBHs, some models predict the formation of PBH clusters~\cite{RubKhlSak00,RubSakKhl01}. The evolution of such clusters and their possible role in astrophysics and cosmology, within the general scenario of structure formation in the Universe, are of considerable interest.

\begin{figure}
\centering
\includegraphics[angle=0,width=0.48\textwidth]{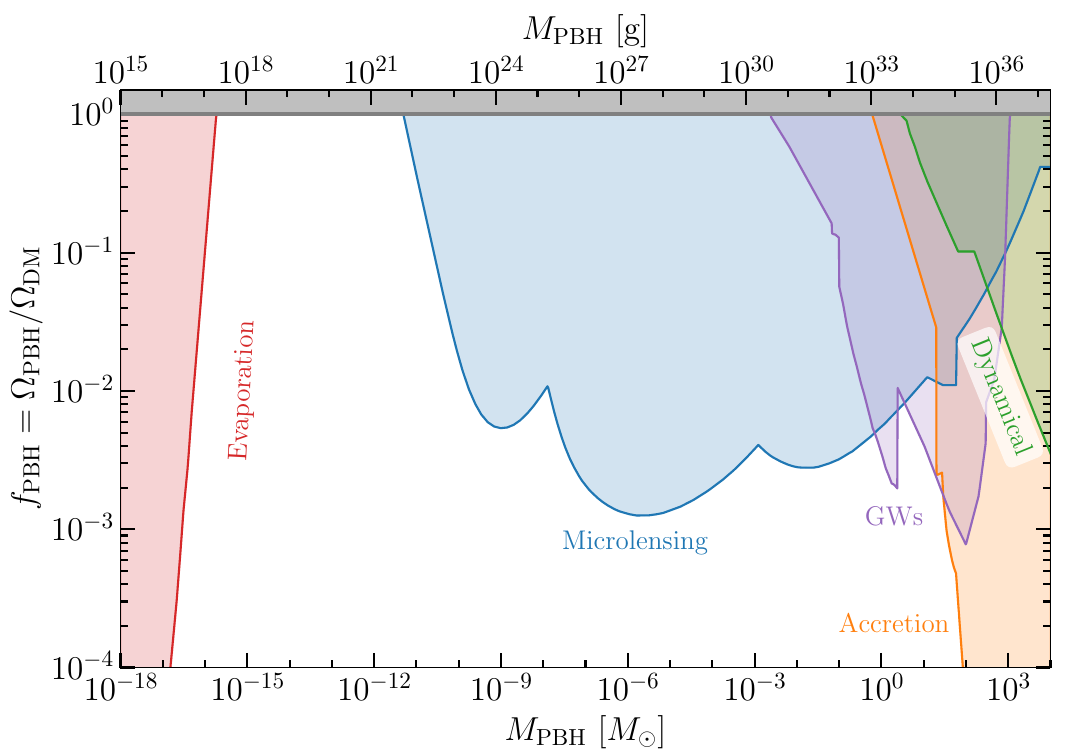}
\caption{Constraints on the mass fraction $f_{\rm PBH}$ of PBHs in cold DM as a function of their mass $M_{\rm PBH}$, derived from various effects for a monochromatic PBH mass function. Shown are the constraints from quantum evaporation of PBHs, from gravitational microlensing, from gravitational-wave (GW) detections, and from gas accretion onto PBHs, as well as dynamical constraints. The figure was produced using the publicly available code PBHbounds~\cite{Kav19}.}
\label{grconstr}
\end{figure}

Suppose that, in one way or another, a population of PBHs has formed during the radiation-dominated stage of the Universe. As the Universe expands, the mean PBH density $\rho_{\rm PBH}$ decreases in the same way as the density of all nonrelativistic particles, $\rho_{\rm PBH}\propto a^{-3}$, where $a(t)$ is the scale factor of the Universe. At the same time, the radiation density decreases as $\rho_r\propto a^{-4}$. This means that the relative contribution of PBHs to the total density of the Universe during the radiation-dominated stage grows as $\propto a(t)$, and this growth continues until the onset of the matter-dominated stage near the moment $t_{\rm eq}$ of matter--radiation equality. After that, PBHs constitute a fixed fraction of the mean DM density, provided that PBH evaporation and accretion-driven mass growth are insignificant. If a fraction $\beta$ of the density of the Universe was converted into PBHs at their formation time $t_{\rm H}$, the present-day cosmological PBH density is
\begin{equation}
\Omega_{\rm PBH}\simeq\beta a(t_{\rm eq})/a(t_{\rm H}).
\label{restr}
\end{equation} 
A similar estimate and a constraint based on it were first given already in~\cite{ZelNov66}. Later, a number of other constraints were obtained on the basis of various effects. A detailed modern review of various PBH constraints is given in~\cite{2021RPPh...84k6902C}. Some of the constraints are very stringent and robust, while others are model-dependent. Importantly, even the absence of any observable PBH effect constrains the PBH abundance in a given mass range and, hence, the parameters of the models that could have produced PBHs. Some of the constraints are shown in Fig.~\ref{grconstr} and explained later in the text.

In this review, we consider various physical mechanisms of PBH formation and possible processes involving PBHs, at different levels of detail. We pay more attention to the effects that have not yet been fully covered in reviews on PBHs but that may prove to be of fundamental importance for astrophysics and cosmology. For most other topics, we explain the basic principles and give bibliographic references. Throughout the review, we use Planck CMB cosmological parameters \cite{Planck:2018vyg}. Unless stated otherwise, we use the Gaussian (CGS) system of units. In some sections, we set $c=1$ or $c=\hbar=1$.

\section{Mechanisms of primordial black hole formation}

\subsection{Formation of primordial black holes from adiabatic perturbations}
\label{adiabsubsec}

We begin with the historically first mechanism of PBH formation from adiabatic density perturbations, also called curvature perturbations. This mechanism was partly described in the Introduction. Adiabatic perturbations are expected to arise during the inflationary stage. On scales larger than the cosmological horizon the amplitudes of adiabatic perturbations in radiation, $\delta_r$, and in matter, $\delta_m$, are of the same order, $\delta_r\sim\delta_m$~\cite{GorRub}.

\medskip  
{\bf Analytical model} 
\medskip

The simplest model of collapse of adiabatic perturbations is the evolution of a spherically symmetric region of space with a uniform density exceeding the mean density of the Universe~\cite{Car75} (see also a brief description of this mechanism in~\cite{Yok97}). This model neglects the inhomogeneities that real perturbations inevitably have. The overdense region can be approximately treated as part of a closed universe evolving against the background of a globally flat Universe. While the background Universe keeps expanding, the closed patch decelerates, turns around, and starts to contract and, under certain conditions, may collapse into a BH. Calculation of the collapse is the simplest and most transparent for a patch with uniform density. We describe the approach used in~\cite{Car75} to study such a collapse, but with a refined value of the PBH formation threshold (see~\cite{Berezin:2020tbk} for details).
   
\begin{figure}
\centering
\includegraphics[angle=0,width=0.35\textwidth]{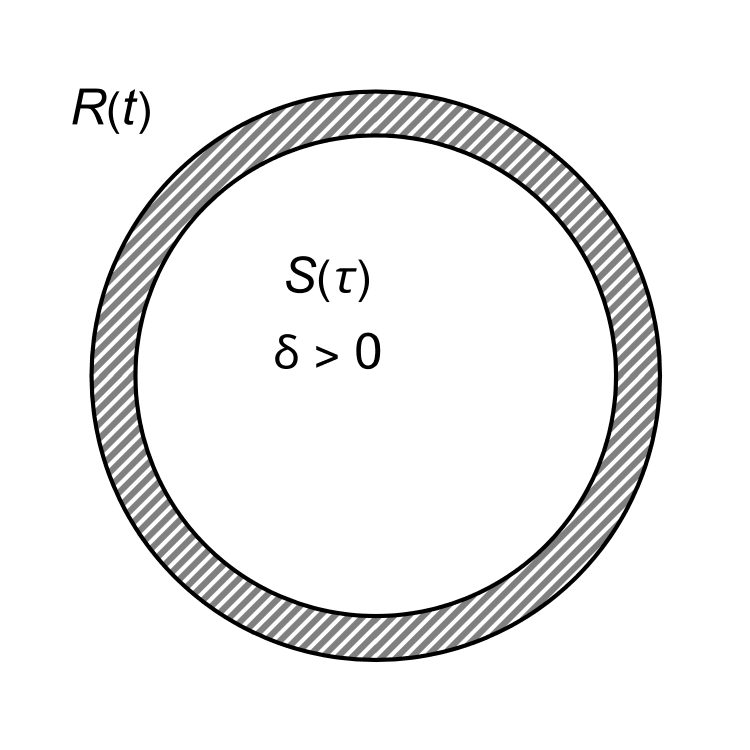}
\caption{The collapsing region is modeled by a part of a closed universe with the scale factor $S(\tau)$. Outside the perturbed region, the Universe is assumed to be flat with the scale factor $R(t)$; transition layers may be present. }
\label{grcoll}
\end{figure}

Following Ref.~\cite{Car75}, we model the perturbed region of space that collapses into a PBH by a patch of a closed universe with the metric
\begin{equation}
ds^2=d\tau^2-S^2\left(\frac{dr^2}{1-r^2}+r^2d\Omega^2\right),
\end{equation}
where $S=S(\tau)$ is the scale factor and $d\Omega^2$ is the metric on the unit 2-sphere. The metric outside the perturbed region is that of the flat Universe,
\begin{equation}
ds^2=dt^2-R^2\left(dr^2+r^2d\Omega^2\right),
\end{equation}
where $R=R(t)$; see Fig.~\ref{grcoll}.

The Friedmann equations for the flat and closed cosmological models read
\begin{equation}
\frac{1}{R^2}\left(\frac{dR}{dt}\right)^2=\frac{8\pi G}{3}\bar\varepsilon,
\label{meqr}
\end{equation}  
\begin{equation}
\frac{1}{S^2}\left(\frac{dS}{d\tau}\right)^2+\frac{1}{S^2}=\frac{8\pi G}{3}\varepsilon.
\label{meqs}
\end{equation}
The initial data are specified at some early time $t_i$ ($\tau_i$) after inflation, when the density perturbations are still small and can be described by linear theory. Quantities at this initial time are denoted by the subscript ``i''. We consider the radiation-dominated cosmological epoch (equation of state $p=\varepsilon/3$), so that 
\begin{equation}
\bar\varepsilon=\bar\varepsilon_i\frac{R_i^4}{R^4}, \qquad \varepsilon=\varepsilon_i\frac{S_i^4}{S^4}.
\end{equation}
The last, closing equation,
\begin{equation}
\frac{d\tau}{dt}=\frac{S}{R(1+\delta_i)^{1/4}},
\label{taut}
\end{equation} 
relates $\tau$ and $t$~\cite{Har70}. This equation follows from the general relations for a comoving coordinate system~\cite{LL-II}. The density perturbation is defined as
\begin{equation}
\delta=\frac{\varepsilon-\bar\varepsilon}{\bar\varepsilon},
\label{defdelta}
\end{equation} 
and the initial constant-time hypersurface is chosen so that at $t=t_i$ ($\tau=\tau_i$) the following conditions hold~\cite{Har70}:
\begin{equation}
S_i=R_i, \qquad \left.dS/d\tau\right|_{\tau_i}=\left.dR/dt\right|_{t_i}. 
\label{hyper}
\end{equation}
Note that some other papers~\cite{Kopp:2010sh, Carr:2014pga} choose the constant-time hypersurface differently (we use the choice adopted in~\cite{Car75}~and~\cite{Har70}).

The condition $dS/d\tau=0$ in (\ref{meqs}) gives the scale factor at the moment $\tau_{\rm max}$ of maximum expansion of the perturbed region, $S_{\rm max}\approx S_i/\delta_i^{1/2}$. The evolution of the scale factor is found in parametric form~\cite{Har70}: $S=S_{\rm max}\sin{\psi}$, where the parameter $\psi$ is related to $\tau$ by $\tau=S_{\rm max}(1-\cos{\psi})$; for the relation between $\psi$ and $t$, see~\cite{Berezin:2020tbk}. The maximum expansion of the perturbed region corresponds to the time $t_{\rm max}\simeq t_i(\pi/2)^2/\delta_i$, with $\tau_{\rm max}=2t_i /\delta_i$.

The criterion for PBH formation was obtained analytically in~\cite{Car75}. Gravitational collapse with PBH formation occurs if the relative amplitude of the fluctuation at the horizon crossing satisfies the conditions
\begin{equation}
\delta_c\le\delta_{\rm H}\le1.
\label{usl0}
\end{equation}
Early works estimated the PBH formation threshold $\delta_c$ using the criterion that a PBH forms if the radius of the perturbed region at the moment its expansion stops exceeds the Jeans radius, $S_{\rm max}>R_J=c_s t$, where the sound speed is $c_s=c/\sqrt{3}$. This corresponds to the left inequality in (\ref{usl0}) and
\begin{equation}
\delta_c\simeq\frac{1}{3},
\label{dc13}
\end{equation}
while the right inequality corresponds to the formation of a BH rather than of a separate universe~\cite{Car75}. The conditions for the formation of a separate universe at $\delta_{\rm H}>1$, and this possibility itself, are discussed in~\cite{Kopp:2010sh} and~\cite{Carr:2014pga}.

As noted in~\cite{HarYooKoh14}, criterion \eqref{dc13} contains an ambiguity in the expression for the Jeans radius and can therefore be used only as an estimate. A refined criterion for PBH formation was derived in~\cite{HarYooKoh14}. According to this criterion, a PBH forms if, by the moment its expansion stops, a sound wave has not had enough time to travel from the periphery of the perturbation to its center. In this case, the Jeans radius is taken to be equal to the radius of the sound horizon.

If the metric of the perturbed region is written as
\begin{equation}
ds^2=d\tau^2-S^2\left(d\chi^2+\sin^2\chi d\Omega^2\right),
\end{equation}
a sound wave obeys $S(d\chi/d\tau)=\pm1/\sqrt{3}$~\cite{HarYooKoh14}. Substituting the expressions given above, we obtain $d\chi/d\psi=\pm1/\sqrt{3}$. Integrating up to the moment of maximum expansion, $\psi=\pi/2$, we find $\chi_{\rm max}=\pi/(2\sqrt{3})$.
The condition that the size of the region equals the Jeans radius (the sound horizon) and the condition of cosmological horizon crossing form the system of equations
\begin{equation}
R_{\rm J}=S_{\rm max}\sin\chi_{\rm max}= S_{\rm max} \sin \chi_a,
\end{equation}
\begin{equation}
R_{\rm H}=2t_H=S_{\rm max}\sin\chi_a\sin\psi_{\rm H},
\end{equation}
where $\chi_a$ is the boundary of the perturbed region and $\psi_{\rm H}$ is the value of the parameter $\psi$ at horizon crossing.
These equations yield a refined value of the PBH formation threshold,
\begin{equation}
\delta_c\simeq0.42.
\label{dcutichn}
\end{equation}

Let us now determine the fraction of PBHs in the total matter density in the Universe. Inflationary theory generally predicts a Gaussian distribution of perturbations, which is supported by the CMB anisotropy observations. The PBH abundance at the time of their formation is then determined from the following estimate~\cite{Car75}. The fraction of the mass of the Universe contained in perturbations whose amplitude exceeds the threshold $\delta_c$ but is less than unity is
\begin{equation}
\begin{array}{l}
{\displaystyle
\beta=
\gamma\int\limits_{\delta_c}^{1}
\frac{\displaystyle d\delta}{\displaystyle\sqrt{2\pi}\sigma_{\rm H}}
\exp(-\frac{\displaystyle\delta^2}{\displaystyle2\sigma_{\rm H}^2})}
\\
\\
{\displaystyle
=\frac{1}{2}\gamma\left[
{\rm erf}\left(\frac{1}{\sqrt{2}\sigma_{\rm H}}\right)-
{\rm erf}\left(\frac{\delta_c}{\sqrt{2}\sigma_{\rm H}}\right)
\right]
}
\\
\\
{\displaystyle
\simeq
\frac{\gamma\sigma_{\rm H}}{\delta_c\sqrt{2\pi}}\exp(-\frac{\delta_
c^2}{2\sigma_{\rm H}^2})},
\label{eq:beta1}
\end{array}
\end{equation}
where ${\rm erf}(x)$ is the error function, $\sigma_{\rm H}$ is the dispersion of density fluctuations on the Hubble-radius scale, and $\gamma=M_\pbh/M_H\leq1$. This expression shows the importance of the PBH formation threshold, because $\delta_c$ enters the exponent. We have also used the asymptotic expression for the error function, since $\sigma_{\rm H} \ll \delta_c$. Otherwise, collapse into PBHs would occur everywhere in the Universe, leading to DM overproduction. Expression (\ref{restr}) immediately gives the relation to the present-day PBH density in the Universe:
\begin{equation}\label{eq:beta2}
f_{\pbh}=\frac{\Omega_{\rm PBH}}{\Omega_{\rm DM}}\sim10^9\beta\left ( \frac{M_\pbh}{M_{\odot}} \right)^{-1/2}.
\end{equation}
If $f_{\pbh}\sim1$, then $\beta\ll1$, and only extremely rare regions in the early Universe could have collapsed into PBHs.

On the other hand, it follows from~\eqref{eq:beta1} that, for PBHs to make a noticeable contribution to DM, the density fluctuations must have an amplitude $\sigma_{\rm H} \simeq 0.06$. This greatly exceeds the value $\sigma_{\rm H} \sim 10^{-5}$ inferred from the CMB and the analysis of the large-scale structure of the Universe. However, the power spectrum of perturbations has been measured only for scales $k\lesssim 10$~Mpc$^{-1}$. From this point of view, PBHs can serve as a tool for probing the power spectrum on very small spatial scales. Thus, PBH formation in this traditional scenario requires an enhancement of the perturbation power spectrum on small scales; possible mechanisms of such an enhancement are discussed below in Section~\ref{nonstandsec}. Note that $\beta$ is exponentially sensitive to $\sigma_{\rm H}$, as can be seen from~\eqref{eq:beta1}; therefore, such models generally require fine-tuning of parameters.

Already in~\cite{Car75}, an attempt was made to calculate the mass function using an approach similar to the Press--Schechter theory. In~\cite{ZabMarNas82}, this method was improved by a mathematically consistent treatment accounting for the effect of correlations in the random field of inhomogeneities on the PBH distribution function.
 

\medskip
{\bf Non-Gaussianity of perturbations} 
\medskip

In most models, PBHs form from large density perturbations --- much larger than the typical perturbation amplitude. For Gaussian perturbations, one says that PBH formation occurs in the tail of the distribution, where the perturbation is much larger than the root-mean-square value. This was demonstrated above in Eq.~(\ref{eq:beta1}). It is therefore clear that deviations of the perturbation distribution function from the standard Gaussian shape can drastically change the quantitative predictions for PBH formation. Indeed, it was shown in~\cite{BulPri97} that the tails of distributions generated during inflation do not follow Gaussian statistics, although this effect is model-dependent. As a result, the number of PBHs formed should be smaller than under the assumption of a strictly Gaussian distribution of perturbations, and the constraint on the perturbation amplitude on the horizon scale, $\sigma_{\rm H}$, can be relaxed by a factor of a few. On the role of non-Gaussianity of perturbations in PBH formation, see also~\cite{1998PhRvD..57.7145I, 2017JCAP...10..046P}.

\medskip
{\bf Numerical simulations}
\medskip

Analytical calculations make it possible to study only fairly simple models of collapse. We have described above the model of evolution of a homogeneous region developed in~\cite{Car75}. In reality, the collapsing region may be inhomogeneous and even non-spherical. The central overdensity may be compensated by an underdense region that either existed initially or arose as a result of evolution of the perturbation. Numerical methods are needed to study such more complex configurations.

An important step in this direction was made in~\cite{NadNovPol78} (see also~\cite{Novetal79}). The authors of~\cite{NadNovPol78} aimed to determine how strongly the perturbed region must differ from a patch of a flat Universe for gravitational collapse to occur in it and a PBH to be born. In~\cite{NadNovPol78}, the collapsing region was initially modeled by a patch of a homogeneous closed universe, but its nonlinear evolution was followed by numerically solving the Einstein equations for the spherically symmetric inhomogeneous case. In addition,~\cite{NadNovPol78} introduced an underdense transition region compensating the mass excess in the perturbed region, while the unperturbed Friedmann model was used outside the transition region. Specifically, the metric in the perturbed region was taken in the form
\begin{equation}
ds^2=dt^2-a^2(t)[dR^2+\sin^2Rd\Omega^2],
\end{equation}
and in the external unperturbed Universe it was given by the same formula with the replacement $\sin^2R\to R^2$. At the initial moment, there is a patch of a homogeneous closed universe at $R<R_1$, a transition region at $R_1<R<R_2$, and a flat Universe at $R>R_2$. Using calculations in a comoving coordinate system, the authors of~\cite{NadNovPol78} studied in detail the role of density gradients in the evolution of the perturbed region. It turned out that for $\Delta=0.5R_1$, where $\Delta=R_2-R_1$, a PBH forms if $R_1/R_{\rm max}\geq0.85-0.9$, where $R_{\rm max}=\pi/2$. The smaller the size $\Delta$ of the transition region, the stronger the density gradient, which hinders PBH formation. According to these calculations, the mass of the newborn PBH is only a few percent of the radiation mass inside the cosmological horizon, which rules out catastrophic accretion of radiation onto the PBH according to the approximate criterion formulated in~\cite{ZelNov66}.

Numerical calculations of PBH formation were later performed in a number of works using various approaches. For example, Ref.~\cite{BicHen79} was the first to study collapse in a Universe without a transition region, i.e., with only a mass excess in the perturbed region. Once the system of equations for the density evolution has been derived, there remains the problem of choosing the initial conditions --- the shape of the initial hypersurface, which was chosen differently in different works. For a modern approach based on the $(3+1)$ decomposition, see~\cite{ShiSas99,Musco:2018rwt}. Interestingly, PBHs can form not only from positive density perturbations but also in deep minima (underdensities, voids)~\cite{JoaYuw25}. As shown by the numerical simulations of~\cite{JoaYuw25}, this is due to the nonlinear recollapse of the perturbation at the center of the void, which gives rise to a positive perturbation that collapses into a PBH. Roughly speaking, this effect is caused by the pressure gradient at the boundaries of a deep void.

\medskip
{\bf Threshold of primordial black hole formation} 
\medskip

The PBH formation threshold $\delta_c$ is an important quantity for calculations because knowing only its value one can study the statistics of PBH formation from a given field of density perturbations using Eqs.~(\ref{eq:beta1}) and (\ref{eq:beta2}) without delving into details of complicated processes of gravitational collapse itself. The analytical estimates described above, based on heuristic arguments, give threshold values $1/3\leq\delta_c\leq0.42$. More detailed studies have also been performed using the theory of perturbation evolution in GR and the results of numerical GR simulations~\cite{2014JCAP...01..037N}. For example, calculations of the critical gravitational collapse yield $\delta_c\simeq0.7$~\cite{NieJed99}. For a general perfect-fluid equation of state $p=w\rho$, the analytical expression $\delta_c=\sin^2[\pi\sqrt{w}/(1+3w)]$ for the threshold was derived in~\cite{HarYooKoh14}; it agrees well with the results of a number of numerical simulations. In particular, for $w=1/3$ this formula gives $\delta_c=0.6203$. As shown in~\cite{2016PhRvD..94f3514K}, taking into account the non-sphericity of the perturbation modifies the PBH formation threshold as $\delta_{ec}\simeq\delta_c(1+3e)$, where $e$ is the eccentricity of the ellipsoid used to model the collapsing region.

In a more precise formulation of the problem, the threshold $\delta_c$ depends on the shape of the density perturbation~\cite{Musco:2018rwt,EscGerShe20,Musetal20}. A perturbation of typical shape can be characterized with reasonable accuracy by a single ``compactness parameter''. We write the metric in the region of the perturbation collapsing into a PBH as
\begin{equation}
ds^2=dt^2-a^2(t)e^{2\zeta(\hat{r})}[d\hat{r}^2+\hat{r}^2 d\Omega^2].
\end{equation}
Let $r_m=R_m/a(t)$ denote the comoving scale at horizon crossing. We also introduce the compaction function $C=2\delta M/R$, which depends on the difference between the Misner--Sharp mass and the mass in a homogeneous Universe inside a sphere of areal radius $R$. It can be rewritten as $C=-(2/3)\hat{r}\zeta'(\hat{r})[2+\hat{r}\zeta'(\hat{r})]$. The compactness parameter is then defined as
\begin{equation}
\alpha=-\frac{C''(r_m)r_m^2}{4C(r_m)},
\end{equation}
and the threshold is given, to within a few percent, by~\cite{EscGerShe20}
\begin{equation}
\delta_c\simeq\frac{4}{15}e^{-1/\alpha}\frac{\alpha^{1-5/(2\alpha)}}{\Gamma\!\left(\frac{5}{2\alpha}\right)-\Gamma\!\left(\frac{5}{2\alpha},\frac{1}{\alpha}\right)},
\end{equation}
where $\Gamma$ denotes the (complete and incomplete) gamma functions. As a result, the threshold was found to lie in the range $0.4\leq\delta_c\leq2/3$. Ref.~\cite{Musetal20} proposed a refined algorithm for calculating $\delta_c$ that takes into account nonlinear effects near horizon crossing, the shape of the perturbation spectrum, and results of numerical simulations. For each specific spectrum, this algorithm involves calculating the characteristic scale of the peak and the characteristic compactness parameter $\alpha$. As a result, the following approximation for the threshold was obtained:
\begin{equation}
\delta_c=
\left\{
\begin{array}{rcl}
\alpha^{0.125}-0.05\,&;&\,0.1\leq\alpha\leq3;
\\
\\
\alpha^{0.06}+0.025\,&;&\,3\leq\alpha\leq8;
\\
\\
1.15\,&;&\,8\leq\alpha\leq30.
\end{array}
\right.
\end{equation}
This corresponds to the range $0.7\leq\delta_c\leq1.15$.

\medskip
{\bf Critical gravitational collapse}
\medskip

Numerical simulations of the gravitational collapse of a scalar field and of a perfect fluid revealed a scaling relation, $M_{\rm BH}\propto (p-p_c)^\gamma$, between the mass of the resulting BH and the difference between the pressure $p$ in the perturbed region and a certain critical value $p_c$, where $\gamma\approx0.37$. The emergence of this relation was called the critical gravitational collapse effect. In~\cite{NieJed98}, a similar relation was applied to calculate the mass function of PBHs forming during the radiation-dominated stage,
\begin{equation}
M_{\rm PBH}(\delta)=kM_{\rm H}(\delta-\delta_c)^{\gamma},
\label{critsoot}
\end{equation}
where $1\leq k\leq10$. On the basis of (\ref{critsoot}), it was shown that for $\delta_c=1/3$ the mean PBH mass relative to the mass inside the horizon at formation, $M_{\rm H}$, is $\langle M_{\rm PBH}\rangle\simeq0.6M_{\rm H}$, i.e., it is of the order of $M_{\rm H}$, as assumed in earlier works, although the mass function also contains PBHs with masses $M_{\rm PBH}\ll M_{\rm H}$. A mass function of similar shape was also obtained in~\cite{GreLid99} using the extended Press--Schechter theory (excursion set formalism)~\cite{LacCol93} and the value $\delta_c\simeq0.7$ derived in~\cite{NieJed99}.

\medskip
{\bf Spins of primordial black holes formed from the collapse of adiabatic perturbations}
\medskip

The question of the angular momentum (spin\footnote{Although, strictly speaking, spin refers to the intrinsic angular momentum of elementary particles that cannot be reduced to their rotation, in astrophysical terminology the angular momentum of a BH is also called spin. This is justified by the fact that the exterior metric of a stationary BH is characterized only by its mass, angular momentum, and charge, regardless of their origin.}) of PBHs is very important for testing the PBH hypothesis observationally using data on gravitational-wave bursts. Measurements of BH masses and spins may help determine whether the BHs were born in astrophysical processes or have a primordial origin of one kind or another.

In general, a perturbation collapsing into a PBH is inhomogeneous and non-spherical. The matter in its surroundings is also distributed in a complex way. This means that the surrounding matter produces (in Newtonian language) tidal gravitational forces that interact with the quadrupole moment of the collapsing region, giving rise to its angular momentum. The situation is, on the whole, similar to the generation of angular momentum of galaxies~\cite{Peebles:1969jm,Dor70,Eisenstein:1994ni}. However, for PBHs these processes take place during the radiation-dominated stage of the evolution of the Universe, and general-relativistic effects must be taken into account in calculating the angular momentum, since the processes occur on scales close to the cosmological horizon scale at the time of collapse.

The dimensionless spin parameter is defined as
\begin{equation}
a_S=\frac{cL}{GM_{\rm PBH}^2}.
\end{equation} 
The results of calculations of the PBH spin differ greatly between different works. In~\cite{ChiYok17}, the constraint $a_S<0.4$ was obtained; in~\cite{MirGruNor19,DeLuca:2019buf}, the value $a_S\sim0.01$ was found; and in~\cite{Haretal20}, a smaller value $a_S\sim O(10^{-3})$ was obtained. The authors of~\cite{ChiYok17} used scaling expressions for the angular momentum to calculate the final PBH spin, whereas the authors of~\cite{MirGruNor19,DeLuca:2019buf,Haretal20} carefully calculated the spin distribution within the theory of cosmological perturbations in the early Universe.

In the model of the collapse of adiabatic perturbations during the radiation-dominated stage, the PBH spin is small primarily because there is a strong correlation between the directions of the principal axes of the quadrupole moment of the collapsing region and of the external matter distribution producing the tidal forces, while the shape of perturbations in the case of high density peaks is close to spherically symmetric.

\subsection{Early matter-dominated stages}

At redshift $z\sim3000$, the Universe underwent a smooth transition from the radiation-dominated stage to the matter-dominated stage (the equation of state changed from $p=\varepsilon/3$ to $p=0$), and at $z\lesssim 1$ the Universe became dominated by a component with an equation of state close to that of vacuum ($p\approx-\varepsilon$). However, it is possible, and even quite likely, that at $z\gg 3000$ the Universe went through early matter-dominated stages~\cite{PolKhl85}, when the density of the Universe became dominated by some hypothetical nonrelativistic particles. The existence of such particles is predicted by Grand Unified Theories. These particles could then decay, and the Universe could return to the radiation-dominated stage.

PBH formation during early matter-dominated stages was first studied by A.G.~Polnarev and M.Yu.~Khlopov~\cite{KhlPol80,PolKhl82}. This mechanism was subsequently developed in~\cite{ZabNasPol87} and considered in detail in a review in {\it Soviet Physics Uspekhi}~\cite{PolKhl85} (see also the review~\cite{Khl10}); therefore, we focus here only on some key points.

The simple estimates given in the previous section show that the lower the pressure, the easier it is for a perturbation to collapse into a PBH. During early matter-dominated stages, the pressure is $p=0$, and in the strictly spherically symmetric case such matter could form PBHs quite easily, since the particles would simply fall onto a single center, where a mass could accumulate within its gravitational radius. However, a complicating factor that considerably reduces the probability of collapse is the non-sphericity of dust-like configurations. Because of non-sphericity, it is very difficult for matter to accumulate at the center; in addition, ``violent relaxation'' (virialization) processes take place and halt the contraction. Matter configurations have a statistical distribution over their parameters, including those describing their shape. Only very rare objects with a high degree of spherical symmetry could collapse into PBHs. But since even a tiny collapsed fraction of matter can eventually come to dominate the density (see (\ref{restr})), even very rare collapses during an early matter-dominated stage could have produced a significant PBH population today. A crucial role in calculating the PBH formation probability in this model is played by the statistical distribution of density perturbations over their non-sphericity parameters, derived by A.G.~Doroshkevich within the theory of Gaussian random fields~\cite{Dor70}.

PBH formation during early matter-dominated stages was also considered in a number of later works. In~\cite{Haretal16}, a calculation that remains valid even when the perturbation amplitude is $\sim1$ was performed using the Zel'dovich approximation, K.S.~Thorne's ``hoop conjecture'', and the above-mentioned distribution of A.G.~Doroshkevich. The semi-analytical dependence for the PBH formation probability obtained numerically in~\cite{Haretal16}, $\beta_0\simeq0.05556\sigma_{\rm H}^5$, where $\sigma_{\rm H}$ is the perturbation amplitude on the horizon scale, is close to that obtained in~\cite{KhlPol80,PolKhl82}. It turned out that, for $\sigma_{\rm H}\leq0.05$, the probability of PBH formation during the matter-dominated stage is higher than during the radiation-dominated stage, whereas for $\sigma_{\rm H}\geq0.05$ the situation is reversed. A hybrid scenario with several generations of PBHs has also been considered~\cite{KotNas98}: a first generation of very light PBHs is born from adiabatic density perturbations, these PBHs come to dominate the density, the Universe enters a matter-dominated stage, and during this stage a second generation of PBHs is born via the mechanism considered in this section. It is also important that large spins are expected for PBHs formed during early matter-dominated stages~\cite{Haretal17,Kuh20}.

\subsection{Phase transitions in the early Universe}

In the 1970s, D.A.~Kirzhnits, A.D.~Linde, and other researchers developed the theory of vacuum phase transitions in the early Universe, associated with the symmetry groups of elementary particles, including proposals for Grand Unified Theories. Based on these ideas, models were later developed for the evolution of islands of the new vacuum inside the old one and, conversely, of remnants of the old vacuum inside the new vacuum dominating by volume in the post-inflationary epoch. It was shown in~\cite{Satetal81,Maeetal82} that bubbles of one vacuum inside another can undergo gravitational collapse, forming BHs and even wormholes. BHs formed in this way are classified as PBHs. A number of works also developed percolation theory, which makes it possible to estimate the probability of various spatial configurations of the new vacuum, in particular, the probability of forming closed domain walls; this in principle allows one to estimate the mass functions of the resulting PBHs.

In~\cite{Satetal81,Maeetal82}, it was assumed that the domain wall --- the boundary between the vacua --- moves at the speed of light. However, V.A.~Berezin, V.A.~Kuzmin, and I.I.~Tkachev~\cite{BerKuzTka83} showed that, when the energy density of the domain wall is taken into account, its velocity only asymptotically approaches the speed of light. This effect significantly influences PBH formation, and~\cite{BerKuzTka83} obtained a refined solution to the problem of the evolution and collapse of bubbles of the new or old vacuum phase. An important achievement of~\cite{BerKuzTka83} is that the equations were solved and the collapse conditions were analyzed within GR, including a study of the global space--time geometry. The only simplifications were that the problem was assumed to be spherically symmetric and the domain walls were treated as infinitely thin (the Israel formalism), which is a good approximation.

In particular, it was shown in~\cite{BerKuzTka83} that the vacuum configuration has two horizons (cosmological and Schwarzschild) and, accordingly, a PBH forms when the mass parameter of the metric (an integral of motion determined by the initial data) satisfies
\begin{equation}
m>\frac{M_{\rm Pl}^3}{3(8\pi \mathcal{E}_{\rm out})^{1/2}},
\end{equation}
and the wall radius satisfies $\rho<\rho_c$, where
\begin{equation}
\rho_c=\frac{M_{\rm Pl}}{4\pi}\left(\frac{S^2}{M_{\rm Pl}^2}+\frac{\mathcal{E}_{\rm in}}{6\pi}\right)^{-1/2},
\end{equation}
$M_{\rm Pl}$ is the Planck mass, $\mathcal{E}_{\rm in}$ and $\mathcal{E}_{\rm out}$ are the energy densities in the regions of space inside and outside the wall, respectively, and $S$ is the surface density of the energy--momentum tensor, with $S=S_0^0$ for a timelike surface and $S=S_1^1$ for a spacelike one. In the opposite case, $\rho>\rho_c$, a wormhole forms.

The theory developed in~\cite{Satetal81,Maeetal82,BerKuzTka83} is fairly general and applicable to many types of vacuum phase transitions. These works demonstrated that PBH formation in such processes is possible in principle. The collapse of bubbles of the new phase and of domain walls into PBHs was also calculated within GR in~\cite{Deng:2017uwc,GarVilZha16,Deng:2016vzb}. We consider the statistics and properties of such PBHs in more detail in Section~\ref{PBH_from_walls}, for a specific realization of the phase transition in an axion-like model with a tilted Mexican-hat potential.

\subsection{Inflationary prerequisites for primordial black hole formation} 
\label{nonstandsec}

Observations of the CMB inhomogeneities and of the large-scale structure of the Universe (galaxies, galaxy clusters, and superclusters) indicate that the initial spectrum of cosmological density perturbations on large scales is very close to a power law. Extrapolating this spectrum to small scales yields an extremely small number of PBHs. Efficient PBH production requires spectra with enhanced power on small scales. In this section, we describe some of the theoretical models that predict such more complex perturbation spectra.

A natural source of density perturbations that later give rise to PBHs is quantum fluctuations during inflation. The inflationary paradigm (a short period of exponential expansion of the Universe) is currently regarded as the best-motivated and most plausible description of the early Universe, although alternatives are not excluded. Here we assume that this paradigm holds. Density perturbations or topological defects that could lead to PBH formation originate during inflation and are determined by the inflationary conditions. Inflation is characterized by the Lagrangian of the physical field or fields that drove inflation (the inflaton) or evolved during inflation.

PBH formation directly during inflation is not forbidden, but it is inefficient because of the exponential expansion of space. However, inflationary perturbations that survive after the end of inflation can generate a sufficient number of massive PBHs. Let us consider several basic mechanisms on which specific models are based. Common to all of them is that rare field fluctuations create conditions for BH formation in the post-inflationary period. The probability of such fluctuations is considered in Appendix I.

\medskip
{\bf Potential with a local plateau} 
\medskip

\begin{figure}[htbp] 
	\centering
	\includegraphics[width=0.9\linewidth]{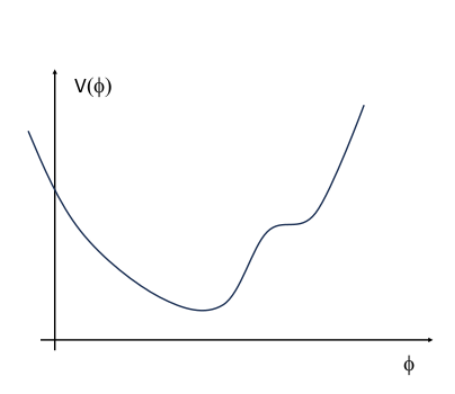}
	\caption{Typical shape of a potential with a local plateau, where, according to \eqref{drhoPhi}, the field fluctuations increase sharply. 
	\label{stepot}}
	\end{figure}
    
The model of PBHs as the dominant DM component (see Section~\ref{dmsec} for details) developed in~\cite{IvaNasNov94} is based on~\cite{Starobinsky:1992ts}, which considered a nonstandard perturbation spectrum arising from features in the inflationary potential, with the power spectrum enhanced at a certain scale; it is this scale that determines the PBH mass in the model of~\cite{IvaNasNov94}. A feature in the spectrum is needed, because a nearly scale-invariant spectrum produces an insufficient abundance of PBHs to explain all of the DM. A feature (a flat segment, or plateau) in the inflationary potential solves the problem of producing massive PBHs. Indeed, matter density fluctuations in the post-inflationary period $t_f$ originated during inflation as quantum fluctuations of the scalar field $\varphi$ at the time $t_{in}$. According to the well-known formula (see, e.g.,~\cite{Khlopov:2004rw})
\begin{equation} \left( \frac{\delta \rho
	}{\rho }\right) _{t_{f}}\simeq\left( \frac{9H^3(\varphi )}{2\pi
		V^{\prime }(\varphi )}\right) _{t_{in}},  \label{drhoPhi}
\end{equation}
the density fluctuations are inversely proportional to the derivative of the inflaton potential $V(\varphi)$. In this equation, $H$ is the Hubble parameter, which depends on the inflaton field $\varphi$ and varies very slowly with time during inflation. Therefore, as follows from \eqref{drhoPhi}, any distortion of the potential like the one shown in Fig.~\ref{stepot} leads to an enhancement of density fluctuations on a certain scale, which depends on the position of the step. This is exactly what is needed for more efficient PBH production. The higher up the potential the flat region is located, the earlier the corresponding modes exit the horizon and the larger their scale in the post-inflationary stage, and the more massive the PBHs that can form. The PBH formation mechanism developed in~\cite{IvaNasNov94,2017PDU....18...47G,Yokoyama:1998pt} is based on this idea.
Calculations within the plateau model must take into account that the presence of a flat segment violates the usually assumed slow-roll condition for the scalar field during inflation. In some models, this leads to a suppression of the generated density perturbations. To avoid a delay of inflation at the flat plateau, potentials with inflection points were considered in~\cite{2017PDU....18...47G}. A well-developed modern model based on a nontrivial shape of the scalar field potential can be found in~\cite{Frolovsky:2025qre}.

\medskip
{\bf Transition of field fluctuations over the potential maximum} 
\medskip

This model is described in detail in Section~\ref{PBH_from_walls}, so here we outline only the idea of the mechanism proposed in~\cite{RubSakKhl01}. Consider a scalar field with a double-well potential that has a maximum at $\varphi=\varphi_0$. Without loss of generality, assume that the field in the patch that became our observable Universe starts at $\varphi>\varphi_0$, and that after inflation, the field rolls down to the right minimum $\varphi_+$. However, the presence of fluctuations during inflation leads to a nontrivial effect. If the classical motion of the field, which decreases the potential energy, is sufficiently slow (as assumed in most inflationary models), then in some causally disconnected regions quantum fluctuations produce the opposite effect: an increase in the potential energy. After several e-folds, the field in such regions may reach the potential maximum at $\varphi=\varphi_0$. From this moment on, at each subsequent e-fold, causally disconnected regions will appear in which the field satisfies $\varphi<\varphi_0$. Later, after inflation, the classical dynamics will bring the field in these regions to the potential minimum $\varphi=\varphi_-$. The regions with the field at the minimum $\varphi_-$, which arose as a result of quantum fluctuations, are much smaller in volume than the regions with the field at the minimum $\varphi_+$, toward which the field was initially rolling as a result of its classical motion.

Thus, after inflation, there are spatial regions with the field $\varphi_-$ surrounded by a region with the field $\varphi_+$. To pass from one region to the other, one has to cross a surface (the boundary of the region) characterized by the potential maximum. Consequently, the formation of a closed field wall is inevitable.

Taking into account, in addition, the classical dynamics of the field in a region where the field potential has two or more maxima makes it possible to vary the final PBH mass spectrum over a wide range, which facilitates fitting it to observational data~\cite{KirRub21}.

The tilted Mexican-hat shape of the potential may seem somewhat artificial, but the main idea appears in various models. For example, all models of PBH formation from field walls based on the transition of field fluctuations over a maximum or a saddle point of the potential are characterized by the condition
$$\frac{dV(\phi)}{d\phi}=0 $$ at some value of the field $\phi$. According to \eqref{drhoPhi}, when the field passes through such an extremum or saddle point, the energy density fluctuations in the post-inflationary epoch increase sharply, which may provide an additional contribution to the PBH mass spectrum.

\medskip
{\bf Models with running parameters} 
\medskip

Models with varying parameters are based on the dependence of the field parameters on the inflaton or on some other scalar field. As an example, consider the potential
\begin{equation}\label{PotD}
    V(\chi,\phi)=\lambda (\chi^2-a^2)^2 + g\chi^2(\phi-\phi_1)^2.
\end{equation}
It is assumed that the field $\chi$, which provides the conditions for PBH formation, does not affect the dynamics of the inflaton $\phi$. At the initial stage of inflation, when $\phi\gg \phi_1$, the field $\chi$ sits at the potential minimum $\chi=0$ for an appropriate choice of parameters, or, more precisely, provided that
\begin{equation}
    -2\lambda a^2+g(\phi-\phi_1)^2 >0.
\end{equation}
The same condition also holds for $\phi\ll \phi_1$. The situation changes drastically when the inflaton field, in the course of its slow motion toward the minimum, approaches the value $\phi\sim \phi_1$. In this case, the second term in potential \eqref{PotD} becomes small, and the field $\chi$ rolls down to the new minimum $\chi=a$. Thus, a rapid motion of the field takes place, which can be regarded as a large-scale fluctuation during inflation and used to construct a PBH model. This effect was first used by Dolgov and Silk~\cite{DolSil93,DolKawKev09,Dol18}. This mechanism is described in more detail in Section~\ref{Dolgov}.

\medskip
{\bf Hybrid inflation}
\medskip

The hybrid inflation model~\cite{Linde1990,Linde:1994eq} is one realization of models with running parameters and involves several physical mechanisms that lead, in particular, to PBH formation~\cite{Randall:1995dj,Garcia-Bellido:1996egv}. Let us consider this idea of generating the corresponding inhomogeneities in more detail.

The hybrid inflation potential has the form
$$V (\chi, \sigma) = \frac{\kappa^2}{2} \left(M^2 - \frac{\chi^2}{4}\right)^2 + \frac{\lambda^2}{4} \chi^2 \sigma^2 + \frac{1}{2} m^2 \sigma^2.$$
The motion of the fields is described by the classical equations
\begin{align}
\label{sigma}
\ddot{\sigma} + 3H\dot{\sigma} + \frac{1}{2}\lambda^2\chi^2\sigma + m^2\sigma &= 0, \\
\label{eq:chi}
\ddot{\chi} + 3H\dot{\chi} - \frac{\kappa^2}{2}\chi \left(M^2 - \frac{\chi^2}{4}\right) + \frac{1}{2}\lambda^2\chi\sigma^2 &= 0,
\end{align}
where $H$ is the Hubble parameter during this period. Inflation proceeds during slow roll along the ``valley''
$$\chi = 0, \quad \sigma > \sigma_c.$$
When the critical point
\begin{equation}\label{sc}
    \sigma_c = {\frac{\kappa M}{\lambda}}
\end{equation}
is reached, the motion along the line $\chi = 0$, $\sigma < \sigma_c$ becomes unstable, and the field $\chi$ rapidly rolls down to one of the minima $\chi_{\pm} = \pm 2M, \sigma = 0$. The choice of the minimum depends on a random fluctuation of the field in a particular spatial domain inside the horizon. The numbers of regions in the two minima are then approximately equal, and as a result we arrive at a Universe dominated by field walls, which contradicts observations.

This picture is correct if the Universe originally emerged with the field $\chi$ exactly equal to zero at the onset of inflation. The situation changes significantly if the field inside the horizon already has a value $\chi=\chi_0 \neq 0$ at the first e-fold. In this case, the dynamical equations \eqref{sigma} and \eqref{eq:chi} bring the field to the point $\chi =\chi_c \neq 0$ by the time the critical point $\sigma_c$ is reached. The symmetry between the vacua $\chi_{\pm}$ is broken, and only a small number of spatial regions will have the field $\chi_+$ (say), while the bulk of the space surrounding them will contain the field $\chi_-$. These regions are separated from the bulk of the space by field walls, which may collapse into PBHs~\cite{2005astro.ph.11181R}. The mass spectrum of such BHs depends strongly on the value of $\chi_c$. Note that at the critical point \eqref{sc} we have the conditions $dV/d\chi =0,\, d^2V/d\chi^2 =0$, which means that the potential has a plateau. In this case, the additional PBH formation mechanism described above is realized.

\medskip
{\bf Cascade relaxation of the gravitating vacuum}
\medskip

A new paradigm for the evolution of the cosmological vacuum was proposed in~\cite{LukMik25}. This model does not assume a potential of a predefined shape; instead, it is based on a picture in which fields in the Lagrangian successively separate out and one of the fields dominates at a certain moment. First, the gravitational field is separated out in the Lagrangian in the form of the curvature scalar $R$ (in standard GR); then, one of the other fields is separated out in the remaining part of the Lagrangian:
\begin{equation}
\mathcal{L} \to \mathcal{L}(\partial\phi,\phi,...)\to (\partial\phi)^2-V(\phi,...)
\end{equation}
Importantly, the field starts from zero. This makes it possible to consider a series expansion and its lowest-order terms without specifying the exact form of the Lagrangian. The dominance of particular fields is accompanied by accelerated expansion of the Universe. At an early stage, this was inflationary expansion; at present, it is the accelerated expansion attributed to dark energy. Other stages of accelerated expansion may have existed in the past or may occur in the future. The model depends on two parameters fixed by observations and is consistent with observational data, including the constraint on the tensor-to-scalar ratio.

In the context of PBHs, this cascade vacuum relaxation model is interesting because it naturally produces enhancements of the power spectrum in the form of bumps. In~\cite{LukMik25}, solutions of the evolution equation for the scalar field were obtained which imply that the field has most likely passed through the point $\dot\phi=0$; in this case, a bump arises on the corresponding scale, since the perturbation is $\propto 1/\dot\phi$. However, the position of the bump is not fixed in the model and is an additional parameter. In particular, the excess of high-redshift galaxies found by the James Webb Space Telescope (JWST) may point to a bump corresponding to galaxy masses of $\sim10^9-10^{10}M_\odot$~\cite{Tkachev:2023acf}.

The mechanisms considered in this section do not exhaust all the inflationary possibilities for generating inhomogeneities that lead to PBH formation. For example, isocurvature perturbations can also be generated in an inflationary model with several scalar fields~\cite{Yok97}.

\subsection{Formation of primordial black holes from the collapse of domain walls}
   \label{PBH_from_walls}

Models based on the collapse of closed domain walls or their fragments, as well as on wall collisions~\cite{Khlopov:2004rw}, are a promising direction in PBH research. As discussed above, the presence of at least one maximum or saddle point in the field potential can create conditions for the formation of field walls and their collapse after the end of inflation. Here we give an example proposed in~\cite{RubKhlSak00,RubSakKhl01}, which can easily be generalized to other models.

Consider a complex scalar field with the Lagrangian
\begin{equation}
\mathcal{L}=\frac{1}{2}\,\partial^\mu\phi\partial_\mu\phi^*-V,
\label{lagr}
\end{equation}
where the potential has the shape of a tilted Mexican hat,
\begin{equation}
V=\lambda\left(\phi\phi^*-\frac{f^2}{2}\right)^2+\Lambda^4(1-\cos\theta),
\label{hateq}
\end{equation}
where $\theta=\arg(\phi)$, $f\gtrsim H$ is the vacuum expectation value, $H\sim10^{13}$~GeV is the Hubble parameter during inflation, and the parameter $\Lambda$ remains free and can be fixed by observational data if some populations of astrophysical black holes are PBHs formed via this mechanism~\cite{RubSakKhl01}.

The presence of a small tilt of the potential, $\Lambda^4(1-\cos \theta)$, is motivated, for example, by instanton effects. The field $\theta(x)$ can form domain walls, since the tilt gives rise to a discrete set of minima of the potential $V(\phi)$ at $\theta =2\pi n$, where $n$ is an integer.

During exponential expansion, the field $\theta$ fluctuation per e-fold is $\delta\theta=H/(2\pi f)$. Over time, in a small fraction of regions, $\theta$ may evolve stochastically from the potential minimum $\theta=0$ across the value $\theta=\pi$ and, in the post-inflationary period, oscillate near the next minimum $\theta=2\pi$. Thus, rare regions with $\theta\simeq2\pi$ appear, surrounded by regions with $\theta\simeq0$. There must then exist a closed boundary surface with $\theta=\pi$, which is a domain wall carrying the surface energy density
\begin{equation}
\sigma=4\Lambda^2f.
\label{sigeq}
\end{equation}
For a detailed description of these processes, see~\cite{RubKhlSak00, RubSakKhl01}. The mass function of PBHs formed from the collapse of domain walls was calculated in~\cite{KirRub21}.

A domain wall has a considerable surface tension, but its contraction begins only after the wall enters the cosmological horizon, when the wall radius is $R\sim c/H=2ct$ (during the radiation-dominated stage). At this moment, the total mass of the wall, $M=4\pi R^2\sigma$, is fixed. The evolution of the wall is significantly affected by its interaction with the plasma~\cite{RubSakKhl01, KurRub20}. Gravitational collapse into a black hole is possible if the wall thickness is smaller than its total gravitational radius. Before horizon entry, the wall consists of causally disconnected parts with a fractal structure, which shed entropy via gravitational radiation, emission of $\theta$ scalar particles, interaction with the surrounding medium, and Hubble friction.

The surface density~\eqref{sigeq} can vary over a fairly wide range. According to~\cite{RubSakKhl01}, PBHs of astrophysically interesting masses $10^{25}-10^{40}$~g can form at $f \approx 1.8 H$ and $\Lambda\sim1-5$~GeV. The surface density of the wall is then $\sigma\sim 7\times10^{13}-2\times10^{15}$~GeV$^3$.

\medskip
{\bf Inflationary fluctuations of the scalar\,field}
\medskip

Let us consider in more detail the evolution of the scalar field during inflation that leads to the formation of domain walls. Quantum fluctuations of the scalar field during inflation can be described as a random walk governed by the Fokker--Planck equation for the field $\phi$~\cite{Linde1990}:
\begin{equation}\label{fokker_planck}
\frac{\partial f(\phi,t)}{\partial t}=\frac{H^3}{8\pi^2}\frac{\partial^2 f(\phi,t)}{\partial \phi^2}+\frac{1}{3H}\frac{\partial}{\partial \phi}\left(f(\phi,t)\,\frac{dV}{d\phi}\right),
\end{equation}
where $f(\phi,t)$ is the distribution of the values of the field $\phi$ at time $t$, $H$ is the Hubble parameter during inflation, and $V(\phi)$ is the potential of the field $\phi$. We solve this equation in the ``flat potential'' approximation, i.e., we neglect the influence of the potential gradient on the quantum fluctuations (the second term on the right-hand side of \eqref{fokker_planck}). For any reasonable realization of the PBH generation mechanism, the parameters of the potential must be sub-inflationary and ensure the validity of this approximation (for more detailed calculations, see~\cite{Linde:1982uu,Starobinsky:1982ee}). The solution of Eq.~\eqref{fokker_planck} then has the form (see Eq.~\eqref{dP} in Appendix I)
\begin{equation}\label{f_ot_t}
f(\phi,N)=\frac{1}{\sqrt{2\pi}\sigma(N)}e^{-\frac{(\phi-\phi_\text{0})^2}{2\sigma^2(N)}}, \quad \sigma(N)=\frac{H}{2\pi}\sqrt{N}\,,
\end{equation}
where the time elapsed since the beginning of inflation is expressed by the number of e-folds $N\equiv Ht$, and $\phi_\text{0}$ is the initial value of the field $\phi$ at the beginning of inflation.

The probability of a critical fluctuation that brings the field to the other vacuum is
\begin{equation}\label{Pt}
    P(N, \phi_\text{cr})=
    \int\limits_{\phi_\text{cr}}^{\infty} f(\phi,N)\, d\phi = 
    \frac{1}{2}\text{erfc} \left(\frac{\phi_\text{cr}-\phi_\text{0}}{\sqrt{2}\sigma(N)}\right)\,,
\end{equation}
where $\phi_\text{cr}$ is the field value at which the potential reaches its maximum.

The spatial size of a quantum fluctuation is limited by the horizon size $H^{-1}$. At the $N$th e-fold of inflation, the space of the observable Universe is divided into $e^{3N}$ causally independent regions. Consequently, the number of critical regions (in which the field has fluctuated into the other vacuum) is
\begin{equation}\label{crit}
n(N)=P(N, \phi_\text{cr})\, e^{3N}\,.
\end{equation}

The number of critical regions is assumed to be much smaller than the total number of causally independent regions; otherwise, the abundance of walls would contradict observational data. After inflation, the size of a critical region (a domain of the other vacuum) is
\begin{equation}\label{rinfott}
r(N)=H^{-1}e^{N_\text{inf} - N}\,,
\end{equation}
where $N_\text{inf}\simeq 60$ is the total number of e-folds required to form the observable Universe. From this, we find the size distribution of the domains right after inflation:
\begin{align}\label{post_inf_spectr}
\frac{dn(r)}{dr} = - \frac{e^{3N_\text{inf}}}{H^3 r^4} &\Big[3 P\big(N_\text{inf} - \ln(Hr), \phi_\text{cr}\big)  \nonumber\\
+P'\big(N_\text{inf}& - \ln(Hr), \phi_\text{cr}\big)\Big]\,.
\end{align}

\medskip
{\bf Mass spectrum of primordial black holes}
\medskip

After the end of inflation, the scalar field $\phi$ ``unfreezes'' and rapidly rolls down to the potential minima through classical evolution (due to the decrease of the friction term $3H\dot{\phi}$ in the field equation of motion). At this moment, field walls possessing mass and surface tension form around the domains of the other vacuum. Immediately after their formation, the walls are not spherical, since the shape of the domains results from quantum fluctuations and is fractal~\cite{RubSakKhl01,KhlRubSa02,Kirillov:2024pcs}.

Due to the surface tension, the parts of the field wall that are inside the horizon rapidly reach near-luminal velocities, shedding entropy into the surrounding medium via emission of $\phi$ particles and gravitational waves and via interaction with the plasma~\cite{Vachaspati:1984yi,Kibble:1976sj}. Causally connected parts of the surface therefore have, at each moment, a characteristic curvature radius of the order of the horizon size, which grows as $r_\text{hor}(t)=2t$ during the radiation-dominated stage. By the time the whole domain enters the horizon, the wall surrounding it is nearly spherical. It is worth mentioning that there are models in which domain walls do not become spherical efficiently and remain irregular at horizon entry~\cite{Friedland:2002qs}. In the rest of this section we assume that domain walls become spherical after entering the horizon.

During the radiation-dominated stage, the domain wall itself expands with the Hubble flow as $\tilde{r}(t)~=~r\sqrt{t/t_\text{inf}}$, where $r$ is the domain size right after inflation and $t_\text{inf} = N_\text{inf}/H$ is the duration of inflation. Therefore, the moment of horizon entry is determined by the equality $r_\text{hor}(t) = \tilde{r}(t)$, which gives the relation between the wall radius right after inflation and after horizon entry:
\begin{equation}\label{r_final}
\tilde{r}\simeq\frac{r^2}{2t_\text{inf}}\,.
\end{equation}
After entering the horizon, the spherical wall starts to contract due to its surface tension, delivering energy into a small volume, which can lead to PBH formation if the mass is sufficient. The PBH mass is determined by the final size of the wall,
\begin{equation}\label{mass_wall}
    M(\tilde{r}) \simeq 4 \pi \tilde{r}^2 \sigma = \frac{\pi \sigma}{t_\text{inf}^2}r^4\,,
\end{equation}
where $\sigma$ is the surface energy density of the wall, determined by the parameters of the potential of the field $\phi$.

Expressing $r(M)$ from \eqref{mass_wall} and substituting it into \eqref{post_inf_spectr}, we obtain the PBH mass spectrum in this model:
\begin{align}\label{tvr}
\frac{dn(M)}{dM} &\approx -\frac{3}{4} \left(\frac{\sqrt{\pi\sigma}}{N_\text{inf}\, H}\right)^{3/2} M^{-7/4}\,e^{3N_\text{inf}}  \nonumber\\
&\times P\left(N_\text{inf} - \frac14 \ln\left(\frac{N_\text{inf}^2\, H^2 M}{\pi\sigma}\right),\, \phi_\text{cr}\right)\,,
\end{align}
where we have neglected the second term in \eqref{post_inf_spectr}, which is small compared with the first one for any reasonable model parameters.

The masses of PBHs formed in this way cannot take arbitrary values. Since during contraction the radius of the domain wall cannot become smaller than its thickness $\delta$ (which is also determined by the parameters of the potential of the field $\phi$), the condition for PBH formation is given by the inequality $r_g = 2 G M > \delta$. Thus, the PBH spectrum is cut off from below by the condition
\begin{equation}
M > \frac{\delta}{2G}\,.
\label{min_mass_dw}
\end{equation}
In addition, the gravitational radius of the wall cannot exceed its size, $r_g = 2GM(\tilde{r}) < \tilde{r}$, since once this condition is violated, the wall collapses immediately (such walls are called ``supercritical'' in the terminology of~\cite{Deng:2016vzb} and lead to the creation of a baby universe), while an external observer sees a BH. Thus, the PBH spectrum is cut off from above by the resulting condition
\begin{equation}
M < \frac{1}{16\pi G^2\sigma}\,.
\label{max_mass_dw}
\end{equation}

The wall energy density $\sigma$ and its thickness $\delta$ depend on the specific shape of the potential. A commonly used potential (although not the only one possible for this mechanism --- see, e.g.,~\cite{Kirillov:2024pcs}) is the tilted ``Mexican hat'' \eqref{hateq}. The field $\theta(x)$ plays the role of the field forming the domain walls. When inflation ends, it rolls down to the potential minima, and neighboring minima (separated in space by a domain wall) are connected through the valley of the ``hat''. The point $\theta = \pi$ forms a potential barrier along the valley and plays the role of the critical point $\phi_\text{cr}$ in \eqref{Pt}. The tilt parameter $\Lambda$ is small, so that the ``flat potential'' approximation holds during inflation.

The one-dimensional ``kink'' solution for such a cosine potential is well known and describes the profile of the domain wall:
\begin{equation}
 \label{kink}
 \theta(z)=-4\arctan\left[e^{\frac{\Lambda^2}{f}(z-z_0)}\right]\!.
\end{equation}
The energy of this field configuration is concentrated in a cross section of thickness $\delta = 2f/\Lambda^2$, and the surface energy density of the wall is $\sigma=4\Lambda^2 f$.

The number of PBHs depends significantly on the vacuum expectation value $f$. In addition, the number of PBHs is strongly affected by the initial value $\theta_0$ of the field for the observable Universe: the closer the initial value is to the critical value $\theta_\text{cr} = \pi$, the more PBHs of all masses subsequently form in the Universe. The wall density, which depends on the parameter $\Lambda$, also affects how massive the PBHs can be, but this effect is less significant than the effect of the initial value $\theta_0$.

\medskip
{\bf Angular momentum of primordial black holes}
\medskip

The spin of a PBH born from the collapse of a domain wall was estimated in~\cite{Ero21} in the Newtonian approximation. The wall was modeled by the surface of an ellipsoid, and the angular momentum acquired by the wall under the action of tidal gravitational forces from external density perturbations was calculated. The tidal forces that spin up the wall are produced by perturbations in the radiation distribution, which propagate away as sound waves after horizon entry; therefore, the collapsing wall acquires its spin within approximately one Hubble time after it enters the cosmological horizon. In~\cite{Ero21}, the mean PBH spin is given as
\begin{equation}
\!\langle a_S \rangle\! \simeq\! 10^{-4} f_{\rm H} \!\left(\frac{\sigma_{\rm H}}{6\times10^{-5}}\!\right)\left(\frac{30M_\odot}{M_{\rm PBH}}\right)
\!\left(\!\frac{10^{14} \text{GeV}^3}{\sigma}\!\right)\!,
\label{spinwalls}
\end{equation}
where $f_{\rm H}=M_{\rm PBH}/M_{\rm H}$ is the ratio of the PBH mass to the total mass inside the horizon at the moment the wall enters it, and the root-mean-square perturbation amplitude at horizon crossing is $\sigma_{\rm H}=6\times10^{-5}$ for $M_{\rm H}=30M_\odot$ (the spectrum on small scales is assumed to be a standard power law). It follows from~\eqref{spinwalls} that the dimensionless spin parameter of the Kerr metric can be small, $a_S<1$, only for PBHs with masses $M>10^{-3}M_\odot$, whereas less massive PBHs, according to this estimate, acquire extremal spins $a_S\simeq1$. Such efficient spin-up at small masses is explained by the fact that, in the collapse of domain walls, unlike in the collapse of adiabatic perturbations, there is no suppression due to the correlation between the directions of the quadrupole moments of the internal and external matter distributions; moreover, the shape of domain walls is highly non-spherical.

For low-mass PBHs in the early Universe, spin can play an important role~\cite{PacSil20}. For example, it is important for their Hawking evaporation, affecting the power and spectrum of the emitted radiation~\cite{ArbAufSil19,ArbAufSil20, DasLahRay20}. As for supermassive PBHs, they can serve as seeds for quasar BHs and substantially increase their mass through disk accretion. The resulting BH may then have a much larger dimensionless spin than the seed. The spin of the BH at the center of the galaxy M87 is $a_S=0.75\pm0.15$~\cite{DokNaz19}, but this BH could have increased its mass by many orders of magnitude through gas accretion or mergers with other BHs. Therefore, a primordial origin of BHs at galactic centers does not contradict their spin values.

\subsection{Formation of primordial black holes in a model with extra dimensions}
\label{adddimsubsec}

Another PBH formation mechanism exists in a class of models based on modified gravity (namely, a multidimensional version of $f(R)$ gravity), in which different gravitational vacua separated by domain walls can arise. These models differ from the purely field-theoretic ones discussed above in that the scalar field responsible for domain walls is effective and is obtained in the low-energy limit from a purely gravitational action. Conceptually, this approach is close to the well-known Starobinsky inflation model~\cite{Starobinsky:1980te}, in which an effective scalar inflaton field arises from adding terms quadratic in curvature to the gravitational Lagrangian.

Adding terms nonlinear in curvature $R$ to the Lagrangian is motivated by two considerations. First, GR is believed to be an effective low-energy theory of gravity; therefore, the Lagrangian should contain curvature corrections due to quantum gravity effects, which are negligible at low energies~\cite{Utiyama:1962sn,Stelle:1976gc,Vilkovisky:1992pb}. It was precisely for such reasons the $R^2$ term was originally introduced in the Starobinsky inflation model in ordinary (3+1)-dimensional space--time. The second consideration is that the nonlinearity of gravity is one of the mechanisms for the compactification of extra dimensions~\cite{Bronnikov:2005iz,Rador:2007gq}. Although other compactification mechanisms are possible, for example, those based on introducing additional matter fields~\cite{Carroll:2001ih} that give a nonzero energy--momentum tensor, purely gravitational models seem aesthetically attractive. Moreover, $f(R)$ theories make it possible to describe cosmological inflation and the compactification of the extra space in a unified way, as manifestations of the common gravitational dynamics in different subspaces~\cite{Fabris:2019ecx, Lyakhova:2018zsr}.

In this section, we are interested in a by-product of multidimensional $f(R)$ models: PBH production. Here we give an example of such a model, considered in~\cite{Nikulin:2022ryt}. It is assumed that at high energies physics is described by a $D=4+n$-dimensional Lagrangian with an action nonlinear in the Ricci curvature:
\begin{align}\label{act0}
S[g_{\mu \nu}] =&
\frac{m_{D}^{D-2}}{2} \int d^{D} x \sqrt{|g_D|}\,
\Big[f(R) + c_1R_{AB}R^{AB} \nonumber\\
+& c_2 R_{ABCD} R^{ABCD} \Big]\,,
\end{align}
\begin{equation}\label{fR}
f(R)=a R^2 + R - 2\Lambda_D\,,
\end{equation}
where $m_D\equiv1$ is the multidimensional Planck mass. The multidimensional space is represented as a direct product $\mathbb{M} = \mathbb{M}_4 \times \mathbb{M}_n$, where $\mathbb{M}_4$ is the four-dimensional space--time and $\mathbb{M}_n$ is a compact extra space with $n$ dimensions:
\begin{equation}
    ds^2 = g_{\mu \nu} dx^\mu dx^\nu - e^{2\beta(t)} d\Omega^2_n\,,  
\end{equation}
where $g_{\mu \nu}$ is the metric of the four-dimensional space $\mathbb{M}_4$, $d\Omega^2_n$ is the metric on the unit n-sphere, and $e^{\beta}$ is its radius.

To obtain the observed four-dimensional dynamics, one has to integrate out the action \eqref{act0} over the extra dimensions and transform to the Einstein frame, which ensures minimal coupling of the curvature to the resulting effective fields. The procedure for obtaining such a theory is described in detail in~\cite{Bronnikov:2005iz}, and the inflation and compactification of the subspaces $\mathbb{M}_4$ and $\mathbb{M}_n$ arising in it are studied in~\cite{Fabris:2019ecx, Bronnikov:2021ptu}.

The extra space $\mathbb{M}_n$ is assumed to be maximally symmetric, which gives its Ricci scalar $R_n = n(n-1)/e^{2\beta}$, where $e^\beta \equiv \rho$ is the radius of the extra space. The effective theory is valid in the regime where the extra space is small compared with the four-dimensional one and varies slowly:
\begin{align}\label{effective}
&R = R_4 + R_n + P_k\,,\nonumber\\
&P_k = 2 n\,\partial^2 \beta + n(n+1) (\partial \beta)^2\,,\nonumber\\
&R_4, P_k \ll R_n\,,
\end{align}
where $R_4$ and $R_n$ are the Ricci scalars of $\mathbb{M}_4$ and $\mathbb{M}_n$.

\medskip
{\bf Effective scalar field}
\medskip

After integrating out the extra space and transforming to the Einstein frame via a conformal transformation~\cite{Bronnikov:2005iz}, the effective Lagrangian becomes four-dimensional with the usual gravitational part $~R_4$, while the Ricci scalar of the extra space is interpreted as a new effective field $\phi$ with the action
\begin{equation}\label{act_eff}
S_{\rm eff} \!=\! \frac{m_4^2}{2}\!\!\int\!\! d^4x \sqrt{-g_4}\  \text{sign}(f') \!\left[R_4\! +\! K(\phi) (\partial \phi)^2 \!-\! 2 V(\phi) \!\right]\!,
\nonumber
\end{equation}
where the effective Planck mass in the Einstein frame is $m_4 = m_D\,\sqrt{2\pi^{\frac{n+1}{2}} / \Gamma(\frac{n+1}{2})}$, and $g_4^{\mu\nu}$ is the observed four-dimensional metric. The emerging scalar field has the meaning of the Ricci scalar of the extra space, $\phi \equiv R_n$, in the original theory~\eqref{act0}. The action $S_{\rm eff}$ contains a potential and a nontrivial kinetic term, which are expressed in terms of the original parameters of action~\eqref{act0} (a detailed derivation is given in~\cite{Fabris:2019ecx}):
\begin{align}\label{Pot}
V(\phi) \!&=\! -\!\frac{\operatorname{sign}f^{\prime}}{2 (f^{\prime})^2}\!\left(\frac{|\phi|}{n(n-1)}\!\!\right)^{\!\!\frac{n}{2}}\!\!\left(\!f(\phi) +\!\frac{c_1\!+\!\frac{2c_2}{n-1}}{n} \phi^{2}\!\right)\!,\\
K(\phi)\! &\!=\! \frac32 \!\!\left(\frac{f^{\prime \prime}}{f^{\prime}}\!\right)^{2} \!-\!\! \frac{n}{2\phi}\frac{f^{\prime \prime}}{f^{\prime}} \!+\! \frac{n(n\!+\!2)}{8\phi^2} \!+\! \frac{c_{1}\!+c\!_{2}}{\phi f^{\prime}}. \label{Kphi}
\end{align}

Since the field $\phi$ represents the curvature $R_n$ of the extra space, its vacua (the minima of potential \eqref{Pot}) correspond to stable configurations of the maximally symmetric extra space with radius $\rho \equiv e^\beta$.

\begin{figure}[h]
    \includegraphics[width=0.48\textwidth]{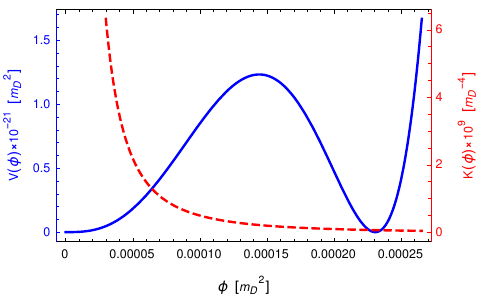}
    \caption{Plots of the potential term and the kinetic factor \eqref{Pot} for the parameters $n=6$, $c_1=-8000$, $c_2=-5000$, $a_2=-500$. Rolling down to the left vacuum $\phi=0$ corresponds to decompactified multidimensional universe; while the right vacuum,
    $\phi_{\rm min} =2.3\cdot10^{-4}$, corresponds to the observed four-dimensional Universe.}
    \label{potential&kinetic}
\end{figure}

There exist combinations of the parameters of action~\eqref{act0} for which potential \eqref{Pot} has two different vacua in $\phi$, one of which corresponds to $\phi_\text{min} = 0$ which signifies decompactification of the extra space (see an example of such a choice of parameters in Fig.~\ref{potential&kinetic}).

Rolling of the field $\phi$ to the right minimum of the potential (Fig.~\ref{potential&kinetic}) corresponds to stabilization of the compact extra space at $R_n\neq 0$ (the extra space is compactified and has some radius $\rho_0$) and leads to the observed four-dimensional cosmology~\cite{Fabris:2019ecx}. The presence of the left minimum, $R_n\approx0$, indicates the possibility of another scenario (studied in~\cite{Bronnikov:2021ptu}), in which the extra space is unstable and expands together with the main space to macroscopic sizes.

Cosmological evolution toward different states of the extra space, either compactified ($\phi_{min}\neq 0$) or decompactified ($\phi_{min}=0$), gives rise to inclusions of one vacuum inside another, i.e., to the formation of ``bubbles'' of decompactified extra space inside the compactified four-dimensional Universe. Such ``bubbles'' must be surrounded by massive domain walls.

\medskip
{\bf Domain walls and primordial black holes}
\medskip

The subsequent mechanism of PBH formation from domain walls is similar to that considered in Section~\ref{PBH_from_walls}. The essential difference is that the domain wall in the extra-dimensional approach is extremely massive. Therefore, its collapse into a PBH occurs right in the course of the rolling of the field $\phi$ toward the potential minima (such walls are called supercritical~\cite{Deng:2016vzb,Deng:2017uwc}), which starts at the moment the scalar field ``unfreezes'', $H \ll \sqrt{V''(\phi_\text{min})}$. This process begins immediately after the end of inflation (because of the large value of $\sqrt{V''(\phi_\text{min})})$; therefore, the domains of the vacuum $\phi=0$ do not have time to undergo significant post-inflationary expansion, unlike in the models of Section~\ref{PBH_from_walls}, and the final wall radius is $\tilde{r}\approx r$, in contrast to Eq.~\eqref{r_final}.

As the field $\phi$ rolls down toward the potential minima, the energy density $\sigma(t)$ of the domain wall and its mass increase. When the wall reaches the Schwarzschild mass for its radius, $4\pi r^2 \sigma(t) = r/2G$, it turns into a black hole. The maximum energy density of the wall, $\sigma_\text{max}$, is so large that it is never reached. In other words, the inequality $4\pi \delta^2 \sigma_\text{max} > \delta/2G$ holds in this model, where $\delta$ is the wall thickness, which is at the same time the minimum possible domain radius. It follows that the PBH mass is determined solely by the size of the domain of the vacuum $\phi=0$ formed during inflation, and the wall parameters play no role in determining the PBH mass.

The PBH mass spectrum $dn(M)/dM$ in this theory corresponds to the size spectrum $dn(r)/dr$ of the domains of field $\phi$ fluctuations generated during inflation. Substituting $r = 2G M$, we obtain the mass spectrum of the resulting PBHs:
\begin{equation}
\frac{dn(M)}{dM} = 2 G\, \frac{dn(2G M)}{dr},
\end{equation}
where $dn(r)/dr$ was calculated in Section~\ref{PBH_from_walls}; see Eq.~\eqref{post_inf_spectr}.

\subsection{Formation of primordial black holes due to baryon charge fluctuations}\label{Dolgov}

One of the first models of massive PBH formation is the Dolgov--Silk model proposed in 1993~\cite{DolSil93}, which developed a mechanism for producing PBHs with a log-normal mass spectrum as a result of baryon charge fluctuations during baryogenesis in the early Universe; see further developments of this theory in~\cite{DolKawKev09} and the review~\cite{Dol18}. The model is based on a generalization of the Affleck--Dine baryogenesis model~\cite{AffDin85}, whose potential is supplemented with a term describing the interaction, with coupling constant $g$, between the inflaton field $\Phi$ and the complex scalar field $\chi$ carrying the baryon number:
\begin{equation}
U(\chi)=m_0^2|\chi|^2+g|\chi|^2(\Phi-\Phi_1)^2+\lambda|\chi|^4\ln\frac{|\chi|^2}{\sigma}.
\end{equation}
In this potential, one can identify an effective mass that includes the coupling between the fields:
\begin{equation}
m_{\rm eff}^2=m_0^2+g(\Phi-\Phi_1)^2+\ldots.
\end{equation}
For simplicity, terms related to temperature and curvature are omitted here. The condensate of the field $\chi$ is responsible for $CP$ violation and the generation of the baryon asymmetry. Baryon charge fluctuations arise during inflation, and~\cite{DolSil93} considered a chaotic inflation model.

Initially, when the field $\Phi$ is large, the mass of $\chi$ is large, and the potential has a minimum at $\chi=0$. As $\Phi$ evolves, decreasing and passing through the value $\Phi_1$, the mass of the field $\chi$ becomes smaller than the Hubble parameter during inflation, and the potential develops a deeper minimum at $\chi\neq0$. Owing to quantum fluctuations, this field can then diffuse far from zero along the valley of the potential and, fluctuating across the valley, pass into a state with an excess of baryon charge of one sign or the other. Pictorially, in the complex plane ($\chi_1,\chi_2$), where $\chi=\chi_1+i\chi_2$, the field ``rotates'' around zero, and the angular momentum of this rotation corresponds to the baryon number (this follows from the expression for the baryon current). A phase transition with the formation of a bubble of the new phase occurs if quantum fluctuations of the field $\chi$ manage to bring it into the region of classical rolling. As $\Phi$ decreases further below $\Phi_1$, baryogenesis ends, and after inflation the field returns to $\chi=0$, conserving its angular momentum and, accordingly, the accumulated baryon charge. When the $\chi$ meson decays into quarks, different ratios of the numbers of quarks and antiquarks arise in different regions of space, and the baryon charge fluctuations turn into density perturbations.

Although the average baryon asymmetry in the Universe may correspond to the observed small value $6\times10^{-10}$, this model predicts compact regions in which the asymmetry reaches $\sim1$. These may include regions made of antimatter~\cite{BamDol07,Blietal15}. Some of the regions with asymmetry $\sim1$ could form compact star-like objects and even collapse into PBHs, if the size of the perturbation exceeded the Jeans radius at that epoch. In~\cite{DolSil93}, a log-normal PBH mass function was obtained,
\begin{equation}
\frac{dN}{dM}=\mu^2\exp\left(-\gamma\ln^2\frac{M}{M_0}\right),
\label{lognor}
\end{equation}
where the parameters $\mu$ and $\gamma$ are not fixed by the theory and can be chosen within wide ranges. They can be chosen so as to describe the available observational data. From the condition of gravitational collapse during the QCD phase transition, the parameter $M_0$ is estimated as $\sim10M_\odot$, which describes the observations well~\cite{DolPos20}.

The Dolgov--Silk model with mass function~\eqref{lognor} can explain several observational facts at once~\cite{Blietal16}. Low-mass PBHs can constitute DM, stellar-mass PBHs can be responsible for gravitational-wave bursts, and more massive PBHs can explain the supermassive BHs in galactic nuclei. In particular, this theory agrees well with recent observations by the James Webb Space Telescope~\cite{Dol23}, which detected an unexpectedly large number of massive galaxies in the pre-reionization epoch. Baryon charge fluctuations can explain the excess of heavy chemical elements observed at redshifts $z\geq10$, as well as the anomalous chemical composition of some stars in the Galaxy. Importantly, despite the presence of antimatter regions, the model does not contradict data on gamma rays from annihilation~\cite{Blietal15}; {moreover, annihilation of antimatter and matter near hypothetical antistars may proceed through the intermediate formation of protonium atoms, with cascade emission of narrow X-ray emission lines before annihilation, which can be tested in future X-ray spectroscopic observations}~\cite{Bonetal22}.

\subsection{QCD phase transition}

According to calculations in quantum chromodynamics (QCD), at temperatures $\sim 100-150$~MeV the early Universe underwent a transition from the quark--gluon plasma to the hadron gas, when quarks combined into baryons and mesons. During this epoch, owing to the nature of quark and gluon interactions, the pressure of the medium drops noticeably and the sound speed $c_s=\sqrt{dp/d\rho}=w^{1/2}$ decreases. According to calculations similar to those presented in Section~\ref{adiabsubsec}, but with a more general equation of state $p=w\rho$, PBHs should then form more efficiently. This is because the PBH formation threshold in the simplest estimates is $\delta_c\simeq w$, while PBH formation (in the model of the collapse of adiabatic perturbations) occurs in the tail of the Gaussian distribution of perturbation amplitudes under the condition $\delta>\delta_c$; therefore, lowering the threshold results in more PBHs. Of course, this requires curvature perturbations on the horizon scale during the QCD transition. This PBH formation mechanism was proposed by Jedamzik~\cite{Jed97}. The mass of radiation inside the cosmological horizon at temperature $T$ of the Universe is
\begin{equation}
M_{\rm H}(T)\simeq15\left(\frac{T}{100\mbox{~MeV}}\right)^{-2}\left(\frac{g_{\rm eff}}{10.75}\right)^{-1/2}M_\odot,
\label{thorkhd}
\end{equation}
where $g_{\rm eff}$ is the effective number of degrees of freedom at the given temperature. A plausible value is $g_{\rm eff}\simeq40$~\cite{DolPos20}; therefore, at $T\sim 100$~MeV one can expect the production of PBHs with masses $\sim8M_\odot$~\cite{DolPos20}. \PK{\footnote{Note that a possible magnetic field and a nonzero chemical potential during the QCD transition lower the temperature $T_{QCD}$~\cite{Areetal23}.} }

Some works have assumed that PBHs formed during the QCD transition may have masses $M_{\rm PBH}\sim1M_\odot$. In particular, masses $\sim1M_\odot$ were considered in~\cite{Jed97}; in~\cite{2018JCAP...08..041B}, a PBH mass range $0.3\leq M_{\rm PBH}/M_\odot\leq1.4$ was obtained, with the maximum of the mass distribution at $M_{\rm PBH}\sim0.7M_\odot$; and in~\cite{Musco:2023dak}, the peak position is given as $\sim1.9M_\odot$. Within the model under consideration, such masses are possible if the temperature of the QCD transition is sufficiently high, or if the PBH masses are a small fraction of the radiation mass inside the horizon at the time of PBH formation~\cite{NadNovPol78}. The latter is also predicted by numerical calculations of critical gravitational collapse, see (\ref{critsoot}), with a broad mass distribution. In this case, low-mass PBHs from the distribution could be of the same order of magnitude as the $\sim0.5M_\odot$ compact objects observed via microlensing events in the 1990s; therefore, the closeness of this mass to the prediction of~\cite{Jed97} was regarded as good agreement with observational data~\cite{Jed98}. Some of the microlensing data were later revised; this is discussed in more detail in Section~\ref{dmsec}. This PBH formation model, for sufficiently small PBH masses, has recently gained relevance in connection with the gravitational-wave event S251112cm observed by the LIGO--Virgo--KAGRA detectors, in which the chirp mass lies in the range $0.1-0.87M_\odot$, and with probability $>99$~\% one of the BHs in the merged pair has a mass $<1M_\odot$~\cite{Vieetal26}. If this estimate is correct, this BH may not be of stellar origin, and its mass may correspond to a PBH born during the QCD transition.


\subsection{Collapse of cosmic strings}

Above, we have already considered models of PBH formation from the collapse of topological defects, namely closed two-dimensional domain walls. Models have also been proposed in which PBHs originate from the evolution and collapse of one-dimensional topological defects, cosmic strings. The first such model was proposed by S.~Hawking in 1989~\cite{Haw89}.

Initially, starting with~\cite{Haw89}, models were developed in which PBHs form from the collapse of approximately circular closed loops~\cite{PolZem91,GarVil93,CalGat93,1996PhRvD..53.3002C}. Collapse occurs if a loop contracts by a factor of about $G\mu/c^2$, where $\mu$ is the linear mass density (tension) of the string, and ends up inside its gravitational radius. The first detailed numerical simulation of loop collapse in GR was performed in~\cite{HelAurLim19}. In~\cite{Jametal20}, PBH constraints were used to obtain a constraint on the fraction of collapsing cosmic string loops, $f\leq7\times10^{-15}[(G\mu/c^2)/10^{-11}]^{-3/2}$.

Closed cosmic string loops make up only a very small fraction of all strings; however,~\cite{JenSak20} extended the possibility of PBH formation to the population of infinite strings. Cosmic strings can rapidly change their shape, forming cusps. The following condition for PBH formation from a string segment of complex shape was justified in~\cite{JenSak20}: if a cosmic string segment with mass $M$ is enclosed within a sphere of radius $R$ and the condition $2GM/(c^2R)\geq1$ holds, this segment can collapse into a BH. In other words, collapse occurs if the sphere contains a mass whose Schwarzschild radius is larger than the radius of the sphere (the ``hoop conjecture''). String segments that can collapse via this mechanism are mainly cusps. Furthermore, it was shown in~\cite{JenSak20} that the PBHs formed in this model have large spins and relativistic initial velocities.


\subsection{Primordial black holes in alternative and modified theories of gravity}

The process of PBH formation depends on the theory of gravity that operates during the formation epoch, and modern theories allow the form of the basic gravitational equations to change when certain symmetries are broken during the expansion of the Universe. In addition to numerous analytical and numerical calculations performed within GR, there are studies of PBH formation in modified theories of gravity. Ref.~\cite{BarCar96} considered the influence of a varying gravitational constant in the Brans--Dicke theory on the number of PBHs formed. The evolution of the PBH population in Brans--Dicke cosmology was also considered in~\cite{NayMajSin10, AliZar20}. The number of PBHs formed is very sensitive to their formation threshold. PBHs form only from density perturbations exceeding a certain threshold value. In GR, this threshold is $\sim1/3$--$0.42$. In~\cite{Che19}, the modification of the PBH formation threshold in Eddington-inspired Born--Infeld gravity was studied, and it was shown that in this theory of gravity the threshold is not constant but depends on the epoch of PBH formation.

In~\cite{Berezin:2020tbk}, corrections to the PBH formation threshold were calculated within the Brans--Dicke theory. The equations for the evolution of spherically symmetric perturbations in the Brans--Dicke theory were written down, and a correction to the PBH formation threshold was derived from them. This correction depends on the initial derivative of the Brans--Dicke scalar field. Owing to this dependence, PBHs form with different probabilities in regions with different derivatives of the scalar field. The scalar field thus modulates the large-scale distribution of PBHs. This influence is similar to the well-known biasing effect and may lead to PBH clustering.


\section{Dark matter around primordial black holes}\label{sec:DM_around_PBH}

PBHs can constitute all of the dark matter (DM) if their masses lie within a certain range. The range $\sim10^{17}-10^{22}$~g remains allowed. PBHs of other masses apparently cannot constitute all of the DM, but only a small fraction of it, owing to a number of constraints on PBHs. In this case, DM is a separate cosmological component, most likely unrelated to PBHs. However, the gravitational field of a PBH can perturb the DM, which should lead to the formation of DM density spikes and DM halos around PBHs. We refer to the DM overdensities that form around PBHs during the radiation-dominated era as density spikes, and to the more extended structures that form during the matter-dominated era as DM halos. Density spikes are of interest because the DM density in them is high, so DM particles can annihilate intensively and produce gamma rays and other types of radiation. These effects can be used to obtain joint constraints on PBHs and on the properties of DM particles.


\subsection{Density spikes}
\label{spikesubsec}

One of the most popular DM models is DM in the form of some elementary particles that have not yet been directly detected in experiments. We will assume that this DM model (new elementary particles) is correct. A number of works have studied the accumulation of DM particles on orbits around PBHs and the formation of density spikes during the radiation-dominated stage of the evolution of the Universe~\cite{LacBea10, SaiShi11-2, SaiShi11,Don11, Ero16,Bouetal18, Adaetal19, Ero19, CarKuhVis21, Ire24}. Before the kinetic decoupling of DM particles from radiation at time $t_d$, they were coupled to radiation and could not move freely; they were only carried along by the common hydrodynamic flow. In particular, they could not accumulate significantly around PBHs, because this was prevented by radiation pressure, which counteracted the gravitational attraction of the PBH. In the model of DM in the form of WIMPs (weakly interacting massive particles), kinetic decoupling occurs at the temperature~\cite{CarKuhVis21}
\begin{equation}
T_d\sim10\left(\frac{m_\chi}{100\mbox{~GeV}}\right)^{5/4}\mbox{~MeV}.
\label{tdbig}
\end{equation}
After kinetic decoupling, DM particles began to move freely, and some of them could remain on bound orbits around PBHs. During the radiation-dominated stage, these particles formed dense spikes around PBHs. For low PBH masses, the residual velocities of DM particles after kinetic decoupling are important, but for large PBH masses the role of residual thermal velocities decreases. After the Universe entered the matter-dominated stage, the formation of DM halos around PBHs continued on larger scales; during this epoch, the halo mass already exceeded the mass of the seed PBH, and the halo grew in the secondary infall regime~\cite{Ber85}.

The condition for a DM particle located at a distance $r_i$ from a PBH to be captured onto a bound orbit is that its initial velocity be smaller than the escape velocity, $v_i<(2GM_{\rm BH}/r_i)^{1/2}$. The second condition is that, at any point of the orbit, the radius $r$ lies between the minimum and maximum distances of the DM particle from the center,
\begin{equation}
r_{\rm min}=a(1-e)\leq r \leq r_{\rm max}=a(1+e),
\label{minmaxcond}
\end{equation}
where $a$ is the semimajor axis of the particle orbit. The particle travels from $r$ to $r+dr$ twice per orbital period $T_{\rm orb}$; therefore, it spends a fraction $2dt/T_{\rm orb}$ of its time at distances between $r$ and $r+dr$, where $dt$ is the time it takes to move from $r$ to $r+dr$. If the initial DM density is $\rho_i(r_i)$, the final density $\rho(r)$ can be written as
\begin{equation}
\rho(r)4\pi r^2dr\!=\!\!\int\!\! 4\pi r_i^2dr_i \rho_i(r_i) \!\!\int\!\! d^3vf(v)\frac{2(dt/dr)}{T_{\rm orb}}dr,
\label{rhoitog}
\end{equation}
where $f(v)$ is the thermal distribution function of DM particles, and the derivative $dt/dr$ has the well-known form from the Kepler problem. This approach makes it possible to calculate the density profile of a DM spike around a PBH in different cases~\cite{CarKuhVis21}:
\[
\rho(r)=
\left\{
\begin{array}{rcl}
\rho_d\left(\frac{r_c}{r}\right)^{3/4} \,&;& \,r<r_c;
\\
\\
\frac{\rho_{\rm eq}}{2}\left(\frac{M_{\rm PBH}}{M_\odot}\right)^{3/2}\left(\frac{\tilde r}{r}\right)^{3/2} \,&;& \,r_c<r<r_K;
\\
\\
\frac{\rho_{\rm eq}}{2}\left(\frac{M_{\rm PBH}}{M_\odot}\right)^{3/4}\left(\frac{\tilde r}{r}\right)^{9/4} \,&;& r>r_K.
\end{array}
\right.,
\]
where $\rho_d$ is the density of the Universe at kinetic decoupling, $\rho_{\rm eq}$ is the density of the Universe at matter--radiation equality, $r_c=2GM_{\rm PBH}m_\chi/(T_d)$, $m_\chi$ is the mass of DM particles, $T_d$ is the temperature of their kinetic decoupling from radiation, $\tilde r=GM_\odot t_{\rm eq}m_\chi/(t_dT_d)$, and $r_K=(2t_dT_d)^2/(2GM_{\rm PBH}m_\chi^2)$.

To calculate the gamma-ray flux from DM particle annihilation, it is important to know the density profile of the spike and the maximum possible density in the central region. The maximum DM density at the halo center can be estimated from the simple condition $\rho\langle\sigma_{\rm ann} v\rangle t_0/m\sim1$~\cite{BerGurZyb92,SilSte93}, where $\langle\sigma_{\rm ann} v\rangle$ is the velocity-averaged annihilation cross section, $m$ is the mass of DM particles, and $t_0$ is the present age of the Universe. This estimate yields a central plateau with constant DM density,
\begin{eqnarray}
&&\rho_{\rm max}\simeq \frac{m}{\langle\sigma_{\rm ann} v\rangle t_0} \simeq10^{-14}\left(\frac{m}{70\mbox{~GeV}}\right) \times 
\label{annwr}
\\
&&\!\left(\!\frac{\langle\sigma_{\rm ann} v\rangle}{3\times10^{-26}\mbox{~cm$^3$s$^{-1}$}}\right)^{-1}
\!\!\!\left(\frac{t_0}{1.36\times10^{10}\mbox{~yr}}\!\right)^{-1}\! 
\mbox{g~cm$^{-3}$}.
\nonumber
\end{eqnarray}

In the plateau region, the DM density should have dropped to $ \rho_{\rm max}$ by the present time. This approach assumes that the density change is a local process that proceeds independently at each radius. Estimate (\ref{annwr}) can be somewhat refined by solving the equation $\partial\rho/\partial t=-\langle\sigma_{\rm ann} v\rangle \rho^2/m$, which gives a smooth transition to the plateau region~\cite{GonSil99} and is exact only for circular orbits~\cite{Vas07}. In general, however, DM particles move on elongated orbits, so annihilation at one radius depletes the density at other radii as well. In~\cite{LacBea10}, the lifetime of a DM particle was estimated taking into account that, in its orbital motion, the particle passes through regions of different density. That work considered an orbit with a single characteristic set of orbital parameters. The next refinement is to take into account the distribution of DM particles over orbital parameters. One can expect that the presence of elongated, mutually overlapping orbits will lead, as particles annihilate, to the formation of a cusp rather than a core in the central region. This was shown in~\cite{Vas07} for astrophysical black holes with a power-law distribution over angular momenta. For PBHs, the calculation showed that the profile differs from a flat plateau and is close to a power-law cusp with an exponent of $\approx-0.7$~\cite{Ero24}. This difference increases the annihilation rate at the halo center, and hence the predicted flux of annihilation radiation, by about 30\%.

Limits on the gamma-ray background yield joint constraints on the PBH abundance and on the parameters of DM particles. These constraints are given in the most complete form in~\cite{CarKuhVis21}. In the mass range $2\times10^{-12}M_\odot(m_\chi/\mbox{1~TeV})^{-3.2}<M_{\rm PBH}<5\times10^{12}M_\odot(m_\chi/\mbox{1~TeV})^{1.1}$, for the thermal cross section of DM production
$\langle\sigma_{\rm ann} v\rangle\sim 3\times10^{-26}\mbox{~cm$^3$ s$^{-1}$}$, the constraint has the form~\cite{CarKuhVis21}
\begin{equation}
f_{\rm PBH}\leq 2\times10^{-9}\left(\frac{m_\chi}{\mbox{1~TeV}}\right)^{1.1}.
\label{fcomogr}
\end{equation}
In particular, it follows that annihilating DM with this cross section and PBHs in the amount required to explain the LIGO/Virgo/KAGRA gravitational-wave bursts are incompatible with each other. Either the PBH abundance must be much smaller than (\ref{fcomogr}), or DM must consist of particles with different properties.

\subsection{Supermassive primordial black holes}

Massive PBHs could accumulate a large amount of DM around themselves. Before turning to this question, let us discuss the status of supermassive PBHs themselves. The observation of quasars at redshifts $z>6$ poses a challenge to astrophysical theories of supermassive BH formation. The point is that the luminosity of the quasar discovered at redshift $z=6.4$ corresponds to accretion onto a BH with a mass of $3\times10^9M_{\odot}$, whereas the formation time of such BHs via ordinary astrophysical mechanisms is rather long (at least, existing models face serious difficulties). Calculations in~\cite{2021ApJ...907L...1W} (Fig.~3 therein) have shown that, even with rapid mass growth through accretion, BH seeds with masses of at least $\sim10^4M_\odot$ are required. PBHs could successfully explain the presence of BHs at high redshifts, since they could form at very early epochs, during the radiation-dominated stage.

Another line of research concerns the seeds of supermassive BHs in galactic nuclei. The James Webb Space Telescope has revealed a larger number of massive galaxies at redshifts~$z>6$ than predicted by the standard $\Lambda$CDM model. Among the proposed explanations are models with an excess of perturbations in the power spectrum. However already early works on PBHs put forward the idea that supermassive PBHs could serve as seeds for the formation of galaxies, or at least of some of them~\cite{Blietal16}. In contrast to the hierarchical picture of galaxy formation, the dark halo around a PBH should grow not through mergers of smaller objects, but as a result of the general infall of DM onto the PBH. This produces centrally concentrated objects with a density profile $\propto r^{-9/4}$, corresponding to the secondary infall mechanism~\cite{Dokuchaev:2001ekr}.

Current observational constraints on the PBH abundance in the Universe in various mass ranges are given in the review~\cite{2021RPPh...84k6902C} in the form of an upper limit on the fraction $f_{\rm PBH}$ of PBHs in dark matter. In the range of interest to us, $10^8-10^{10}M_\odot$, there are constraints from CMB $\mu$-distortions, from millilensing of compact radio sources, as well as a number of dynamical constraints described below in Section~\ref{dmsec}.

The strongest of the dynamical constraints comes from the dynamical friction of PBHs in the Galactic halo. If the halo contained PBHs of such large mass, dynamical friction would have made them sink to the Galactic center over the lifetime of the Galaxy. However, the black hole at the Galactic center is known to have a mass of only $\sim 4\times 10^6M_\odot$, much smaller than the masses under consideration.

The constraint from CMB $\mu$-distortions may be several orders of magnitude stronger than the dynamical constraints. This constraint is related to Silk damping, i.e., the dissipation of density perturbations, which produces CMB $\mu$-distortions~\cite{1993PhRvD..48..543C}. As a result, the number of PBHs in the mass range $(6\times10^4 - 5\times10^{13})M_\odot$ is strongly constrained~\cite{2014PhRvD..90h3514K}. However, as pointed out in~\cite{Nakama:2016kfq,2021RPPh...84k6902C}, the CMB $\mu$-distortion constraint applies only if PBHs formed from high density peaks of Gaussian perturbations, i.e., as in the historically first model of PBH formation. In other PBH formation models, including those considered in Sections~\ref{PBH_from_walls} and~\ref{adddimsubsec}, this constraint does not apply. Nor does it apply in the presence of non-Gaussianity, which is expected in the region of high peaks. Thus, the constraint from CMB $\mu$-distortions is model-dependent and may not hold.

Let us assume that the constraint from CMB $\mu$-distortions plays no role and adopt, as the most conservative one, the constraint $f_{\rm PBH}\leq10^{-5}$ that follows from dynamical friction. Let us estimate the mean separation between PBHs under this constraint. Their number density is
\begin{equation}
n_{\rm PBH}(z)=\frac{\rho_{c,0}\Omega_m f_{\rm PBH}(1+z)^3}{M_{\rm PBH}},
\end{equation}
where $\rho_{c,0}$ is the present critical density of the Universe and $\Omega_m$ is the cosmological density parameter of DM. The mean separation between PBHs, $l(z)=(n_{\rm PBH}(z))^{-1/3}$, is
\begin{equation}
l(z)\simeq4\left(\frac{1+z}{21}\right)^{-1}
\!\!\left(\frac{f_{\rm PBH}}{10^{-5}}\right)^{\!\!-1/3}
\!\!\left(\frac{M_{\rm PBH}}{10^{10}M_\odot}\right)^{\!1/3}\mbox{Mpc}. \nonumber
\end{equation}
In particular, this shows that in the present-day Universe, at $z=0$, the separation between such objects is $\geq84$~Mpc, i.e., they are quite rare.

\medskip
{\bf Primordial black holes of extremely large mass}  
\medskip

In most models, the maximum PBH mass is formally limited by the mass of matter inside the Hubble radius, although such massive PBHs should be very rare. Among the known supermassive BHs in quasars, the largest estimated mass, $\sim7\times10^{10}M_\odot$, belongs to the BH in the quasar TON~618 at redshift $z=2.2$~\cite{Shemmer:2004ph}. In~\cite{CarKuhVis21}, a rather extreme case was considered in which PBHs have masses $\sim(10^{11}-10^{18})M_\odot$. Such PBHs, with masses exceeding the known masses of BHs in galaxies, were called Stupendously Large Black Holes (SLABs). Interestingly, these PBHs could constitute a noticeable fraction of the DM in the Universe, although, of course, they cannot explain the presence of DM inside galaxies. It was shown in~\cite{CarKuhVis21} that PBHs of extremely large mass could have acquired a significant part of their mass through accretion; that work also obtained constraints on such PBHs from several effects: dynamical constraints, gravitational lensing, background radiation from accretion onto PBHs, and gamma rays from the annihilation of DM around PBHs.

\subsection{Large-scale structure}     
\label{krupnosubsec}

If supermassive PBHs were born in the Universe, each such PBH could have produced density perturbations in the baryons and DM around it, leading to the formation of a galaxy~\cite{Rya72,CarRee84}. Consider a PBH formed during the radiation-dominated cosmological stage. Before the onset of the matter-dominated stage, infall of DM onto the PBH, let alone of baryons, is inefficient. Only in the immediate vicinity of the PBH does a DM density spike form. However, after the end of the radiation-dominated stage, secondary infall becomes efficient, and a virialized DM halo forms around the PBH, with a universal density profile $\rho(r)\propto r^{-9/4}$~\cite{Gun77} and a universal self-similar profile outside the virial radius~\cite{FilGol84,Ber85}. The growth of these halos ends when the ordinary inflationary density perturbations surrounding the PBH enter the nonlinear stage and become competing centers of attraction for DM, stopping its flow onto the PBH~\cite{Dokuchaev:2001ekr,DokEro03}. Alternatively, if the number of PBHs in the Universe is sufficiently large, halos around neighboring PBHs could compete with each other~\cite{CarRee84}.

Such objects may constitute a subclass of galaxies with a steep density increase toward the center. Let us describe their structure. One of the characteristic scales is the size $r_c$ of the region in which the gravitational influence of the PBH dominates at the time $t_{\rm eq}$ when the Universe enters the matter-dominated stage. Consider a PBH of mass $M_{\rm PBH}$. Let $\rho_{\rm eq}$ be the DM density in the Universe at matter--radiation equality ($z=z_{\rm eq}\simeq3119$); then
\begin{equation}
r_c=\left(\frac{3M_{\rm PBH}}{4\pi\rho_{\rm eq}}\right)^{1/3}=0.05\left(\frac{M_{\rm PBH}}{M_\odot}\right)^{1/3}~\mbox{pc.}
\end{equation}
The region of radius $r_c$ contains a DM density spike formed during the radiation-dominated cosmological stage, as well as a cloud of hot ionized gas.

\begin{figure}[t]
	\begin{center}
\includegraphics[angle=0,width=0.45\textwidth]{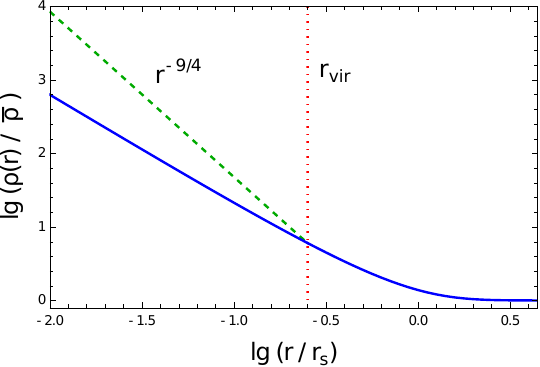}
\label{figrho}
	\end{center}
\caption{The self-similar DM density profile around a BH obtained in~\cite{Ber85} is shown by the solid line. The density is given in units of the mean cosmological density. The shell turns around at radius $r=r_s(z)$, i.e., at $r/r_s=1$. The vertical dotted line marks the virial radius $r_{\rm vir}$. To the left of this line is the virialized, phase-mixed region with density (\ref{rhovir}); to the right is the single-stream flow region. The dashed line shows the actual density in the virialized region, which differs from the density given by the self-similar solution.}
\label{auto}
\end{figure}  

The next characteristic scale at each given moment is set by the radius of the DM shell that has turned around~\cite{Ber85}:
\begin{equation}
r_s(z)=6.7\times10^{21}\left(\frac{M_{\rm PBH}}{10^6M_\odot}\right)^{1/3}\left(\frac{1+z}{18}\right)^{-4/3}~\mbox{cm}.
\end{equation}
Within the virial radius $r_{\rm vir}\simeq r_s/4$, i.e., at $r<r_{\rm vir}$, DM has undergone virialization and phase mixing; this region contains a stationary DM distribution with the density profile obtained in the secondary infall model:
\begin{equation}
\rho(r)=1.3\times10^{-19}\!\left(\!\frac{M_{\rm PBH}}{10^6M_\odot}\!\right)\!^{3/4}\!\!\left(\frac{r}{1\mbox{~pc}}\right)^{\!\!-9/4}\!\mbox{g~cm$^{-3}$}.
\label{rhovir}
\end{equation}
At $r>r_{\rm vir}$, the density is given by the self-similar law found in~\cite{FilGol84, Ber85}. This density is shown in Fig.~\ref{auto}, where the self-similar density profile is matched to the virial density profile (\ref{rhovir}). DM halos forming around PBHs should have such a structure. In~\cite{DubGla12}, it was proposed to call supermassive PBHs with DM halos around them ``cosmological dinosaurs''.

In~\cite{Mes74,Mes75}, attempts were made to explain the long-wavelength perturbation spectrum on the scales of galaxies and galaxy clusters by statistical fluctuations in the PBH medium. In a region of space containing $N$ PBHs, the relative fluctuation of their number is $\sim1/\sqrt{N}$. As shown in~\cite{Car77}, this mechanism is inefficient on large scales. In~\cite{Mes74}, initial (deterministic) long-wavelength fluctuations in the mixture of PBHs and radiation were also considered. However, fluctuations in the number of PBHs may prove important for low-mass halos, as discussed in detail in the next section. Statistical fluctuations in the number of PBHs could also affect absorption in the neutral hydrogen line (the Ly-$\alpha$ forest)~\cite{Afshordi:2003zb}. This effect imposes the constraint $f_\pbh M_\pbh \lesssim 100~M_{\odot}$ on the PBH contribution to the DM density~\cite{Murgia:2019duy}.

\medskip
{\bf Absorption in the 21~cm line}
\medskip

As shown above, objects made of DM and baryonic gas can form around supermassive PBHs; at present, they may constitute a subclass of rare spheroidal galaxies. Regions with an enhanced optical depth of the gas to absorption in the 21~cm line may form at the periphery of these galaxies. These regions may now be observable as absorption rings around galaxies at redshifts $z\simeq20$. The position relative to the center and the parameters of the main absorption feature depend only on the mass of the central PBH and on time. The existence of such rings was predicted in~\cite{VasShc12,DubrovichGra19} for the first galaxies and in~\cite{DubEroGra21} for halos around supermassive PBHs. The rings arise because of codirected peculiar velocities of the gas along the line of sight near the turnaround radius of the object. The angular size of the absorption ring is rather small, $\sim9''$, but a radio telescope with a collecting area of more than 1~km$^2$ would make it possible to search for such objects. Thus, observations of such absorption rings in the radio band at $z>15-20$ may become feasible for the planned next-generation radio telescopes. The heating and ionization of the cosmic medium by accreting PBHs and the influence of this process on emission and absorption in the 21~cm line were considered earlier in~\cite{2013MNRAS.435.3001T} for a wide range of possible PBH masses, $(10-10^8)M_\odot$. Compared with~\cite{2013MNRAS.435.3001T}, the work~\cite{DubEroGra21} took into account, in particular, the motion of the gas in the evolving DM halo. This effect leads to the deep ring-shaped absorption regions described above.

\medskip
{\bf ``Little red dot'' galaxies}
\medskip

JWST observations have also revealed, at redshifts $z\geq6$, compact (with radii $\simeq150$~pc) bright galaxies called ``little red dots''~\cite{2025ApJ...989L...7T,2025A&A...704A.313D}. It is quite likely that ``little red dots'' host supermassive BHs whose masses are 2--3 orders of magnitude larger than implied by the correlations observed in the present-day Universe. In~\cite{Zhang:2025oyl}, it was hypothesized that supermassive PBHs are responsible for the formation of ``little red dots''. One may further hypothesize that ``little red dots'' formed not around individual PBHs but around PBH clusters. Radiation from accretion onto multiple PBHs could have suppressed star formation, which would explain the relatively small observed stellar mass compared with the mass of the central black hole in ``little red dots''.

\section{Clustering of primordial black holes}

As some models predict, PBHs in the Universe may be distributed not uniformly (more precisely, not according to a random Poisson distribution) but gathered in overdense regions or in gravitationally bound clusters. PBH clustering may be of interest because PBHs in clusters would have an increased probability of pairwise encounters. Accordingly, this would increase the rate of gravitational-wave bursts, which is important for their observation.

\subsection{Clustering of primordial black holes in low-mass halos} 
\label{sec:51}  

In this section, we consider a model in which PBHs are initially formed unclustered (isolated) in the Universe. In this case, their spatial distribution follows Poisson statistics, which gives rise to relative density fluctuations with amplitude $\delta_p \sim f_\pbh / \sqrt{\langle N_\pbh \rangle}$~\cite{Carr:2018rid}, where $\langle N_\pbh \rangle$ is the mean number of PBHs in the volume under consideration. These fluctuations are isocurvature modes; therefore, during the matter-dominated stage, they are related to their initial values by $\delta = 3s/(2s_\eq) \delta_i$~\cite{GorRub}, where $s$ is the scale factor of the Universe. On sufficiently small scales, these fluctuations exceed the conventional inflationary predictions, which leads to early structure formation~\cite{Afshordi:2003zb,Inman:2019wvr,Delos:2024poq}. We define the variance of the fluctuations, $\sigma^2_p =\langle \delta_p^2 \rangle$:
\begin{equation}\label{eq:sp1}
\sigma^2_p (M,z) = \frac{9 \, M_\pbh f_\pbh}{4 \, M}  \left ( \frac{1 + z_{\eq}}{1 + z} \right)^{2},
\end{equation}
where $M$ is the mass scale under consideration. A similar result could be obtained by introducing the power spectrum $P_\pbh = f^2_\pbh/n_\pbh$ and using expression~\eqref{eq:s_inf}. According to the Press--Schechter formalism~\cite{Press:1973iz}, the halo mass function has the form
\begin{equation}\label{eq:dNdM}
    \frac{d n}{d \ln M} = \frac{1}{\sqrt{2 \pi}} \frac{\rho_{\dm}}{M} \sqrt{\frac{M}{M_{\rm ch}}} \exp{\left ( -\frac{M}{2 M_{\rm ch}} \right)},
\end{equation}
where we have defined the characteristic halo mass by the standard relation $\sigma^2(M_{\rm ch},z) = \delta_c^2$:
\begin{equation} \label{eq:M_char}
    M_{\rm ch} = \frac{9 \, M_\pbh f_\pbh }{4 \, \delta_c^2} \left ( \frac{1 + z_{\rm eq}}{1 + z} \right)^{2},
\end{equation}
and $\delta_c = 1.69$ is the critical linear overdensity for spherical collapse. After formation, the mean halo density is $\overline{\rho}_h = 18 \pi^2 \rho_M(z)$, where $\rho_M(z)$ is the background matter density; the virial radius of the halo is $R_{\rm vir} = (3M/4 \pi \overline{\rho}_h)^{1/3}$, the number of PBHs in it is $\simeq f_\pbh M_{\rm ch}/M_\pbh$, and the characteristic velocities in the halo are estimated as $v\simeq \sqrt{GM/R_{\rm vir}}$:
\begin{equation}\label{eq:vch}
    v \simeq 1.4 \left ( \frac{f_\pbh M_\pbh }{\, M_{\odot}} \right)^{1/3} \left (\frac{1+z}{20} \right)^{-1/6} \mbox{km~s$^{-1}$}.
\end{equation}
Here, we have used Eq.~\eqref{eq:M_char} for the halo mass.

\begin{figure}
	\begin{center}
\includegraphics[angle=0,width=0.48\textwidth]{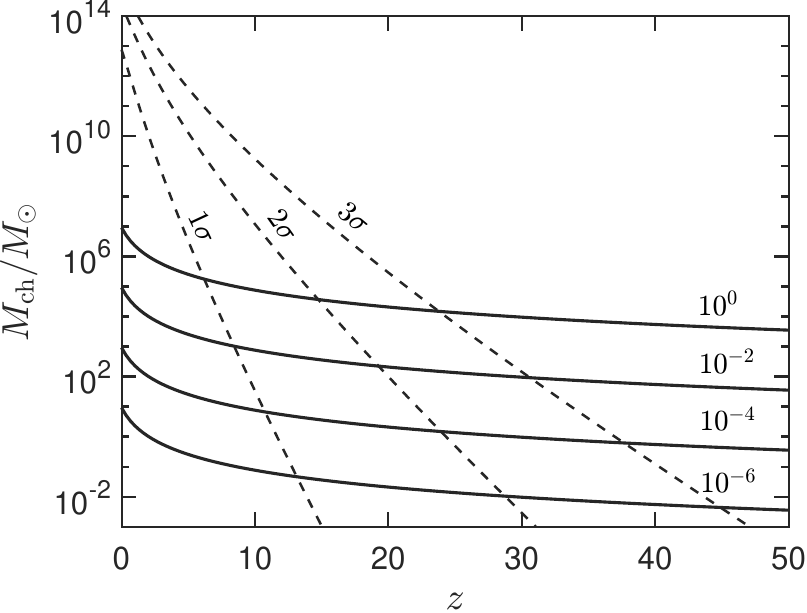}
	\end{center}
\caption{Characteristic halo mass as a function of redshift. The solid lines are given by expression~\eqref{eq:M_char} and are labeled with the value of the parameter $M_\pbh f_\pbh/M_{\odot}$. The dotted lines show $N\sigma$ halos from inflationary perturbations, determined by solving the equation $N\sigma_{\rm inf}(M,z) = \delta_c$, where $N = 1,2,3$ is also indicated in the plot.}
	\label{gr:Mch} 
\end{figure}

To illustrate the influence of PBHs on early structure formation, let us compare these results with the predictions of the generally accepted standard cosmological model, $\Lambda$CDM, with a nearly scale-invariant power spectrum $P_{\rm inf} \propto k^{n_s} T^2_{\rm ad}(k)$, where $n_s = 0.97$ and $T_{\rm ad}$ is the transfer function for adiabatic perturbations, an analytical approximation for which is given in~\cite{Bardeen:1985tr}. The variance of fluctuations is related to the power spectrum by
\begin{equation} \label{eq:s_inf}
	\sigma^2_{\rm inf}(M,z) = D^2(z)\int \frac{dk}{2 \pi^2} \, k^2 P_{\rm inf}(k) W^2 (kR),
\end{equation}
where the spatial scale is related to the mass scale by $M =  4 \pi R^3 \rho_{M} / 3$, the function $D(z) = g(z) / [g(0) (1 + z)]$ describes the growth of fluctuations in the linear regime~\cite{Mo:2002ft}, and $W(kR) = 3(\sin kR - kR \cos kR) / (kR)^3$ is the window function in Fourier space. We now define the so-called $N\sigma$ halos as solutions of the equation $N\sigma_{\rm inf}(M,z) = \delta_c$. The redshift dependence of the mass of $N\sigma$ halos is shown by the dotted lines in Fig.~\ref{gr:Mch} ($N = 1$ corresponds to halos of characteristic mass). One can see that at $z \gtrsim 20$ the structures induced by Poisson noise can be more massive than rare ($\lesssim 5\%$) $2\sigma$ halos from inflationary perturbations. However, the mass of typical inflationary halos grows very rapidly with decreasing redshift. Therefore, in the present epoch, the large-scale structure of the Universe is determined by density perturbations of inflationary origin, and the Poisson noise of PBHs plays no role, at least at the level of isolated halos rather than substructures. The moment of this transition can be estimated from the intersections of the solid lines with the dotted $1\sigma$ halo line for inflationary perturbations. One can see that at redshifts $z \gtrsim 10$ structures still form owing to Poisson noise in the spatial distribution of PBHs. We use this to estimate the $f_\pbh$ for which this mechanism is at all viable. As a rough criterion, we require that a halo contain at least $3$ PBHs; then, using expression~\eqref{eq:M_char} evaluated at $z \sim 10$, we conclude that the PBH fraction must be $f_\pbh > 0.006$.

\begin{figure}
	\begin{center}
\includegraphics[angle=0,width=0.49\textwidth]{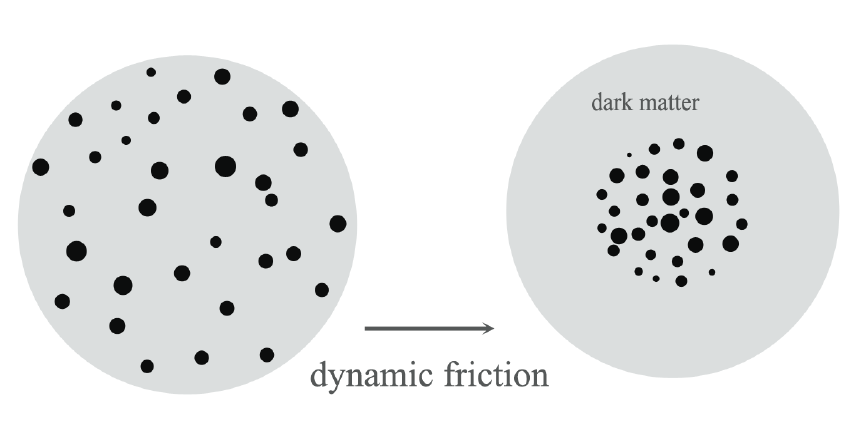}
	\end{center}
\caption{Formation of a PBH subsystem at the center of a DM halo due to dynamical friction.}
	\label{grdfr1} 
\end{figure}

\medskip
{\bf Evolution of PBH clusters as self-gravitating systems} 
\medskip

Let us now discuss the internal dynamics of early structures. Owing to dynamical friction against DM particles, PBHs will lose energy and sink toward the halo center. This process is shown schematically in Fig.~\ref{grdfr1}. Moreover, this process is most efficient in the dense central region of the halo. The dynamical friction timescale is~\cite{Merritt2013}
\begin{align}\label{eq:t_df}
    t_{\rm df} &= \frac{3}{8} \sqrt{\frac{2}{\pi}} \frac{v^3}{G^2 M_\pbh \rho_c \ln{\Lambda}} \nonumber \\
    & \simeq 8.5 \,  f_{\pbh} k^{-1} \left ( \frac{1+z}{40} \right)^{-7/2}10^9~\mbox{yr}.
\end{align}
In this expression, $\rho_c = k \overline{\rho}_h$ is the central density of the halo, where $k \sim 10-100$ is a numerical constant that we determined from the analysis of halo density profiles in numerical simulations of structure formation with PBHs~\cite{Inman:2019wvr,Delos:2024poq}; $v$ is the one-dimensional velocity dispersion, which can be estimated by its characteristic value~\eqref{eq:vch}; $\ln{\Lambda} \sim 10$ is the Coulomb logarithm; and for the halo mass we have used its typical value from Eq.~\eqref{eq:M_char}. Thus, within a time much shorter than the present age of the Universe, the PBH density in the central region of early structures greatly exceeds the DM density: a PBH cluster forms. These processes and the evolution of the resulting clusters were considered in detail in~\cite{Stasenko:2023zmf,Stasenko:2024pzd}. Note also that we consider here a model in which PBHs are a subdominant DM component. In this case, each PBH should be surrounded by a DM halo that forms around it via secondary infall (see Section~\ref{krupnosubsec}). Therefore, early structures will contain not just PBHs, but compact halos several times more massive, with PBHs at their centers. This, in turn, should effectively reduce the dynamical friction time~\eqref{eq:t_df}, at least for the very first structures.

This picture is valid only for an isolated halo. In reality, the halo should be absorbed into larger halos (or absorb smaller ones, increasing its mass) in the course of hierarchical structure formation. The halo may be absorbed and gradually disrupted within a larger structure before the PBHs have time to sink to its center and form a cluster. Therefore, these two competing processes must be compared. As a characteristic structure formation time, one can take the halo mass-doubling time (for the justification and theoretical details of hierarchical halo mergers, see~\cite{LacCol93}), which amounts to several Hubble times $t_H = H^{-1}$. Then, from~\eqref{eq:t_df}, we obtain the ratio
\begin{equation}\label{eq:tdftH}
    \frac{t_{\rm df}}{t_H} \simeq 80\, f_\pbh k^{-1} \left ( \frac{1+z}{40} \right)^{-2}.
\end{equation}
One can see that at sufficiently high $z$, for $f_\pbh \lesssim 0.1$ and $k \sim 10$, PBH clusters can indeed form in the central regions of halos, in the sense that this process is faster than structure formation. Note also that the dynamical friction time~\eqref{eq:t_df} does not depend on the PBH mass; therefore, this clustering effect inevitably arises when PBHs constitute a noticeable fraction $f_\pbh \gtrsim 0.01$ of all DM. Although such dense relic structures would be very hard to detect today, they may have left imprints in the younger Universe. In particular, PBH clustering will be discussed below in the context of gravitational-wave observations.

Let us now discuss the scenario in which PBHs are the dominant DM component. We first consider the case of a broad mass distribution. Current constraints allow dark matter to consist of PBHs with masses $\sim 10^{17}-10^{22}$~g (the so-called asteroid-mass window). In this case, the situation is no different from that considered above, provided there is a sufficient fraction, $f_\pbh \simeq 0.01 - 0.1$, of massive PBHs whose Poisson noise leads to the formation of early halos. The situation is different if PBHs have a monochromatic mass spectrum. In this case, the halo evolves due to two-body interactions between PBHs. This is actually somewhat different from the situation considered above. The characteristic timescale of this process is the relaxation time~\cite{BinTre2008}
\begin{equation}
    t_r = \frac{0.34 \, v^3}{G^2 M_\pbh \rho_c \ln{\Lambda}},
\end{equation}
where the Coulomb logarithm is $\ln{\Lambda} \sim \ln{N_\pbh}$. In the course of evolution, PBHs diffuse in velocity space, and some of them acquire velocities high enough to leave the central core. This leads to contraction of the core and an increase in its density. Asymptotically, this process leads to complete ``evaporation'' of the core and its contraction to zero radius, with the density tending to infinity; this phenomenon is known as core collapse, or the gravothermal catastrophe~\cite{1968MNRAS.138..495L,1980ApJ...242..765C}. The name reflects the fact that self-gravitating systems have negative heat capacity, so that when heat flows out of the core (fast particles escape), the core heats up even more. In real astrophysical systems, however, such as globular star clusters, core collapse is eventually halted by the formation of hard binaries~\cite{HegHut06}. What matters here is that the core-collapse time is $t_{\rm cc}\sim 10^2 \ t_r$~\cite{Quinlan:1996bw,BinTre2008}, which is much longer than the Hubble time (see the discussion before~\eqref{eq:tdftH}). Consequently, in this model, Poisson halos do not have time to undergo core collapse, and compact PBH clusters do not form in them. Instead, PBHs keep clustering hierarchically, so the internal dynamics of individual halos is unimportant compared with the build-up of ever larger structures.


   \subsection{Clusters of primordial black holes in the domain wall model}
   \label{clustdomensubs}

Above, we have already considered a model in which PBHs form from the collapse of domain walls created by a scalar field. We discussed a collapse that leads to the formation of a single main PBH. A more detailed analysis shows that PBHs should form not individually but within hierarchical clusters~\cite{KhlRubSa02,2005APh....23..265K,Belotsky:2018wph} in this model. In such clusters, the most massive PBH is located at the center and is surrounded by less massive ones.

\begin{figure}[b]
	\begin{center}
\includegraphics[angle=0,width=0.49\textwidth]{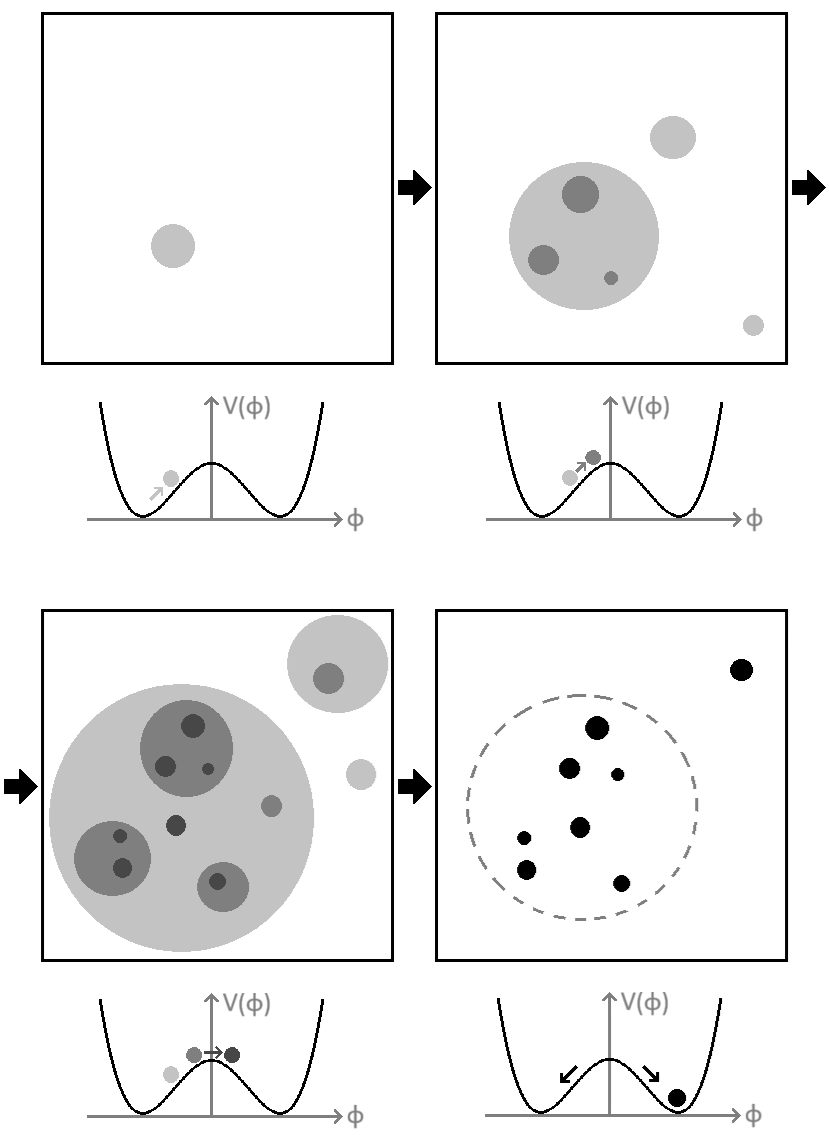}
	\end{center}
\caption{Initial clustering of PBHs arising in the domain wall mechanism due to the spatial correlation of quantum field fluctuations during inflation. Fluctuations at successive moments during inflation (frames 1--3) push the scalar field over the potential maximum (frame 3). After the end of inflation, the field rolls down to the potential minima, and a cluster of domain walls forms (frame 4), which then collapse into PBHs.}
	\label{cluster_walls} 
\end{figure}

The clustering mechanism is illustrated in Fig.~\ref{cluster_walls} and is based on the following consideration. During inflation, the scalar field $\phi$ undergoes repeated quantum fluctuations in the exponentially expanding Universe. If, in some spatial region, the value of the field $\phi$ has come close to the critical value $\phi_\text{cr}$, then at the next moment (when this region has been expanded by inflation) the probability of a critical fluctuation forming a domain wall is increased in this region. The fluctuations turn out to be spatially correlated, as are the resulting domain walls (from which PBHs form). Therefore, the probability of forming new PBHs around an already existing one increases. As a result, PBHs form clusters with a fractal-like structure, in which each PBH is surrounded by a population of smaller PBHs. Accordingly, each smaller PBH is surrounded by a population of even smaller PBHs, and so on. Further evolution of the cluster leads to mergers of black holes into more massive ones.

To obtain the spatial and mass structure of a cluster, we choose a ``seed'' (parent) PBH of mass $M_0$ and consider the cluster of smaller PBHs around it. We fix the size $r_\text{cl}$ of the PBH cluster at the end of inflation. Then, according to Eq.~\eqref{rinfott}, the e-fold at which the region containing the cluster was formed has the number $N_\text{cl}=N(r_\text{cl})$. We also need the e-fold at which the seed PBH was formed. It can be determined from Eq.~\eqref{mass_wall}: $N_0=N(r(M_0))$. Since the ``seed'' PBH is, by definition, the only one in the cluster, only one critical fluctuation must occur during the time interval $N_\text{0}-N_\text{cl}$. Using~\eqref{crit}, we write this condition as
\begin{equation}
n\big(N_\text{0}-N_\text{cl}, \phi_\text{cl} \big)=1\,.
\end{equation}
This equation relates the size $r_\text{cl}$ of the chosen region to the initial field value $\phi_\text{cl}$ for this region in such a way that only one PBH (the seed) of mass $M_0$ forms in this region. Solving this equation for the initial field value, we obtain the dependence $\phi_\text{cl}(M_\text{0},r_\text{cl})$. It is now possible to obtain the spatial and mass distribution of PBHs within a given region of space. It is described by the expression obtained from Eq.~\eqref{crit} by substituting the initial field value $\phi_0(M_\text{0},r_\text{cl})$ and the time $N-N_\text{cl}$:
\begin{equation}
n\big(N(r(m))-N(r_\text{cl}), \phi_\text{cl}(M_\text{0},r_\text{cl})\big)=n(M_\text{0},r_\text{cl},m)\,.
\end{equation}
The function on the right-hand side of this equation gives the number of PBHs with masses above $m$ within a region of radius $r_\text{cl}$ around the ``seed'' PBH of mass $M_0$. This distribution is cumulative; the mass spectrum and the spatial distribution of PBHs inside the cluster can be obtained by differentiating it with respect to $m$ and $r_\text{cl}$. The physical properties of such clusters are discussed in the review~\cite{Belotsky:2018wph}.

\subsection{Early formation of quasars around clusters of primordial black holes}

As described above, in the model with collapsing domain walls, PBHs form not individually but within clusters. The presence of a cluster makes the DM halo around the central PBH even more compact than in the case of an isolated seed PBH. The formation of such compact DM halos around PBH clusters was considered in a series of papers~\cite{Dokuchaev:2004kr,Dokuchaev:2008hz,2009AstL...35..143D,2010AstL...36..773D}. It is important to note that this model can explain the appearance of early galaxies implied by JWST observations~\cite{Dokuchaev:2008hz}, as well as the observed early quasar activity~\cite{Dokuchaev:2004kr}. In addition, this model predicts gravitational-wave bursts from collisions of PBHs in the cluster, detectable by future space-based gravitational-wave detectors (LISA, etc.)~\cite{2009AstL...35..143D}.

\begin{figure}[htbp]
\centering
\includegraphics[width=0.9\linewidth]{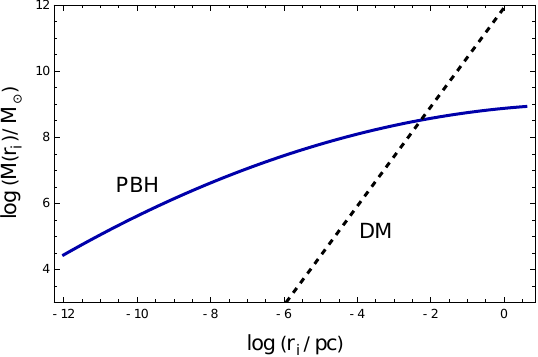}
\caption{Initial mass profile $M_{\rm h}(r_i)$ of a PBH cluster (solid curve) and DM mass profile $M_{\rm DM}(r_i)$ (dashed curve).  } 
\label{massprof1} 
\end{figure}

Let us consider sufficiently dense clusters that evolve early and in which quasars can ignite. The most massive BH at the center of the cluster forms almost immediately after its gravitational radius crosses the cosmological horizon; then, the central dense region of the cluster decouples from the cosmological expansion, contracts, and virializes. Quasar activity is initiated by gas accretion onto the central BH. As an example, consider a model in which a fraction $\sim10^{-3}$ of the mass of the Universe is in PBH clusters. An example of the initial mass profile $M_{\rm h}(r_i)$ of a PBH cluster in the domain wall model is shown in Fig.~\ref{massprof1}. For comparison, the same figure shows the mass of DM, $M_{\rm DM}(r_i)$, enclosed within spherical shells of the same radii. The radii $r_i$ are given at different times $t_i$ when the spherical shells cross the cosmological horizon (this explains the deviation from the law $M_{\rm DM}\propto r^3$). Only the central mass of $4.3\times10^7M_{\odot}$ ends up inside its gravitational radius; thus, the full range of masses and radii shown in Fig.~\ref{massprof1} is not actually realized.

We now turn to the gravitational dynamics of the cluster of PBHs and DM, starting from the radiation-dominated stage. Consider a system of radius $r<ct$ consisting of PBHs with total mass $M_{\rm h}$ inside a sphere of radius $r$, radiation (a mixture of all relativistic particles) with density $\rho_r$, and DM with density $\rho_{\rm DM}$, in the presence of a cosmological constant with the corresponding density $\rho_{\Lambda}$. We introduce the parametrization $r=ab$, in which the evolution equation for the shell with parameter $b$ takes the form
\begin{equation}
\frac{d^2b}{dz^2}+\frac{db}{dz}S(z)+
\left(\frac{1+\delta_h}{b^2}-b\right)\frac{\Omega_{m,0}(1+z)}{2E^2(z)}=0,
\label{d2bdz1}
\end{equation}
where we have introduced the function $S(z)=E^{-1}(z)dE(z)/dz-1/(1+z)$ and $\delta_h=M_{\rm h}/M_{\rm DM}$. In the limiting case $\varepsilon_{\Lambda}=0$, Eq.~(\ref{d2bdz1}) is equivalent to the corresponding equation in~\cite{Kolb:1994fi}. We solve Eq.~(\ref{d2bdz1}) starting from the redshift $z_i$ at which the shell under consideration crosses the horizon, $r\sim ct$. The initial conditions are given in Fig.~\ref{massprof1}.

The moment when the shell stops expanding, $\dot r=0$, at some radius $r_s$ corresponds to the equation $db/dz=b/(1+z)$. As usual, we assume that after contracting from $r_s$ to $r_c=r_s/2$, the shell virializes and stays at radius $r_c$. Therefore, the mean density $\rho$ of the virialized object is 8 times its density at maximum expansion, $\rho=8\rho_{m,0}(1+z_s)^3b_s^{-3}$, and its virial radius is
$r_c=[3M_{\rm DM}/(4\pi\rho)]^{1/3}$. Inner shells stop expanding earlier than outer ones. A BH of mass $M_c=4.3\times10^7M_{\odot}$ forms at the center of the cluster at time $t_i$. The subsequent shells with $\delta_h>1$ (and $M_{\rm DM}<M_{\rm h}$) also decouple from the cosmological expansion during the radiation-dominated stage. These shells are accreted by the central BH as they contract owing to two-body relaxation. The boundary value $\delta_h=1$ corresponds to the mass $M_{\rm h}=M_{\rm DM}=3.3\times10^8M_{\odot}$.

\begin{figure}[t]
\includegraphics[angle=0,width=0.9\linewidth]{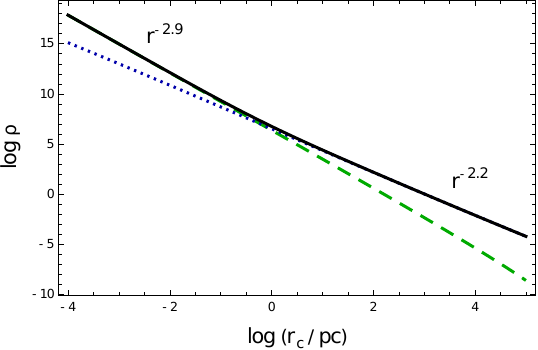}
\caption{\label{massprof2} Final density profile (\ref{dprofeq}) of the induced halo ($\rho$ in units of $M_{\odot}$~pc$^{-3}$) as a function of the distance $r_c$ from the cluster center for DM (dotted line), for PBHs (dashed line), and for the total density (solid line). The asymptotic power laws are also shown. }
\end{figure}

For early-forming PBHs, a process similar to secondary infall takes place. As a result, PBHs or their clusters end up surrounded by extended DM halos. Their density profile will not exactly follow the secondary infall profile $\rho\propto r^{-9/4}$, because the central mass is not compact. In the course of galaxy formation, after virialization, the DM distribution has the form
\begin{equation}
\rho_{\rm DM}(r)=\frac{1}{4\pi r^2}\frac{dM_{\rm DM}(r)}{dr},
\label{dprofeq}
\end{equation}
where the function $M_{\rm DM}(r_c)$ is determined by solving Eq.~(\ref{d2bdz1}). The BH density profile, as well as the total density profile, can be determined in the same way as the DM profile. The results of this calculation are shown in Fig.~\ref{massprof2}. As can be seen from this figure, the total density at $r=8$~kpc is $0.7$~GeV~cm$^{-3}$. Thus, in some of their characteristics, the objects under consideration partly resemble our Galaxy, but they are denser in the central part and contain supermassive BHs.

A DM halo stops growing when the gravitational attraction of neighboring density perturbations of inflationary origin becomes strong. Comparing the amplitudes of inflationary perturbations and of perturbations from BH clusters, we obtain the mass of the DM halo at the end of its growth, $M_{\rm
DM}=1.8\times10^{12}M_{\odot}$. Halo growth stops completely at redshift $z=1.64$, although the quasar activity of the BH began much earlier, at high $z$. Quasar activity in a galaxy is probably triggered by tidal gas flows onto the central BH during mergers of the galaxy with neighboring objects. The resulting induced halos look like large spheroidal galaxies or elliptical galaxies with small ellipticity. They have a central DM density spike, shown in Fig.~\ref{massprof2}, and central supermassive BHs. Within a radius of $r=10$~pc from the center, the total mass of PBHs (including the mass of the central BH) is $2.9\times10^8M_{\odot}$. Within the same radius, the DM mass is $3.4\times10^8M_{\odot}$.

\subsection{Biasing effect for primordial black holes}

Besides the formation of dense PBH clusters, it is also possible that either very ``loose'' clusters or even non-virialized regions with an enhanced PBH concentration form. One case in which the enhancement of the PBH density is most likely not very significant is the biasing effect. In general terms, the biasing effect means that large-scale density perturbations influence the formation of objects on small scales and thereby modulate their distribution. The biasing effect is known, for example, in the theory of galaxy formation. The possibility of a similar biasing effect in PBH formation from Gaussian perturbations within GR was demonstrated in~\cite{Chi06}. However, more accurate calculations later showed that this effect is extremely small~\cite{DesRio18}. This is related to subtleties in summing long-wavelength perturbation modes~\cite{YouByrSas14}. At the same time, weak PBH clustering may indeed occur in the case of non-Gaussian perturbations~\cite{2015PhRvD..91h3521Y,2015PhRvD..91l3534T, SuyYok19}.

It was shown in~\cite{Berezin:2020tbk} that the biasing effect operates efficiently in the Brans--Dicke theory in the presence of independent fluctuations in the scalar field distribution, which may lead to the formation of PBH clusters. Regions with an enhanced PBH number density are in fact entropy perturbations (isocurvature perturbations). In~\cite{Starobinsky:1994mh}, curvature perturbations and perturbations in the Brans--Dicke scalar field, distinct from the inflaton, were calculated. The calculations of~\cite{Starobinsky:1994mh} show that, in the Brans--Dicke theory, the contribution of entropy perturbations generated during inflation to the total perturbation amplitude is negligible. However, it was shown in~\cite{Berezin:2020tbk} that this contribution can be enhanced by about three orders of magnitude owing to PBH formation. Such a contribution of entropy perturbations will not contradict observations of the CMB anisotropy if the perturbation spectrum of the Brans--Dicke scalar field is enhanced only on small, subgalactic scales. On large scales, probed by the CMB, deviations from the standard picture may then be very weak, below the level accessible to current observations.

\section{Primordial black holes and gravitational-wave sources}

The direct detection of gravitational waves from a binary BH merger in 2015~\cite{LIGOScientific:2016aoc} gave a powerful impetus to the rapid development of PBH theory for the following decade. The idea that some of the detected gravitational-wave events have a primordial rather than stellar origin immediately became extremely popular~\cite{Bird:2016dcv,Sasaki:2016jop,Clesse:2016vqa,Ali-Haimoud:2017rtz,Blietal16,Raidal:2018bbj}. PBH binaries can form both in the very early Universe, during the radiation-dominated stage, owing to the decoupling of PBH pairs from the Hubble flow~\cite{1997ApJ...487L.139N,Ioka:1998nz,Sasaki:2016jop}, and at a later epoch, in subgalactic structures, as a result of dynamical interactions: two-body scatterings of PBHs accompanied by gravitational-wave emission~\cite{Bird:2016dcv,Clesse:2016vqa} and three-body interactions~\cite{Franciolini:2022ewd}. The PBH merger rate depends nontrivially on their fraction $f_\pbh$ in DM, which in turn determines the relative contribution of each binary formation channel. Below, we review the main mechanisms of PBH binary formation and the evolution of the binaries in the course of clustering.

\subsection{Primordial black hole binaries and LIGO/Virgo/KAGRA observations}
\label{sec:61}

We first consider the simplest case in which there is no initial PBH clustering, i.e., PBHs are born at random points in space (a Poisson process). This holds to good accuracy, for example, in the model of PBH formation from Gaussian curvature perturbations. Since PBHs are initially distributed randomly, some of them could by chance have formed quite close to other PBHs. As the Universe expands, the PBHs in such close pairs first move apart, but from some moment on they can start approaching each other and form a gravitationally bound binary. If the Universe contained no objects other than the two PBHs under consideration, these PBHs would move exactly toward each other under their mutual gravitational attraction. In this idealized situation, they would eventually collide and merge into a single BH. In reality, however, there is a third, nearest PBH next to the pair, and other PBHs farther away. In addition, inhomogeneities of matter (DM, baryons, and radiation) should be present nearby. All these surrounding objects gravitationally perturb the trajectories of the PBHs in the pair while they move apart and first approach each other. Under the action of tidal forces, the PBHs deviate from motion exactly toward each other, and their orbit eventually becomes not strictly radial but elliptical, with eccentricity $e<1$. As a result, the two PBHs merge not immediately but after a long time, during which their orbit shrinks owing to gravitational-wave emission or other processes. Below, we explain how to calculate the orbital distribution of PBH pairs and the resulting rate of their mergers (and hence of gravitational-wave bursts), taking into account the effects described here qualitatively.
  
Two PBHs randomly distributed in space
may form quite close to each other. Deep in the radiation-dominated stage, the local density of these two PBHs decreases owing to cosmological expansion as $\rho \propto s^{-3}$, while the radiation density decreases as $\rho_r \propto s^{-4}$ \PK{(in this section, $s=1/(1+z)$ is the dimensionless scale factor)}. At some moment, the mean \PK{energy} density of the two PBHs exceeds the \PK{energy} density of radiation; a gravitationally bound pair then forms \PK{and decouples from the Hubble flow}. Let us determine the time $t_\dec$ when this happens; here and below, we follow the formalism of~\cite{Sasaki:2016jop}. We introduce the mean separation between PBHs at matter--radiation \PK{energy} density equality, $t_\eq$:
\begin{equation}\label{eq:x_av}
\overline{x} = \left ( \frac{M_\pbh}{\rho_\pbh(t_\eq)}  \right)^{1/3} = \frac{s_\eq}{ f_\pbh^{1/3}} \left ( \frac{M_\pbh}{\Omega_\dm \rho_{\rm crit}}  \right)^{1/3}
\end{equation}
\PK{(in the second equality, $\rho_\mathrm{crit}=3H_0^2/8\pi G$ is the present critical density of the Universe).}
Consider two PBHs separated by a distance \PK{$R=(s/s_\mathrm{eq})x$, which equals $x$ at $s_\eq$}. At the moment determined by the condition \PK{that the energy density inside a sphere of radius $R$ equals the background energy density},
\begin{equation}\label{eq:s_dec1}
\frac{M_\pbh}{x^3} \left ( \frac{s_\eq}{s_\dec} \right)^3 \simeq \rho_r(t_\dec) = \rho_{r, \eq} \left ( \frac{s_\eq}{s_\dec} \right)^4,
\end{equation}
these two PBHs decouple from the cosmological expansion of the Universe and form a gravitationally bound binary. Using expression~\eqref{eq:x_av}, we obtain the relation between the scale factor $s_\dec$ and the separation $x$ between the PBHs:
\begin{equation}
    s_\dec \simeq \frac{s_\eq}{f_\pbh} \left ( \frac{x}{\overline{x}} \right)^3. 
\end{equation}
Only those PBH pairs whose separation satisfies $x < \overline{x} f_\pbh^{1/3}$ form binaries.

After decoupling from the expansion of the Universe, the two PBHs avoid a head-on collision and move on elliptical orbits. The angular momentum of the binary is generated by tidal forces from the surrounding PBHs and inflationary perturbations. To simplify the analysis, we consider only the influence of the nearest, third PBH. We denote by $y$ the distance between the third PBH and the binary at $s_\eq$, with the condition $x < y < \overline{x}$. Since PBHs are born at random points in space, the probability distribution over $x$ and $y$ is Poissonian, which for our purposes can be taken to be uniform to good accuracy:
\begin{equation} \label{eq:dp_dxdy}
    dP = \frac{9}{\overline{x}^6} x^2 y^2 dx dy.
\end{equation}
The generated angular momentum can be estimated as $L \sim RF_t\tau$, where $F_t \simeq GM_\pbh^2R/d^3$ is the tidal force from the third PBH at the time of binary formation, where $d = y s_\dec/s_\eq$ is the distance to the third PBH at that time, and $\tau \sim 1 /\sqrt{G\rho_r(t_\dec)}$ is the duration of the tidal force action, which is of the order of the Hubble time $t_{\rm H}$. Thus, the angular momentum is estimated as
\begin{equation}
    L  \sim M_\pbh \left ( \frac{R}{d} \right)^3 \sqrt{GM_\pbh R},
\end{equation}
where we have used relation~\eqref{eq:s_dec1} for the radiation density at $t_\dec$. For the subsequent analysis, we need the orbital parameters of the forming binaries. The semimajor axis is determined by the separation between the two PBHs at decoupling, $a\sim R$. The semiminor axis is $b = L/\sqrt{M_\pbh |E|}$, where $|E| \sim GM^2_\pbh/R$ is the binding energy of the binary. Therefore, up to factors of order unity, the orbital parameters are
\begin{equation}
	a = \frac{x}{f_\pbh} \left( \frac{x}{\overline{x}} \right)^3, \quad b = a \left ( \frac{x}{y} \right)^3.
\end{equation}
It is convenient to introduce the dimensionless \PK{orbital} angular momentum $j = b/a$; then, using~\eqref{eq:dp_dxdy}, we obtain the distribution over the parameters of PBH binaries:
\begin{equation} \label{eq:dp_dadj}
dP = \frac{3}{4} \left ( \frac{f_\pbh}{\overline{x}} \right)^{3/2} \frac{\sqrt{a}}{j^2} \, da dj,
\end{equation}
where the variables range over $0<a<\overline{x}f_\pbh^{1/3}$ and $(a f_\pbh/\overline{x})^{3/4} < j < 1$, and the angular momentum is related to the eccentricity of the binary by $j = \sqrt{1 -e^2}$.

After formation, the PBH binary gradually \PK{loses energy} through gravitational-wave emission and merges within the time~\cite{Peters:1964zz}\footnote{This estimate neglects the possible growth of PBH masses due to accretion.}
\begin{equation} \label{eq:tmer}
t_{\rm mer}=\frac{3 \, c^5 a^4 j^7}{170 \, G^3M_\pbh^3}.
\end{equation}
To obtain the explicit time dependence of the merger rate, it is simplest to change variables, $a = a(j,t)$, using the merger time~\eqref{eq:tmer}. The differential probability of PBH binary mergers per unit time is then
\begin{align}\label{eq:dPdt1}
\frac{dP}{dt} &= \frac{3}{16} \frac{f^2_\pbh}{T} \left ( \frac{t}{T} \right)^{-5/8} \int \limits_{j_{\rm min}}^{1} j^{-37/8}  \, dj \nonumber \\
&\approx \frac{3}{58} \frac{f^{16/37}_\pbh}{T} \left ( \frac{t}{T} \right)^{-34/37},
\end{align}
where the main contribution to the integral comes from the lower limit $j_{\rm min} = (t/T)^{3/37} f^{16/37}_\pbh \ll 1$, and
\begin{equation}
T = \frac{3 \, c^5 \overline{x}^4 f_\pbh^{4/3}}{170 \, G^3 M_\pbh^3} \sim 10^{34} \left ( \frac{M_\pbh}{M_{\odot}} \right)^{-5/3} \, \text{yr}
\end{equation}
is the merger time of a PBH binary on a circular orbit with the largest possible semimajor axis for $f_\pbh = 1$. The merger rate of PBH binaries is given by
\begin{align}\label{eq:mr1}
&\mathcal{R} = \frac{f_\pbh\Omega_\dm}{M_\pbh} \frac{3 H_0^2}{8 \pi G}  \frac{dP}{dt} S_f  \\
&\approx \frac{1.3 \times 10^6}{\text{Gpc}^{3} \, \text{yr}} f_\pbh^{53/37} S_f  \left ( \frac{M_\pbh}{M_{\odot}} \right)^{-32/37} \left ( \frac{t}{t_U} \right)^{-34/37}, \nonumber
\end{align}
where $t_U$ is the present age of the Universe, and the factor $S_f \approx 0.6 \big(1 + 2.3\,\sigma^2_{M, \eq}/f^2_\pbh \big)^{-21/74}$ corrects the leading contribution and accounts for the influence of inflationary perturbations and of the other PBHs on the binary orbit~\cite{Eroshenko:2016hmn,Ali-Haimoud:2017rtz,Raidal:2018bbj}.

Let us now estimate the characteristic orbital parameters of binaries merging in the present epoch. Consider binaries with a fixed semimajor axis $a$. As can be seen from distribution~\eqref{eq:dp_dadj}, the angular momenta must be small, so a reasonable estimate for them is $j_{\rm min} = (a f_\pbh/\overline{x})^{3/4}$. Now expressing the angular momentum through the merger time~\eqref{eq:tmer}, we obtain the characteristic value of the semimajor axis:
\begin{align}
    a_{\rm ch} &\simeq \left ( \frac{170\, G^3M_\pbh^{19/4} t}{3\, c^5 f_\pbh^7  \rho^{7/4}_\dm(t_\eq)} \right)^{4/37} \nonumber \\
    &\approx 43 \, f_\pbh^{-28/37} \left ( \frac{M_\pbh}{M_{\odot}} \right)^{19/37} \left ( \frac{t}{t_U} \right)^{4/37} \, \text{au}.
\end{align}
The characteristic angular momentum, in turn, can be estimated as $j_{\rm ch} \simeq (a_{\rm ch} f_\pbh/\overline{x})^{3/4}$:
\begin{equation}\label{eq:j_ch}
    j_{\rm ch} \approx 0.011 \, f_\pbh^{16/37} \left (\frac{M_\pbh}{M_{\odot}} \right)^{5/37} \left ( \frac{t}{t_U} \right)^{3/37}.
\end{equation}
These estimates show that PBH binaries have orbital angular momenta $j \ll 1$. This fact turns out to be important when perturbations of binaries by the surrounding PBHs in clusters are taken into account, as discussed below.

PBHs that may be responsible for the LIGO/Virgo/KAGRA events constitute only a small fraction of DM. The rest of the DM is distributed around the PBHs as a uniform background and as overdensities of various kinds. As discussed in Section~\ref{sec:DM_around_PBH}, density spikes and induced halos form around PBHs. Before the binary forms, DM accretes onto each PBH separately; after it forms, DM accretes onto the binary as a whole. The interaction of the PBHs with DM particles in such a halo affects the orbital evolution of the binary and introduces a correction to expression~\eqref{eq:mr1}, which, however, is difficult to include in analytical calculations. Various aspects of these processes were considered in~\cite{Hayasaki:2009ug,Eroshenko:2016hmn,Kavanagh:2018ggo,Pilipenko:2022emp,Jangra:2023mqp,Stasenko:2024dui}.

\medskip
{\bf LIGO/Virgo/KAGRA observations}
\medskip

Observations of gravitational-wave bursts by the \PK{ground-based laser interferometers} LIGO/Virgo/KAGRA have revealed two interesting features of the population of merging objects: relatively large masses and small spins of the individual objects~\cite{FerPro19}. In many cases, the measured masses significantly exceeded the average masses of compact remnants of stellar evolution. This feature is easily explained by a selection effect: mergers of the most massive objects are more readily observed. This effect had been predicted even before the LIGO/Virgo detections~\cite{TutYun93,Lipetal97,Grietal01}, including in the context of the formation of massive stars in low-metallicity regions~\cite{chris1}.

Until recently, the record mass of a BH formed in a merger was $\sim140M_\odot$. The international LIGO/Virgo/KAGRA Collaboration reported in~\cite{LIGOScientific:2025rsn} the observation of the event GW231123, in which the final BH mass is $190-265~M_\odot$, and the masses of the merging BHs are estimated as $137^{+22}_{-17}~M_\odot$ and $103^{+20}_{-52}M_\odot$. The merger occurred at a distance of 0.7--4.1~Gpc from Earth (redshift $z=0.39^{+0.27}_{-0.24}$); five cycles of the gravitational wave were observed, and the signal-to-noise ratio of the detection by the two LIGO detectors was $\sim22.5$. An unusual property of GW231123 is that one or both of the merging BHs had masses falling into the so-called mass gap, $60-130M_\odot$. As follows from calculations, BHs within this gap cannot form in supernova explosions because of electron--positron pair production in the hot stellar core \PK{already at the stage of carbon burning into oxygen}, and the associated oscillations of the stellar radius and mass loss~\cite{Wooetal07}. However, the boundaries of the gap are known imprecisely owing to theoretical uncertainties~\cite{WooHeg21}. A BH mass within the forbidden gap may be explained by more complex processes of stellar evolution and by mergers accompanied by mass growth. Either stars could have merged before collapsing into BHs, or BHs could have merged in dense star clusters after their formation. The resulting massive BHs then merged, producing the observed event GW231123. Both scenarios have their difficulties; for example, when two BHs merge, the resulting massive BH usually receives a large recoil velocity and escapes from the cluster or the host galaxy unless the velocity dispersion in the cluster is very high ($\geq10^3-10^4$~km/s). At the same time, the large dimensionless spins of the BHs before the merger ($\sim0.9$ and $\sim0.8$) are well explained by this hierarchical mechanism of BH formation in previous mergers. A model with merging PBHs could naturally explain the observed features, since PBHs could have had masses inside the forbidden mass gap from the outset~\cite{DeLuca:2025fln}. A primordial origin of the BHs observed by the LIGO/Virgo/KAGRA gravitational-wave interferometers may be indicated not only by their large masses but also by the shape of their mass function, which is close to log-normal~\cite{Blietal16,Doletal20}.

It cannot be ruled out, however, that some unusual gravitational-wave events with extreme parameters are related not to BH mergers in pairs but to processes of an entirely different nature. In~\cite{Boybeyi:2026lnq}, an attempt was made to explain the events GW190521 and GW231123 by the passage of a field configuration in the form of a cosmic domain wall through the detector. Such walls could constitute a fraction of the DM in the Universe. Although the BH model was found to be preferable in terms of the signal shape, the domain wall option cannot be completely ruled out either; moreover, the domain wall parameters obtained for these two independent events agree well with each other.

The BH spin measured in some gravitational-wave events turned out to be much smaller than the spin expected from calculations of stellar core collapse~\cite{FerPro19,GarSilMor21,Beletal20}. However, small BH spins can also be explained by processes of ordinary stellar evolution~\cite{Beletal20,PosKurMit19}. The most popular PBH formation mechanisms produce black holes with small spins, whose masses can be distributed over a wide range. Observations of gravitational-wave bursts make it possible to determine the ``effective spin'' of a pair of BHs with masses $m_1$ and $m_2$, defined as the mass-weighted sum of the spin projections,
\begin{equation}
\chi_{\rm eff}=\frac{(m_1\vec a_1+m_2\vec a_2)\,\vec n}{m_1+m_2},
\label{spin}
\end{equation}
where $\vec a_1$ and $\vec a_2$ are the dimensionless spin parameters of the merging BHs, $\vec a=c\vec L/(Gm^2)$, and $\vec n$ is the unit vector along the orbital angular momentum. The spins of the two BHs individually cannot be determined from observations of gravitational-wave bursts.

When the first gravitational-wave events were observed, it was often noted that for BH pairs of astrophysical origin the case $\chi_{\rm eff}>0$ should be preferred, since the two stars in a pair have a common origin, and their angular momenta should therefore be approximately aligned with the orbital angular momentum. In contrast, for PBHs, it was assumed that $\chi_{\rm eff}>0$ and $\chi_{\rm eff}<0$ should be equally probable, and the PBH spin should be close to zero (as in the historically first model of their formation). It later turned out that the situation is more complex and diverse, both from the point of view of stellar evolution theory and from the point of view of the possibilities for PBH formation and pairing.

\PK{The data of the latest catalog of gravitational-wave events, GWTC-5, which contains about 300 binary BH mergers, clearly reveal distinct populations characterized by masses of 10--20~$M_\odot$ and $\sim 45 M_\odot$~\cite{GWTC5}. The effective spin is $\chi_{\rm eff}<0.25$ for $\sim80\%$ of the events, although there is a subpopulation of events with large spins reaching $\sim 0.7$. The latter is most likely formed by dynamical BH mergers in dense clusters.}

The small value of $\mu_\chi$ in GW merger events can be explained in stellar evolution models~\cite{PosKur19,PosKurMit19}, \PK{especially in the model of BH formation in the collapse of a massive star, in which the BH mass is directly related to the mass of the collapsing CO core, $M_{BH}\approx 0.9 M_{CO}$. Note that this relation is independently supported by the analysis of the mass distributions of Galactic BHs in close binaries and of the masses of Wolf--Rayet stars, which are the progenitors of massive stars collapsing into BHs~\cite{Posetal26}.} It was also shown in~\cite{PosKurMit19} that PBHs in pairs can slightly increase their spins through gas accretion. This effect may explain a small nonzero effective spin of PBH pairs, even if the initial PBH spin was close to zero.

The LIGO--Virgo--KAGRA detectors have recently recorded the event S251112cm, in which the so-called chirp mass, $\mathcal{M}_c=(M_1M_2)^{3/5}/(M_1+M_2)^{1/5}$, lies in the range $0.1-0.87M_\odot$, and with probability $>99$~\% one of the BHs in the merged pair has a subsolar mass $<1M_\odot$~\cite{Vieetal26}. A subsolar BH mass may be evidence of a primordial origin~\cite{HaqIocVis26,MagCap26}. However, this event is so far unique, and further statistics are needed to confirm the reality of such events.

It is usually believed that the detection of BHs with subsolar masses would prove their primordial origin. Indeed, this would be a very strong argument in favor of PBHs. Still, some exotic possibilities for producing subsolar-mass BHs in the present-day Universe remain. Suppose, for example, that $P$-parity~\cite{BerDolMoh96} is broken in a model with baryonic mirror matter~\cite{Oku07,Bli14} constituting a fraction of the dark matter of the Universe. Suppose also that the nucleon masses in the mirror world are three times larger than in ours (see the details of such models in~\cite{BerDolMoh96}), while the proton--neutron mass difference remains small, so that stellar evolution with supernova explosions and BH formation is possible. Then, from the simplest estimate of the maximum neutron star mass (the Oppenheimer--Volkoff limit),
\begin{equation}
M_{\rm OV}\propto\frac{1}{m_n^2}\left(\frac{\hbar c}{G}\right)^{3/2},
\end{equation}
one can conclude that supernova explosions in the mirror world should produce BHs with masses starting from $\sim0.6M_\odot$, which are subsolar masses from the point of view of our world.

Another model selection criterion is related to the dependence of the gravitational-wave signal on the orbital eccentricity before the BH merger. If a BH pair formed long ago (for example, at the end of the radiation-dominated stage), it has had time to circularize ($e\approx0$), and the signal shape within each period will be more symmetric in amplitude. In contrast, for PBH pairs formed dynamically in clusters, the orbit remains highly elongated before the merger ($e\approx1$), which should manifest itself in a characteristic asymmetric signal shape~\cite{1999PhRvD..60l4008M,2018CQGra..35f3001S}.

On the whole, it can be said that gravitational-wave observations by LIGO/Virgo/KAGRA do not yet provide evidence for the primordial origin of even some of the BHs. The event with a subsolar mass requires verification, and BH masses within the mass gap, as noted above, may be of astrophysical origin. The shape of the BH mass function is also not a decisive argument at this stage, owing to uncertainties in the evolution of ordinary stars. The possible contribution of PBHs to the statistics of LIGO/Virgo/KAGRA events, taking into account all events including the first part of the fourth observing run, O4a, was studied quantitatively in~\cite{2026arXiv260515749A}. That work showed that the LIGO/Virgo/KAGRA observations are consistent with an astrophysical origin (stellar core collapse) of the BHs in binaries, and obtained improved constraints on the PBH abundance (on their fraction $f_{\rm PBH}$ in DM) in the mass range $0.6M_\odot\leq M_{\rm PBH}\leq 100M_\odot$ for a monochromatic mass function. For an extended mass function, the constraints at the ends of this range are somewhat weaker, but they also exist at $\sim 10^3M_\odot$.

\subsection{Effect of primordial black hole clustering on their mergers} 
\label{sec:62}

\begin{figure}
	\begin{center}
\includegraphics[angle=0,width=0.48\textwidth]{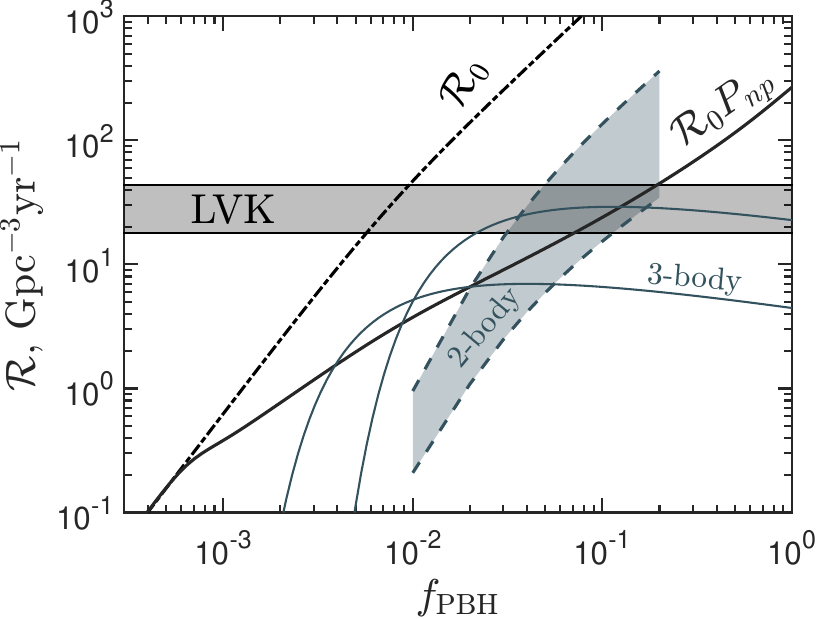}
	\end{center}
\caption{Present-day merger rate of PBHs with mass $M_\pbh = 30 \, M_{\odot}$ as a function of their fraction in DM. The gray region bounded by the dotted lines shows the range of merger rates of late binaries depending on the efficiency of cluster destruction during structure formation; the upper boundary corresponds to the limiting case $w = 1$ (see expression~\eqref{eq:mr_late}), in which cluster destruction is neglected. The horizontal band shows the LIGO/Virgo/KAGRA observations from the O3 run, $ \mathcal{R} = 17.9  -44 $~Gpc$^{-3}$~yr$^{-1}$~\cite{KAGRA:2021duu}. The solid line corresponds to the merger rate of early binaries including the suppression factor $P_{\rm np}$ (mergers of unperturbed binaries); the dash-dotted line, without suppression.}
	\label{gr:mrLVK} 
\end{figure}

Clustering of PBHs into early structures has two key consequences: (1) early binaries are subject to gravitational perturbations from single PBHs, which effectively ``exclude'' them from the merger process; and (2) a channel of binary formation through dynamical processes in clusters opens up. We begin with the former. Early binaries have extremely small angular momenta (see expression~\eqref{eq:j_ch}), and scattering off a single PBH predominantly increases the angular momentum of the binary. This, in turn, leads to a significant increase in the merger time of the binaries, according to expression~\eqref{eq:tmer}. Most importantly, the resulting merger time exceeds the present age of the Universe. It is precisely because of these gravitational perturbations that early binaries can be regarded as excluded from the population of merging PBHs. This process cannot be described analytically because of the complexity of the three-body problem, but the physical considerations are as follows: owing to interactions with other PBHs, the initial angular momentum distribution~\eqref{eq:dp_dadj} tends toward a thermal one, $dP/dj \propto j$. Numerical simulations of gravitational scattering between early binaries and single PBHs indeed confirm this expectation~\cite{Jedamzik:2020ypm,Delos:2024poq}; the formation and disruption of PBH binaries in the first gravitationally bound structures were also studied in~\cite{Tkachev:2020uin}. However, not all PBHs are clustered, so only the binaries that end up in virialized structures will be perturbed. This leads to the following modification of the merger rate: $\mathcal{R} = \mathcal{R}_0 P_{\rm np}$, where $\mathcal{R}_0$ is given by expression~\eqref{eq:mr1}, and the factor $P_{\rm np}$ is the fraction of unperturbed binaries, calculated in~\cite{Stasenko:2024pzd, Vaskonen:2019jpv}. The merger rate of early PBH binaries as a function of the PBH fraction in dark matter is shown in Fig.~\ref{gr:mrLVK} by the black solid and dash-dotted lines.

\medskip
{\bf Formation of primordial black hole binaries via gravitational-wave emission} 
\medskip

We now turn to the formation of late binaries in clusters, for which there are two possibilities: two-body and three-body interactions between PBHs. We first consider the two-body channel. When two PBHs scatter with a sufficiently small impact parameter, gravitational waves are emitted, and if the energy they carry away exceeds the initial kinetic energy, a binary forms. The cross section of this process is~\cite{QuinShap89}
\begin{equation} \label{eq:cs_bbh}
	 \Sigma_{2b} = \pi R^2_s \left( \frac{85 \pi}{3} \right )^{2/7} \left ( \frac{v_{\rm rel}}{c} \right)^{-18/7},
\end{equation}
and, as a rule, the binary that forms has a high binding energy and such a short lifetime that it merges almost immediately~\cite{Stasenko:2024pzd}. The merger rate per cluster is~\cite{Stasenko:2021vmm}
\begin{align}\label{eq:g2b}
    &\gamma_{2b} = \frac{1}{2} \int d\vec{v}\, d\vec{r} f(\vec{r}, \vec{v}) \langle \Sigma_{2b} v_{\rm rel} \rangle  \\  
    &= \frac{14 \pi^2}{3} \left ( \frac{85 \pi}{14} \right)^{2/7} R_s^2 \int dr \, r^2 \overline{v} \left ( \frac{\rho_\pbh}{M_\pbh} \right)^2  \left ( \frac{\overline{v}}{c} \right)^{-18/7}, \nonumber
\end{align}
where $f$ is the PBH distribution function in phase space, and the angular brackets denote averaging over relative velocities. To evaluate the integral, we adopt the delta-function approximation for the distribution function, $ f(\bm{r}, \bm{v}) \propto \rho (r) \delta(v - \overline{v})$~\cite{QuinShap89}, where $ \sqrt{\langle v^2(r) \rangle} = \overline{v}(r)$ denotes the root-mean-square velocity. Since the cluster evolves with time, the merger rate $\gamma_{2b}$ is a function of time. The total PBH merger rate is obtained by integrating expression~\eqref{eq:g2b} over the cluster mass function. The difficulty here lies in determining this mass function. The point is that clusters form continuously in the young Universe, but some of them are subsequently destroyed during the formation of larger structures. Here we follow the procedure for calculating the merger rate of such binaries given in~\cite{Stasenko:2024pzd}. We assume that clusters form at redshift $z_f$ and that some of them are subsequently destroyed; we denote by $w$ the fraction of clusters that survive. The formation of new clusters is not taken into account, which makes this estimate conservative (a lower bound). The merger rate of late binaries is then
\begin{equation} \label{eq:mr_late}
    \mathcal{R}_{\rm late, 2b} = \int w(M,z_f) \frac{dn(z_f)}{dM} \gamma_{2b} \, dM.
\end{equation}
The functions $w(M,z)$ and $dn/dM$ are calculated within the Press--Schechter formalism. Numerical analysis shows that the merger rate depends only weakly on the redshift $z_f$ at which the mass function is ``frozen''~\cite{Stasenko:2024pzd}. The shaded region in Fig.~\ref{gr:mrLVK} bounded by the dotted lines corresponds to the merger rate of late two-body binaries; the uncertainty is related to the destruction of clusters during the formation of large-scale structure. The upper dotted line corresponds to the limiting case $w = 1$ in expression~\eqref{eq:mr_late}, in which there is no cluster destruction.

\medskip
{\bf Formation of primordial black hole binaries via three-body interactions} 
\medskip

Let us now consider binaries that form in three-body interactions. With some probability, the outcome of a gravitational encounter of three PBHs in a cluster is the formation of a hard and, generally speaking, long-lived binary. This binary then predominantly hardens in interactions with single PBHs~\cite{1975AJ.....80..809H,1975MNRAS.173..729H} and eventually acquires a large recoil energy $\sim kM_{\pbh} v^2$ and leaves the cluster (for globular star clusters, $k \sim 100$). These issues are discussed in detail in the book~\cite{HegHut06}. As a zeroth approximation, we assume that the binding energy of the binaries is determined by their recoil energy and that the angular momentum distribution is isotropic:
\begin{equation}\label{eq:3bdadj}
	 dP_{3b} = 2 \, \delta(a - a_*) j \, dj da,
\end{equation}
where $a_* = GM_{\pbh} / (k \, v^2)$, and the velocity depends on the halo mass, $v\simeq \sqrt{GM/R}$. Proceeding in the same way as in the derivation of~\eqref{eq:dPdt1}, we obtain
\begin{align} \label{eq:dp3bdt}
    &\frac{dP_{3b}}{dt} = \left ( \frac{170 \, k^4 v^8 t}{3 \, c^5 G M_\pbh} \right)^{2/7} \frac{2}{7t}.  
\end{align}
It is necessary to determine the fraction of three-body binaries in a cluster. A difficulty similar to that encountered for mergers of two-body binaries arises here: binaries form from the onset of PBH clustering until the onset of the formation of the present-day large structures of the Universe at $z \lesssim 10$. Therefore, binaries will end up in newly forming Poisson halos, in which further binaries will form and existing ones will be perturbed (or even disrupted). These effects can be followed only in numerical $N$-body simulations that take into account both structure formation and the dynamics of PBHs in halos, but such simulations are computationally expensive. Therefore, to estimate the merger rate, we assume that a certain fraction $\alpha$ of PBHs in DM halos formed at some redshift $z_f$ are in binaries:
\begin{align}\label{eq:mr3b1}
    &\mathcal{R}_{\rm 3b} = \int \alpha(M) \, N_\pbh \frac{dP_{\rm 3b}}{dt} \frac{dn}{dM} dM,
\end{align}
where $N_\pbh = f_\pbh M/M_\pbh$ is the number of PBHs in a halo, and we set $k = 100$. A reasonable lower estimate for $\alpha$ is one binary per halo: $\alpha = 2/N_\pbh$, with $\alpha = 0$ for $N_\pbh < 2$. The merger rate of three-body binaries is shown by the thin solid lines in Fig.~\ref{gr:mrLVK}. The sharp cutoff of this dependence at $f_\pbh \lesssim 0.01$ is due to the fact that Poisson clustering of PBHs becomes inefficient at such a contribution to DM, as discussed in Section~\ref{sec:51}. An analytical estimate of the merger rate can be obtained from the integral:
\begin{align}
    \mathcal{R}_{\rm 3b} \simeq \frac{23 \, f_\pbh^{-5/21}}{\text{yr Gpc}^{3}} \left ( \frac{M_\pbh}{M_{\odot}} \right)^{-\frac{11}{21}} \left ( \frac{1+z_f}{10} \right)^{\frac{34}{21}} \left ( \frac{t}{t_U} \right)^{-\frac{5}{7}}. \nonumber
\end{align}
The integral $\mathcal{R}_{\rm 3b}$ is easily evaluated if one takes into account that the main contribution comes from $M < M_{\rm ch}$. For the estimate to be valid, $M_{\rm ch} > 3 M_{\pbh}/f_\pbh$ is required.

The results of this section are summarized in Fig.~\ref{gr:mrLVK}, which shows the present-day PBH merger rate as a function of their fraction $f_\pbh$ in DM. Interestingly, in the range $f_\pbh = 0.01 - 0.1$, early binaries may even make a subdominant contribution to the total PBH merger rate compared with binaries that form deep in the matter-dominated stage. The region $f_\pbh \gtrsim 0.1$ is constrained by the analysis of gravitational microlensing~\cite{Mroz:2024mse}. As can be seen, to avoid contradicting the LIGO/Virgo/KAGRA observations, the PBH fraction in DM must also be $f_\pbh \lesssim 0.1$. It is important to note that the formation of clusters, in which PBHs interact gravitationally with each other intensely, reduces the merger rate of early binaries. On the other hand, the probability of formation and merger of late binaries increases in clusters. Clustering thus suppresses the early channel but inevitably enhances the late one.

\subsection{Mergers in clusters}

Let us now consider PBH binary mergers in models with initial clustering, following the formalism of~\cite{Stasenko:2025vqz}. PBH binaries form in the same way as in the model of Section~\ref{sec:61}, via pairwise decoupling from the Hubble flow. However, owing to gravitational perturbations from the surrounding PBHs, the angular momentum of the binaries increases, which leads to a substantial increase in the binary merger time. As in the discussion in Section~\ref{sec:62}, these perturbations lead to a gradual suppression of the binary merger rate with time. The probability of gravitational perturbations is determined by the mean free time, $\tau^{-1} = \rho_{\pbh} \langle \Sigma_p v \rangle /M_\pbh$, where $\Sigma_p$ is the gravitational cross section for scattering of a binary by a single PBH, which we define by the condition that the pericenter distance equal the semimajor axis of the binary, $r_p = a$. After averaging over relative velocities, we obtain the characteristic timescale for perturbations of the orbital parameters of PBH binaries:
\begin{equation} \label{eq:tau}
    \tau^{-1} = \frac{2 \sqrt{2 \pi}}{M_\pbh} 140 \, \rho_{\eq} \delta_\pbh^4 \sigma a^2 \left (1 + \frac{3\, G M_\pbh}{a \sigma^2} \right),
\end{equation}
where $\delta_\pbh (\bf x) = \rho_{\pbh}(\bf x) / \rho_{\dm}$ is the local PBH density contrast, which for simplicity is assumed to be constant on some scale, and $\sigma$ is the velocity dispersion of the Maxwellian distribution. To obtain the merger rate of PBH binaries, one has to introduce into distribution~\eqref{eq:dp_dadj} a factor describing the depletion of binaries due to perturbations and make the replacement $f_\pbh \rightarrow \delta_\pbh$. For simplicity, we choose this factor in the exponential form $e^{-t/\tau}$, which gives
\begin{equation}\label{eq:dPdt_Sf}
    \frac{dP}{dt} = \frac{3}{16} \frac{\delta_\pbh^2}{T} \left ( \frac{t}{T} \right)^{-5/8} \int \limits_{j_{\rm min}}^{1} j^{-37/8} \, e^{-t/\tau(t,j)} \, dj,
\end{equation}
where the lower limit is $j_{\rm min} = \left ( t/T \right)^{3/37} \delta_\pbh^{16/37}$, and the transition $\tau(a) \rightarrow \tau(t,j)$ is made using the merger time~\eqref{eq:tmer}.

Let us now discuss the consequences of taking perturbations into account. Consider a binary with a given lifetime $t$. The widest binaries, with angular momenta $j \ll 1$, are efficiently perturbed. In contrast, binaries with $j \sim 1$ are very tight systems with a small interaction cross section. Therefore, owing to perturbations, the fraction of binaries with small angular momenta is effectively excluded from the merging population. This cuts off integral~\eqref{eq:dPdt_Sf} from below at some critical angular momentum $j_p > j_{\rm min}$, which can be found from the equation $\tau(t,j_p) =t$. With this effect taken into account, the merger rate is estimated as
\begin{align}\label{eq:mr_cl}
&\mathcal{R}_{\rm np} = \frac{f_\pbh \Omega_{\dm}}{M_\pbh} \frac{3 H_0^2}{8 \pi G}  \frac{dP}{dt} \simeq \frac{0.24 \, f_\pbh}{\text{yr Gpc}^{3}}  \delta_\pbh^{-44/7}   \\
&\times \left ( \frac{t}{t_U} \right)^{-45/14} \left ( \frac{M_\pbh}{M_{\odot}} \right)^{-27/14} \left ( \frac{\sigma}{10 \,\text{km/s}} \right)^{29/14}. \nonumber
\end{align}
One can see that, in models with initial clustering, the time dependence of the merger rate differs significantly from the unclustered case~\eqref{eq:mr1}.

It is important to note that perturbed binaries are not completely excluded from the merger process; they merely acquire different orbital parameters. The contribution of such binaries must be taken into account separately. Within a rather simplified model, one can assume that for perturbed binaries the initial angular momentum distribution~\eqref{eq:dp_dadj} becomes thermal. In this case, the merger rate can be written as
\begin{equation}\label{eq:mr_per_cl}
\mathcal{R}_{\rm per} \!= \!\frac{f_\pbh \rho_\dm}{M_\pbh} \!\!\int \limits_{0}^{1}  \!\frac{3}{2}\! \left (\! \frac{\delta_\pbh}{\overline{x}} \!\right)^{3/2} \!\!\!\!\sqrt{a}\,\theta \big(a_h\! -\! a \big) j \, \frac{\partial a}{\partial t}  \,dj.
\end{equation}
In this expression, $a_h = GM_\pbh/2\sigma^2$; the dependence of the semimajor axis on time and angular momentum, $a(t,j)$, is determined, as before, using~\eqref{eq:tmer}; and $\theta \big(a_h -a \big)$ is the Heaviside step function, which separates soft binaries, disrupted by perturbations, from hard ones. Since interactions with single PBHs make hard binaries even tighter, expression~\eqref{eq:mr_per_cl} should be regarded as a lower bound on the merger rate.

\subsection{Primordial black hole mergers at high redshifts} 

One of the future observational tests for the existence of PBHs with masses of the order of a solar mass is the detection of gravitational-wave signals from cosmological distances. Unlike astrophysical BHs, which form only after the onset of star formation at $z \lesssim 30$~\cite{Koushiappas:2017kqm}, PBHs merge throughout the entire cosmic history, starting from the moment of their formation. This is why the observation of gravitational waves from BH mergers at high $z$ would point to their primordial nature. In this section, we compare the astrophysical and primordial channels.

\begin{figure}
\begin{center}
\includegraphics[angle=0,width=0.48\textwidth]{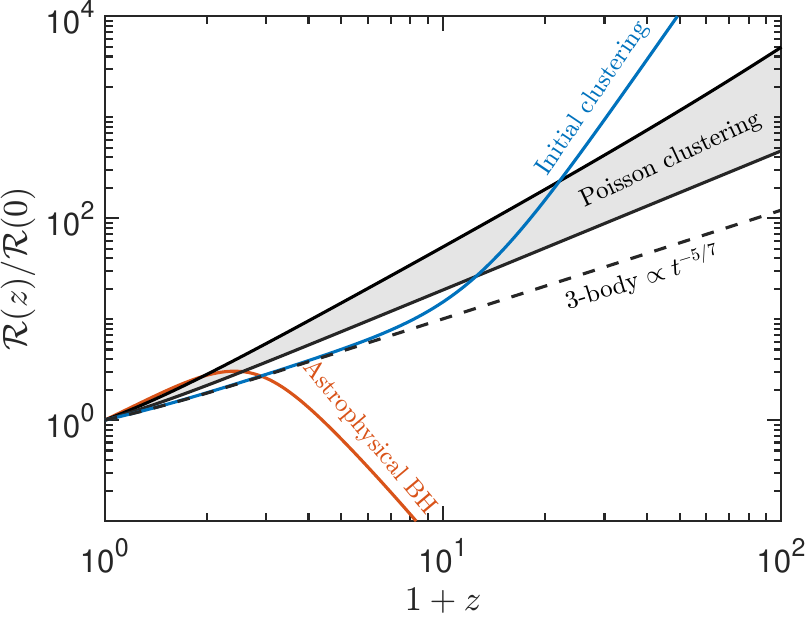}
\end{center}
\caption{Redshift dependence of the merger rates of primordial and astrophysical BHs. The gray hatched region shows the merger rate of PBH binaries formed in the early Universe, taking into account their Poisson clustering; the upper boundary corresponds to $f_\pbh = 0.1$, and the lower boundary corresponds to the absence of clustering effects, $f_\pbh < 10^{-3}$. The dashed line shows binaries formed in three-body interactions in a cluster~\eqref{eq:mr3b1}. The blue line shows the scenario with initial PBH clustering for the parameters $\delta_\pbh = 1$, $\sigma$ = 10~km/s, and $M_\pbh = 30M_{\odot}$ (see expressions~\eqref{eq:tau}~and~\eqref{eq:dPdt_Sf}).}
\label{gr:mrz} 
\end{figure} 

In the astrophysical scenario, the merger rate of stellar BHs is given by
\begin{align}\label{eq:mrabh}
&\mathcal{R}_{\rm{ABH}}(z) \propto \int_z^{\infty} dz_f \, \frac{dt_f}{dz_f} \, dt_{\rm del} \, \mathcal{R}_{\rm{SFR}}(z_f) \nonumber \\
&\times P_{\rm del}(t_{\rm del}) \delta \Big (t(z) - t(z_f) - t_{\rm del} \Big),
\end{align}
where $\mathcal{R}_{\rm{SFR}}$ is the star formation rate, given by the empirical formula~\cite{Madau:2014bja}
\begin{equation}\label{eq:sfr}
	\mathcal{R}_{\rm{SFR}}(z) \propto \frac{(1+z)^{2.7}}{1 + \big ((1 + z)/2.9 \big)^{5.6}}, 
\end{equation}
and $P_{\rm del} \propto t^{-1}$ is the distribution of delay times between star formation and merger, with $P_{\rm del} = 0$ for $t < 50\times 10^6$~yr~\cite{Belczynski:2016obo}. The redshift dependence of the stellar BH merger rate is shown in Fig.~\ref{gr:mrz} by the red line. The redshift evolution of the PBH merger rate depends on their clustering. If PBHs are distributed uniformly (clustering arises only from Poisson noise), the merger rate varies with redshift as $\mathcal{R}_\pbh \propto \big (t(z)/t_U \big)^{-34/37} P_{\rm np}(z)$ (see expression~\eqref{eq:mr1} and the discussion in Section~\ref{sec:62}). This dependence is shown by the gray shaded region in Fig.~\ref{gr:mrz}, where the lower boundary corresponds to the absence of clustering, i.e., to the case in which the PBH fraction in dark matter is $f_\pbh < 10^{-3}$. If PBHs are born clustered, the merger rate is the sum of the contributions of unperturbed and perturbed binaries, $\mathcal{R} = \mathcal{R}_{\rm np} + \mathcal{R}_{\rm per}$, given by expressions~\eqref{eq:mr_cl} and~\eqref{eq:mr_per_cl}. This dependence is shown by the blue line. The sharp change of slope to $\propto t^{-5/7}$ occurs when a significant fraction of binaries in this scenario have already been perturbed.

The curves in Fig.~\ref{gr:mrz} clearly demonstrate the differences between the astrophysical and primordial BH merger channels. Measuring the redshift dependence of the merger rate would make it possible to identify the PBH population and to test various scenarios of their formation, in particular, the features of their clustering. Conversely, the absence of the corresponding signal would lead to new constraints on the parameters of such models. Future gravitational-wave detectors, such as the Einstein Telescope~\cite{ET:2025xjr} and Cosmic Explorer~\cite{CE:2021} (see also~\cite{LIGOScientific:2016wof}), are expected to have sufficient sensitivity to explore this possibility.

\subsection{Gravitational-wave background}

Individual BH merger events at high redshifts may be inaccessible to observation. Nevertheless, they contribute to the energy density of gravitational waves in the Universe, which is conventionally described by the fractional spectral energy density~\cite{Phinney:2001di}
\begin{align}\label{eq:Ogw}
\Omega_{\rm gw} &= \frac{1}{\rho_{\rm crit} c^2} \frac{d \rho_{\rm gw}}{d \ln \nu} \nonumber \\
&= \frac{1}{\rho_{\rm crit} c^2} \int_0^{\infty} dz \, \frac{\mathcal{R}(z)}{H(z) (1 + z)^2} \frac{d E_{\rm gw}}{d \ln \nu_s},
\end{align}
where $\nu_s = (1 + z) \nu$ is the gravitational-wave frequency in the source frame, and $dE_{\rm gw}/d \ln \nu_s$ is the gravitational-wave energy emitted in a binary BH merger per logarithmic frequency interval. Let us consider how the gravitational-wave background can provide evidence for the existence of PBHs. To this end, we parametrize the redshift evolution of the merger rate as $\mathcal{R} = \mathcal{R}_0 (1 + z)^{\kappa}$ and consider only the inspiral phase, whose energy spectrum is~\cite{Magg2007}
\begin{equation} \label{eq:dEgwdlnf}
\frac{d E}{d \ln \nu_s} = \frac{\pi^{2/3}}{3 G} (G M_c)^{5/3} \nu_s^{2/3},
\end{equation}
where $M_c = (M_1 M_2)^{3/5}/(M_1 + M_2)^{1/5}$ is the chirp mass. For simplicity, we assume that the inspiral phase continues up to some maximum frequency $\nu_{\rm max}$, which can be defined via the innermost stable circular orbit, after which $dE / d\ln \nu_s$ drops to zero. The amplitude of the gravitational-wave background then depends on frequency as
\begin{align} \label{eq:Ogw2}
\Omega_{\rm gw} &= \nu^{2/3} \frac{\mathcal{R}_0 \pi^{2/3} (G M_c)^{5/3}}{3 G \rho_{\rm crit} c^2} \!\!\!\int\limits_{0}^{(\nu_{\rm max}/\nu)-1} \!\!\!\frac{(1 + z)^{\kappa +2/3}}{(1+z)^2H(z)} dz \nonumber \\
&\propto \frac{ \nu^{2/3}}{\kappa -11/6}  \left [ \left ( \frac{\nu_{\rm max}}{\nu}  \right)^{\kappa - 11/6} -1  \right],
\end{align}
where we have used the matter-dominated approximation for the Hubble parameter, $H(z) \approx H_0 \sqrt{\Omega_M (1 + z)^3}$. One can see that, for $\kappa < 11/6$, the second term in the brackets in expression~\eqref{eq:Ogw2} dominates, which leads to the classical power-law slope $\Omega_{\rm gw} \propto \nu^{2/3}$ characteristic of BH mergers. However, for $\kappa > 11/6$, the spectral slope deviates from the canonical one and takes the form $\Omega_{\rm gw} \propto \nu^{5/2 - \kappa}$. This is precisely the behavior that arises in scenarios with PBH clustering.

\begin{figure}
	\begin{center}
\includegraphics[angle=0,width=0.47\textwidth]{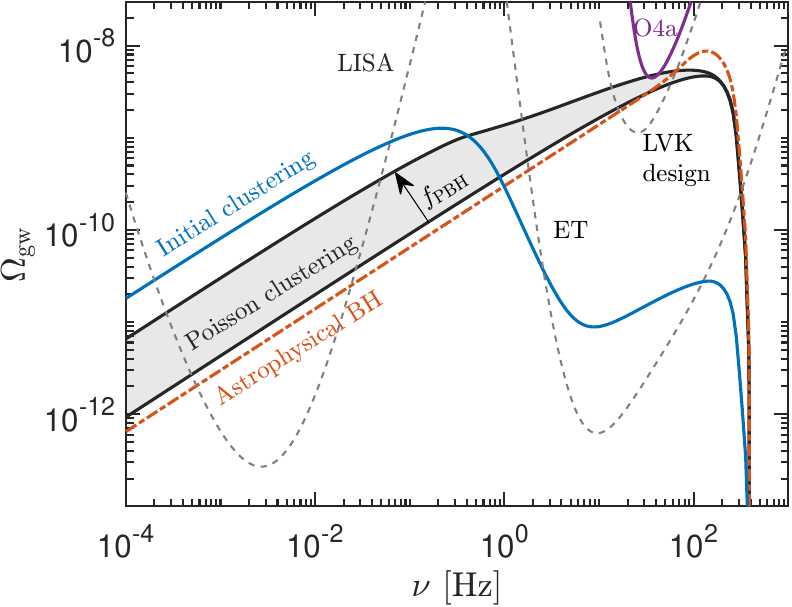}
	\end{center}
\caption{Gravitational-wave background from mergers of PBH binaries with mass $M_\pbh = 30 \, M_{\odot}$ for various clustering scenarios. The gray shaded region corresponds to PBH clustering due to Poisson noise for $f_\pbh = 0.007-0.1$. The blue solid line shows the model with initial PBH clustering with parameters $\delta_\pbh = 1$ and $f_\pbh = 0.1$, where the merger rate is given by the sum of expressions~\eqref{eq:mr_cl} and~\eqref{eq:mr_per_cl}. The red dash-dotted line shows the contribution of astrophysical BHs \eqref{eq:mrabh}, whose merger rate is normalized to $\mathcal{R}_{\rm ABH}(t_0) = 40$~Gpc$^{-3}$~yr$^{-1}$. The dashed lines show the projected sensitivities of some future gravitational-wave detectors; the upper limit from the LIGO/Virgo/KAGRA O4a observing run~\cite{LIGOScientific:2025kry} is also shown.}
	\label{gr:Ogw} 
\end{figure}

Figure~\ref{gr:Ogw} shows the gravitational-wave background calculated for primordial and astrophysical BHs. One can see that PBH clustering leads to deviations of the spectral characteristics of the gravitational-wave energy density from the standard behavior $\Omega_{\rm gw} \propto \nu^{2/3}$. This effect is especially pronounced in models with initial clustering, where the merger rate increases rapidly with redshift, $\mathcal{R}_{\rm np}\propto (1 + z)^{135/28}$, according to~\eqref{eq:mr_cl}, which, in accordance with~\eqref{eq:Ogw2}, leads to the spectral slope $\Omega_{\rm gw} \propto \nu^{-65/28}$. A detailed analysis of these effects is presented in~\cite{Stasenko:2025vqz}.

\subsection{Gravitational waves generated by primordial black hole formation}

Gravitational waves can be generated not only by PBH mergers but also in connection with the processes of PBH formation in the early Universe. The wavelength of such gravitational waves has, of course, increased by many orders of magnitude by now. The formation of a significant number of PBHs implies the presence of large scalar curvature perturbations, $\mathcal{P}_\zeta\simeq10^{-2}$, on some scale. Such perturbations simultaneously generate gravitational waves at second order in perturbation theory, whereas at first order scalar and tensor perturbations evolve independently~\cite{MatPanSae94,BugKli10, KohTer18, PiSas20, InoTer20, Dom21, VasVee21}. Interestingly, the frequency of such gravitational waves generated during the formation of PBHs with masses $\sim30M_\odot$ (as in the first LIGO event) is~\cite{2009PhRvL.102p1101S}
\begin{equation}
\nu_{gw}\simeq1\times10^{-9}\left(\frac{M_{\rm PBH}}{30M_\odot}\right)^{-1/2}\mbox{~Hz},
\label{gwnano}
\end{equation}
i.e., it falls into the nanohertz band. A stochastic background of such gravitational waves can be searched for using pulsar timing~\cite{PosPorPsh25}.

Conversely, if the existence of a cosmological gravitational-wave background is known from pulsar timing observations or other experiments, and if these gravitational waves are assumed to have been produced in the very early Universe (for example, during inflation), then calculations show that scalar density perturbations can be generated from the gravitational waves at second order in perturbation theory. These perturbations, in turn, can lead to the formation of a significant number of PBHs~\cite{2015PhRvD..92l1304N,2016PhRvD..94d3507N,Kumar:2025jfi}. In fact, this is yet another PBH formation mechanism, in addition to those considered at the beginning of the review.

Another possible mechanism of gravitational-wave generation accompanying PBH formation is the collapse of closed domain walls. It was shown in~\cite{SakEroRub21} that the gravitational-wave spectrum arising in this model, $h\propto f^{\alpha}$ with $\alpha=-5/4$, agrees well with the spectrum observed with pulsar timing by the NANOGrav project, since for $\alpha=-5/4$ we have $\gamma=3-2\alpha=5.5$\footnote{Here, $\gamma$ is the power-law index of the power spectrum of pulse time-of-arrival residuals in pulsar timing, $S(f)\sim h^2/f^3$~\cite{PosPorPsh25}.}, while NANOGrav gives the constraint $4.5<\gamma<6.5$ for the signal in the frequency range from 2.5~nHz to 12~nHz.

\section{Gas accretion onto primordial black holes}

Like BHs of astrophysical origin, PBHs can accrete matter. Accreting gas heats up and radiates.

\medskip
{\bf Accretion during the radiation-dominated stage}
\medskip

Although accretion of matter onto PBHs during the radiation-dominated stage was considered as early as~\cite{ZelNov66}, it remained unclear whether accretion leads to a catastrophic growth of the PBH mass, in which its gravitational radius grows as fast as the cosmological horizon. Indeed,~\cite{ZelNov66} considered the capture of particles by a PBH. The capture cross section of ultrarelativistic particles by a Schwarzschild sphere of radius $R_g=2GM_{\rm PBH}/c^2$ is $\sigma=(27/4)\pi R_g^2$~\cite{ZelNov65, LL-II}; therefore, the PBH mass grows at the rate
\begin{equation}
\frac{dM_{\rm PBH}}{dt}=\frac{27}{4}\pi R_g^2c\rho_r,
\label{dmdteq1}
\end{equation}
where the radiation density during the radiation-dominated stage is $\rho_r=3/(32\pi Gt^2)$. The solution of Eq.~(\ref{dmdteq1}) on the time interval from $t_0$ to $t_1$ is
\begin{equation}
M_{\rm PBH}=\frac{M(t_0)}{1-\frac{81}{32}\frac{GM_0}{c^3}\left(\frac{1}{t_0}-\frac{1}{t_1}\right)}.
\label{dmdtsol1}
\end{equation}
Formally, for sufficiently small $t_0$, there is a moment when the mass tends to infinity. However, as noted in~\cite{ZelNov66}, the conclusion about catastrophic growth cannot be drawn because of uncertainties in the initial density distribution near the moment of PBH formation. The quasi-stationary accretion approximation (\ref{dmdteq1}) is also insufficiently justified. These uncertainties should give rise to an additional coefficient in the denominator of (\ref{dmdtsol1}), which fundamentally affects the result. For example, if at the moment of PBH formation its gravitational radius is much smaller than the radius of the cosmological horizon, this relation will persist later on, and the mass accreted over the lifetime of the Universe will not exceed $\sim M_0$.

To clarify this question, the equations of relativistic hydrodynamics must be studied. This question was first resolved in~\cite{1974MNRAS.168..399C}, where self-consistent self-similar solutions for accretion were obtained and used to show that there is no catastrophic accretion, regardless of how close the gravitational radius is to the radius of the cosmological horizon at the initial moment. Self-similarity meant that the quantities depend only on the ratio $r/t$. Thus,~\cite{1974MNRAS.168..399C} showed that the PBH model is self-consistent with respect to catastrophic accretion. Had this not been the case, PBH models would effectively have been ruled out already in 1974; after the important result of~\cite{1974MNRAS.168..399C}, the way was opened for more detailed studies of many other aspects of PBH physics. It was also pointed out in~\cite{1974MNRAS.168..399C} that there may exist solutions containing PBHs that are not exactly but only asymptotically Friedmann at infinity. However, this requires special initial conditions under which the flow does not pass through a sonic point, remaining supersonic everywhere.

The study of accretion onto PBHs during the radiation-dominated stage, with $p=\rho/3$, and, more generally, of accretion of matter with a barotropic equation of state $p=(\gamma-1)\rho$ was continued in a number of later works (see the review~\cite{CarHarMae10}). Attention was focused, in particular, on the search for self-similar accretion solutions and on the consistent embedding of the PBH metric into the cosmological background at large distances. It turned out that for $1<\gamma<2$ there are no self-similar solutions connected to the Friedmann background through a sonic point, while for $2/3\leq\gamma<1$ there are asymptotically Friedmann solutions with a solid-angle deficit at infinity, which may nevertheless correspond to a real physical situation. Finally, for dark energy, $0<\gamma<2/3$, asymptotically Friedmann solutions were found. Interestingly, under some assumptions about the equation of state that differ from those adopted in~\cite{1974MNRAS.168..399C} and correspond to accretion of a massive scalar field, the size of a PBH may indeed grow as fast as the cosmological horizon, and this growth may explain the large masses of BHs in galactic nuclei.

Another noteworthy effect is that accretion and the corresponding increase in PBH mass can delay the quantum evaporation of PBHs. The constraints related to PBH evaporation in the early Universe should then be somewhat modified.

Theoretically, in addition to primordial BHs, there could also exist relic BHs left over from previous epochs in a cyclic model of the Universe, if the Universe follows such a model~\cite{GorTul21,Gaz26}. We do not consider such models in this review. The specific angular momentum and masses of relic BHs could change strongly due to accretion of the surrounding matter during the contraction phase of the Universe, and the enormous mass growth may severely constrain or rule out this class of models.

\medskip
{\bf Accretion during the Dark Ages}
\medskip

In recent years, there has been considerable interest in studying the cosmological epoch of the Dark Ages, which lasted from recombination until the reionization of the Universe, i.e., at redshifts $z\sim6-1100$. This is largely due to JWST observations of early galaxies at the end of the Dark Ages. Unexpectedly, there turned out to be about an order of magnitude more galaxies at that time than expected. The study of the physical properties of the baryonic medium in this epoch is of great interest. In particular, radiation from accretion onto PBHs could have fundamentally affected the properties of the medium, which yields additional constraints on PBHs.

The first detailed studies of gas accretion onto PBHs in the pregalactic epoch were carried out in~\cite{Car79,Car81}. According to a number of calculations, the spectrum of accretion radiation is a power law with a high-energy cutoff,
\begin{equation}
L_\omega\propto \omega^{-a}\exp\left(-\omega/T_s\right).
\end{equation}
In~\cite{Car81}, emission at the Eddington limit was assumed and a flat spectrum with $a=0$ was used, whereas the later work~\cite{Pouetal17} adopted $T_s\sim 200$~keV, $|a|\leq0.5$, and $\omega_{\rm min}\leq10$~eV. The uncertainty in the emission spectrum is related, in particular, to uncertainties in the accretion model (spherically symmetric or disk). In~\cite{Car81}, the structure of the gas ionization region around a PBH was calculated as a function of distance. It turned out that the transition from a fully ionized region to a weakly ionized one is rather sharp. PBHs, if present in sufficiently large numbers, could have changed the thermal history of the Universe and kept it ionized. From the requirement that the Universe was not fully ionized much earlier than $z\sim6$, one can derive constraints on the number of accreting PBHs in various mass ranges; this effect is very strong for PBHs of stellar and larger masses. A detailed calculation of accretion onto PBHs and its observational consequences was performed in~\cite{AliKam17}, which (see also~\cite{Ser24, Casetal24}) gives modern constraints on accreting PBHs during the Dark Ages; in addition to the reionization history of the Universe, the influence of accretion radiation on the CMB spectrum and polarization was also considered. The resulting constraints are shown in Fig.~\ref{grconstr}. In particular, comparison of the calculations with Planck data rules out PBHs with masses $\geq10^2M_\odot$ as the dominant DM component.

\medskip
{\bf Accretion in the present-day Universe}
\medskip

If the Universe contains a sufficiently large number of PBHs of stellar and higher masses, then after recombination these PBHs could accrete gas and emit X-rays, most likely from an accretion disk. Unidentified ultraluminous X-ray sources are observed in some nearby galaxies. It is possible that these sources are accreting PBHs, although the primordial nature of these BHs has not been proven, and they may well be astrophysical BHs rapidly growing in mass. In~\cite{Taketal22}, the heating of interstellar gas in galaxies by PBH accretion radiation was studied, and this effect was used to constrain PBHs in the mass range $\sim(1-10^7)M_\odot$. It was noted in~\cite{2017PhRvL.118x1101G} that gas accretion onto BHs usually produces relativistic jets of charged particles, which generate radio emission. Therefore, one can try to search for accreting PBHs with large radio telescopes.

\medskip
{\bf Primordial black holes and neutron stars}
\medskip

For sufficiently small masses, the gravitational radius of a PBH becomes comparable to or smaller than the size of a nucleon. Therefore, quantum effects should manifest themselves in accretion of matter onto low-mass PBHs. For larger PBH masses, however, steady accretion onto a PBH inside a neutron star is possible, and over a long time the neutron star may be completely converted into a BH~\cite{Haw71}. The capture of PBHs by neutron stars and the accretion of matter onto PBHs inside these objects were considered in~\cite{GidMan08}, taking into account the quantum effects mentioned above as well as a possible influence of extra dimensions. It was found that, if a PBH with a mass of $\sim10^{15}$~g is captured inside a neutron star, the neutron star will survive for $\sim 3\times10^5$~yr. Under high-density conditions, the PBH rapidly increases its mass, eventually leading to the collapse of the whole neutron star into a black hole; however, constraints of this kind are model-dependent~\cite{Monetal19}.

\section{Evaporation of primordial black holes}

Perhaps the most striking and fundamentally interesting effect associated with PBHs is the possibility of their quantum evaporation, predicted by S.~Hawking~\cite{1974Natur.248...30H}. Earlier works by Ya.B.~Zel'dovich and A.A.~Starobinsky on particle creation in gravitational fields and on quantum radiation from rotating BHs made a major contribution to the discovery of this effect. It was quantum evaporation that drew increased attention to PBHs in the mid-1970s. Only PBHs can have masses of $\sim5\times10^{14}$~g, at which their evaporation is active at the present time; therefore, the observation of an evaporating BH would prove its primordial origin. Stellar BHs are 19 orders of magnitude more massive, and no mechanisms are known for forming astrophysical BHs with small, subsolar masses.

The evaporation of a sufficiently nearby PBH could be distinguished with high confidence from other radiation processes owing to the unique shape of the spectrum and of the signal time profile. In addition, a nearby evaporating PBH would appear as a moving gamma-ray source. Detection of PBH evaporation would make it possible, in particular, to test quantum gravity effects. Even the mere absence of evaporating PBHs in observations, as well as the influence of PBH evaporation on processes in the early Universe, have already yielded very stringent constraints on the shape of the spectrum of cosmological perturbations on small scales and, accordingly, on the inflationary models used to describe the generation of perturbations.

\medskip
{\bf Emission spectra}
\medskip

S.~Hawking showed that a BH undergoing quantum evaporation radiates as a black body with temperature $T=\hbar c^3/(8\pi k_{\rm B}G M)\sim10^{-7}(M/M_\odot)^{-1}$~K (for a BH without spin and charge). The evaporation process can be described both as the creation of pairs of virtual particles near the BH horizon~\cite{Fro76} and in the quantum tunneling formalism. The emission rate of particles with energy $E$ is
\begin{equation}
\frac{d^2N}{dEdt}=\frac{1}{2\pi}\frac{\Gamma_{s}(M,E)}{e^{E/T}-(-1)^{2s}},
\end{equation}
where $\Gamma_{s}(M,E)$ is the so-called greybody factor, obtained from rather complicated calculations, and $s$ is the particle spin. The particles carry away energy, the BH mass gradually decreases, and at each moment particles with masses (or energies) $\sim T$ are emitted most efficiently. For PBH masses $M_{\rm PBH}\geq10^{17}$~g, mainly only photons, neutrinos, and possibly other light particles are emitted. As the mass decreases below $\sim10^{17}$~g, electrons and positrons begin to make up a significant fraction of the emission, and as the PBH mass decreases further, heavy leptons and various hadrons appear among the evaporation products. The particles emitted by a PBH interact with each other in the region of space closest to the PBH, developing cascades of secondary particles. As a result, the initial emission spectrum is modified, which is important for searches for signals from evaporating PBHs. In particular, the secondary spectrum has a maximum at $\sim100$~MeV due to pion decays. The final spectra of various particles produced by PBH evaporation are calculated accurately with a number of numerical codes (see the review~\cite{2023PrPNP.13104040A}). Calculating the mass evolution yields the following dependence of the PBH lifetime on its mass~\cite{2021RPPh...84k6902C}:
\begin{equation}
\tau(M)\simeq 1.4\times10^{10}\left(\!\frac{M}{5.1\times10^{14}\mbox{~g}}\right)^{\!3}\!\mbox{yr}. 
\end{equation}

\medskip
{\bf Searches for signals from primordial black hole evaporation and evaporation constraints}
\medskip

At the very last stage, evaporation looks like an explosion accompanied by a gamma-ray flash. If a sufficiently large number of PBHs were evaporating in the vicinity of the Solar System, the corresponding gamma-ray flashes could be observed. We would detect such flashes from space, or the background radiation from many flashes if there were many evaporating PBHs~\cite{Car76}. So far, no such flashes have been reliably observed. It has been hypothesized that the 220~PeV neutrino detected by the KM3NeT telescope in 2023 may be a product of the evaporation of a nearby PBH~\cite{KliKai25}, but this hypothesis has not been proven. Even the very absence of flashes, however, yields stringent constraints on the number of evaporating PBHs.

More massive PBHs that have not yet reached the final stage of evaporation should also contribute to the gamma-ray background. Since we do not see an excess of such radiation, there should be few PBHs in the corresponding mass range, $\sim10^{15}-10^{17}$~g. As noted above, this yields a strong constraint on inhomogeneities in the early Universe on the scales corresponding to these masses. A number of detectors have provided limits on the gamma-ray and X-ray backgrounds that constrain the number of nearby evaporating PBHs. These imply constraints on the fraction of matter that collapsed into PBHs in the early Universe. Such a constraint is shown in Fig.~\ref{grconstr}. Recently, a new constraint from the neutrino flux measured by the IceCube neutrino telescope in Antarctica has been added, since the evaporation products could include high-energy neutrinos. In~\cite{BugKon02}, the extragalactic neutrino background from evaporating PBHs was calculated. Comparison with data from neutrino detectors, primarily Kamiokande and LSD, made it possible to constrain the possible spectral indices of the initial density perturbations.

Interestingly, radiation from PBH evaporation during the Dark Ages could have prevented the formation of hydrogen molecules responsible for gas cooling in the first DM halos, which, in turn, creates conditions for the direct collapse of gas clouds into BHs~\cite{ZacPicKus24}. Thus, small PBHs could have helped give birth to much more massive BHs that became the seeds of supermassive BHs in galactic nuclei.

\medskip
{\bf Evaporating primordial black holes in the early Universe}
\medskip

PBH evaporation at earlier cosmological epochs cannot be observed directly, but evaporating PBHs could have influenced processes in the Universe.

If PBHs had initial masses (at formation) of $\sim10^9-10^{12}$~g, their active evaporation took place during the epoch of big bang nucleosynthesis. At that time, when the Universe was $\sim2-20$~minutes old, the first atomic nuclei up to lithium were formed. PBH radiation during this epoch could have affected the element yields by changing the proton-to-neutron ratio, as well as by disintegrating the newly formed nuclei with the emitted high-energy particles~\cite{Zeletal77,Caretal10}. Calculations of this process take into account a large number of reactions and interactions~\cite{2021RPPh...84k6902C}. The neutron diffusion length exceeds the mean separation between neighboring PBHs (for PBH abundances at which their role is noticeable); therefore, a uniform distribution of PBH evaporation products is a good approximation in such calculations.

PBHs with masses $\leq10^9$~g should evaporate before big bang nucleosynthesis. In~\cite{ZelSta76}, the increase in the entropy of the Universe due to PBH evaporation was calculated, and constraints on the number of PBHs were obtained from this effect. PBH evaporation could produce DM particles if such particles are among the evaporation products~\cite{KhlBarGra06,ChaChaKhl25}. Models have also been proposed for generating the baryon asymmetry, i.e., the predominance of matter over antimatter in the observable Universe, through PBH evaporation~\cite{Haw75, Zel76,TurSch79, Dol81, Dol92, Nag99, Fujetal14}. Note, however, that baryogenesis may be related to PBHs not only through quantum evaporation but also through common inflationary mechanisms~\cite{DolSil93,2022PhRvL.128c1102W}. It is also important that Planck-mass remnants of PBH evaporation, if evaporation stops at this mass, could constitute all of the DM in the Universe; this is discussed below in Section~\ref{dmsec}. In models with extra spatial dimensions, for example in brane-world cosmology, PBH evaporation proceeds differently than in the ordinary Friedmann model, which modifies the evaporation constraints on PBHs~\cite{GueClaLid02}.

\section{Primordial black holes as dark matter}
\label{dmsec}

The nature of DM is one of the main unsolved puzzles of cosmology. One of the main candidates is new elementary particles, but attempts to detect them with existing detectors have so far been unsuccessful. PBHs are also a plausible DM candidate. The idea that PBHs may constitute the DM in the Universe was put forward in~\cite{Haw71, Cha75}, even before the concept of DM became established in cosmology, and was later developed in a model with features in the inflationary potential~\cite{IvaNasNov94}.

Above, we have considered a large number of different effects in which PBHs could participate. Although the existence of PBHs has not yet been proven, these effects yield upper limits on the PBH abundance in various mass ranges (see the reviews~\cite{2020ARNPS..70..355C,Carr:2021bzv,2021JPhG...48d3001G}). These limits, usually expressed in terms of the mass fraction $f_{\rm PBH}$ of PBHs in DM, are rather stringent at all masses except for the ``window'' $10^{17}-10^{22}$~g. In this range, PBHs may still constitute a significant fraction of DM ($f_{\rm PBH}\sim1$), although part of this range, $\sim10^{19}-10^{20}$~g, is probably already excluded by white dwarf explosions (see below). Outside this allowed window, PBHs can constitute only a certain fraction of all DM, and the bulk of DM consists of something else, for example, elementary particles not yet detected experimentally.

Above, it was assumed that all PBHs have a single characteristic mass. However, the constraints on PBHs are somewhat relaxed if PBHs are ``spread'' over masses, i.e., if they have a sufficiently broad mass function. Then, in the ranges where the constraints are strong, PBHs can make up only a small fraction of the critical density of the Universe, but integrally (summed over all masses) they can constitute all of the DM. Clearly, in this case the constraints obtained under the assumption of a monochromatic mass function do not constrain the entire PBH population, but only part of it.

\medskip
{\bf Dynamical constraints on the abundance of primordial black holes}
\medskip

First of all, there are stringent dynamical constraints on PBHs as DM. PBHs are massive compact objects. Therefore, if their mass is sufficiently large and they are located inside or near other astrophysical objects, their gravitational field can noticeably perturb the motion of stars or other bodies and even disrupt them during close passages. Several possible situations in which PBHs could manifest themselves in this way have been studied, and the absence of a visible influence yields constraints on PBHs in the corresponding mass ranges. Such constraints based on the gravitational influence of PBHs are called dynamical constraints. Dynamical constraints on PBHs have been discussed in detail in reviews (see, e.g.,~\cite{2020ARNPS..70..355C,Carr:2021bzv,2021RPPh...84k6902C}); therefore, here we only briefly list the main ones. These dynamical constraints are shown in Fig.~\ref{grconstr}.

Passages of PBHs through degenerate stars, i.e., white dwarfs, could cause local temperature increases along the trajectory and thereby trigger thermonuclear explosions of white dwarfs (type Ia supernovae). The existence of unexploded white dwarfs in space and the measured rates of type Ia supernovae constrain the number of PBHs with masses $\sim10^{19}-10^{22}$~g, ruling out $f_{\rm PBH}=1$ in the range $\sim10^{19}-10^{20}$~g~\cite{2015PhRvD..92f3007G}. Capture of PBHs by neutron stars should lead to accretion of matter, instability, and gravitational collapse of the neutron star into a BH~\cite{CapPshTin13-2,2014PhRvD..90d3512K}. Wide binary stars are observed in our Galaxy and in nearby dwarf galaxies. If PBHs were sufficiently numerous, these binaries could be disrupted, which constrains the number of PBHs. Sufficiently massive PBHs could also disrupt globular star clusters, and the observed number of surviving clusters yields constraints on PBHs. PBHs could heat the stellar population of the disk, i.e., increase the velocity dispersion of disk stars as sufficiently massive PBHs pass through the disk. Very massive PBHs, $M\geq7\times10^{10}M_\odot$, could disrupt entire galaxies in galaxy clusters~\cite{2021RPPh...84k6902C}.

\medskip
{\bf Microlensing}
\medskip

One of the most important methods for searching for PBHs and constraining their abundance is the observation of gravitational lensing events (see the review~\cite{Gre26}). This effect, predicted by A.~Einstein in 1936, consists in the magnification of a source's light when light rays are deflected by a massive object on the line of sight. For example, if a star is being observed and a PBH passes across the line of sight between the star and the observer, the light flux from the star will first smoothly increase and then equally smoothly return to its original level. The mass of the gravitational lens can be inferred from the distances and the characteristics of the lensing event. B.~Paczy\'nski proposed in 1986 to use stars in the Magellanic Clouds to search for compact objects in the Galactic halo. The source stars in microlensing observations are usually stars of the Large and Small Magellanic Clouds, of the Galactic bulge, or of the galaxy M31 (the Andromeda Galaxy). In these observations, the individual images produced by the lens cannot be resolved; only a time-symmetric change in the brightness of the star is observed. The magnification factor is limited by the angular size of the source star.

To date, microlensing observations of stars have been carried out mainly by three collaborations: MACHO, EROS, and OGLE. Each of these collaborations has conducted several observing campaigns, gradually rejecting false signals and refining the constraints on the abundance of lensing objects (their mass fraction $f$ in dark matter) in various mass ranges. The statistics of microlensing events depend on the model of the Galactic dark halo, and the results are usually presented for a chosen standard halo model in the form of an isothermal sphere (see the discussion of the influence of this factor in~\cite{GarHaw24,HawGar25, Mroetal25}). Supernova explosions in distant galaxies and the variability of source stars are sometimes mistaken for microlensing by compact objects. For several events initially identified as gravitational microlensing, further observations revealed a repeated brightening of the source stars, which indicates that these stars are variable and the events merely mimic gravitational lensing. Microlensing constraints have also been obtained from observations with the Kepler space telescope and the Subaru HSC (Hyper Suprime-Cam) telescope.

In the history of microlensing searches, there have been reports of positive results that were later excluded on the basis of larger data sets. Thus, in the early 1990s, the MACHO collaboration reported the detection of gravitational microlensing of stars in the Large Magellanic Cloud by unknown compact objects on the line of sight with masses $\sim0.5M_\odot$, constituting $\sim20$~\% of the DM density ($f\sim0.2$). Such a large fraction cannot be explained by baryonic objects because of the constraint on the amount of baryons produced in big bang nucleosynthesis. In~\cite{IvaNasNov94}, it was hypothesized that these compact objects are PBHs, and a model of ``neutralino stars'' was considered in~\cite{GurZybSir97}. However, subsequent EROS and OGLE observations ruled out $f\sim0.2$ and established more stringent constraints on the abundance of such objects. Recently, the OGLE collaboration reported the possible observation of six microlensing events that could be caused by planetary-mass objects. However, more stringent constraints on the number of such objects were later obtained, which cast doubt on the reliability of the detection of these six objects. Nevertheless, this effect has already been used to detect a free-floating exoplanet with a mass of the order of that of Saturn.

One can speak with more confidence not about detections but about constraints on the abundance of compact objects in various mass ranges.
The existing constraints are shown in Fig.~\ref{grconstr} for a delta-function PBH mass function. It should be noted that for an extended mass function the constraints are modified; a method for refining the constraints in this case was developed in~\cite{Yok98, Caretal17}. One can already conclude that PBHs in the mass range $10^{-11}M_\odot\leq M_{\rm PBH}\leq10^4M_\odot$ cannot constitute all of the DM in the Galaxy, and PBHs in the range $10^{-10}M_\odot\leq M_{\rm PBH}\leq1M_\odot$ constitute no more than 1~\%~\cite{Gre26}. Microlensing observations of stars in the Andromeda galaxy (M31) were carried out with the Subaru HSC telescope~\cite{2019NatAs...3..524N}. The identification of only one candidate event made it possible to obtain the most stringent constraint at that time on PBHs in the wide mass range $10^{-11}M_\odot\leq M_{\rm PBH}\leq10^{-6}M_\odot$. Observations with the Vera C.~Rubin Observatory are expected to yield constraints several orders of magnitude stronger. For masses $M_{\rm PBH}\leq10^{-11}M_\odot$, the wavelength of light already exceeds the Schwarzschild radius of the lens; therefore, the number of PBHs can be constrained by stellar microlensing only at larger masses.

It was pointed out in~\cite{2016PhRvL.117i1301M} that future observations of gravitational lensing of fast radio bursts may constrain PBHs with masses $\geq20M_\odot$. Fast radio bursts (powerful single pulses of radio emission with durations of the order of milliseconds and large dispersion measures, see~\cite{PopPosPsh18}) most likely originate in processes on magnetars~\cite{2010vaoa.conf..129P,2013arXiv1307.4924P}. The ``femtolensing'' of the emission of cosmic gamma-ray bursts has also been considered~\cite{Gou92}. The terms micro-, femto-, and picolensing correspond to the characteristic angular separations between the images of the lensed source.

The statistics of microlensing events may be substantially different if PBHs are gathered in clusters~\cite{GorGre22,Tosetal26}. If such clusters are not very dense, the PBHs in a cluster lens light as individual objects; however, clusters with a high PBH number density (for example, those considered in Section~\ref{clustdomensubs}) act as a single lens~\cite{CleGar17}.

The main conclusion from gravitational lensing observations is that PBHs are ruled out as the dominant DM component over a wide mass range, except for the so-called ``asteroid-mass window'' (masses of the order of asteroid masses), $10^{17}-10^{22}$~g. Note, however, that this range may also be closed by the neutron star stability argument~\cite{CapPshTin13-1,CapPshTin13-2, PanLoe14,CapPshTin14}, although this constraint is model-dependent and is not yet regarded as final~\cite{Monetal19}.

\medskip
{\bf Remnants of primordial black hole evaporation (planckeons)}
\medskip

As the mass of an evaporating PBH decreases, more and more massive and energetic particles begin to be emitted. Already at energies approaching the Grand Unification scale, calculating the emission spectrum is difficult because of theoretical uncertainties in particle physics models, but qualitatively the evaporation picture probably does not change much. However, as the PBH mass decreases and approaches the Planck mass, quantum gravity effects should come into play, and the evaporation process may change qualitatively. In particular, evaporation may stop altogether. It has been hypothesized that a PBH does not evaporate completely but leaves behind a stable remnant, a ``planckeon''.

Several reasons why PBH evaporation may stop near the Planck mass have been discussed~\cite{CarGilLid94}. For example, according to the argument of M.A.~Markov, this follows from the quantum uncertainty principle for the momentum uncertainty $c\Delta M$ and the position uncertainty $\Delta l$, where $l\sim r_g=2GM/c^2$, in the form $c\Delta M\Delta l>\hbar/2$. This yields an estimate of the minimum BH mass, $\sim M_{\rm Pl}=(\hbar c/G)^{1/2}\simeq10^{-5}$~g. Ya.B.~Zel'dovich noted that BHs of sub-Planckian mass may represent a new class of stable elementary particles. Other possibilities considered include quadratic corrections to the gravitational Lagrangian, a new axion charge of BHs, arguments based on symmetry conservation, effects of the dilaton field, the information paradox, and others (for references to works on these topics, see~\cite{CarGilLid94}).

In theories with extra dimensions, the masses of PBH evaporation remnants may exceed the usual Planck mass of $10^{-5}$~g. In~\cite{CheAdl03}, the generalized uncertainty principle (GUP) was considered as the reason for the cessation of evaporation. This principle, as in M.A.~Markov's argument, implies that gravitational interaction gives rise to an absolute minimum uncertainty in the position of a particle, of the order of the Planck length, and to a corresponding minimum black hole mass.

It has been suggested that planckeons may be DM particles in the Universe~\cite{MacGib87,Barrow:1992hq}. Experimental detection of planckeons appears to be a challenging task because of their small interaction cross section. However, the existence of PBH evaporation remnants constrains the perturbation spectra on small scales generated, for example, in the hybrid inflation model~\cite{CarGilLid94}.

\section{Primordial black holes in the Solar System}

If PBHs formed in sufficiently large numbers in the Universe, there is a chance that PBHs have entered the Solar System in the past and may do so in the present epoch, perturbing planetary orbits, and may even have been captured onto bound orbits around the Sun. The possibility of PBHs inside the Earth and other celestial bodies has also been considered, and the observational consequences of such a situation have been studied.

\medskip
{\bf  Primordial black holes inside the Earth}
\medskip

In~\cite{DokVol}, the hypothesis that there is a PBH inside the Earth was studied. Such a PBH should accrete matter and emit neutrinos. From the constraints on neutrino fluxes, the presence of a PBH inside the Earth was ruled out because of the unacceptably large neutrino emission produced by accretion. In~\cite{Luoetal12}, constraints from seismic events excited by PBHs were considered.

\medskip
{\bf  A primordial black hole as Planet Nine}
\medskip

The discovery of certain correlations in the motion of trans-Neptunian objects underlies the hypothesis of a ninth planet with a mass of 5--15 Earth masses at a distance of 300--700~AU from the Sun~\cite{Batetal19,LykIto23}. This planet could perturb the motion of trans-Neptunian objects. In~\cite{Scholtz:2019csj}, it was hypothesized that this object is not a rocky or icy planet but a PBH. Although the PBH option looks more exotic than a planet, the probability of capturing a PBH into the Solar System may be comparable to the probability of capturing a free-floating planet~\cite{Scholtz:2019csj}.

A number of effects have been considered that could be used to detect a PBH in the inner Oort cloud. Ref.~\cite{SirLoe20} discussed future constraints from X-ray flares in the periphery of the Solar System. These flares should arise when a PBH collides with icy bodies. In~\cite{Arbey:2020urq}, it was proposed to search for radio-frequency Hawking radiation using microprobes with radio detectors flying past the PBH. Ref.~\cite{Scholtz:2019csj} also derived a constraint from DM annihilation in the density spike around such a PBH (see Section~\ref{spikesubsec}). Such a constraint applies if DM particles can annihilate and the annihilation products are in principle observable.

It was shown in~\cite{EroPop23} that, if a PBH is located at the periphery of the Solar System, its gravitational field perturbs the orbits of dust particles. As a result, dust may fall into the inner Solar System, penetrating inside the Earth's orbit. It was shown that, if the PBH mass is of the order of $10M_\oplus$ and the total mass of dust in the inner part of the so-called ``Oort cloud'' is of the same order of magnitude, the dust flux is comparable to that actually observed by counting dust grains in Antarctic ice and by direct measurements on spacecraft, $\sim 3.0-5.6$~$\mu$g~m$^{-2}$~yr$^{-1}$~\cite{Rojetal21}. Similar calculations~\cite{EroPop23} apply to all dust or meteoroid particles larger than about 10~micrometers and to iron- and graphite-free particles smaller than 0.02~micrometers, which are not blown out by radiation pressure. It is known that a significant part, or all, of the observed dust is brought to Earth by passing comets or originates from collisions and fragmentation of objects in the asteroid belt. A captured PBH or planet at the periphery of the Solar System could produce a dust flux of the same order of magnitude if a significant amount of dust is present there. Even a small contribution of such high-velocity dust from the periphery of the Solar System to the total flux may be of interest for understanding the structure of the inner Oort cloud, if its chemical composition differs from that of asteroid-belt dust.

\medskip
{\bf  Passages of primordial black holes through the Solar System}
\medskip

In~\cite{SirLoe21}, an attempt was made to constrain the rate of passages of low-mass PBHs, $\sim10^{-10}M_\odot$, through the Solar System on the basis of the growth of the orbital eccentricities of trans-Neptunian objects, and it was shown that this effect is very small. For PBHs of larger masses (from an Earth mass to a solar mass), the perturbations of Neptune's orbit were obtained in~\cite{SirLoe21-2}. However, this effect has not yet yielded significant constraints on PBHs.

\section{Conclusion}

Despite more than half a century of research, PBHs remain among the most intriguing objects of modern cosmology. Their distinctive feature is that they can arise from a wide variety of physical processes in the early Universe: amplification of primordial density perturbations, phase transitions, collapse of topological defects, collisions of true-vacuum bubbles, early matter-dominated stages, and a number of other mechanisms. Therefore, the study of PBHs is of interest not only in its own right but also as a unique tool for investigating physical processes that took place at energies inaccessible to modern particle accelerators.

The current observational situation is characterized by increasingly stringent constraints combined with the emergence of new opportunities for PBH searches. Data on the CMB, gravitational lensing, the dynamics of stellar systems, gravitational-wave events, gamma-ray signals, and the large-scale structure of the Universe substantially constrain the allowed fraction of PBHs in the DM density in various mass ranges. At the same time, none of the existing observational results can yet definitively exclude the existence of PBH populations that could make some contribution, however small, to DM or carry information about physical processes in the early Universe.

Further progress in this field is expected to come from the development of multimessenger astronomy, combining cosmological observations, gravitational-wave experiments, and particle physics research. A new generation of observational facilities will significantly increase the sensitivity to PBH signatures and related signals. Regardless of whether their role as a DM component and as gravitational-wave sources is confirmed, PBHs already represent an important theoretical and observational tool for studying inflation, phase transitions, topological defects, and other manifestations of physics beyond the Standard Model. In this sense, PBHs are not only possible cosmological relics but also among the most promising physical objects for understanding the origin and early evolution of the Universe. The analysis of PBH formation mechanisms presented in our review highlights the need to move from qualitative estimates to precise modeling of PBH mass spectra and clustering parameters. Systematizing the various scenarios, from the collapse of adiabatic perturbations to the evolution of cosmic strings, allows one to identify the most promising ``observational windows'' for future experiments.

The topic of PBHs is very broad, and this review covers only some of the research directions. In particular, discussions of other possible astrophysical and cosmological properties of PBHs can be found in the reviews~\cite{Khlopov:2004rw,Caretal10,Dol18,2018CQGra..35f3001S,2020ARNPS..70..355C,Carr:2021bzv,2021JPhG...48d3001G,2021RPPh...84k6902C,2023PrPNP.13104040A,2024bheg.book..261E,2024Univ...10..444H,2024PhR..1054....1C,2025LRR....28....1B,Green2025,Gre26,2026NCimR..49..225C}. The breadth and importance of research on PBHs are illustrated by the recent publication of a 732-page book written by 25 authors~\cite{PBHs}.

The most remarkable milestone in PBH research would, of course, be a reliable confirmation of their actual existence in the Universe. As noted by A.~Green~\cite{Gre25}, in searching for PBHs through astronomical observations, one should distinguish between (1) constraints on PBHs, (2) possible evidence for PBHs, and (3) a definitive detection of PBHs. In the first case, if PBHs could influence some process in cosmology or astrophysics but this influence has not yet been detected, the absence of a detection at the achieved level of precision yields constraints on the PBH abundance, such as those shown in Fig.~\ref{grconstr}. On the other hand, some known observed phenomena can be explained both by ordinary astrophysical processes and by effects related to PBHs. If PBH models are admissible in such cases, and all the more so if they are preferable, one can speak of possible evidence for PBHs. Finally, if it is convincingly shown that only a PBH model can describe a certain process, while an astrophysical explanation is ruled out with high confidence, this, as A.~Green pointed out~\cite{Gre25}, will constitute a detection of PBHs. At present, there is no evidence that any particular process can be explained only by PBHs. A decisive event could be the detection of radiation from a nearby evaporating BH or the detection of gravitational waves from mergers of binary BHs of subsolar mass.

{\textbf{Acknowledgments}. The work of V.D.S., S.G.R., and V.V.N. was supported by the Ministry of Science and Higher Education of the Russian Federation within project FSWU-2026-0010. The authors are grateful to A.A.~Kirillov for useful discussions and valuable comments.}

\setcounter{section}{0} 
\renewcommand{\thesection}{\Alph{section}} 

\section*{Appendix I: Quantum Fluctuations of a Field with Classical Evolution during Inflation}\label{Classmo}

Quantum fluctuations are usually considered under the assumption that the classical field has a fixed value. However, many PBH models assume that the classical field evolves during inflation; this applies, in particular, to some of the mechanisms considered above. In this case, it is convenient to have an analytical expression for the probability of field fluctuations on a dynamical background~\cite{KirRub21}. In this section, we derive an expression for the probability of a field fluctuation at a given spatial point for an arbitrary scalar field potential $V$, assuming its slow rolling.
For our purposes, it is sufficient to know the value of the field inside the horizon averaged over spatial coordinates, which makes it possible to obtain the required result.

Consider a scalar field $\Phi$ with the standard action
\begin{equation}
\label{action4}
S = \frac{M_\text{Pl}^{2}}{2}\int d^{4}x \, \sqrt{g_{4}} \Big[R+
(\partial_\mu\Phi)^2 - 2V(\Phi) \Big].
\end{equation}
Here, $M_\text{Pl}$ is the Planck mass. The scalar field may be either the inflaton or a spectator field. Following the reasoning of~\cite{KirRub21, Dokuchaev:2008hz,RubKhlSak00}, we write the scalar field as
\begin{equation}
   \Phi=\Phi_0 + Q.
\end{equation}
Here, $Q(x,t)$ is the ``short-wavelength'' part of the Fourier expansion of the field $\Phi$. The field equation in de Sitter space can then be written as~\cite{Khlopov:2004rw}
\begin{gather}
    \frac{\partial \Phi_0 }{\partial t}-\frac{1}{3H}\left[e^{-2Ht}\Delta \Phi_0 - \frac{\partial V(\Phi_0 )}{\partial \Phi_0}\right] = y(\mathbf{x},t),
    \label{randEq} 
    \\ 
    y(\mathbf{x},t) \equiv \left( -\frac{1}{3H}\frac{\partial ^{2}}{\partial t^{2}}-\frac{\partial }{\partial t}+\frac{1}{3H}e^{-2Ht}\Delta\right) Q( \mathbf{x},t).
    \nonumber
\end{gather}
This equation has been simplified: the second time derivative has been dropped in the slow-roll approximation, which is standard during inflation, and higher powers of the function $y(\mathbf{x} ,t)$ have been neglected.
Equation~\eqref{randEq} describes the classical motion of the field $\Phi_0$ under the action of a random ``force'' $y$.
The latter is assumed to be small, so we seek a solution of the equation in the form
\begin{equation}
    \label{clquant}
    \Phi_{0} =\Phi_\text{cl}+\phi.
\end{equation}
The inequality $\Phi_\text{cl} \gg \phi $ is assumed to hold, which is valid if the random ``force'' $y(\mathbf{x},t)$ is small.
We are interested in fluctuation scales exceeding the horizon $H^{-1}$ during inflation, where the fluctuations do not depend on the spatial coordinates. In this case, the homogeneous distribution $\Phi =\Phi (t)$ is determined by simpler equations:
\begin{align}
& \frac{\partial \Phi_\text{cl} }{\partial t} + \frac{1}{3H}\frac{\partial
V(\Phi_\text{cl} )}{ \partial \Phi_\text{cl} }=0,  
\label{lineart} 
\\ 
& \frac{\partial \phi }{\partial t}+\mu(t)\phi =y(t);\quad  \mu(t) \equiv \frac{m^{2}(t)}{3H},
\label{lineart2}
\end{align}
where we have neglected the spatial derivatives. The parameter
\begin{equation}
    m^2(t) \equiv V^{\prime \prime }(\Phi_\text{cl})
\end{equation}
depends on time through the classical field $\Phi_{cl}$.
It is therefore natural to assume that the function $m(t)$ varies slowly during inflation.

The probability density of the random function $y(t)$ is known to be Gaussian,
\begin{eqnarray}
\label{Wy}
W(y) &=& C_1 \cdot \exp \left[ -\frac 1{2\sigma^2}\!\int\! y^2(t) \, d{t} \right]\!\!, \\  
C_1&=&const, \quad \sigma=\frac{H^{3/2}}{2\pi},
\end{eqnarray}
since its correlator is proportional to a $\delta$-function~\cite{Rey:1986zk}:
\begin{equation}
    \big\langle y(t_{1}) \, y(t_{2}) \big\rangle 
       = \frac{H^3}{4\pi ^{2}}\delta(t_{1}-t_{2}).
\end{equation}
Since $y(t)$ and $\phi(t)$ are linearly related, the probability distribution of the function $\phi$ can be obtained by substituting relation \eqref{lineart2} into expression \eqref{Wy}. This means that the probability of finding a certain value of $\phi(t)$ within a small interval is~\cite{FeynHibb}
\begin{equation}
    d{P}(\phi) = C_2 \, d\phi \,
    \exp \left[ -\frac{1}{2\sigma^{2}}\int 
    \left[ \frac{\partial \phi }{\partial t}+\mu(t)\phi \right]^{2} \!\!dt \right]\!, 
\end{equation}
where $C_2=const$.
 
We can now obtain the probability of finding the field value $\phi_{2}$ at time $t_{2}$, given its value $\phi _{1}$ at time $t_{1}$ (see~\cite{FeynHibb}). To do so, it suffices to integrate over all paths $\phi(t)$ within the interval $(t_{1},t_{2})$:
\begin{eqnarray} \label{fundistr}
&&   \frac{d{P}(\phi_{2},t_{2};\phi _{1},t_{1})}{ d\phi_{2}}= \\
&&        C_2\int\limits_{\phi
_{1}}^{\phi _{2}}\mathcal{D}\phi \, 
\exp\left[ -\frac{1}{2\sigma^{2}} \int\limits_{t_{1}}^{t_{2}}
\left[\frac{\partial \phi}{\partial t}+\mu(t) \phi\right]^{2} d{t}\right]. \nonumber
\end{eqnarray}
The constant factor $C_2$ in this expression is determined by the normalization condition
\begin{equation}
    \int\limits_{-\infty }^{\infty } \frac{dP(\phi_{2},t_{2}; \phi_{1},t_{1})}{d\phi_2} \, d\phi_{2}
        = 1.
\end{equation}
The functional integral \eqref{fundistr} can be evaluated by the standard saddle-point method, by finding the extremal trajectory for the integral in the exponent, i.e., by solving the equation
\begin{equation}
\ddot{\phi} - \mu^{2}(t)\phi = 0,
\label{eq:DDotPhi}
\end{equation}
where the term with $\dot{\mu}$ is assumed to be small, since the parameter $\mu(t)$ depends only weakly on time during inflation. Note that the squared ``mass'' $\mu^2\propto m^4$ is therefore positive both when the field is near a maximum and when it is near a minimum of the potential.

Imposing the boundary conditions
\begin{equation}
\phi (t_{1})=\phi _{1},  \quad   \phi (t_{2})=\phi _{2},
\end{equation}
we find the solution of Eq.~\eqref{eq:DDotPhi} in the form
\begin{align}
&\phi(t) = A\exp\big( M(t) \big) + B\exp\big( -M(t) \big); \\
&M(t)\equiv \int\limits_{t_1}^{t}\mu(t') \,d{t'}; \nonumber \\
&A = \frac{\phi_{2}-\phi_{1}e^{-M(t_2)}}{2\sinh\big(M(t_2)\big)},
\quad
B = -\frac{\phi_{2}-\phi_{1}e^{M(t_2)}}{2\sinh\big(M(t_2)\big)}. \label{MAB}
\end{align}
Note that $M(t_1)=0$ by definition.

Substituting this solution into the integral in the exponent of expression \eqref{fundistr}, we obtain the desired probability:
\begin{align}
&d{P}(\phi_{2}, t_2; \phi_{1}, t_{1}) = C_3 \, d\phi_{2} \, \exp\left[ -\frac{2A^2}{\sigma^2}\!\int\limits_{t_1}^{t_2}\! \mu^2(t) e^{2M(t)} \, d{t} \right] \nonumber \\ 
&= C_3 \, d\phi_{2} \, \exp\bigg[-q^2 \Big(\phi_2-\phi_1 e^{-M(t_2)}\Big)^2\bigg], 
\label{q2}
\end{align}
where $C_3=const$ and
\begin{equation}
q^2\equiv \frac{1}{2\sigma^2\sinh^2\big(M(t_2)\big)}\int\limits_{t_1}^{t_2} \mu^2(t)e^{2M(t)}\,d{t}.
\end{equation}
Formula \eqref{q2} gives the probability of finding a specific value of the fluctuating part of the field~\eqref{clquant}. The ``classical'', slowly varying part of the field, $\Phi_\text{cl}$, enters implicitly through the definition of the function $M(t)$. The probability of a given value of the field $\Phi$ (the distribution function $f$) is easily obtained by substituting $\phi(t)=\Phi(t)-\Phi_\text{cl}(t)$ into the above formula:
\begin{eqnarray}
f(\Phi_2, t) &=& 
\frac{dP(\Phi_2, t; \Phi_1, t_1)}{d\Phi_2}\nonumber \\
&=&
\sqrt{\cfrac{q^2}{\pi}} 
\exp\left[-q^2 \left(\Phi_2 - \Phi_\text{cl}(t) \right)^2\right]\!\!.
\label{dP}
\end{eqnarray}
Here we have taken into account that at the initial moment there is no fluctuation at the chosen spatial point: $\phi_1= \Phi_1 - \Phi_\text{cl}(t_1)=0$.

Let us calculate the variance that plays an important role in the study of fluctuations during inflation
\begin{eqnarray}\label{f2}
 &&   \langle \varphi^2\rangle= \int_{-\infty}^{\infty}d\varphi \varphi^2f(\varphi,t)=\frac{1}{2q^2}.
\end{eqnarray}
This formula, which is the main goal of this section, looks rather compact; however, determining $q^2$ requires the nontrivial chain of formulas \eqref{q2}, \eqref{MAB}, \eqref{Wy}, and \eqref{lineart2}. In addition, obtaining the result requires knowledge of the classical trajectory of the scalar field, which is the solution of Eq.~\eqref{lineart}. As shown in~\cite{KirRub21}, taking into account the classical motion in PBH formation strongly affects the resulting PBH mass spectrum.

Let us compare this formula with the standard results obtained under the assumption that the classical component is stationary, i.e., $\Phi_\text{cl}(t_2)= \Phi_\text{cl}(t_1)$.
In this case, $M(t)=\mu t$, with the initial time $t_1=0$ and $\mu (t)=m^2/3H=const$. Then, formula \eqref{f2} simplifies considerably:
\begin{equation}
\langle \varphi^2\rangle=\frac{\sigma^2}{2\mu}(1-e^{-2\mu t}),
\end{equation}
which coincides with the result obtained in~\cite{Linde1990}.

\bigskip
{\bf List of abbreviations:}
\bigskip

BH --- black hole

CMB --- cosmic microwave background

DM --- dark matter

GR --- General Relativity

JWST --- James Webb Space Telescope

PBH --- primordial black hole

\bibliography{pbh-en}

\begin{thebibliography}{359}%
\makeatletter
\providecommand \@ifxundefined [1]{%
 \@ifx{#1\undefined}
}%
\providecommand \@ifnum [1]{%
 \ifnum #1\expandafter \@firstoftwo
 \else \expandafter \@secondoftwo
 \fi
}%
\providecommand \@ifx [1]{%
 \ifx #1\expandafter \@firstoftwo
 \else \expandafter \@secondoftwo
 \fi
}%
\providecommand \natexlab [1]{#1}%
\providecommand \enquote  [1]{``#1''}%
\providecommand \bibnamefont  [1]{#1}%
\providecommand \bibfnamefont [1]{#1}%
\providecommand \citenamefont [1]{#1}%
\providecommand \href@noop [0]{\@secondoftwo}%
\providecommand \href [0]{\begingroup \@sanitize@url \@href}%
\providecommand \@href[1]{\@@startlink{#1}\@@href}%
\providecommand \@@href[1]{\endgroup#1\@@endlink}%
\providecommand \@sanitize@url [0]{\catcode `\\12\catcode `\$12\catcode `\&12\catcode `\#12\catcode `\^12\catcode `\_12\catcode `\%12\relax}%
\providecommand \@@startlink[1]{}%
\providecommand \@@endlink[0]{}%
\providecommand \url  [0]{\begingroup\@sanitize@url \@url }%
\providecommand \@url [1]{\endgroup\@href {#1}{\urlprefix }}%
\providecommand \urlprefix  [0]{URL }%
\providecommand \Eprint [0]{\href }%
\providecommand \doibase [0]{http://dx.doi.org/}%
\providecommand \selectlanguage [0]{\@gobble}%
\providecommand \bibinfo  [0]{\@secondoftwo}%
\providecommand \bibfield  [0]{\@secondoftwo}%
\providecommand \translation [1]{[#1]}%
\providecommand \BibitemOpen [0]{}%
\providecommand \bibitemStop [0]{}%
\providecommand \bibitemNoStop [0]{.\EOS\space}%
\providecommand \EOS [0]{\spacefactor3000\relax}%
\providecommand \BibitemShut  [1]{\csname bibitem#1\endcsname}%
\let\auto@bib@innerbib\@empty
\bibitem [{\citenamefont {Reitze}(2017)}]{Rei17}%
  \BibitemOpen
  \bibfield  {author} {\bibinfo {author} {\bibfnamefont {D.~H.}\ \bibnamefont {Reitze}},\ }\href {\doibase 10.3367/UFNe.2016.11.038176} {\bibfield  {journal} {\bibinfo  {journal} {Phys. Usp.}\ }\textbf {\bibinfo {volume} {60}},\ \bibinfo {pages} {823} (\bibinfo {year} {2017})}\BibitemShut {NoStop}%
\bibitem [{\citenamefont {Dokuchaev}\ and\ \citenamefont {Nazarova}(2020)}]{DokNaz20}%
  \BibitemOpen
  \bibfield  {author} {\bibinfo {author} {\bibfnamefont {V.~I.}\ \bibnamefont {Dokuchaev}}\ and\ \bibinfo {author} {\bibfnamefont {N.~O.}\ \bibnamefont {Nazarova}},\ }\href {\doibase 10.3367/UFNe.2020.01.038717} {\bibfield  {journal} {\bibinfo  {journal} {Phys. Usp.}\ }\textbf {\bibinfo {volume} {63}},\ \bibinfo {pages} {583} (\bibinfo {year} {2020})}\BibitemShut {NoStop}%
\bibitem [{\citenamefont {Cherepashchuk}(2014)}]{Cherepashchuk:2014}%
  \BibitemOpen
  \bibfield  {author} {\bibinfo {author} {\bibfnamefont {A.~M.}\ \bibnamefont {Cherepashchuk}},\ }\href {\doibase 10.3367/UFNe.0184.201404d.0387} {\bibfield  {journal} {\bibinfo  {journal} {Phys. Usp.}\ }\textbf {\bibinfo {volume} {57}},\ \bibinfo {pages} {359} (\bibinfo {year} {2014})}\BibitemShut {NoStop}%
\bibitem [{\citenamefont {Dokuchaev}(1991)}]{Dok91}%
  \BibitemOpen
  \bibfield  {author} {\bibinfo {author} {\bibfnamefont {V.~I.}\ \bibnamefont {Dokuchaev}},\ }\href {\doibase 10.1070/PU1991v034n06ABEH002383} {\bibfield  {journal} {\bibinfo  {journal} {Sov. Phys. Usp.}\ }\textbf {\bibinfo {volume} {34}},\ \bibinfo {pages} {447} (\bibinfo {year} {1991})}\BibitemShut {NoStop}%
\bibitem [{\citenamefont {Dolgov}(2018)}]{Dol18}%
  \BibitemOpen
  \bibfield  {author} {\bibinfo {author} {\bibfnamefont {A.~D.}\ \bibnamefont {Dolgov}},\ }\href {\doibase 10.3367/UFNe.2017.06.038153} {\bibfield  {journal} {\bibinfo  {journal} {Phys. Usp.}\ }\textbf {\bibinfo {volume} {61}},\ \bibinfo {pages} {115} (\bibinfo {year} {2018})}\BibitemShut {NoStop}%
\bibitem [{\citenamefont {Carr}\ \emph {et~al.}(2024)\citenamefont {Carr}, \citenamefont {Clesse}, \citenamefont {Garcia-Bellido}, \citenamefont {Hawkins},\ and\ \citenamefont {Kuhnel}}]{2024PhR..1054....1C}%
  \BibitemOpen
  \bibfield  {author} {\bibinfo {author} {\bibfnamefont {B.}~\bibnamefont {Carr}}, \bibinfo {author} {\bibfnamefont {S.}~\bibnamefont {Clesse}}, \bibinfo {author} {\bibfnamefont {J.}~\bibnamefont {Garcia-Bellido}}, \bibinfo {author} {\bibfnamefont {M.}~\bibnamefont {Hawkins}}, \ and\ \bibinfo {author} {\bibfnamefont {F.}~\bibnamefont {Kuhnel}},\ }\href {\doibase 10.1016/j.physrep.2023.11.005} {\bibfield  {journal} {\bibinfo  {journal} {Phys. Rep.}\ }\textbf {\bibinfo {volume} {1054}},\ \bibinfo {pages} {1} (\bibinfo {year} {2024})},\ \Eprint {http://arxiv.org/abs/2306.03903} {arXiv:2306.03903 [astro-ph.CO]} \BibitemShut {NoStop}%
\bibitem [{\citenamefont {{Novikov}}(1965)}]{Nov64}%
  \BibitemOpen
  \bibfield  {author} {\bibinfo {author} {\bibfnamefont {I.~D.}\ \bibnamefont {{Novikov}}},\ }\href@noop {} {\bibfield  {journal} {\bibinfo  {journal} {Sov. Astron.}\ }\textbf {\bibinfo {volume} {8}},\ \bibinfo {pages} {857} (\bibinfo {year} {1965})}\BibitemShut {NoStop}%
\bibitem [{\citenamefont {{Ne'eman}}(1965)}]{1965ApJ...141.1303N}%
  \BibitemOpen
  \bibfield  {author} {\bibinfo {author} {\bibfnamefont {Y.}~\bibnamefont {{Ne'eman}}},\ }\href {\doibase 10.1086/148218} {\bibfield  {journal} {\bibinfo  {journal} {Astrophys. J.}\ }\textbf {\bibinfo {volume} {141}},\ \bibinfo {pages} {1303} (\bibinfo {year} {1965})}\BibitemShut {NoStop}%
\bibitem [{\citenamefont {{Ambartsumyan}}(1962)}]{1962PrCmg...8....2A}%
  \BibitemOpen
  \bibfield  {author} {\bibinfo {author} {\bibfnamefont {V.~A.}\ \bibnamefont {{Ambartsumyan}}},\ }\href@noop {} {\bibfield  {journal} {\bibinfo  {journal} {Problems of Cosmogeny}\ }\textbf {\bibinfo {volume} {8}},\ \bibinfo {pages} {2} (\bibinfo {year} {1962})}\BibitemShut {NoStop}%
\bibitem [{\citenamefont {{Zel'dovich}}\ and\ \citenamefont {{Novikov}}(1967)}]{ZelNov66}%
  \BibitemOpen
  \bibfield  {author} {\bibinfo {author} {\bibfnamefont {Y.~B.}\ \bibnamefont {{Zel'dovich}}}\ and\ \bibinfo {author} {\bibfnamefont {I.~D.}\ \bibnamefont {{Novikov}}},\ }\href@noop {} {\bibfield  {journal} {\bibinfo  {journal} {Sov. Astron.}\ }\textbf {\bibinfo {volume} {10}},\ \bibinfo {pages} {602} (\bibinfo {year} {1967})}\BibitemShut {NoStop}%
\bibitem [{\citenamefont {Hawking}(1971)}]{Haw71}%
  \BibitemOpen
  \bibfield  {author} {\bibinfo {author} {\bibfnamefont {S.}~\bibnamefont {Hawking}},\ }\href {\doibase 10.1093/mnras/152.1.75} {\bibfield  {journal} {\bibinfo  {journal} {Mon. Not. R. Astron. Soc.}\ }\textbf {\bibinfo {volume} {152}},\ \bibinfo {pages} {75} (\bibinfo {year} {1971})}\BibitemShut {NoStop}%
\bibitem [{\citenamefont {{Carr}}\ and\ \citenamefont {{Hawking}}(1974)}]{1974MNRAS.168..399C}%
  \BibitemOpen
  \bibfield  {author} {\bibinfo {author} {\bibfnamefont {B.~J.}\ \bibnamefont {{Carr}}}\ and\ \bibinfo {author} {\bibfnamefont {S.~W.}\ \bibnamefont {{Hawking}}},\ }\href {\doibase 10.1093/mnras/168.2.399} {\bibfield  {journal} {\bibinfo  {journal} {Mon. Not. R. Astron. Soc.}\ }\textbf {\bibinfo {volume} {168}},\ \bibinfo {pages} {399} (\bibinfo {year} {1974})}\BibitemShut {NoStop}%
\bibitem [{\citenamefont {{Hawking}}(1974)}]{1974Natur.248...30H}%
  \BibitemOpen
  \bibfield  {author} {\bibinfo {author} {\bibfnamefont {S.~W.}\ \bibnamefont {{Hawking}}},\ }\href {\doibase 10.1038/248030a0} {\bibfield  {journal} {\bibinfo  {journal} {Nature}\ }\textbf {\bibinfo {volume} {248}},\ \bibinfo {pages} {30} (\bibinfo {year} {1974})}\BibitemShut {NoStop}%
\bibitem [{\citenamefont {{Carr}}(1976)}]{Car76}%
  \BibitemOpen
  \bibfield  {author} {\bibinfo {author} {\bibfnamefont {B.~J.}\ \bibnamefont {{Carr}}},\ }\href {\doibase 10.1086/154351} {\bibfield  {journal} {\bibinfo  {journal} {Astrophys. J.}\ }\textbf {\bibinfo {volume} {206}},\ \bibinfo {pages} {8} (\bibinfo {year} {1976})}\BibitemShut {NoStop}%
\bibitem [{\citenamefont {{Carr}}\ \emph {et~al.}(2021)\citenamefont {{Carr}}, \citenamefont {{Kohri}}, \citenamefont {{Sendouda}},\ and\ \citenamefont {{Yokoyama}}}]{2021RPPh...84k6902C}%
  \BibitemOpen
  \bibfield  {author} {\bibinfo {author} {\bibfnamefont {B.}~\bibnamefont {{Carr}}}, \bibinfo {author} {\bibfnamefont {K.}~\bibnamefont {{Kohri}}}, \bibinfo {author} {\bibfnamefont {Y.}~\bibnamefont {{Sendouda}}}, \ and\ \bibinfo {author} {\bibfnamefont {J.}~\bibnamefont {{Yokoyama}}},\ }\href {\doibase 10.1088/1361-6633/ac1e31} {\bibfield  {journal} {\bibinfo  {journal} {Rept. Prog. Phys.}\ }\textbf {\bibinfo {volume} {84}},\ \bibinfo {eid} {116902} (\bibinfo {year} {2021})},\ \Eprint {http://arxiv.org/abs/2002.12778} {arXiv:2002.12778 [astro-ph.CO]} \BibitemShut {NoStop}%
\bibitem [{\citenamefont {{Auffinger}}(2023)}]{2023PrPNP.13104040A}%
  \BibitemOpen
  \bibfield  {author} {\bibinfo {author} {\bibfnamefont {J.}~\bibnamefont {{Auffinger}}},\ }\href {\doibase 10.1016/j.ppnp.2023.104040} {\bibfield  {journal} {\bibinfo  {journal} {Progress in Particle and Nuclear Physics}\ }\textbf {\bibinfo {volume} {131}},\ \bibinfo {eid} {104040} (\bibinfo {year} {2023})},\ \Eprint {http://arxiv.org/abs/2206.02672} {arXiv:2206.02672 [astro-ph.CO]} \BibitemShut {NoStop}%
\bibitem [{\citenamefont {Klipfel}\ and\ \citenamefont {Kaiser}(2025)}]{KliKai25}%
  \BibitemOpen
  \bibfield  {author} {\bibinfo {author} {\bibfnamefont {A.~P.}\ \bibnamefont {Klipfel}}\ and\ \bibinfo {author} {\bibfnamefont {D.~I.}\ \bibnamefont {Kaiser}},\ }\href {\doibase 10.1103/vnm4-7wdc} {\bibfield  {journal} {\bibinfo  {journal} {Phys. Rev. Lett.}\ }\textbf {\bibinfo {volume} {135}},\ \bibinfo {pages} {121003} (\bibinfo {year} {2025})},\ \Eprint {http://arxiv.org/abs/2503.19227} {arXiv:2503.19227 [hep-ph]} \BibitemShut {NoStop}%
\bibitem [{\citenamefont {{Nakamura}}\ \emph {et~al.}(1997)\citenamefont {{Nakamura}}, \citenamefont {{Sasaki}}, \citenamefont {{Tanaka}},\ and\ \citenamefont {{Thorne}}}]{1997ApJ...487L.139N}%
  \BibitemOpen
  \bibfield  {author} {\bibinfo {author} {\bibfnamefont {T.}~\bibnamefont {{Nakamura}}}, \bibinfo {author} {\bibfnamefont {M.}~\bibnamefont {{Sasaki}}}, \bibinfo {author} {\bibfnamefont {T.}~\bibnamefont {{Tanaka}}}, \ and\ \bibinfo {author} {\bibfnamefont {K.~S.}\ \bibnamefont {{Thorne}}},\ }\href {\doibase 10.1086/310886} {\bibfield  {journal} {\bibinfo  {journal} {Astrophys. J. Lett.}\ }\textbf {\bibinfo {volume} {487}},\ \bibinfo {pages} {L139} (\bibinfo {year} {1997})},\ \Eprint {http://arxiv.org/abs/astro-ph/9708060} {arXiv:astro-ph/9708060 [astro-ph]} \BibitemShut {NoStop}%
\bibitem [{\citenamefont {Ioka}\ \emph {et~al.}(1998)\citenamefont {Ioka}, \citenamefont {Chiba}, \citenamefont {Tanaka},\ and\ \citenamefont {Nakamura}}]{Ioka:1998nz}%
  \BibitemOpen
  \bibfield  {author} {\bibinfo {author} {\bibfnamefont {K.}~\bibnamefont {Ioka}}, \bibinfo {author} {\bibfnamefont {T.}~\bibnamefont {Chiba}}, \bibinfo {author} {\bibfnamefont {T.}~\bibnamefont {Tanaka}}, \ and\ \bibinfo {author} {\bibfnamefont {T.}~\bibnamefont {Nakamura}},\ }\href {\doibase 10.1103/PhysRevD.58.063003} {\bibfield  {journal} {\bibinfo  {journal} {Phys. Rev. D}\ }\textbf {\bibinfo {volume} {58}},\ \bibinfo {pages} {063003} (\bibinfo {year} {1998})},\ \Eprint {http://arxiv.org/abs/astro-ph/9807018} {arXiv:astro-ph/9807018} \BibitemShut {NoStop}%
\bibitem [{\citenamefont {Sasaki}\ \emph {et~al.}(2016)\citenamefont {Sasaki}, \citenamefont {Suyama}, \citenamefont {Tanaka},\ and\ \citenamefont {Yokoyama}}]{Sasaki:2016jop}%
  \BibitemOpen
  \bibfield  {author} {\bibinfo {author} {\bibfnamefont {M.}~\bibnamefont {Sasaki}}, \bibinfo {author} {\bibfnamefont {T.}~\bibnamefont {Suyama}}, \bibinfo {author} {\bibfnamefont {T.}~\bibnamefont {Tanaka}}, \ and\ \bibinfo {author} {\bibfnamefont {S.}~\bibnamefont {Yokoyama}},\ }\href {\doibase 10.1103/PhysRevLett.117.061101} {\bibfield  {journal} {\bibinfo  {journal} {Phys. Rev. Lett.}\ }\textbf {\bibinfo {volume} {117}},\ \bibinfo {pages} {061101} (\bibinfo {year} {2016})},\ \bibinfo {note} {[Erratum: Phys.Rev.Lett. 121, 059901 (2018)]},\ \Eprint {http://arxiv.org/abs/1603.08338} {arXiv:1603.08338 [astro-ph.CO]} \BibitemShut {NoStop}%
\bibitem [{\citenamefont {Postnov}\ \emph {et~al.}(2019)\citenamefont {Postnov}, \citenamefont {Kuranov},\ and\ \citenamefont {Mitichkin}}]{PosKurMit19}%
  \BibitemOpen
  \bibfield  {author} {\bibinfo {author} {\bibfnamefont {K.~A.}\ \bibnamefont {Postnov}}, \bibinfo {author} {\bibfnamefont {A.~G.}\ \bibnamefont {Kuranov}}, \ and\ \bibinfo {author} {\bibfnamefont {N.~A.}\ \bibnamefont {Mitichkin}},\ }\href {\doibase 10.3367/UFNe.2019.04.038593} {\bibfield  {journal} {\bibinfo  {journal} {Phys. Usp.}\ }\textbf {\bibinfo {volume} {62}},\ \bibinfo {pages} {1153} (\bibinfo {year} {2019})}\BibitemShut {NoStop}%
\bibitem [{\citenamefont {Carr}\ and\ \citenamefont {Kuhnel}(2022)}]{Carr:2021bzv}%
  \BibitemOpen
  \bibfield  {author} {\bibinfo {author} {\bibfnamefont {B.}~\bibnamefont {Carr}}\ and\ \bibinfo {author} {\bibfnamefont {F.}~\bibnamefont {Kuhnel}},\ }\href {\doibase 10.21468/SciPostPhysLectNotes.48} {\bibfield  {journal} {\bibinfo  {journal} {SciPost Phys. Lect. Notes}\ }\textbf {\bibinfo {volume} {48}},\ \bibinfo {pages} {1} (\bibinfo {year} {2022})},\ \Eprint {http://arxiv.org/abs/2110.02821} {arXiv:2110.02821 [astro-ph.CO]} \BibitemShut {NoStop}%
\bibitem [{\citenamefont {Green}(2026)}]{Gre26}%
  \BibitemOpen
  \bibfield  {author} {\bibinfo {author} {\bibfnamefont {A.~M.}\ \bibnamefont {Green}},\ }\href {\doibase 10.3390/universe12050123} {\bibfield  {journal} {\bibinfo  {journal} {Universe}\ }\textbf {\bibinfo {volume} {12}},\ \bibinfo {pages} {123} (\bibinfo {year} {2026})},\ \Eprint {http://arxiv.org/abs/2602.15974} {arXiv:2602.15974 [astro-ph.GA]} \BibitemShut {NoStop}%
\bibitem [{\citenamefont {Blinnikov}\ \emph {et~al.}(2016)\citenamefont {Blinnikov}, \citenamefont {Dolgov}, \citenamefont {Porayko},\ and\ \citenamefont {Postnov}}]{Blietal16}%
  \BibitemOpen
  \bibfield  {author} {\bibinfo {author} {\bibfnamefont {S.}~\bibnamefont {Blinnikov}}, \bibinfo {author} {\bibfnamefont {A.}~\bibnamefont {Dolgov}}, \bibinfo {author} {\bibfnamefont {N.~K.}\ \bibnamefont {Porayko}}, \ and\ \bibinfo {author} {\bibfnamefont {K.}~\bibnamefont {Postnov}},\ }\href {\doibase 10.1088/1475-7516/2016/11/036} {\bibfield  {journal} {\bibinfo  {journal} {J. Cosmol. Astropart. Phys.}\ }\textbf {\bibinfo {volume} {11}},\ \bibinfo {pages} {036} (\bibinfo {year} {2016})},\ \Eprint {http://arxiv.org/abs/1611.00541} {arXiv:1611.00541 [astro-ph.HE]} \BibitemShut {NoStop}%
\bibitem [{\citenamefont {Ziparo}\ \emph {et~al.}(2025)\citenamefont {Ziparo}, \citenamefont {Gallerani},\ and\ \citenamefont {Ferrara}}]{Zipetal25}%
  \BibitemOpen
  \bibfield  {author} {\bibinfo {author} {\bibfnamefont {F.}~\bibnamefont {Ziparo}}, \bibinfo {author} {\bibfnamefont {S.}~\bibnamefont {Gallerani}}, \ and\ \bibinfo {author} {\bibfnamefont {A.}~\bibnamefont {Ferrara}},\ }\href {\doibase 10.1088/1475-7516/2025/04/040} {\bibfield  {journal} {\bibinfo  {journal} {J. Cosmol. Astropart. Phys.}\ }\textbf {\bibinfo {volume} {04}},\ \bibinfo {pages} {040} (\bibinfo {year} {2025})},\ \Eprint {http://arxiv.org/abs/2411.03448} {arXiv:2411.03448 [astro-ph.CO]} \BibitemShut {NoStop}%
\bibitem [{\citenamefont {Khlopov}\ and\ \citenamefont {Polnarev}(1980)}]{KhlPol80}%
  \BibitemOpen
  \bibfield  {author} {\bibinfo {author} {\bibfnamefont {M.~Y.}\ \bibnamefont {Khlopov}}\ and\ \bibinfo {author} {\bibfnamefont {A.~G.}\ \bibnamefont {Polnarev}},\ }\href {\doibase 10.1016/0370-2693(80)90624-3} {\bibfield  {journal} {\bibinfo  {journal} {Phys. Lett. B}\ }\textbf {\bibinfo {volume} {97}},\ \bibinfo {pages} {383} (\bibinfo {year} {1980})}\BibitemShut {NoStop}%
\bibitem [{\citenamefont {{Polnarev}}\ and\ \citenamefont {{Khlopov}}(1982)}]{PolKhl82}%
  \BibitemOpen
  \bibfield  {author} {\bibinfo {author} {\bibfnamefont {A.~G.}\ \bibnamefont {{Polnarev}}}\ and\ \bibinfo {author} {\bibfnamefont {M.~Y.}\ \bibnamefont {{Khlopov}}},\ }\href@noop {} {\bibfield  {journal} {\bibinfo  {journal} {Sov. Astron.}\ }\textbf {\bibinfo {volume} {26}},\ \bibinfo {pages} {391} (\bibinfo {year} {1982})}\BibitemShut {NoStop}%
\bibitem [{\citenamefont {Polnarev}\ and\ \citenamefont {Khlopov}(1985)}]{PolKhl85}%
  \BibitemOpen
  \bibfield  {author} {\bibinfo {author} {\bibfnamefont {A.~G.}\ \bibnamefont {Polnarev}}\ and\ \bibinfo {author} {\bibfnamefont {M.~Y.}\ \bibnamefont {Khlopov}},\ }\href {\doibase 10.1070/PU1985v028n03ABEH003858} {\bibfield  {journal} {\bibinfo  {journal} {Sov. Phys. Usp.}\ }\textbf {\bibinfo {volume} {28}},\ \bibinfo {pages} {213} (\bibinfo {year} {1985})}\BibitemShut {NoStop}%
\bibitem [{\citenamefont {Sato}\ \emph {et~al.}(1981)\citenamefont {Sato}, \citenamefont {Sasaki}, \citenamefont {Kodama},\ and\ \citenamefont {Maeda}}]{Satetal81}%
  \BibitemOpen
  \bibfield  {author} {\bibinfo {author} {\bibfnamefont {K.}~\bibnamefont {Sato}}, \bibinfo {author} {\bibfnamefont {M.}~\bibnamefont {Sasaki}}, \bibinfo {author} {\bibfnamefont {H.}~\bibnamefont {Kodama}}, \ and\ \bibinfo {author} {\bibfnamefont {K.-i.}\ \bibnamefont {Maeda}},\ }\href {\doibase 10.1143/PTP.65.1443} {\bibfield  {journal} {\bibinfo  {journal} {Prog. Theor. Phys.}\ }\textbf {\bibinfo {volume} {65}},\ \bibinfo {pages} {1443} (\bibinfo {year} {1981})}\BibitemShut {NoStop}%
\bibitem [{\citenamefont {Maeda}\ \emph {et~al.}(1982)\citenamefont {Maeda}, \citenamefont {Sato}, \citenamefont {Sasaki},\ and\ \citenamefont {Kodama}}]{Maeetal82}%
  \BibitemOpen
  \bibfield  {author} {\bibinfo {author} {\bibfnamefont {K.-i.}\ \bibnamefont {Maeda}}, \bibinfo {author} {\bibfnamefont {K.}~\bibnamefont {Sato}}, \bibinfo {author} {\bibfnamefont {M.}~\bibnamefont {Sasaki}}, \ and\ \bibinfo {author} {\bibfnamefont {H.}~\bibnamefont {Kodama}},\ }\href {\doibase 10.1016/0370-2693(82)91151-0} {\bibfield  {journal} {\bibinfo  {journal} {Phys. Lett. B}\ }\textbf {\bibinfo {volume} {108}},\ \bibinfo {pages} {98} (\bibinfo {year} {1982})}\BibitemShut {NoStop}%
\bibitem [{\citenamefont {Berezin}\ \emph {et~al.}(1983)\citenamefont {Berezin}, \citenamefont {Kuzmin},\ and\ \citenamefont {Tkachev}}]{BerKuzTka83}%
  \BibitemOpen
  \bibfield  {author} {\bibinfo {author} {\bibfnamefont {V.~A.}\ \bibnamefont {Berezin}}, \bibinfo {author} {\bibfnamefont {V.~A.}\ \bibnamefont {Kuzmin}}, \ and\ \bibinfo {author} {\bibfnamefont {I.~I.}\ \bibnamefont {Tkachev}},\ }\href {\doibase 10.1016/0370-2693(83)90630-5} {\bibfield  {journal} {\bibinfo  {journal} {Phys. Lett. B}\ }\textbf {\bibinfo {volume} {120}},\ \bibinfo {pages} {91} (\bibinfo {year} {1983})}\BibitemShut {NoStop}%
\bibitem [{\citenamefont {Rubin}\ \emph {et~al.}(2000)\citenamefont {Rubin}, \citenamefont {Khlopov},\ and\ \citenamefont {Sakharov}}]{RubKhlSak00}%
  \BibitemOpen
  \bibfield  {author} {\bibinfo {author} {\bibfnamefont {S.~G.}\ \bibnamefont {Rubin}}, \bibinfo {author} {\bibfnamefont {M.~Y.}\ \bibnamefont {Khlopov}}, \ and\ \bibinfo {author} {\bibfnamefont {A.~S.}\ \bibnamefont {Sakharov}},\ }\href@noop {} {\bibfield  {journal} {\bibinfo  {journal} {Grav. Cosmol.}\ }\textbf {\bibinfo {volume} {6}},\ \bibinfo {pages} {51} (\bibinfo {year} {2000})},\ \Eprint {http://arxiv.org/abs/hep-ph/0005271} {arXiv:hep-ph/0005271} \BibitemShut {NoStop}%
\bibitem [{\citenamefont {Rubin}\ \emph {et~al.}(2001)\citenamefont {Rubin}, \citenamefont {Sakharov},\ and\ \citenamefont {Khlopov}}]{RubSakKhl01}%
  \BibitemOpen
  \bibfield  {author} {\bibinfo {author} {\bibfnamefont {S.~G.}\ \bibnamefont {Rubin}}, \bibinfo {author} {\bibfnamefont {A.~S.}\ \bibnamefont {Sakharov}}, \ and\ \bibinfo {author} {\bibfnamefont {M.~Y.}\ \bibnamefont {Khlopov}},\ }\href {\doibase 10.1134/1.1385631} {\bibfield  {journal} {\bibinfo  {journal} {J. Exp. Theor. Phys.}\ }\textbf {\bibinfo {volume} {91}},\ \bibinfo {pages} {921} (\bibinfo {year} {2001})},\ \Eprint {http://arxiv.org/abs/hep-ph/0106187} {arXiv:hep-ph/0106187} \BibitemShut {NoStop}%
\bibitem [{\citenamefont {{Dolgov}}\ and\ \citenamefont {{Silk}}(1993)}]{DolSil93}%
  \BibitemOpen
  \bibfield  {author} {\bibinfo {author} {\bibfnamefont {A.}~\bibnamefont {{Dolgov}}}\ and\ \bibinfo {author} {\bibfnamefont {J.}~\bibnamefont {{Silk}}},\ }\href {\doibase 10.1103/PhysRevD.47.4244} {\bibfield  {journal} {\bibinfo  {journal} {Phys. Rev.~D}\ }\textbf {\bibinfo {volume} {47}},\ \bibinfo {pages} {4244} (\bibinfo {year} {1993})}\BibitemShut {NoStop}%
\bibitem [{\citenamefont {{Kavanagh}}(2019)}]{Kav19}%
  \BibitemOpen
  \bibfield  {author} {\bibinfo {author} {\bibfnamefont {B.~J.}\ \bibnamefont {{Kavanagh}}},\ }\href {\doibase 10.5281/zenodo.3538999} {\enquote {\bibinfo {title} {bradkav/pbhbounds: Release version v1.0},}\ } (\bibinfo {year} {2019}),\ \bibinfo {note} {doi: \url{10.5281/zenodo.3538999}}\BibitemShut {NoStop}%
\bibitem [{\citenamefont {Aghanim}\ \emph {et~al.}(2020)\citenamefont {Aghanim} \emph {et~al.}}]{Planck:2018vyg}%
  \BibitemOpen
  \bibfield  {author} {\bibinfo {author} {\bibfnamefont {N.}~\bibnamefont {Aghanim}} \emph {et~al.} (\bibinfo {collaboration} {Planck}),\ }\href {\doibase 10.1051/0004-6361/201833910} {\bibfield  {journal} {\bibinfo  {journal} {Astron. Astrophys.}\ }\textbf {\bibinfo {volume} {641}},\ \bibinfo {pages} {A6} (\bibinfo {year} {2020})},\ \bibinfo {note} {[Erratum: Astron.Astrophys. 652, C4 (2021)]},\ \Eprint {http://arxiv.org/abs/1807.06209} {arXiv:1807.06209 [astro-ph.CO]} \BibitemShut {NoStop}%
\bibitem [{\citenamefont {{Gorbunov}}\ and\ \citenamefont {{Rubakov}}(2011)}]{GorRub}%
  \BibitemOpen
  \bibfield  {author} {\bibinfo {author} {\bibfnamefont {D.~S.}\ \bibnamefont {{Gorbunov}}}\ and\ \bibinfo {author} {\bibfnamefont {V.~A.}\ \bibnamefont {{Rubakov}}},\ }\href {\doibase 10.1142/7873} {\emph {\bibinfo {title} {{Introduction to the Theory of the Early Universe: Cosmological Perturbations and Inflationary Theory}}}}\ (\bibinfo {year} {2011})\BibitemShut {NoStop}%
\bibitem [{\citenamefont {Carr}(1975)}]{Car75}%
  \BibitemOpen
  \bibfield  {author} {\bibinfo {author} {\bibfnamefont {B.~J.}\ \bibnamefont {Carr}},\ }\href {\doibase 10.1086/153853} {\bibfield  {journal} {\bibinfo  {journal} {Astrophys. J.}\ }\textbf {\bibinfo {volume} {201}},\ \bibinfo {pages} {1} (\bibinfo {year} {1975})}\BibitemShut {NoStop}%
\bibitem [{\citenamefont {Yokoyama}(1997)}]{Yok97}%
  \BibitemOpen
  \bibfield  {author} {\bibinfo {author} {\bibfnamefont {J.}~\bibnamefont {Yokoyama}},\ }\href@noop {} {\bibfield  {journal} {\bibinfo  {journal} {Astron. Astrophys.}\ }\textbf {\bibinfo {volume} {318}},\ \bibinfo {pages} {673} (\bibinfo {year} {1997})},\ \Eprint {http://arxiv.org/abs/astro-ph/9509027} {arXiv:astro-ph/9509027} \BibitemShut {NoStop}%
\bibitem [{\citenamefont {Berezin}\ \emph {et~al.}(2020)\citenamefont {Berezin}, \citenamefont {Dokuchaev}, \citenamefont {Eroshenko},\ and\ \citenamefont {Smirnov}}]{Berezin:2020tbk}%
  \BibitemOpen
  \bibfield  {author} {\bibinfo {author} {\bibfnamefont {V.~A.}\ \bibnamefont {Berezin}}, \bibinfo {author} {\bibfnamefont {V.~I.}\ \bibnamefont {Dokuchaev}}, \bibinfo {author} {\bibfnamefont {Y.~N.}\ \bibnamefont {Eroshenko}}, \ and\ \bibinfo {author} {\bibfnamefont {A.~L.}\ \bibnamefont {Smirnov}},\ }\href {\doibase 10.3390/universe6100158} {\bibfield  {journal} {\bibinfo  {journal} {Universe}\ }\textbf {\bibinfo {volume} {6}},\ \bibinfo {pages} {158} (\bibinfo {year} {2020})},\ \Eprint {http://arxiv.org/abs/2007.15347} {arXiv:2007.15347 [astro-ph.CO]} \BibitemShut {NoStop}%
\bibitem [{\citenamefont {Harrison}(1970)}]{Har70}%
  \BibitemOpen
  \bibfield  {author} {\bibinfo {author} {\bibfnamefont {E.~R.}\ \bibnamefont {Harrison}},\ }\href {\doibase 10.1103/PhysRevD.1.2726} {\bibfield  {journal} {\bibinfo  {journal} {Phys. Rev. D}\ }\textbf {\bibinfo {volume} {1}},\ \bibinfo {pages} {2726} (\bibinfo {year} {1970})}\BibitemShut {NoStop}%
\bibitem [{\citenamefont {Landau}\ and\ \citenamefont {Lifshitz}(1975)}]{LL-II}%
  \BibitemOpen
  \bibfield  {author} {\bibinfo {author} {\bibfnamefont {L.~D.}\ \bibnamefont {Landau}}\ and\ \bibinfo {author} {\bibfnamefont {E.~M.}\ \bibnamefont {Lifshitz}},\ }\href@noop {} {\emph {\bibinfo {title} {The classical theory of fields}}}\ (\bibinfo  {publisher} {Pergamon Press},\ \bibinfo {year} {1975})\BibitemShut {NoStop}%
\bibitem [{\citenamefont {Kopp}\ \emph {et~al.}(2011)\citenamefont {Kopp}, \citenamefont {Hofmann},\ and\ \citenamefont {Weller}}]{Kopp:2010sh}%
  \BibitemOpen
  \bibfield  {author} {\bibinfo {author} {\bibfnamefont {M.}~\bibnamefont {Kopp}}, \bibinfo {author} {\bibfnamefont {S.}~\bibnamefont {Hofmann}}, \ and\ \bibinfo {author} {\bibfnamefont {J.}~\bibnamefont {Weller}},\ }\href {\doibase 10.1103/PhysRevD.83.124025} {\bibfield  {journal} {\bibinfo  {journal} {Phys. Rev. D}\ }\textbf {\bibinfo {volume} {83}},\ \bibinfo {pages} {124025} (\bibinfo {year} {2011})},\ \Eprint {http://arxiv.org/abs/1012.4369} {arXiv:1012.4369 [astro-ph.CO]} \BibitemShut {NoStop}%
\bibitem [{\citenamefont {Carr}\ and\ \citenamefont {Harada}(2015)}]{Carr:2014pga}%
  \BibitemOpen
  \bibfield  {author} {\bibinfo {author} {\bibfnamefont {B.~J.}\ \bibnamefont {Carr}}\ and\ \bibinfo {author} {\bibfnamefont {T.}~\bibnamefont {Harada}},\ }\href {\doibase 10.1103/PhysRevD.91.084048} {\bibfield  {journal} {\bibinfo  {journal} {Phys. Rev. D}\ }\textbf {\bibinfo {volume} {91}},\ \bibinfo {pages} {084048} (\bibinfo {year} {2015})},\ \Eprint {http://arxiv.org/abs/1405.3624} {arXiv:1405.3624 [astro-ph.CO]} \BibitemShut {NoStop}%
\bibitem [{\citenamefont {Harada}\ \emph {et~al.}(2013)\citenamefont {Harada}, \citenamefont {Yoo},\ and\ \citenamefont {Kohri}}]{HarYooKoh14}%
  \BibitemOpen
  \bibfield  {author} {\bibinfo {author} {\bibfnamefont {T.}~\bibnamefont {Harada}}, \bibinfo {author} {\bibfnamefont {C.-M.}\ \bibnamefont {Yoo}}, \ and\ \bibinfo {author} {\bibfnamefont {K.}~\bibnamefont {Kohri}},\ }\href {\doibase 10.1103/PhysRevD.88.084051} {\bibfield  {journal} {\bibinfo  {journal} {Phys. Rev. D}\ }\textbf {\bibinfo {volume} {88}},\ \bibinfo {pages} {084051} (\bibinfo {year} {2013})},\ \bibinfo {note} {[Erratum: Phys.Rev.D 89, 029903 (2014)]},\ \Eprint {http://arxiv.org/abs/1309.4201} {arXiv:1309.4201 [astro-ph.CO]} \BibitemShut {NoStop}%
\bibitem [{\citenamefont {Zabotin}\ \emph {et~al.}(1982)\citenamefont {Zabotin}, \citenamefont {Marochnik},\ and\ \citenamefont {Nasel'skii}}]{ZabMarNas82}%
  \BibitemOpen
  \bibfield  {author} {\bibinfo {author} {\bibfnamefont {N.~A.}\ \bibnamefont {Zabotin}}, \bibinfo {author} {\bibfnamefont {L.~S.}\ \bibnamefont {Marochnik}}, \ and\ \bibinfo {author} {\bibfnamefont {P.~D.}\ \bibnamefont {Nasel'skii}},\ }\href {\doibase 10.1007/BF01004475} {\bibfield  {journal} {\bibinfo  {journal} {Astrophysics}\ }\textbf {\bibinfo {volume} {18}},\ \bibinfo {pages} {97} (\bibinfo {year} {1982})}\BibitemShut {NoStop}%
\bibitem [{\citenamefont {Bullock}\ and\ \citenamefont {Primack}(1997)}]{BulPri97}%
  \BibitemOpen
  \bibfield  {author} {\bibinfo {author} {\bibfnamefont {J.~S.}\ \bibnamefont {Bullock}}\ and\ \bibinfo {author} {\bibfnamefont {J.~R.}\ \bibnamefont {Primack}},\ }\href {\doibase 10.1103/PhysRevD.55.7423} {\bibfield  {journal} {\bibinfo  {journal} {Phys. Rev. D}\ }\textbf {\bibinfo {volume} {55}},\ \bibinfo {pages} {7423} (\bibinfo {year} {1997})},\ \Eprint {http://arxiv.org/abs/astro-ph/9611106} {arXiv:astro-ph/9611106} \BibitemShut {NoStop}%
\bibitem [{\citenamefont {{Ivanov}}(1998)}]{1998PhRvD..57.7145I}%
  \BibitemOpen
  \bibfield  {author} {\bibinfo {author} {\bibfnamefont {P.}~\bibnamefont {{Ivanov}}},\ }\href {\doibase 10.1103/PhysRevD.57.7145} {\bibfield  {journal} {\bibinfo  {journal} {Phys. Rev. D}\ }\textbf {\bibinfo {volume} {57}},\ \bibinfo {pages} {7145} (\bibinfo {year} {1998})},\ \Eprint {http://arxiv.org/abs/astro-ph/9708224} {arXiv:astro-ph/9708224 [astro-ph]} \BibitemShut {NoStop}%
\bibitem [{\citenamefont {Pattison}\ \emph {et~al.}(2017)\citenamefont {Pattison}, \citenamefont {Vennin}, \citenamefont {Assadullahi},\ and\ \citenamefont {Wands}}]{2017JCAP...10..046P}%
  \BibitemOpen
  \bibfield  {author} {\bibinfo {author} {\bibfnamefont {C.}~\bibnamefont {Pattison}}, \bibinfo {author} {\bibfnamefont {V.}~\bibnamefont {Vennin}}, \bibinfo {author} {\bibfnamefont {H.}~\bibnamefont {Assadullahi}}, \ and\ \bibinfo {author} {\bibfnamefont {D.}~\bibnamefont {Wands}},\ }\href {\doibase 10.1088/1475-7516/2017/10/046} {\bibfield  {journal} {\bibinfo  {journal} {J. Cosmol. Astropart. Phys.}\ }\textbf {\bibinfo {volume} {10}},\ \bibinfo {pages} {046} (\bibinfo {year} {2017})},\ \Eprint {http://arxiv.org/abs/1707.00537} {arXiv:1707.00537 [hep-th]} \BibitemShut {NoStop}%
\bibitem [{\citenamefont {{Nadezhin}}\ \emph {et~al.}(1978)\citenamefont {{Nadezhin}}, \citenamefont {{Novikov}},\ and\ \citenamefont {{Polnarev}}}]{NadNovPol78}%
  \BibitemOpen
  \bibfield  {author} {\bibinfo {author} {\bibfnamefont {D.~K.}\ \bibnamefont {{Nadezhin}}}, \bibinfo {author} {\bibfnamefont {I.~D.}\ \bibnamefont {{Novikov}}}, \ and\ \bibinfo {author} {\bibfnamefont {A.~G.}\ \bibnamefont {{Polnarev}}},\ }\href@noop {} {\bibfield  {journal} {\bibinfo  {journal} {Sov. Astron.}\ }\textbf {\bibinfo {volume} {22}},\ \bibinfo {pages} {129} (\bibinfo {year} {1978})}\BibitemShut {NoStop}%
\bibitem [{\citenamefont {{Novikov}}\ \emph {et~al.}(1979)\citenamefont {{Novikov}}, \citenamefont {{Polnarev}}, \citenamefont {{Starobinskii}},\ and\ \citenamefont {{Zel'dovich}}}]{Novetal79}%
  \BibitemOpen
  \bibfield  {author} {\bibinfo {author} {\bibfnamefont {I.~D.}\ \bibnamefont {{Novikov}}}, \bibinfo {author} {\bibfnamefont {A.~G.}\ \bibnamefont {{Polnarev}}}, \bibinfo {author} {\bibfnamefont {A.~A.}\ \bibnamefont {{Starobinskii}}}, \ and\ \bibinfo {author} {\bibfnamefont {Y.~B.}\ \bibnamefont {{Zel'dovich}}},\ }\href@noop {} {\bibfield  {journal} {\bibinfo  {journal} {Astron. \& Astrophys.}\ }\textbf {\bibinfo {volume} {80}},\ \bibinfo {pages} {104} (\bibinfo {year} {1979})}\BibitemShut {NoStop}%
\bibitem [{\citenamefont {{Bicknell}}\ and\ \citenamefont {{Henriksen}}(1979)}]{BicHen79}%
  \BibitemOpen
  \bibfield  {author} {\bibinfo {author} {\bibfnamefont {G.~V.}\ \bibnamefont {{Bicknell}}}\ and\ \bibinfo {author} {\bibfnamefont {R.~N.}\ \bibnamefont {{Henriksen}}},\ }\href {\doibase 10.1086/157325} {\bibfield  {journal} {\bibinfo  {journal} {\apj}\ }\textbf {\bibinfo {volume} {232}},\ \bibinfo {pages} {670} (\bibinfo {year} {1979})}\BibitemShut {NoStop}%
\bibitem [{\citenamefont {Shibata}\ and\ \citenamefont {Sasaki}(1999)}]{ShiSas99}%
  \BibitemOpen
  \bibfield  {author} {\bibinfo {author} {\bibfnamefont {M.}~\bibnamefont {Shibata}}\ and\ \bibinfo {author} {\bibfnamefont {M.}~\bibnamefont {Sasaki}},\ }\href {\doibase 10.1103/PhysRevD.60.084002} {\bibfield  {journal} {\bibinfo  {journal} {Phys. Rev. D}\ }\textbf {\bibinfo {volume} {60}},\ \bibinfo {pages} {084002} (\bibinfo {year} {1999})},\ \Eprint {http://arxiv.org/abs/gr-qc/9905064} {arXiv:gr-qc/9905064} \BibitemShut {NoStop}%
\bibitem [{\citenamefont {Musco}(2019)}]{Musco:2018rwt}%
  \BibitemOpen
  \bibfield  {author} {\bibinfo {author} {\bibfnamefont {I.}~\bibnamefont {Musco}},\ }\href {\doibase 10.1103/PhysRevD.100.123524} {\bibfield  {journal} {\bibinfo  {journal} {Phys. Rev. D}\ }\textbf {\bibinfo {volume} {100}},\ \bibinfo {pages} {123524} (\bibinfo {year} {2019})},\ \Eprint {http://arxiv.org/abs/1809.02127} {arXiv:1809.02127 [gr-qc]} \BibitemShut {NoStop}%
\bibitem [{\citenamefont {Joana}\ and\ \citenamefont {Yuwen}(2026)}]{JoaYuw25}%
  \BibitemOpen
  \bibfield  {author} {\bibinfo {author} {\bibfnamefont {C.}~\bibnamefont {Joana}}\ and\ \bibinfo {author} {\bibfnamefont {Z.-Y.}\ \bibnamefont {Yuwen}},\ }\href {\doibase 10.1103/j3hw-d5cx} {\bibfield  {journal} {\bibinfo  {journal} {Phys. Rev. D}\ }\textbf {\bibinfo {volume} {113}},\ \bibinfo {pages} {023518} (\bibinfo {year} {2026})},\ \Eprint {http://arxiv.org/abs/2510.11611} {arXiv:2510.11611 [astro-ph.CO]} \BibitemShut {NoStop}%
\bibitem [{\citenamefont {Nakama}\ \emph {et~al.}(2014)\citenamefont {Nakama}, \citenamefont {Harada}, \citenamefont {Polnarev},\ and\ \citenamefont {Yokoyama}}]{2014JCAP...01..037N}%
  \BibitemOpen
  \bibfield  {author} {\bibinfo {author} {\bibfnamefont {T.}~\bibnamefont {Nakama}}, \bibinfo {author} {\bibfnamefont {T.}~\bibnamefont {Harada}}, \bibinfo {author} {\bibfnamefont {A.~G.}\ \bibnamefont {Polnarev}}, \ and\ \bibinfo {author} {\bibfnamefont {J.}~\bibnamefont {Yokoyama}},\ }\href {\doibase 10.1088/1475-7516/2014/01/037} {\bibfield  {journal} {\bibinfo  {journal} {J. Cosmol. Astropart. Phys.}\ }\textbf {\bibinfo {volume} {01}},\ \bibinfo {pages} {037} (\bibinfo {year} {2014})},\ \Eprint {http://arxiv.org/abs/1310.3007} {arXiv:1310.3007 [gr-qc]} \BibitemShut {NoStop}%
\bibitem [{\citenamefont {Niemeyer}\ and\ \citenamefont {Jedamzik}(1999)}]{NieJed99}%
  \BibitemOpen
  \bibfield  {author} {\bibinfo {author} {\bibfnamefont {J.~C.}\ \bibnamefont {Niemeyer}}\ and\ \bibinfo {author} {\bibfnamefont {K.}~\bibnamefont {Jedamzik}},\ }\href {\doibase 10.1103/PhysRevD.59.124013} {\bibfield  {journal} {\bibinfo  {journal} {Phys. Rev. D}\ }\textbf {\bibinfo {volume} {59}},\ \bibinfo {pages} {124013} (\bibinfo {year} {1999})},\ \Eprint {http://arxiv.org/abs/astro-ph/9901292} {arXiv:astro-ph/9901292} \BibitemShut {NoStop}%
\bibitem [{\citenamefont {{K{\"u}hnel}}\ and\ \citenamefont {{Sandstad}}(2016)}]{2016PhRvD..94f3514K}%
  \BibitemOpen
  \bibfield  {author} {\bibinfo {author} {\bibfnamefont {F.}~\bibnamefont {{K{\"u}hnel}}}\ and\ \bibinfo {author} {\bibfnamefont {M.}~\bibnamefont {{Sandstad}}},\ }\href {\doibase 10.1103/PhysRevD.94.063514} {\bibfield  {journal} {\bibinfo  {journal} {Phys. Rev. D}\ }\textbf {\bibinfo {volume} {94}},\ \bibinfo {eid} {063514} (\bibinfo {year} {2016})},\ \Eprint {http://arxiv.org/abs/1602.04815} {arXiv:1602.04815 [astro-ph.CO]} \BibitemShut {NoStop}%
\bibitem [{\citenamefont {Escriv{\`a}}\ \emph {et~al.}(2020)\citenamefont {Escriv{\`a}}, \citenamefont {Germani},\ and\ \citenamefont {Sheth}}]{EscGerShe20}%
  \BibitemOpen
  \bibfield  {author} {\bibinfo {author} {\bibfnamefont {A.}~\bibnamefont {Escriv{\`a}}}, \bibinfo {author} {\bibfnamefont {C.}~\bibnamefont {Germani}}, \ and\ \bibinfo {author} {\bibfnamefont {R.~K.}\ \bibnamefont {Sheth}},\ }\href {\doibase 10.1103/PhysRevD.101.044022} {\bibfield  {journal} {\bibinfo  {journal} {Phys. Rev. D}\ }\textbf {\bibinfo {volume} {101}},\ \bibinfo {pages} {044022} (\bibinfo {year} {2020})},\ \Eprint {http://arxiv.org/abs/1907.13311} {arXiv:1907.13311 [gr-qc]} \BibitemShut {NoStop}%
\bibitem [{\citenamefont {Musco}\ \emph {et~al.}(2021)\citenamefont {Musco}, \citenamefont {De~Luca}, \citenamefont {Franciolini},\ and\ \citenamefont {Riotto}}]{Musetal20}%
  \BibitemOpen
  \bibfield  {author} {\bibinfo {author} {\bibfnamefont {I.}~\bibnamefont {Musco}}, \bibinfo {author} {\bibfnamefont {V.}~\bibnamefont {De~Luca}}, \bibinfo {author} {\bibfnamefont {G.}~\bibnamefont {Franciolini}}, \ and\ \bibinfo {author} {\bibfnamefont {A.}~\bibnamefont {Riotto}},\ }\href {\doibase 10.1103/PhysRevD.103.063538} {\bibfield  {journal} {\bibinfo  {journal} {Phys. Rev. D}\ }\textbf {\bibinfo {volume} {103}},\ \bibinfo {pages} {063538} (\bibinfo {year} {2021})},\ \Eprint {http://arxiv.org/abs/2011.03014} {arXiv:2011.03014 [astro-ph.CO]} \BibitemShut {NoStop}%
\bibitem [{\citenamefont {Niemeyer}\ and\ \citenamefont {Jedamzik}(1998)}]{NieJed98}%
  \BibitemOpen
  \bibfield  {author} {\bibinfo {author} {\bibfnamefont {J.~C.}\ \bibnamefont {Niemeyer}}\ and\ \bibinfo {author} {\bibfnamefont {K.}~\bibnamefont {Jedamzik}},\ }\href {\doibase 10.1103/PhysRevLett.80.5481} {\bibfield  {journal} {\bibinfo  {journal} {Phys. Rev. Lett.}\ }\textbf {\bibinfo {volume} {80}},\ \bibinfo {pages} {5481} (\bibinfo {year} {1998})},\ \Eprint {http://arxiv.org/abs/astro-ph/9709072} {arXiv:astro-ph/9709072} \BibitemShut {NoStop}%
\bibitem [{\citenamefont {Green}\ and\ \citenamefont {Liddle}(1999)}]{GreLid99}%
  \BibitemOpen
  \bibfield  {author} {\bibinfo {author} {\bibfnamefont {A.~M.}\ \bibnamefont {Green}}\ and\ \bibinfo {author} {\bibfnamefont {A.~R.}\ \bibnamefont {Liddle}},\ }\href {\doibase 10.1103/PhysRevD.60.063509} {\bibfield  {journal} {\bibinfo  {journal} {Phys. Rev. D}\ }\textbf {\bibinfo {volume} {60}},\ \bibinfo {pages} {063509} (\bibinfo {year} {1999})},\ \Eprint {http://arxiv.org/abs/astro-ph/9901268} {arXiv:astro-ph/9901268} \BibitemShut {NoStop}%
\bibitem [{\citenamefont {{Lacey}}\ and\ \citenamefont {{Cole}}(1993)}]{LacCol93}%
  \BibitemOpen
  \bibfield  {author} {\bibinfo {author} {\bibfnamefont {C.}~\bibnamefont {{Lacey}}}\ and\ \bibinfo {author} {\bibfnamefont {S.}~\bibnamefont {{Cole}}},\ }\href {\doibase 10.1093/mnras/262.3.627} {\bibfield  {journal} {\bibinfo  {journal} {Mon. Not. R. Astron. Soc.}\ }\textbf {\bibinfo {volume} {262}},\ \bibinfo {pages} {627} (\bibinfo {year} {1993})}\BibitemShut {NoStop}%
\bibitem [{\citenamefont {Peebles}(1969)}]{Peebles:1969jm}%
  \BibitemOpen
  \bibfield  {author} {\bibinfo {author} {\bibfnamefont {P.~J.~E.}\ \bibnamefont {Peebles}},\ }\href {\doibase 10.1086/149876} {\bibfield  {journal} {\bibinfo  {journal} {Astrophys. J.}\ }\textbf {\bibinfo {volume} {155}},\ \bibinfo {pages} {393} (\bibinfo {year} {1969})}\BibitemShut {NoStop}%
\bibitem [{\citenamefont {{Doroshkevich}}(1970)}]{Dor70}%
  \BibitemOpen
  \bibfield  {author} {\bibinfo {author} {\bibfnamefont {A.~G.}\ \bibnamefont {{Doroshkevich}}},\ }\href {\doibase 10.1007/BF01001625} {\bibfield  {journal} {\bibinfo  {journal} {Astrophysics}\ }\textbf {\bibinfo {volume} {6}},\ \bibinfo {pages} {320} (\bibinfo {year} {1970})}\BibitemShut {NoStop}%
\bibitem [{\citenamefont {Eisenstein}\ and\ \citenamefont {Loeb}(1995)}]{Eisenstein:1994ni}%
  \BibitemOpen
  \bibfield  {author} {\bibinfo {author} {\bibfnamefont {D.~J.}\ \bibnamefont {Eisenstein}}\ and\ \bibinfo {author} {\bibfnamefont {A.}~\bibnamefont {Loeb}},\ }\href {\doibase 10.1086/175193} {\bibfield  {journal} {\bibinfo  {journal} {Astrophys. J.}\ }\textbf {\bibinfo {volume} {439}},\ \bibinfo {pages} {520} (\bibinfo {year} {1995})},\ \Eprint {http://arxiv.org/abs/astro-ph/9405012} {arXiv:astro-ph/9405012} \BibitemShut {NoStop}%
\bibitem [{\citenamefont {Chiba}\ and\ \citenamefont {Yokoyama}(2017)}]{ChiYok17}%
  \BibitemOpen
  \bibfield  {author} {\bibinfo {author} {\bibfnamefont {T.}~\bibnamefont {Chiba}}\ and\ \bibinfo {author} {\bibfnamefont {S.}~\bibnamefont {Yokoyama}},\ }\href {\doibase 10.1093/ptep/ptx087} {\bibfield  {journal} {\bibinfo  {journal} {PTEP}\ }\textbf {\bibinfo {volume} {2017}},\ \bibinfo {pages} {083E01} (\bibinfo {year} {2017})},\ \Eprint {http://arxiv.org/abs/1704.06573} {arXiv:1704.06573 [gr-qc]} \BibitemShut {NoStop}%
\bibitem [{\citenamefont {Mirbabayi}\ \emph {et~al.}(2020)\citenamefont {Mirbabayi}, \citenamefont {Gruzinov},\ and\ \citenamefont {Nore{\~n}a}}]{MirGruNor19}%
  \BibitemOpen
  \bibfield  {author} {\bibinfo {author} {\bibfnamefont {M.}~\bibnamefont {Mirbabayi}}, \bibinfo {author} {\bibfnamefont {A.}~\bibnamefont {Gruzinov}}, \ and\ \bibinfo {author} {\bibfnamefont {J.}~\bibnamefont {Nore{\~n}a}},\ }\href {\doibase 10.1088/1475-7516/2020/03/017} {\bibfield  {journal} {\bibinfo  {journal} {J. Cosmol. Astropart. Phys.}\ }\textbf {\bibinfo {volume} {03}},\ \bibinfo {pages} {017} (\bibinfo {year} {2020})},\ \Eprint {http://arxiv.org/abs/1901.05963} {arXiv:1901.05963 [astro-ph.CO]} \BibitemShut {NoStop}%
\bibitem [{\citenamefont {De~Luca}\ \emph {et~al.}(2019)\citenamefont {De~Luca}, \citenamefont {Desjacques}, \citenamefont {Franciolini}, \citenamefont {Malhotra},\ and\ \citenamefont {Riotto}}]{DeLuca:2019buf}%
  \BibitemOpen
  \bibfield  {author} {\bibinfo {author} {\bibfnamefont {V.}~\bibnamefont {De~Luca}}, \bibinfo {author} {\bibfnamefont {V.}~\bibnamefont {Desjacques}}, \bibinfo {author} {\bibfnamefont {G.}~\bibnamefont {Franciolini}}, \bibinfo {author} {\bibfnamefont {A.}~\bibnamefont {Malhotra}}, \ and\ \bibinfo {author} {\bibfnamefont {A.}~\bibnamefont {Riotto}},\ }\href {\doibase 10.1088/1475-7516/2019/05/018} {\bibfield  {journal} {\bibinfo  {journal} {J. Cosmol. Astropart. Phys.}\ }\textbf {\bibinfo {volume} {05}},\ \bibinfo {pages} {018} (\bibinfo {year} {2019})},\ \Eprint {http://arxiv.org/abs/1903.01179} {arXiv:1903.01179 [astro-ph.CO]} \BibitemShut {NoStop}%
\bibitem [{\citenamefont {Harada}\ \emph {et~al.}(2021)\citenamefont {Harada}, \citenamefont {Yoo}, \citenamefont {Kohri}, \citenamefont {Koga},\ and\ \citenamefont {Monobe}}]{Haretal20}%
  \BibitemOpen
  \bibfield  {author} {\bibinfo {author} {\bibfnamefont {T.}~\bibnamefont {Harada}}, \bibinfo {author} {\bibfnamefont {C.-M.}\ \bibnamefont {Yoo}}, \bibinfo {author} {\bibfnamefont {K.}~\bibnamefont {Kohri}}, \bibinfo {author} {\bibfnamefont {Y.}~\bibnamefont {Koga}}, \ and\ \bibinfo {author} {\bibfnamefont {T.}~\bibnamefont {Monobe}},\ }\href {\doibase 10.3847/1538-4357/abd9b9} {\bibfield  {journal} {\bibinfo  {journal} {Astrophys. J.}\ }\textbf {\bibinfo {volume} {908}},\ \bibinfo {pages} {140} (\bibinfo {year} {2021})},\ \Eprint {http://arxiv.org/abs/2011.00710} {arXiv:2011.00710 [astro-ph.CO]} \BibitemShut {NoStop}%
\bibitem [{\citenamefont {Zabotin}\ \emph {et~al.}(1987)\citenamefont {Zabotin}, \citenamefont {Naselskii},\ and\ \citenamefont {Polnarev}}]{ZabNasPol87}%
  \BibitemOpen
  \bibfield  {author} {\bibinfo {author} {\bibfnamefont {N.~A.}\ \bibnamefont {Zabotin}}, \bibinfo {author} {\bibfnamefont {P.~D.}\ \bibnamefont {Naselskii}}, \ and\ \bibinfo {author} {\bibfnamefont {A.~G.}\ \bibnamefont {Polnarev}},\ }\href@noop {} {\bibfield  {journal} {\bibinfo  {journal} {Sov. Astron.}\ }\textbf {\bibinfo {volume} {31}},\ \bibinfo {pages} {353} (\bibinfo {year} {1987})}\BibitemShut {NoStop}%
\bibitem [{\citenamefont {Khlopov}(2010)}]{Khl10}%
  \BibitemOpen
  \bibfield  {author} {\bibinfo {author} {\bibfnamefont {M.~Y.}\ \bibnamefont {Khlopov}},\ }\href {\doibase 10.1088/1674-4527/10/6/001} {\bibfield  {journal} {\bibinfo  {journal} {Res. Astron. Astrophys.}\ }\textbf {\bibinfo {volume} {10}},\ \bibinfo {pages} {495} (\bibinfo {year} {2010})},\ \Eprint {http://arxiv.org/abs/0801.0116} {arXiv:0801.0116 [astro-ph]} \BibitemShut {NoStop}%
\bibitem [{\citenamefont {Harada}\ \emph {et~al.}(2016)\citenamefont {Harada}, \citenamefont {Yoo}, \citenamefont {Kohri}, \citenamefont {Nakao},\ and\ \citenamefont {Jhingan}}]{Haretal16}%
  \BibitemOpen
  \bibfield  {author} {\bibinfo {author} {\bibfnamefont {T.}~\bibnamefont {Harada}}, \bibinfo {author} {\bibfnamefont {C.-M.}\ \bibnamefont {Yoo}}, \bibinfo {author} {\bibfnamefont {K.}~\bibnamefont {Kohri}}, \bibinfo {author} {\bibfnamefont {K.-i.}\ \bibnamefont {Nakao}}, \ and\ \bibinfo {author} {\bibfnamefont {S.}~\bibnamefont {Jhingan}},\ }\href {\doibase 10.3847/1538-4357/833/1/61} {\bibfield  {journal} {\bibinfo  {journal} {Astrophys. J.}\ }\textbf {\bibinfo {volume} {833}},\ \bibinfo {pages} {61} (\bibinfo {year} {2016})},\ \Eprint {http://arxiv.org/abs/1609.01588} {arXiv:1609.01588 [astro-ph.CO]} \BibitemShut {NoStop}%
\bibitem [{\citenamefont {Kotok}\ and\ \citenamefont {Naselsky}(1998)}]{KotNas98}%
  \BibitemOpen
  \bibfield  {author} {\bibinfo {author} {\bibfnamefont {E.}~\bibnamefont {Kotok}}\ and\ \bibinfo {author} {\bibfnamefont {P.}~\bibnamefont {Naselsky}},\ }\href {\doibase 10.1103/PhysRevD.58.103517} {\bibfield  {journal} {\bibinfo  {journal} {Phys. Rev. D}\ }\textbf {\bibinfo {volume} {58}},\ \bibinfo {pages} {103517} (\bibinfo {year} {1998})},\ \Eprint {http://arxiv.org/abs/astro-ph/9806139} {arXiv:astro-ph/9806139} \BibitemShut {NoStop}%
\bibitem [{\citenamefont {Harada}\ \emph {et~al.}(2017)\citenamefont {Harada}, \citenamefont {Yoo}, \citenamefont {Kohri},\ and\ \citenamefont {Nakao}}]{Haretal17}%
  \BibitemOpen
  \bibfield  {author} {\bibinfo {author} {\bibfnamefont {T.}~\bibnamefont {Harada}}, \bibinfo {author} {\bibfnamefont {C.-M.}\ \bibnamefont {Yoo}}, \bibinfo {author} {\bibfnamefont {K.}~\bibnamefont {Kohri}}, \ and\ \bibinfo {author} {\bibfnamefont {K.-I.}\ \bibnamefont {Nakao}},\ }\href {\doibase 10.1103/PhysRevD.96.083517} {\bibfield  {journal} {\bibinfo  {journal} {Phys. Rev. D}\ }\textbf {\bibinfo {volume} {96}},\ \bibinfo {pages} {083517} (\bibinfo {year} {2017})},\ \bibinfo {note} {[Erratum: Phys.Rev.D 99, 069904 (2019)]},\ \Eprint {http://arxiv.org/abs/1707.03595} {arXiv:1707.03595 [gr-qc]} \BibitemShut {NoStop}%
\bibitem [{\citenamefont {Kuhnel}(2020)}]{Kuh20}%
  \BibitemOpen
  \bibfield  {author} {\bibinfo {author} {\bibfnamefont {F.}~\bibnamefont {Kuhnel}},\ }\href {\doibase 10.1140/epjc/s10052-020-7807-z} {\bibfield  {journal} {\bibinfo  {journal} {Eur. Phys. J. C}\ }\textbf {\bibinfo {volume} {80}},\ \bibinfo {pages} {243} (\bibinfo {year} {2020})},\ \Eprint {http://arxiv.org/abs/1909.04742} {arXiv:1909.04742 [astro-ph.CO]} \BibitemShut {NoStop}%
\bibitem [{\citenamefont {Deng}\ and\ \citenamefont {Vilenkin}(2017)}]{Deng:2017uwc}%
  \BibitemOpen
  \bibfield  {author} {\bibinfo {author} {\bibfnamefont {H.}~\bibnamefont {Deng}}\ and\ \bibinfo {author} {\bibfnamefont {A.}~\bibnamefont {Vilenkin}},\ }\href {\doibase 10.1088/1475-7516/2017/12/044} {\bibfield  {journal} {\bibinfo  {journal} {J. Cosmol. Astropart. Phys.}\ }\textbf {\bibinfo {volume} {12}},\ \bibinfo {pages} {044} (\bibinfo {year} {2017})},\ \Eprint {http://arxiv.org/abs/1710.02865} {arXiv:1710.02865 [gr-qc]} \BibitemShut {NoStop}%
\bibitem [{\citenamefont {Garriga}\ \emph {et~al.}(2016)\citenamefont {Garriga}, \citenamefont {Vilenkin},\ and\ \citenamefont {Zhang}}]{GarVilZha16}%
  \BibitemOpen
  \bibfield  {author} {\bibinfo {author} {\bibfnamefont {J.}~\bibnamefont {Garriga}}, \bibinfo {author} {\bibfnamefont {A.}~\bibnamefont {Vilenkin}}, \ and\ \bibinfo {author} {\bibfnamefont {J.}~\bibnamefont {Zhang}},\ }\href {\doibase 10.1088/1475-7516/2016/02/064} {\bibfield  {journal} {\bibinfo  {journal} {J. Cosmol. Astropart. Phys.}\ }\textbf {\bibinfo {volume} {02}},\ \bibinfo {pages} {064} (\bibinfo {year} {2016})},\ \Eprint {http://arxiv.org/abs/1512.01819} {arXiv:1512.01819 [hep-th]} \BibitemShut {NoStop}%
\bibitem [{\citenamefont {Deng}\ \emph {et~al.}(2017)\citenamefont {Deng}, \citenamefont {Garriga},\ and\ \citenamefont {Vilenkin}}]{Deng:2016vzb}%
  \BibitemOpen
  \bibfield  {author} {\bibinfo {author} {\bibfnamefont {H.}~\bibnamefont {Deng}}, \bibinfo {author} {\bibfnamefont {J.}~\bibnamefont {Garriga}}, \ and\ \bibinfo {author} {\bibfnamefont {A.}~\bibnamefont {Vilenkin}},\ }\href {\doibase 10.1088/1475-7516/2017/04/050} {\bibfield  {journal} {\bibinfo  {journal} {J. Cosmol. Astropart. Phys.}\ }\textbf {\bibinfo {volume} {04}},\ \bibinfo {pages} {050} (\bibinfo {year} {2017})},\ \Eprint {http://arxiv.org/abs/1612.03753} {arXiv:1612.03753 [gr-qc]} \BibitemShut {NoStop}%
\bibitem [{\citenamefont {Ivanov}\ \emph {et~al.}(1994)\citenamefont {Ivanov}, \citenamefont {Naselsky},\ and\ \citenamefont {Novikov}}]{IvaNasNov94}%
  \BibitemOpen
  \bibfield  {author} {\bibinfo {author} {\bibfnamefont {P.}~\bibnamefont {Ivanov}}, \bibinfo {author} {\bibfnamefont {P.}~\bibnamefont {Naselsky}}, \ and\ \bibinfo {author} {\bibfnamefont {I.}~\bibnamefont {Novikov}},\ }\href {\doibase 10.1103/PhysRevD.50.7173} {\bibfield  {journal} {\bibinfo  {journal} {Phys. Rev. D}\ }\textbf {\bibinfo {volume} {50}},\ \bibinfo {pages} {7173} (\bibinfo {year} {1994})}\BibitemShut {NoStop}%
\bibitem [{\citenamefont {Starobinsky}(1992)}]{Starobinsky:1992ts}%
  \BibitemOpen
  \bibfield  {author} {\bibinfo {author} {\bibfnamefont {A.~A.}\ \bibnamefont {Starobinsky}},\ }\href@noop {} {\bibfield  {journal} {\bibinfo  {journal} {JETP Lett.}\ }\textbf {\bibinfo {volume} {55}},\ \bibinfo {pages} {489} (\bibinfo {year} {1992})}\BibitemShut {NoStop}%
\bibitem [{\citenamefont {Khlopov}\ and\ \citenamefont {Rubin}(2004)}]{Khlopov:2004rw}%
  \BibitemOpen
  \bibfield  {author} {\bibinfo {author} {\bibfnamefont {M.~Y.}\ \bibnamefont {Khlopov}}\ and\ \bibinfo {author} {\bibfnamefont {S.~G.}\ \bibnamefont {Rubin}},\ }\href {\doibase 10.1007/978-1-4020-2650-8} {\emph {\bibinfo {title} {{Cosmological pattern of microphysics in the inflationary universe}}}}\ (\bibinfo  {publisher} {Springer},\ \bibinfo {year} {2004})\BibitemShut {NoStop}%
\bibitem [{\citenamefont {{Garc{\'\i}a-Bellido}}\ and\ \citenamefont {{Ruiz Morales}}(2017)}]{2017PDU....18...47G}%
  \BibitemOpen
  \bibfield  {author} {\bibinfo {author} {\bibfnamefont {J.}~\bibnamefont {{Garc{\'\i}a-Bellido}}}\ and\ \bibinfo {author} {\bibfnamefont {E.}~\bibnamefont {{Ruiz Morales}}},\ }\href {\doibase 10.1016/j.dark.2017.09.007} {\bibfield  {journal} {\bibinfo  {journal} {Physics of the Dark Universe}\ }\textbf {\bibinfo {volume} {18}},\ \bibinfo {pages} {47} (\bibinfo {year} {2017})},\ \Eprint {http://arxiv.org/abs/1702.03901} {arXiv:1702.03901 [astro-ph.CO]} \BibitemShut {NoStop}%
\bibitem [{\citenamefont {Yokoyama}(1998{\natexlab{a}})}]{Yokoyama:1998pt}%
  \BibitemOpen
  \bibfield  {author} {\bibinfo {author} {\bibfnamefont {J.}~\bibnamefont {Yokoyama}},\ }\href {\doibase 10.1103/PhysRevD.58.083510} {\bibfield  {journal} {\bibinfo  {journal} {Phys. Rev. D}\ }\textbf {\bibinfo {volume} {58}},\ \bibinfo {pages} {083510} (\bibinfo {year} {1998}{\natexlab{a}})},\ \Eprint {http://arxiv.org/abs/astro-ph/9802357} {arXiv:astro-ph/9802357} \BibitemShut {NoStop}%
\bibitem [{\citenamefont {Frolovsky}\ and\ \citenamefont {Ketov}(2025)}]{Frolovsky:2025qre}%
  \BibitemOpen
  \bibfield  {author} {\bibinfo {author} {\bibfnamefont {D.}~\bibnamefont {Frolovsky}}\ and\ \bibinfo {author} {\bibfnamefont {S.~V.}\ \bibnamefont {Ketov}},\ }\href {\doibase 10.1103/PhysRevD.111.083533} {\bibfield  {journal} {\bibinfo  {journal} {Phys. Rev. D}\ }\textbf {\bibinfo {volume} {111}},\ \bibinfo {pages} {083533} (\bibinfo {year} {2025})},\ \Eprint {http://arxiv.org/abs/2502.00628} {arXiv:2502.00628 [gr-qc]} \BibitemShut {NoStop}%
\bibitem [{\citenamefont {Kirillov}\ and\ \citenamefont {Rubin}(2021)}]{KirRub21}%
  \BibitemOpen
  \bibfield  {author} {\bibinfo {author} {\bibfnamefont {A.~A.}\ \bibnamefont {Kirillov}}\ and\ \bibinfo {author} {\bibfnamefont {S.~G.}\ \bibnamefont {Rubin}},\ }\href {\doibase 10.3389/fspas.2021.777661} {\bibfield  {journal} {\bibinfo  {journal} {Front. Astron. Space Sci.}\ }\textbf {\bibinfo {volume} {8}},\ \bibinfo {pages} {777661} (\bibinfo {year} {2021})},\ \Eprint {http://arxiv.org/abs/2109.02446} {arXiv:2109.02446 [astro-ph.CO]} \BibitemShut {NoStop}%
\bibitem [{\citenamefont {Dolgov}\ \emph {et~al.}(2009)\citenamefont {Dolgov}, \citenamefont {Kawasaki},\ and\ \citenamefont {Kevlishvili}}]{DolKawKev09}%
  \BibitemOpen
  \bibfield  {author} {\bibinfo {author} {\bibfnamefont {A.~D.}\ \bibnamefont {Dolgov}}, \bibinfo {author} {\bibfnamefont {M.}~\bibnamefont {Kawasaki}}, \ and\ \bibinfo {author} {\bibfnamefont {N.}~\bibnamefont {Kevlishvili}},\ }\href {\doibase 10.1016/j.nuclphysb.2008.08.029} {\bibfield  {journal} {\bibinfo  {journal} {Nucl. Phys. B}\ }\textbf {\bibinfo {volume} {807}},\ \bibinfo {pages} {229} (\bibinfo {year} {2009})},\ \Eprint {http://arxiv.org/abs/0806.2986} {arXiv:0806.2986 [hep-ph]} \BibitemShut {NoStop}%
\bibitem [{\citenamefont {Linde}(1990)}]{Linde1990}%
  \BibitemOpen
  \bibfield  {author} {\bibinfo {author} {\bibfnamefont {A.}~\bibnamefont {Linde}},\ }\href@noop {} {\emph {\bibinfo {title} {Inflation and Quantum Cosmology}}},\ edited by\ \bibinfo {editor} {\bibfnamefont {R.~H.}\ \bibnamefont {Brandenberger}}\ (\bibinfo  {publisher} {Academic Press},\ \bibinfo {address} {Boston},\ \bibinfo {year} {1990})\BibitemShut {NoStop}%
\bibitem [{\citenamefont {Linde}(1994)}]{Linde:1994eq}%
  \BibitemOpen
  \bibfield  {author} {\bibinfo {author} {\bibfnamefont {A.~D.}\ \bibnamefont {Linde}},\ }\href {\doibase 10.1038/scientificamerican1194-48} {\bibfield  {journal} {\bibinfo  {journal} {Sci. Am.}\ }\textbf {\bibinfo {volume} {271}},\ \bibinfo {pages} {32} (\bibinfo {year} {1994})}\BibitemShut {NoStop}%
\bibitem [{\citenamefont {Randall}\ \emph {et~al.}(1996)\citenamefont {Randall}, \citenamefont {Soljacic},\ and\ \citenamefont {Guth}}]{Randall:1995dj}%
  \BibitemOpen
  \bibfield  {author} {\bibinfo {author} {\bibfnamefont {L.}~\bibnamefont {Randall}}, \bibinfo {author} {\bibfnamefont {M.}~\bibnamefont {Soljacic}}, \ and\ \bibinfo {author} {\bibfnamefont {A.~H.}\ \bibnamefont {Guth}},\ }\href {\doibase 10.1016/0550-3213(96)00174-5} {\bibfield  {journal} {\bibinfo  {journal} {Nucl. Phys. B}\ }\textbf {\bibinfo {volume} {472}},\ \bibinfo {pages} {377} (\bibinfo {year} {1996})},\ \Eprint {http://arxiv.org/abs/hep-ph/9512439} {arXiv:hep-ph/9512439} \BibitemShut {NoStop}%
\bibitem [{\citenamefont {Garcia-Bellido}\ and\ \citenamefont {Linde}(1997)}]{Garcia-Bellido:1996egv}%
  \BibitemOpen
  \bibfield  {author} {\bibinfo {author} {\bibfnamefont {J.}~\bibnamefont {Garcia-Bellido}}\ and\ \bibinfo {author} {\bibfnamefont {A.~D.}\ \bibnamefont {Linde}},\ }\href {\doibase 10.1016/S0370-2693(97)00211-6} {\bibfield  {journal} {\bibinfo  {journal} {Phys. Lett. B}\ }\textbf {\bibinfo {volume} {398}},\ \bibinfo {pages} {18} (\bibinfo {year} {1997})},\ \Eprint {http://arxiv.org/abs/astro-ph/9612141} {arXiv:astro-ph/9612141} \BibitemShut {NoStop}%
\bibitem [{\citenamefont {Rubin}(2005)}]{2005astro.ph.11181R}%
  \BibitemOpen
  \bibfield  {author} {\bibinfo {author} {\bibfnamefont {S.~G.}\ \bibnamefont {Rubin}},\ }in\ \href@noop {} {\emph {\bibinfo {booktitle} {{International Conference: I.Ya. Pomeranchuk and Physics at the Turn of Centuries}}}}\ (\bibinfo {year} {2005})\ \Eprint {http://arxiv.org/abs/astro-ph/0511181} {arXiv:astro-ph/0511181} \BibitemShut {NoStop}%
\bibitem [{\citenamefont {Lukash}\ and\ \citenamefont {Mikheeva}(2025)}]{LukMik25}%
  \BibitemOpen
  \bibfield  {author} {\bibinfo {author} {\bibfnamefont {V.~N.}\ \bibnamefont {Lukash}}\ and\ \bibinfo {author} {\bibfnamefont {E.~V.}\ \bibnamefont {Mikheeva}},\ }\href {\doibase 10.1134/S0021364024604718} {\bibfield  {journal} {\bibinfo  {journal} {JETP Lett.}\ }\textbf {\bibinfo {volume} {121}},\ \bibinfo {pages} {399} (\bibinfo {year} {2025})},\ \Eprint {http://arxiv.org/abs/2506.03226} {arXiv:2506.03226 [gr-qc]} \BibitemShut {NoStop}%
\bibitem [{\citenamefont {Tkachev}\ \emph {et~al.}(2023)\citenamefont {Tkachev}, \citenamefont {Pilipenko}, \citenamefont {Mikheeva},\ and\ \citenamefont {Lukash}}]{Tkachev:2023acf}%
  \BibitemOpen
  \bibfield  {author} {\bibinfo {author} {\bibfnamefont {M.~V.}\ \bibnamefont {Tkachev}}, \bibinfo {author} {\bibfnamefont {S.~V.}\ \bibnamefont {Pilipenko}}, \bibinfo {author} {\bibfnamefont {E.~V.}\ \bibnamefont {Mikheeva}}, \ and\ \bibinfo {author} {\bibfnamefont {V.~N.}\ \bibnamefont {Lukash}},\ }\href {\doibase 10.1093/mnras/stad3279} {\bibfield  {journal} {\bibinfo  {journal} {Mon. Not. Roy. Astron. Soc.}\ }\textbf {\bibinfo {volume} {527}},\ \bibinfo {pages} {1381} (\bibinfo {year} {2023})},\ \Eprint {http://arxiv.org/abs/2307.13774} {arXiv:2307.13774 [astro-ph.CO]} \BibitemShut {NoStop}%
\bibitem [{\citenamefont {{Kurakin}}\ and\ \citenamefont {{Rubin}}(2020)}]{KurRub20}%
  \BibitemOpen
  \bibfield  {author} {\bibinfo {author} {\bibfnamefont {A.~A.}\ \bibnamefont {{Kurakin}}}\ and\ \bibinfo {author} {\bibfnamefont {S.~G.}\ \bibnamefont {{Rubin}}},\ }\href@noop {} {\  (\bibinfo {year} {2020})},\ \Eprint {http://arxiv.org/abs/2011.01757} {arXiv:2011.01757 [physics.gen-ph]} \BibitemShut {NoStop}%
\bibitem [{\citenamefont {Linde}(1982)}]{Linde:1982uu}%
  \BibitemOpen
  \bibfield  {author} {\bibinfo {author} {\bibfnamefont {A.~D.}\ \bibnamefont {Linde}},\ }\href {\doibase 10.1016/0370-2693(82)90293-3} {\bibfield  {journal} {\bibinfo  {journal} {Phys. Lett. B}\ }\textbf {\bibinfo {volume} {116}},\ \bibinfo {pages} {335} (\bibinfo {year} {1982})}\BibitemShut {NoStop}%
\bibitem [{\citenamefont {Starobinsky}(1982)}]{Starobinsky:1982ee}%
  \BibitemOpen
  \bibfield  {author} {\bibinfo {author} {\bibfnamefont {A.~A.}\ \bibnamefont {Starobinsky}},\ }\href {\doibase 10.1016/0370-2693(82)90541-X} {\bibfield  {journal} {\bibinfo  {journal} {Phys. Lett. B}\ }\textbf {\bibinfo {volume} {117}},\ \bibinfo {pages} {175} (\bibinfo {year} {1982})}\BibitemShut {NoStop}%
\bibitem [{\citenamefont {Khlopov}\ \emph {et~al.}(2002)\citenamefont {Khlopov}, \citenamefont {Rubin},\ and\ \citenamefont {Sakharov}}]{KhlRubSa02}%
  \BibitemOpen
  \bibfield  {author} {\bibinfo {author} {\bibfnamefont {M.~Y.}\ \bibnamefont {Khlopov}}, \bibinfo {author} {\bibfnamefont {S.~G.}\ \bibnamefont {Rubin}}, \ and\ \bibinfo {author} {\bibfnamefont {A.~S.}\ \bibnamefont {Sakharov}},\ }\href@noop {} {\bibfield  {journal} {\bibinfo  {journal} {Grav. Cosmol.}\ }\textbf {\bibinfo {volume} {8}},\ \bibinfo {pages} {57} (\bibinfo {year} {2002})},\ \Eprint {http://arxiv.org/abs/astro-ph/0202505} {astro-ph/0202505} \BibitemShut {NoStop}%
\bibitem [{\citenamefont {Kirillov}\ \emph {et~al.}(2025)\citenamefont {Kirillov}, \citenamefont {Murygin},\ and\ \citenamefont {Nikulin}}]{Kirillov:2024pcs}%
  \BibitemOpen
  \bibfield  {author} {\bibinfo {author} {\bibfnamefont {A.~A.}\ \bibnamefont {Kirillov}}, \bibinfo {author} {\bibfnamefont {B.~S.}\ \bibnamefont {Murygin}}, \ and\ \bibinfo {author} {\bibfnamefont {V.~V.}\ \bibnamefont {Nikulin}},\ }\href {\doibase 10.1016/j.physletb.2024.139201} {\bibfield  {journal} {\bibinfo  {journal} {Phys. Lett. B}\ }\textbf {\bibinfo {volume} {860}},\ \bibinfo {pages} {139201} (\bibinfo {year} {2025})},\ \Eprint {http://arxiv.org/abs/2412.18997} {arXiv:2412.18997 [hep-th]} \BibitemShut {NoStop}%
\bibitem [{\citenamefont {Vachaspati}\ \emph {et~al.}(1984)\citenamefont {Vachaspati}, \citenamefont {Everett},\ and\ \citenamefont {Vilenkin}}]{Vachaspati:1984yi}%
  \BibitemOpen
  \bibfield  {author} {\bibinfo {author} {\bibfnamefont {T.}~\bibnamefont {Vachaspati}}, \bibinfo {author} {\bibfnamefont {A.~E.}\ \bibnamefont {Everett}}, \ and\ \bibinfo {author} {\bibfnamefont {A.}~\bibnamefont {Vilenkin}},\ }\href {\doibase 10.1103/PhysRevD.30.2046} {\bibfield  {journal} {\bibinfo  {journal} {Phys. Rev. D}\ }\textbf {\bibinfo {volume} {30}},\ \bibinfo {pages} {2046} (\bibinfo {year} {1984})}\BibitemShut {NoStop}%
\bibitem [{\citenamefont {Kibble}(1976)}]{Kibble:1976sj}%
  \BibitemOpen
  \bibfield  {author} {\bibinfo {author} {\bibfnamefont {T.~W.~B.}\ \bibnamefont {Kibble}},\ }\href {\doibase 10.1088/0305-4470/9/8/029} {\bibfield  {journal} {\bibinfo  {journal} {J. Phys. A}\ }\textbf {\bibinfo {volume} {9}},\ \bibinfo {pages} {1387} (\bibinfo {year} {1976})}\BibitemShut {NoStop}%
\bibitem [{\citenamefont {Friedland}\ \emph {et~al.}(2003)\citenamefont {Friedland}, \citenamefont {Murayama},\ and\ \citenamefont {Perelstein}}]{Friedland:2002qs}%
  \BibitemOpen
  \bibfield  {author} {\bibinfo {author} {\bibfnamefont {A.}~\bibnamefont {Friedland}}, \bibinfo {author} {\bibfnamefont {H.}~\bibnamefont {Murayama}}, \ and\ \bibinfo {author} {\bibfnamefont {M.}~\bibnamefont {Perelstein}},\ }\href {\doibase 10.1103/PhysRevD.67.043519} {\bibfield  {journal} {\bibinfo  {journal} {Phys. Rev. D}\ }\textbf {\bibinfo {volume} {67}},\ \bibinfo {pages} {043519} (\bibinfo {year} {2003})},\ \Eprint {http://arxiv.org/abs/astro-ph/0205520} {arXiv:astro-ph/0205520} \BibitemShut {NoStop}%
\bibitem [{\citenamefont {Eroshenko}(2021)}]{Ero21}%
  \BibitemOpen
  \bibfield  {author} {\bibinfo {author} {\bibfnamefont {Y.~N.}\ \bibnamefont {Eroshenko}},\ }\href {\doibase 10.1088/1475-7516/2021/12/041} {\bibfield  {journal} {\bibinfo  {journal} {J. Cosmol. Astropart. Phys.}\ }\textbf {\bibinfo {volume} {12}},\ \bibinfo {pages} {041} (\bibinfo {year} {2021})},\ \Eprint {http://arxiv.org/abs/2111.03403} {arXiv:2111.03403 [astro-ph.CO]} \BibitemShut {NoStop}%
\bibitem [{\citenamefont {de~Freitas~Pacheco}\ and\ \citenamefont {Silk}(2020)}]{PacSil20}%
  \BibitemOpen
  \bibfield  {author} {\bibinfo {author} {\bibfnamefont {J.~A.}\ \bibnamefont {de~Freitas~Pacheco}}\ and\ \bibinfo {author} {\bibfnamefont {J.}~\bibnamefont {Silk}},\ }\href {\doibase 10.1103/PhysRevD.101.083022} {\bibfield  {journal} {\bibinfo  {journal} {Phys. Rev. D}\ }\textbf {\bibinfo {volume} {101}},\ \bibinfo {pages} {083022} (\bibinfo {year} {2020})},\ \Eprint {http://arxiv.org/abs/2003.12072} {arXiv:2003.12072 [astro-ph.CO]} \BibitemShut {NoStop}%
\bibitem [{\citenamefont {Arbey}\ \emph {et~al.}(2020)\citenamefont {Arbey}, \citenamefont {Auffinger},\ and\ \citenamefont {Silk}}]{ArbAufSil19}%
  \BibitemOpen
  \bibfield  {author} {\bibinfo {author} {\bibfnamefont {A.}~\bibnamefont {Arbey}}, \bibinfo {author} {\bibfnamefont {J.}~\bibnamefont {Auffinger}}, \ and\ \bibinfo {author} {\bibfnamefont {J.}~\bibnamefont {Silk}},\ }\href {\doibase 10.1093/mnras/staa765} {\bibfield  {journal} {\bibinfo  {journal} {Mon. Not. Roy. Astron. Soc.}\ }\textbf {\bibinfo {volume} {494}},\ \bibinfo {pages} {1257} (\bibinfo {year} {2020})},\ \Eprint {http://arxiv.org/abs/1906.04196} {arXiv:1906.04196 [astro-ph.CO]} \BibitemShut {NoStop}%
\bibitem [{\citenamefont {Arbey}\ \emph {et~al.}(2021)\citenamefont {Arbey}, \citenamefont {Auffinger},\ and\ \citenamefont {Silk}}]{ArbAufSil20}%
  \BibitemOpen
  \bibfield  {author} {\bibinfo {author} {\bibfnamefont {A.}~\bibnamefont {Arbey}}, \bibinfo {author} {\bibfnamefont {J.}~\bibnamefont {Auffinger}}, \ and\ \bibinfo {author} {\bibfnamefont {J.}~\bibnamefont {Silk}},\ }\href {\doibase 10.22323/1.390.0585} {\bibfield  {journal} {\bibinfo  {journal} {PoS}\ }\textbf {\bibinfo {volume} {ICHEP2020}},\ \bibinfo {pages} {585} (\bibinfo {year} {2021})},\ \Eprint {http://arxiv.org/abs/2012.14767} {arXiv:2012.14767 [astro-ph.CO]} \BibitemShut {NoStop}%
\bibitem [{\citenamefont {Dasgupta}\ \emph {et~al.}(2020)\citenamefont {Dasgupta}, \citenamefont {Laha},\ and\ \citenamefont {Ray}}]{DasLahRay20}%
  \BibitemOpen
  \bibfield  {author} {\bibinfo {author} {\bibfnamefont {B.}~\bibnamefont {Dasgupta}}, \bibinfo {author} {\bibfnamefont {R.}~\bibnamefont {Laha}}, \ and\ \bibinfo {author} {\bibfnamefont {A.}~\bibnamefont {Ray}},\ }\href {\doibase 10.1103/PhysRevLett.125.101101} {\bibfield  {journal} {\bibinfo  {journal} {Phys. Rev. Lett.}\ }\textbf {\bibinfo {volume} {125}},\ \bibinfo {pages} {101101} (\bibinfo {year} {2020})},\ \Eprint {http://arxiv.org/abs/1912.01014} {arXiv:1912.01014 [hep-ph]} \BibitemShut {NoStop}%
\bibitem [{\citenamefont {Dokuchaev}\ and\ \citenamefont {Nazarova}(2019)}]{DokNaz19}%
  \BibitemOpen
  \bibfield  {author} {\bibinfo {author} {\bibfnamefont {V.~I.}\ \bibnamefont {Dokuchaev}}\ and\ \bibinfo {author} {\bibfnamefont {N.~O.}\ \bibnamefont {Nazarova}},\ }\href {\doibase 10.3390/universe5080183} {\bibfield  {journal} {\bibinfo  {journal} {Universe}\ }\textbf {\bibinfo {volume} {5}},\ \bibinfo {pages} {183} (\bibinfo {year} {2019})},\ \Eprint {http://arxiv.org/abs/1906.07171} {arXiv:1906.07171 [astro-ph.HE]} \BibitemShut {NoStop}%
\bibitem [{\citenamefont {Starobinsky}(1980)}]{Starobinsky:1980te}%
  \BibitemOpen
  \bibfield  {author} {\bibinfo {author} {\bibfnamefont {A.~A.}\ \bibnamefont {Starobinsky}},\ }\href {\doibase 10.1016/0370-2693(80)90670-X} {\bibfield  {journal} {\bibinfo  {journal} {Phys. Lett. B}\ }\textbf {\bibinfo {volume} {91}},\ \bibinfo {pages} {99} (\bibinfo {year} {1980})}\BibitemShut {NoStop}%
\bibitem [{\citenamefont {Utiyama}\ and\ \citenamefont {DeWitt}(1962)}]{Utiyama:1962sn}%
  \BibitemOpen
  \bibfield  {author} {\bibinfo {author} {\bibfnamefont {R.}~\bibnamefont {Utiyama}}\ and\ \bibinfo {author} {\bibfnamefont {B.~S.}\ \bibnamefont {DeWitt}},\ }\href {\doibase 10.1063/1.1724264} {\bibfield  {journal} {\bibinfo  {journal} {J. Math. Phys.}\ }\textbf {\bibinfo {volume} {3}},\ \bibinfo {pages} {608} (\bibinfo {year} {1962})}\BibitemShut {NoStop}%
\bibitem [{\citenamefont {Stelle}(1977)}]{Stelle:1976gc}%
  \BibitemOpen
  \bibfield  {author} {\bibinfo {author} {\bibfnamefont {K.~S.}\ \bibnamefont {Stelle}},\ }\href {\doibase 10.1103/PhysRevD.16.953} {\bibfield  {journal} {\bibinfo  {journal} {Phys. Rev. D}\ }\textbf {\bibinfo {volume} {16}},\ \bibinfo {pages} {953} (\bibinfo {year} {1977})}\BibitemShut {NoStop}%
\bibitem [{\citenamefont {Vilkovisky}(1992)}]{Vilkovisky:1992pb}%
  \BibitemOpen
  \bibfield  {author} {\bibinfo {author} {\bibfnamefont {G.~A.}\ \bibnamefont {Vilkovisky}},\ }\href {\doibase 10.1088/0264-9381/9/4/008} {\bibfield  {journal} {\bibinfo  {journal} {Class. Quant. Grav.}\ }\textbf {\bibinfo {volume} {9}},\ \bibinfo {pages} {895} (\bibinfo {year} {1992})}\BibitemShut {NoStop}%
\bibitem [{\citenamefont {Bronnikov}\ and\ \citenamefont {Rubin}(2006)}]{Bronnikov:2005iz}%
  \BibitemOpen
  \bibfield  {author} {\bibinfo {author} {\bibfnamefont {K.~A.}\ \bibnamefont {Bronnikov}}\ and\ \bibinfo {author} {\bibfnamefont {S.~G.}\ \bibnamefont {Rubin}},\ }\href {\doibase 10.1103/PhysRevD.73.124019} {\bibfield  {journal} {\bibinfo  {journal} {Phys. Rev. D}\ }\textbf {\bibinfo {volume} {73}},\ \bibinfo {pages} {124019} (\bibinfo {year} {2006})},\ \Eprint {http://arxiv.org/abs/gr-qc/0510107} {arXiv:gr-qc/0510107} \BibitemShut {NoStop}%
\bibitem [{\citenamefont {Rador}(2007)}]{Rador:2007gq}%
  \BibitemOpen
  \bibfield  {author} {\bibinfo {author} {\bibfnamefont {T.}~\bibnamefont {Rador}},\ }\href {\doibase 10.1103/PhysRevD.75.064033} {\bibfield  {journal} {\bibinfo  {journal} {Phys. Rev. D}\ }\textbf {\bibinfo {volume} {75}},\ \bibinfo {pages} {064033} (\bibinfo {year} {2007})},\ \Eprint {http://arxiv.org/abs/hep-th/0701267} {arXiv:hep-th/0701267} \BibitemShut {NoStop}%
\bibitem [{\citenamefont {Carroll}\ \emph {et~al.}(2002)\citenamefont {Carroll}, \citenamefont {Geddes}, \citenamefont {Hoffman},\ and\ \citenamefont {Wald}}]{Carroll:2001ih}%
  \BibitemOpen
  \bibfield  {author} {\bibinfo {author} {\bibfnamefont {S.~M.}\ \bibnamefont {Carroll}}, \bibinfo {author} {\bibfnamefont {J.}~\bibnamefont {Geddes}}, \bibinfo {author} {\bibfnamefont {M.~B.}\ \bibnamefont {Hoffman}}, \ and\ \bibinfo {author} {\bibfnamefont {R.~M.}\ \bibnamefont {Wald}},\ }\href {\doibase 10.1103/PhysRevD.66.024036} {\bibfield  {journal} {\bibinfo  {journal} {Phys. Rev. D}\ }\textbf {\bibinfo {volume} {66}},\ \bibinfo {pages} {024036} (\bibinfo {year} {2002})},\ \Eprint {http://arxiv.org/abs/hep-th/0110149} {arXiv:hep-th/0110149} \BibitemShut {NoStop}%
\bibitem [{\citenamefont {Fabris}\ \emph {et~al.}(2020)\citenamefont {Fabris}, \citenamefont {Popov},\ and\ \citenamefont {Rubin}}]{Fabris:2019ecx}%
  \BibitemOpen
  \bibfield  {author} {\bibinfo {author} {\bibfnamefont {J.~C.}\ \bibnamefont {Fabris}}, \bibinfo {author} {\bibfnamefont {A.~A.}\ \bibnamefont {Popov}}, \ and\ \bibinfo {author} {\bibfnamefont {S.~G.}\ \bibnamefont {Rubin}},\ }\href {\doibase 10.1016/j.physletb.2020.135458} {\bibfield  {journal} {\bibinfo  {journal} {Phys. Lett. B}\ }\textbf {\bibinfo {volume} {806}},\ \bibinfo {pages} {135458} (\bibinfo {year} {2020})},\ \Eprint {http://arxiv.org/abs/1911.03695} {arXiv:1911.03695 [gr-qc]} \BibitemShut {NoStop}%
\bibitem [{\citenamefont {Lyakhova}\ \emph {et~al.}(2018)\citenamefont {Lyakhova}, \citenamefont {Popov},\ and\ \citenamefont {Rubin}}]{Lyakhova:2018zsr}%
  \BibitemOpen
  \bibfield  {author} {\bibinfo {author} {\bibfnamefont {Y.}~\bibnamefont {Lyakhova}}, \bibinfo {author} {\bibfnamefont {A.~A.}\ \bibnamefont {Popov}}, \ and\ \bibinfo {author} {\bibfnamefont {S.~G.}\ \bibnamefont {Rubin}},\ }\href {\doibase 10.1140/epjc/s10052-018-6251-9} {\bibfield  {journal} {\bibinfo  {journal} {Eur. Phys. J. C}\ }\textbf {\bibinfo {volume} {78}},\ \bibinfo {pages} {764} (\bibinfo {year} {2018})},\ \Eprint {http://arxiv.org/abs/1807.06235} {arXiv:1807.06235 [gr-qc]} \BibitemShut {NoStop}%
\bibitem [{\citenamefont {Nikulin}\ \emph {et~al.}(2022)\citenamefont {Nikulin}, \citenamefont {Krasnov},\ and\ \citenamefont {Rubin}}]{Nikulin:2022ryt}%
  \BibitemOpen
  \bibfield  {author} {\bibinfo {author} {\bibfnamefont {V.~V.}\ \bibnamefont {Nikulin}}, \bibinfo {author} {\bibfnamefont {M.~A.}\ \bibnamefont {Krasnov}}, \ and\ \bibinfo {author} {\bibfnamefont {S.~G.}\ \bibnamefont {Rubin}},\ }\href {\doibase 10.3389/fspas.2022.927144} {\bibfield  {journal} {\bibinfo  {journal} {Front. Astron. Space Sci.}\ }\textbf {\bibinfo {volume} {9}},\ \bibinfo {pages} {927144} (\bibinfo {year} {2022})},\ \Eprint {http://arxiv.org/abs/2204.06360} {arXiv:2204.06360 [gr-qc]} \BibitemShut {NoStop}%
\bibitem [{\citenamefont {Bronnikov}\ and\ \citenamefont {Rubin}(2021)}]{Bronnikov:2021ptu}%
  \BibitemOpen
  \bibfield  {author} {\bibinfo {author} {\bibfnamefont {K.~A.}\ \bibnamefont {Bronnikov}}\ and\ \bibinfo {author} {\bibfnamefont {S.~G.}\ \bibnamefont {Rubin}},\ }\href {\doibase 10.3390/physics3030048} {\bibfield  {journal} {\bibinfo  {journal} {MDPI Physics}\ }\textbf {\bibinfo {volume} {3}},\ \bibinfo {pages} {781} (\bibinfo {year} {2021})},\ \Eprint {http://arxiv.org/abs/2107.13893} {arXiv:2107.13893 [gr-qc]} \BibitemShut {NoStop}%
\bibitem [{\citenamefont {{Affleck}}\ and\ \citenamefont {{Dine}}(1985)}]{AffDin85}%
  \BibitemOpen
  \bibfield  {author} {\bibinfo {author} {\bibfnamefont {I.}~\bibnamefont {{Affleck}}}\ and\ \bibinfo {author} {\bibfnamefont {M.}~\bibnamefont {{Dine}}},\ }\href {\doibase 10.1016/0550-3213(85)90021-5} {\bibfield  {journal} {\bibinfo  {journal} {Nuclear Physics B}\ }\textbf {\bibinfo {volume} {249}},\ \bibinfo {pages} {361} (\bibinfo {year} {1985})}\BibitemShut {NoStop}%
\bibitem [{\citenamefont {Bambi}\ and\ \citenamefont {Dolgov}(2007)}]{BamDol07}%
  \BibitemOpen
  \bibfield  {author} {\bibinfo {author} {\bibfnamefont {C.}~\bibnamefont {Bambi}}\ and\ \bibinfo {author} {\bibfnamefont {A.~D.}\ \bibnamefont {Dolgov}},\ }\href {\doibase 10.1016/j.nuclphysb.2007.06.010} {\bibfield  {journal} {\bibinfo  {journal} {Nucl. Phys.~B}\ }\textbf {\bibinfo {volume} {784}},\ \bibinfo {pages} {132} (\bibinfo {year} {2007})},\ \Eprint {http://arxiv.org/abs/astro-ph/0702350} {arXiv:astro-ph/0702350} \BibitemShut {NoStop}%
\bibitem [{\citenamefont {{Blinnikov}}\ \emph {et~al.}(2015)\citenamefont {{Blinnikov}}, \citenamefont {{Dolgov}},\ and\ \citenamefont {{Postnov}}}]{Blietal15}%
  \BibitemOpen
  \bibfield  {author} {\bibinfo {author} {\bibfnamefont {S.~I.}\ \bibnamefont {{Blinnikov}}}, \bibinfo {author} {\bibfnamefont {A.~D.}\ \bibnamefont {{Dolgov}}}, \ and\ \bibinfo {author} {\bibfnamefont {K.~A.}\ \bibnamefont {{Postnov}}},\ }\href {\doibase 10.1103/PhysRevD.92.023516} {\bibfield  {journal} {\bibinfo  {journal} {Phys. Rev.~D}\ }\textbf {\bibinfo {volume} {92}},\ \bibinfo {eid} {023516} (\bibinfo {year} {2015})},\ \Eprint {http://arxiv.org/abs/1409.5736} {arXiv:1409.5736 [astro-ph.HE]} \BibitemShut {NoStop}%
\bibitem [{\citenamefont {Dolgov}\ and\ \citenamefont {Postnov}(2020)}]{DolPos20}%
  \BibitemOpen
  \bibfield  {author} {\bibinfo {author} {\bibfnamefont {A.}~\bibnamefont {Dolgov}}\ and\ \bibinfo {author} {\bibfnamefont {K.}~\bibnamefont {Postnov}},\ }\href {\doibase 10.1088/1475-7516/2020/07/063} {\bibfield  {journal} {\bibinfo  {journal} {J. Cosmol. Astropart. Phys.}\ }\textbf {\bibinfo {volume} {07}},\ \bibinfo {pages} {063} (\bibinfo {year} {2020})},\ \Eprint {http://arxiv.org/abs/2004.11669} {arXiv:2004.11669 [astro-ph.CO]} \BibitemShut {NoStop}%
\bibitem [{\citenamefont {Dolgov}(2024)}]{Dol23}%
  \BibitemOpen
  \bibfield  {author} {\bibinfo {author} {\bibfnamefont {A.}~\bibnamefont {Dolgov}},\ }\href {\doibase 10.22323/1.447.0010} {\bibfield  {journal} {\bibinfo  {journal} {PoS}\ }\textbf {\bibinfo {volume} {MULTIF2023}},\ \bibinfo {pages} {010} (\bibinfo {year} {2024})}\BibitemShut {NoStop}%
\bibitem [{\citenamefont {Bondar}\ \emph {et~al.}(2022)\citenamefont {Bondar}, \citenamefont {Blinnikov}, \citenamefont {Bykov}, \citenamefont {Dolgov},\ and\ \citenamefont {Postnov}}]{Bonetal22}%
  \BibitemOpen
  \bibfield  {author} {\bibinfo {author} {\bibfnamefont {A.~E.}\ \bibnamefont {Bondar}}, \bibinfo {author} {\bibfnamefont {S.~I.}\ \bibnamefont {Blinnikov}}, \bibinfo {author} {\bibfnamefont {A.~M.}\ \bibnamefont {Bykov}}, \bibinfo {author} {\bibfnamefont {A.~D.}\ \bibnamefont {Dolgov}}, \ and\ \bibinfo {author} {\bibfnamefont {K.~A.}\ \bibnamefont {Postnov}},\ }\href {\doibase 10.1088/1475-7516/2022/03/009} {\bibfield  {journal} {\bibinfo  {journal} {J. Cosmol. Astropart. Phys.}\ }\textbf {\bibinfo {volume} {03}},\ \bibinfo {pages} {009} (\bibinfo {year} {2022})},\ \Eprint {http://arxiv.org/abs/2109.12699} {arXiv:2109.12699 [astro-ph.HE]} \BibitemShut {NoStop}%
\bibitem [{\citenamefont {Jedamzik}(1997)}]{Jed97}%
  \BibitemOpen
  \bibfield  {author} {\bibinfo {author} {\bibfnamefont {K.}~\bibnamefont {Jedamzik}},\ }\href {\doibase 10.1103/PhysRevD.55.R5871} {\bibfield  {journal} {\bibinfo  {journal} {Phys. Rev. D}\ }\textbf {\bibinfo {volume} {55}},\ \bibinfo {pages} {5871} (\bibinfo {year} {1997})},\ \Eprint {http://arxiv.org/abs/astro-ph/9605152} {arXiv:astro-ph/9605152} \BibitemShut {NoStop}%
\bibitem [{\citenamefont {{Aref'eva}}\ \emph {et~al.}(2023)\citenamefont {{Aref'eva}}, \citenamefont {{Hajilou}}, \citenamefont {{Rannu}},\ and\ \citenamefont {{Slepov}}}]{Areetal23}%
  \BibitemOpen
  \bibfield  {author} {\bibinfo {author} {\bibfnamefont {I.~Y.}\ \bibnamefont {{Aref'eva}}}, \bibinfo {author} {\bibfnamefont {A.}~\bibnamefont {{Hajilou}}}, \bibinfo {author} {\bibfnamefont {K.}~\bibnamefont {{Rannu}}}, \ and\ \bibinfo {author} {\bibfnamefont {P.}~\bibnamefont {{Slepov}}},\ }\href {\doibase 10.1140/epjc/s10052-023-12309-w} {\bibfield  {journal} {\bibinfo  {journal} {European Physical Journal C}\ }\textbf {\bibinfo {volume} {83}},\ \bibinfo {eid} {1143} (\bibinfo {year} {2023})},\ \Eprint {http://arxiv.org/abs/2305.06345} {arXiv:2305.06345 [hep-th]} \BibitemShut {NoStop}%
\bibitem [{\citenamefont {Byrnes}\ \emph {et~al.}(2018)\citenamefont {Byrnes}, \citenamefont {Hindmarsh}, \citenamefont {Young},\ and\ \citenamefont {Hawkins}}]{2018JCAP...08..041B}%
  \BibitemOpen
  \bibfield  {author} {\bibinfo {author} {\bibfnamefont {C.~T.}\ \bibnamefont {Byrnes}}, \bibinfo {author} {\bibfnamefont {M.}~\bibnamefont {Hindmarsh}}, \bibinfo {author} {\bibfnamefont {S.}~\bibnamefont {Young}}, \ and\ \bibinfo {author} {\bibfnamefont {M.~R.~S.}\ \bibnamefont {Hawkins}},\ }\href {\doibase 10.1088/1475-7516/2018/08/041} {\bibfield  {journal} {\bibinfo  {journal} {J. Cosmol. Astropart. Phys.}\ }\textbf {\bibinfo {volume} {08}},\ \bibinfo {pages} {041} (\bibinfo {year} {2018})},\ \Eprint {http://arxiv.org/abs/1801.06138} {arXiv:1801.06138 [astro-ph.CO]} \BibitemShut {NoStop}%
\bibitem [{\citenamefont {Musco}\ \emph {et~al.}(2024)\citenamefont {Musco}, \citenamefont {Jedamzik},\ and\ \citenamefont {Young}}]{Musco:2023dak}%
  \BibitemOpen
  \bibfield  {author} {\bibinfo {author} {\bibfnamefont {I.}~\bibnamefont {Musco}}, \bibinfo {author} {\bibfnamefont {K.}~\bibnamefont {Jedamzik}}, \ and\ \bibinfo {author} {\bibfnamefont {S.}~\bibnamefont {Young}},\ }\href {\doibase 10.1103/PhysRevD.109.083506} {\bibfield  {journal} {\bibinfo  {journal} {Phys. Rev. D}\ }\textbf {\bibinfo {volume} {109}},\ \bibinfo {pages} {083506} (\bibinfo {year} {2024})},\ \Eprint {http://arxiv.org/abs/2303.07980} {arXiv:2303.07980 [astro-ph.CO]} \BibitemShut {NoStop}%
\bibitem [{\citenamefont {Jedamzik}(1998)}]{Jed98}%
  \BibitemOpen
  \bibfield  {author} {\bibinfo {author} {\bibfnamefont {K.}~\bibnamefont {Jedamzik}},\ }\href {\doibase 10.1016/S0370-1573(98)00067-2} {\bibfield  {journal} {\bibinfo  {journal} {Phys. Rep.}\ }\textbf {\bibinfo {volume} {307}},\ \bibinfo {pages} {155} (\bibinfo {year} {1998})},\ \Eprint {http://arxiv.org/abs/astro-ph/9805147} {arXiv:astro-ph/9805147} \BibitemShut {NoStop}%
\bibitem [{\citenamefont {Vieira}\ \emph {et~al.}(2026)\citenamefont {Vieira} \emph {et~al.}}]{Vieetal26}%
  \BibitemOpen
  \bibfield  {author} {\bibinfo {author} {\bibfnamefont {N.}~\bibnamefont {Vieira}} \emph {et~al.},\ }\href {\doibase 10.3847/1538-4357/ae877c} {\bibfield  {journal} {\bibinfo  {journal} {Astrophys. J.}\ }\textbf {\bibinfo {volume} {1007}},\ \bibinfo {pages} {116} (\bibinfo {year} {2026})},\ \Eprint {http://arxiv.org/abs/2603.17009} {arXiv:2603.17009 [astro-ph.HE]} \BibitemShut {NoStop}%
\bibitem [{\citenamefont {Hawking}(1989)}]{Haw89}%
  \BibitemOpen
  \bibfield  {author} {\bibinfo {author} {\bibfnamefont {S.~W.}\ \bibnamefont {Hawking}},\ }\href {\doibase 10.1016/0370-2693(89)90206-2} {\bibfield  {journal} {\bibinfo  {journal} {Phys. Lett. B}\ }\textbf {\bibinfo {volume} {231}},\ \bibinfo {pages} {237} (\bibinfo {year} {1989})}\BibitemShut {NoStop}%
\bibitem [{\citenamefont {Polnarev}\ and\ \citenamefont {Zembowicz}(1991)}]{PolZem91}%
  \BibitemOpen
  \bibfield  {author} {\bibinfo {author} {\bibfnamefont {A.}~\bibnamefont {Polnarev}}\ and\ \bibinfo {author} {\bibfnamefont {R.}~\bibnamefont {Zembowicz}},\ }\href {\doibase 10.1103/PhysRevD.43.1106} {\bibfield  {journal} {\bibinfo  {journal} {Phys. Rev. D}\ }\textbf {\bibinfo {volume} {43}},\ \bibinfo {pages} {1106} (\bibinfo {year} {1991})}\BibitemShut {NoStop}%
\bibitem [{\citenamefont {Garriga}\ and\ \citenamefont {Vilenkin}(1993)}]{GarVil93}%
  \BibitemOpen
  \bibfield  {author} {\bibinfo {author} {\bibfnamefont {J.}~\bibnamefont {Garriga}}\ and\ \bibinfo {author} {\bibfnamefont {A.}~\bibnamefont {Vilenkin}},\ }\href {\doibase 10.1103/PhysRevD.47.3265} {\bibfield  {journal} {\bibinfo  {journal} {Phys. Rev. D}\ }\textbf {\bibinfo {volume} {47}},\ \bibinfo {pages} {3265} (\bibinfo {year} {1993})},\ \Eprint {http://arxiv.org/abs/hep-ph/9208212} {arXiv:hep-ph/9208212} \BibitemShut {NoStop}%
\bibitem [{\citenamefont {Caldwell}\ and\ \citenamefont {Gates}(1993)}]{CalGat93}%
  \BibitemOpen
  \bibfield  {author} {\bibinfo {author} {\bibfnamefont {R.~R.}\ \bibnamefont {Caldwell}}\ and\ \bibinfo {author} {\bibfnamefont {E.}~\bibnamefont {Gates}},\ }\href {\doibase 10.1103/PhysRevD.48.2581} {\bibfield  {journal} {\bibinfo  {journal} {Phys. Rev. D}\ }\textbf {\bibinfo {volume} {48}},\ \bibinfo {pages} {2581} (\bibinfo {year} {1993})},\ \Eprint {http://arxiv.org/abs/hep-ph/9306221} {arXiv:hep-ph/9306221} \BibitemShut {NoStop}%
\bibitem [{\citenamefont {{Caldwell}}\ and\ \citenamefont {{Casper}}(1996)}]{1996PhRvD..53.3002C}%
  \BibitemOpen
  \bibfield  {author} {\bibinfo {author} {\bibfnamefont {R.~R.}\ \bibnamefont {{Caldwell}}}\ and\ \bibinfo {author} {\bibfnamefont {P.}~\bibnamefont {{Casper}}},\ }\href {\doibase 10.1103/PhysRevD.53.3002} {\bibfield  {journal} {\bibinfo  {journal} {Phys. Rev. D}\ }\textbf {\bibinfo {volume} {53}},\ \bibinfo {pages} {3002} (\bibinfo {year} {1996})},\ \Eprint {http://arxiv.org/abs/gr-qc/9509012} {arXiv:gr-qc/9509012 [astro-ph]} \BibitemShut {NoStop}%
\bibitem [{\citenamefont {Helfer}\ \emph {et~al.}(2019)\citenamefont {Helfer}, \citenamefont {Aurrekoetxea},\ and\ \citenamefont {Lim}}]{HelAurLim19}%
  \BibitemOpen
  \bibfield  {author} {\bibinfo {author} {\bibfnamefont {T.}~\bibnamefont {Helfer}}, \bibinfo {author} {\bibfnamefont {J.~C.}\ \bibnamefont {Aurrekoetxea}}, \ and\ \bibinfo {author} {\bibfnamefont {E.~A.}\ \bibnamefont {Lim}},\ }\href {\doibase 10.1103/PhysRevD.99.104028} {\bibfield  {journal} {\bibinfo  {journal} {Phys. Rev. D}\ }\textbf {\bibinfo {volume} {99}},\ \bibinfo {pages} {104028} (\bibinfo {year} {2019})},\ \Eprint {http://arxiv.org/abs/1808.06678} {arXiv:1808.06678 [gr-qc]} \BibitemShut {NoStop}%
\bibitem [{\citenamefont {James-Turner}\ \emph {et~al.}(2020)\citenamefont {James-Turner}, \citenamefont {Weil}, \citenamefont {Green},\ and\ \citenamefont {Copeland}}]{Jametal20}%
  \BibitemOpen
  \bibfield  {author} {\bibinfo {author} {\bibfnamefont {C.}~\bibnamefont {James-Turner}}, \bibinfo {author} {\bibfnamefont {D.~P.~B.}\ \bibnamefont {Weil}}, \bibinfo {author} {\bibfnamefont {A.~M.}\ \bibnamefont {Green}}, \ and\ \bibinfo {author} {\bibfnamefont {E.~J.}\ \bibnamefont {Copeland}},\ }\href {\doibase 10.1103/PhysRevD.101.123526} {\bibfield  {journal} {\bibinfo  {journal} {Phys. Rev. D}\ }\textbf {\bibinfo {volume} {101}},\ \bibinfo {pages} {123526} (\bibinfo {year} {2020})},\ \Eprint {http://arxiv.org/abs/1911.12658} {arXiv:1911.12658 [astro-ph.CO]} \BibitemShut {NoStop}%
\bibitem [{\citenamefont {Jenkins}\ and\ \citenamefont {Sakellariadou}(2020)}]{JenSak20}%
  \BibitemOpen
  \bibfield  {author} {\bibinfo {author} {\bibfnamefont {A.~C.}\ \bibnamefont {Jenkins}}\ and\ \bibinfo {author} {\bibfnamefont {M.}~\bibnamefont {Sakellariadou}},\ }\href@noop {} {\  (\bibinfo {year} {2020})},\ \Eprint {http://arxiv.org/abs/2006.16249} {arXiv:2006.16249 [astro-ph.CO]} \BibitemShut {NoStop}%
\bibitem [{\citenamefont {Barrow}\ and\ \citenamefont {Carr}(1996)}]{BarCar96}%
  \BibitemOpen
  \bibfield  {author} {\bibinfo {author} {\bibfnamefont {J.~D.}\ \bibnamefont {Barrow}}\ and\ \bibinfo {author} {\bibfnamefont {B.~J.}\ \bibnamefont {Carr}},\ }\href {\doibase 10.1103/PhysRevD.54.3920} {\bibfield  {journal} {\bibinfo  {journal} {Phys. Rev. D}\ }\textbf {\bibinfo {volume} {54}},\ \bibinfo {pages} {3920} (\bibinfo {year} {1996})}\BibitemShut {NoStop}%
\bibitem [{\citenamefont {Nayak}\ \emph {et~al.}(2010)\citenamefont {Nayak}, \citenamefont {Majumdar},\ and\ \citenamefont {Singh}}]{NayMajSin10}%
  \BibitemOpen
  \bibfield  {author} {\bibinfo {author} {\bibfnamefont {B.}~\bibnamefont {Nayak}}, \bibinfo {author} {\bibfnamefont {A.~S.}\ \bibnamefont {Majumdar}}, \ and\ \bibinfo {author} {\bibfnamefont {L.~P.}\ \bibnamefont {Singh}},\ }\href {\doibase 10.1088/1475-7516/2010/08/039} {\bibfield  {journal} {\bibinfo  {journal} {J. Cosmol. Astropart. Phys.}\ }\textbf {\bibinfo {volume} {08}},\ \bibinfo {pages} {039} (\bibinfo {year} {2010})},\ \Eprint {http://arxiv.org/abs/1002.3472} {arXiv:1002.3472 [gr-qc]} \BibitemShut {NoStop}%
\bibitem [{\citenamefont {Aliferis}\ and\ \citenamefont {Zarikas}(2021)}]{AliZar20}%
  \BibitemOpen
  \bibfield  {author} {\bibinfo {author} {\bibfnamefont {G.}~\bibnamefont {Aliferis}}\ and\ \bibinfo {author} {\bibfnamefont {V.}~\bibnamefont {Zarikas}},\ }\href {\doibase 10.1103/PhysRevD.103.023509} {\bibfield  {journal} {\bibinfo  {journal} {Phys. Rev. D}\ }\textbf {\bibinfo {volume} {103}},\ \bibinfo {pages} {023509} (\bibinfo {year} {2021})},\ \Eprint {http://arxiv.org/abs/2006.13621} {arXiv:2006.13621 [gr-qc]} \BibitemShut {NoStop}%
\bibitem [{\citenamefont {Chen}(2021)}]{Che19}%
  \BibitemOpen
  \bibfield  {author} {\bibinfo {author} {\bibfnamefont {C.-Y.}\ \bibnamefont {Chen}},\ }\href {\doibase 10.1142/S0218271821500103} {\bibfield  {journal} {\bibinfo  {journal} {Int. J. Mod. Phys. D}\ }\textbf {\bibinfo {volume} {30}},\ \bibinfo {pages} {02} (\bibinfo {year} {2021})},\ \Eprint {http://arxiv.org/abs/1912.10690} {arXiv:1912.10690 [gr-qc]} \BibitemShut {NoStop}%
\bibitem [{\citenamefont {Lacki}\ and\ \citenamefont {Beacom}(2010)}]{LacBea10}%
  \BibitemOpen
  \bibfield  {author} {\bibinfo {author} {\bibfnamefont {B.~C.}\ \bibnamefont {Lacki}}\ and\ \bibinfo {author} {\bibfnamefont {J.~F.}\ \bibnamefont {Beacom}},\ }\href {\doibase 10.1088/2041-8205/720/1/L67} {\bibfield  {journal} {\bibinfo  {journal} {Astrophys. J. Lett.}\ }\textbf {\bibinfo {volume} {720}},\ \bibinfo {pages} {L67} (\bibinfo {year} {2010})},\ \Eprint {http://arxiv.org/abs/1003.3466} {arXiv:1003.3466 [astro-ph.CO]} \BibitemShut {NoStop}%
\bibitem [{\citenamefont {{Saito R, Matsumoto S, Shirai S, Yanagida T T}}(2011)}]{SaiShi11-2}%
  \BibitemOpen
  \bibfield  {author} {\bibinfo {author} {\bibnamefont {{Saito R, Matsumoto S, Shirai S, Yanagida T T}}},\ }in\ \href@noop {} {\emph {\bibinfo {booktitle} {Proceedings of the 21st Workshop on General Relativity and Gravitation in Japan (JGRG21)}}}\ (\bibinfo {address} {Sendai, Japan},\ \bibinfo {year} {2011})\BibitemShut {NoStop}%
\bibitem [{\citenamefont {Saito}\ and\ \citenamefont {Shirai}(2011)}]{SaiShi11}%
  \BibitemOpen
  \bibfield  {author} {\bibinfo {author} {\bibfnamefont {R.}~\bibnamefont {Saito}}\ and\ \bibinfo {author} {\bibfnamefont {S.}~\bibnamefont {Shirai}},\ }\href {\doibase 10.1016/j.physletb.2011.01.038} {\bibfield  {journal} {\bibinfo  {journal} {Phys. Lett. B}\ }\textbf {\bibinfo {volume} {697}},\ \bibinfo {pages} {95} (\bibinfo {year} {2011})},\ \Eprint {http://arxiv.org/abs/1009.1947} {arXiv:1009.1947 [hep-ph]} \BibitemShut {NoStop}%
\bibitem [{\citenamefont {{Zhang}}(2011)}]{Don11}%
  \BibitemOpen
  \bibfield  {author} {\bibinfo {author} {\bibfnamefont {D.}~\bibnamefont {{Zhang}}},\ }\href {\doibase 10.1111/j.1365-2966.2011.19602.x} {\bibfield  {journal} {\bibinfo  {journal} {Mon. Not. R. Astron. Soc.}\ }\textbf {\bibinfo {volume} {418}},\ \bibinfo {pages} {1850} (\bibinfo {year} {2011})},\ \Eprint {http://arxiv.org/abs/1011.1935} {arXiv:1011.1935 [astro-ph.CO]} \BibitemShut {NoStop}%
\bibitem [{\citenamefont {Eroshenko}(2016)}]{Ero16}%
  \BibitemOpen
  \bibfield  {author} {\bibinfo {author} {\bibfnamefont {Y.~N.}\ \bibnamefont {Eroshenko}},\ }\href {\doibase 10.1134/S1063773716060013} {\bibfield  {journal} {\bibinfo  {journal} {Astron. Lett.}\ }\textbf {\bibinfo {volume} {42}},\ \bibinfo {pages} {347} (\bibinfo {year} {2016})},\ \Eprint {http://arxiv.org/abs/1607.00612} {arXiv:1607.00612 [astro-ph.HE]} \BibitemShut {NoStop}%
\bibitem [{\citenamefont {Boucenna}\ \emph {et~al.}(2018)\citenamefont {Boucenna}, \citenamefont {Kuhnel}, \citenamefont {Ohlsson},\ and\ \citenamefont {Visinelli}}]{Bouetal18}%
  \BibitemOpen
  \bibfield  {author} {\bibinfo {author} {\bibfnamefont {S.~M.}\ \bibnamefont {Boucenna}}, \bibinfo {author} {\bibfnamefont {F.}~\bibnamefont {Kuhnel}}, \bibinfo {author} {\bibfnamefont {T.}~\bibnamefont {Ohlsson}}, \ and\ \bibinfo {author} {\bibfnamefont {L.}~\bibnamefont {Visinelli}},\ }\href {\doibase 10.1088/1475-7516/2018/07/003} {\bibfield  {journal} {\bibinfo  {journal} {J. Cosmol. Astropart. Phys.}\ }\textbf {\bibinfo {volume} {07}},\ \bibinfo {pages} {003} (\bibinfo {year} {2018})},\ \Eprint {http://arxiv.org/abs/1712.06383} {arXiv:1712.06383 [hep-ph]} \BibitemShut {NoStop}%
\bibitem [{\citenamefont {Adamek}\ \emph {et~al.}(2019)\citenamefont {Adamek}, \citenamefont {Byrnes}, \citenamefont {Gosenca},\ and\ \citenamefont {Hotchkiss}}]{Adaetal19}%
  \BibitemOpen
  \bibfield  {author} {\bibinfo {author} {\bibfnamefont {J.}~\bibnamefont {Adamek}}, \bibinfo {author} {\bibfnamefont {C.~T.}\ \bibnamefont {Byrnes}}, \bibinfo {author} {\bibfnamefont {M.}~\bibnamefont {Gosenca}}, \ and\ \bibinfo {author} {\bibfnamefont {S.}~\bibnamefont {Hotchkiss}},\ }\href {\doibase 10.1103/PhysRevD.100.023506} {\bibfield  {journal} {\bibinfo  {journal} {Phys. Rev. D}\ }\textbf {\bibinfo {volume} {100}},\ \bibinfo {pages} {023506} (\bibinfo {year} {2019})},\ \Eprint {http://arxiv.org/abs/1901.08528} {arXiv:1901.08528 [astro-ph.CO]} \BibitemShut {NoStop}%
\bibitem [{\citenamefont {Eroshenko}(2020)}]{Ero19}%
  \BibitemOpen
  \bibfield  {author} {\bibinfo {author} {\bibfnamefont {Y.~N.}\ \bibnamefont {Eroshenko}},\ }\href {\doibase 10.1142/S0217751X20400461} {\bibfield  {journal} {\bibinfo  {journal} {Int. J. Mod. Phys. A}\ }\textbf {\bibinfo {volume} {35}},\ \bibinfo {pages} {2040046} (\bibinfo {year} {2020})},\ \Eprint {http://arxiv.org/abs/1910.01564} {arXiv:1910.01564 [astro-ph.CO]} \BibitemShut {NoStop}%
\bibitem [{\citenamefont {Carr}\ \emph {et~al.}(2021)\citenamefont {Carr}, \citenamefont {Kuhnel},\ and\ \citenamefont {Visinelli}}]{CarKuhVis21}%
  \BibitemOpen
  \bibfield  {author} {\bibinfo {author} {\bibfnamefont {B.}~\bibnamefont {Carr}}, \bibinfo {author} {\bibfnamefont {F.}~\bibnamefont {Kuhnel}}, \ and\ \bibinfo {author} {\bibfnamefont {L.}~\bibnamefont {Visinelli}},\ }\href {\doibase 10.1093/mnras/staa3651} {\bibfield  {journal} {\bibinfo  {journal} {Mon. Not. Roy. Astron. Soc.}\ }\textbf {\bibinfo {volume} {501}},\ \bibinfo {pages} {2029} (\bibinfo {year} {2021})},\ \Eprint {http://arxiv.org/abs/2008.08077} {arXiv:2008.08077 [astro-ph.CO]} \BibitemShut {NoStop}%
\bibitem [{\citenamefont {Ireland}(2025)}]{Ire24}%
  \BibitemOpen
  \bibfield  {author} {\bibinfo {author} {\bibfnamefont {A.}~\bibnamefont {Ireland}},\ }\href {\doibase 10.1103/PhysRevD.111.023513} {\bibfield  {journal} {\bibinfo  {journal} {Phys. Rev. D}\ }\textbf {\bibinfo {volume} {111}},\ \bibinfo {pages} {023513} (\bibinfo {year} {2025})},\ \Eprint {http://arxiv.org/abs/2406.07624} {arXiv:2406.07624 [astro-ph.CO]} \BibitemShut {NoStop}%
\bibitem [{\citenamefont {Bertschinger}(1985)}]{Ber85}%
  \BibitemOpen
  \bibfield  {author} {\bibinfo {author} {\bibfnamefont {E.}~\bibnamefont {Bertschinger}},\ }\href {\doibase 10.1086/191028} {\bibfield  {journal} {\bibinfo  {journal} {Astrophys. J. Suppl.}\ }\textbf {\bibinfo {volume} {58}},\ \bibinfo {pages} {39} (\bibinfo {year} {1985})}\BibitemShut {NoStop}%
\bibitem [{\citenamefont {{Berezinsky}}\ \emph {et~al.}(1992)\citenamefont {{Berezinsky}}, \citenamefont {{Gurevich}},\ and\ \citenamefont {{Zybin}}}]{BerGurZyb92}%
  \BibitemOpen
  \bibfield  {author} {\bibinfo {author} {\bibfnamefont {V.~S.}\ \bibnamefont {{Berezinsky}}}, \bibinfo {author} {\bibfnamefont {A.~V.}\ \bibnamefont {{Gurevich}}}, \ and\ \bibinfo {author} {\bibfnamefont {K.~P.}\ \bibnamefont {{Zybin}}},\ }\href {\doibase 10.1016/0370-2693(92)90686-X} {\bibfield  {journal} {\bibinfo  {journal} {Physics Letters B}\ }\textbf {\bibinfo {volume} {294}},\ \bibinfo {pages} {221} (\bibinfo {year} {1992})}\BibitemShut {NoStop}%
\bibitem [{\citenamefont {{Silk}}\ and\ \citenamefont {{Stebbins}}(1993)}]{SilSte93}%
  \BibitemOpen
  \bibfield  {author} {\bibinfo {author} {\bibfnamefont {J.}~\bibnamefont {{Silk}}}\ and\ \bibinfo {author} {\bibfnamefont {A.}~\bibnamefont {{Stebbins}}},\ }\href {\doibase 10.1086/172846} {\bibfield  {journal} {\bibinfo  {journal} {Astrophys. J.}\ }\textbf {\bibinfo {volume} {411}},\ \bibinfo {pages} {439} (\bibinfo {year} {1993})}\BibitemShut {NoStop}%
\bibitem [{\citenamefont {Gondolo}\ and\ \citenamefont {Silk}(1999)}]{GonSil99}%
  \BibitemOpen
  \bibfield  {author} {\bibinfo {author} {\bibfnamefont {P.}~\bibnamefont {Gondolo}}\ and\ \bibinfo {author} {\bibfnamefont {J.}~\bibnamefont {Silk}},\ }\href {\doibase 10.1103/PhysRevLett.83.1719} {\bibfield  {journal} {\bibinfo  {journal} {Phys. Rev. Lett.}\ }\textbf {\bibinfo {volume} {83}},\ \bibinfo {pages} {1719} (\bibinfo {year} {1999})},\ \Eprint {http://arxiv.org/abs/astro-ph/9906391} {arXiv:astro-ph/9906391} \BibitemShut {NoStop}%
\bibitem [{\citenamefont {Vasiliev}(2007)}]{Vas07}%
  \BibitemOpen
  \bibfield  {author} {\bibinfo {author} {\bibfnamefont {E.}~\bibnamefont {Vasiliev}},\ }\href {\doibase 10.1103/PhysRevD.76.103532} {\bibfield  {journal} {\bibinfo  {journal} {Phys. Rev. D}\ }\textbf {\bibinfo {volume} {76}},\ \bibinfo {pages} {103532} (\bibinfo {year} {2007})},\ \Eprint {http://arxiv.org/abs/0707.3334} {arXiv:0707.3334 [astro-ph]} \BibitemShut {NoStop}%
\bibitem [{\citenamefont {Eroshenko}(2024)}]{Ero24}%
  \BibitemOpen
  \bibfield  {author} {\bibinfo {author} {\bibfnamefont {Y.~N.}\ \bibnamefont {Eroshenko}},\ }\href {\doibase 10.1088/1475-7516/2024/08/019} {\bibfield  {journal} {\bibinfo  {journal} {J. Cosmol. Astropart. Phys.}\ }\textbf {\bibinfo {volume} {08}},\ \bibinfo {pages} {019} (\bibinfo {year} {2024})},\ \Eprint {http://arxiv.org/abs/2407.10225} {arXiv:2407.10225 [astro-ph.CO]} \BibitemShut {NoStop}%
\bibitem [{\citenamefont {{Wang}}\ \emph {et~al.}(2021)\citenamefont {{Wang}} \emph {et~al.}}]{2021ApJ...907L...1W}%
  \BibitemOpen
  \bibfield  {author} {\bibinfo {author} {\bibfnamefont {F.}~\bibnamefont {{Wang}}} \emph {et~al.},\ }\href {\doibase 10.3847/2041-8213/abd8c6} {\bibfield  {journal} {\bibinfo  {journal} {Astrophys. J. Lett.}\ }\textbf {\bibinfo {volume} {907}},\ \bibinfo {eid} {L1} (\bibinfo {year} {2021})},\ \Eprint {http://arxiv.org/abs/2101.03179} {arXiv:2101.03179 [astro-ph.GA]} \BibitemShut {NoStop}%
\bibitem [{\citenamefont {Dokuchaev}\ and\ \citenamefont {Eroshenko}(2001)}]{Dokuchaev:2001ekr}%
  \BibitemOpen
  \bibfield  {author} {\bibinfo {author} {\bibfnamefont {V.~I.}\ \bibnamefont {Dokuchaev}}\ and\ \bibinfo {author} {\bibfnamefont {Y.~N.}\ \bibnamefont {Eroshenko}},\ }\href {\doibase 10.1134/1.1424357} {\bibfield  {journal} {\bibinfo  {journal} {Astron. Lett.}\ }\textbf {\bibinfo {volume} {27}},\ \bibinfo {pages} {759} (\bibinfo {year} {2001})},\ \Eprint {http://arxiv.org/abs/astro-ph/0202019} {arXiv:astro-ph/0202019} \BibitemShut {NoStop}%
\bibitem [{\citenamefont {{Carr}}\ and\ \citenamefont {{Lidsey}}(1993)}]{1993PhRvD..48..543C}%
  \BibitemOpen
  \bibfield  {author} {\bibinfo {author} {\bibfnamefont {B.~J.}\ \bibnamefont {{Carr}}}\ and\ \bibinfo {author} {\bibfnamefont {J.~E.}\ \bibnamefont {{Lidsey}}},\ }\href {\doibase 10.1103/PhysRevD.48.543} {\bibfield  {journal} {\bibinfo  {journal} {Phys. Rev. D}\ }\textbf {\bibinfo {volume} {48}},\ \bibinfo {pages} {543} (\bibinfo {year} {1993})}\BibitemShut {NoStop}%
\bibitem [{\citenamefont {{Kohri}}\ \emph {et~al.}(2014)\citenamefont {{Kohri}}, \citenamefont {{Nakama}},\ and\ \citenamefont {{Suyama}}}]{2014PhRvD..90h3514K}%
  \BibitemOpen
  \bibfield  {author} {\bibinfo {author} {\bibfnamefont {K.}~\bibnamefont {{Kohri}}}, \bibinfo {author} {\bibfnamefont {T.}~\bibnamefont {{Nakama}}}, \ and\ \bibinfo {author} {\bibfnamefont {T.}~\bibnamefont {{Suyama}}},\ }\href {\doibase 10.1103/PhysRevD.90.083514} {\bibfield  {journal} {\bibinfo  {journal} {Phys. Rev. D}\ }\textbf {\bibinfo {volume} {90}},\ \bibinfo {eid} {083514} (\bibinfo {year} {2014})},\ \Eprint {http://arxiv.org/abs/1405.5999} {arXiv:1405.5999 [astro-ph.CO]} \BibitemShut {NoStop}%
\bibitem [{\citenamefont {Nakama}\ \emph {et~al.}(2016)\citenamefont {Nakama}, \citenamefont {Suyama},\ and\ \citenamefont {Yokoyama}}]{Nakama:2016kfq}%
  \BibitemOpen
  \bibfield  {author} {\bibinfo {author} {\bibfnamefont {T.}~\bibnamefont {Nakama}}, \bibinfo {author} {\bibfnamefont {T.}~\bibnamefont {Suyama}}, \ and\ \bibinfo {author} {\bibfnamefont {J.}~\bibnamefont {Yokoyama}},\ }\href {\doibase 10.1103/PhysRevD.94.103522} {\bibfield  {journal} {\bibinfo  {journal} {Phys. Rev. D}\ }\textbf {\bibinfo {volume} {94}},\ \bibinfo {pages} {103522} (\bibinfo {year} {2016})},\ \Eprint {http://arxiv.org/abs/1609.02245} {arXiv:1609.02245 [gr-qc]} \BibitemShut {NoStop}%
\bibitem [{\citenamefont {Shemmer}\ \emph {et~al.}(2004)\citenamefont {Shemmer}, \citenamefont {Netzer}, \citenamefont {Maiolino}, \citenamefont {Oliva}, \citenamefont {Croom}, \citenamefont {Corbett},\ and\ \citenamefont {di~Fabrizio}}]{Shemmer:2004ph}%
  \BibitemOpen
  \bibfield  {author} {\bibinfo {author} {\bibfnamefont {O.}~\bibnamefont {Shemmer}}, \bibinfo {author} {\bibfnamefont {H.}~\bibnamefont {Netzer}}, \bibinfo {author} {\bibfnamefont {R.}~\bibnamefont {Maiolino}}, \bibinfo {author} {\bibfnamefont {E.}~\bibnamefont {Oliva}}, \bibinfo {author} {\bibfnamefont {S.}~\bibnamefont {Croom}}, \bibinfo {author} {\bibfnamefont {E.}~\bibnamefont {Corbett}}, \ and\ \bibinfo {author} {\bibfnamefont {L.}~\bibnamefont {di~Fabrizio}},\ }\href {\doibase 10.1086/423607} {\bibfield  {journal} {\bibinfo  {journal} {Astrophys. J.}\ }\textbf {\bibinfo {volume} {614}},\ \bibinfo {pages} {547} (\bibinfo {year} {2004})},\ \Eprint {http://arxiv.org/abs/astro-ph/0406559} {arXiv:astro-ph/0406559} \BibitemShut {NoStop}%
\bibitem [{\citenamefont {{Ryan}}(1972)}]{Rya72}%
  \BibitemOpen
  \bibfield  {author} {\bibinfo {author} {\bibfnamefont {M.~P.}\ \bibnamefont {{Ryan}}, \bibfnamefont {Jr.}},\ }\href {\doibase 10.1086/181056} {\bibfield  {journal} {\bibinfo  {journal} {Astrophys. J. Lett.}\ }\textbf {\bibinfo {volume} {177}},\ \bibinfo {pages} {L79} (\bibinfo {year} {1972})}\BibitemShut {NoStop}%
\bibitem [{\citenamefont {{Carr}}\ and\ \citenamefont {{Rees}}(1984)}]{CarRee84}%
  \BibitemOpen
  \bibfield  {author} {\bibinfo {author} {\bibfnamefont {B.~J.}\ \bibnamefont {{Carr}}}\ and\ \bibinfo {author} {\bibfnamefont {M.~J.}\ \bibnamefont {{Rees}}},\ }\href {\doibase 10.1093/mnras/206.4.801} {\bibfield  {journal} {\bibinfo  {journal} {Mon. Not. R. Astron. Soc.}\ }\textbf {\bibinfo {volume} {206}},\ \bibinfo {pages} {801} (\bibinfo {year} {1984})}\BibitemShut {NoStop}%
\bibitem [{\citenamefont {{Gunn}}(1977)}]{Gun77}%
  \BibitemOpen
  \bibfield  {author} {\bibinfo {author} {\bibfnamefont {J.~E.}\ \bibnamefont {{Gunn}}},\ }\href {\doibase 10.1086/155715} {\bibfield  {journal} {\bibinfo  {journal} {Astrophys. J.}\ }\textbf {\bibinfo {volume} {218}},\ \bibinfo {pages} {592} (\bibinfo {year} {1977})}\BibitemShut {NoStop}%
\bibitem [{\citenamefont {{Fillmore}}\ and\ \citenamefont {{Goldreich}}(1984)}]{FilGol84}%
  \BibitemOpen
  \bibfield  {author} {\bibinfo {author} {\bibfnamefont {J.~A.}\ \bibnamefont {{Fillmore}}}\ and\ \bibinfo {author} {\bibfnamefont {P.}~\bibnamefont {{Goldreich}}},\ }\href {\doibase 10.1086/162070} {\bibfield  {journal} {\bibinfo  {journal} {Astrophys. J.}\ }\textbf {\bibinfo {volume} {281}},\ \bibinfo {pages} {1} (\bibinfo {year} {1984})}\BibitemShut {NoStop}%
\bibitem [{\citenamefont {Dokuchaev}\ and\ \citenamefont {Eroshenko}(2003)}]{DokEro03}%
  \BibitemOpen
  \bibfield  {author} {\bibinfo {author} {\bibfnamefont {V.~I.}\ \bibnamefont {Dokuchaev}}\ and\ \bibinfo {author} {\bibfnamefont {Y.~N.}\ \bibnamefont {Eroshenko}},\ }\href {\doibase 10.1080/10556790310001609115} {\bibfield  {journal} {\bibinfo  {journal} {Astron. Astrophys. Trans.}\ }\textbf {\bibinfo {volume} {22}},\ \bibinfo {pages} {727} (\bibinfo {year} {2003})},\ \Eprint {http://arxiv.org/abs/astro-ph/0209324} {arXiv:astro-ph/0209324} \BibitemShut {NoStop}%
\bibitem [{\citenamefont {{Dubrovich}}\ and\ \citenamefont {{Glazyrin}}(2012)}]{DubGla12}%
  \BibitemOpen
  \bibfield  {author} {\bibinfo {author} {\bibfnamefont {V.~K.}\ \bibnamefont {{Dubrovich}}}\ and\ \bibinfo {author} {\bibfnamefont {S.~I.}\ \bibnamefont {{Glazyrin}}},\ }\href@noop {} {\  (\bibinfo {year} {2012})},\ \Eprint {http://arxiv.org/abs/1208.3999} {arXiv:1208.3999 [astro-ph.CO]} \BibitemShut {NoStop}%
\bibitem [{\citenamefont {{Meszaros}}(1974)}]{Mes74}%
  \BibitemOpen
  \bibfield  {author} {\bibinfo {author} {\bibfnamefont {P.}~\bibnamefont {{Meszaros}}},\ }\href@noop {} {\bibfield  {journal} {\bibinfo  {journal} {Astron. \& Astrophys.}\ }\textbf {\bibinfo {volume} {37}},\ \bibinfo {pages} {225} (\bibinfo {year} {1974})}\BibitemShut {NoStop}%
\bibitem [{\citenamefont {{Meszaros}}(1975)}]{Mes75}%
  \BibitemOpen
  \bibfield  {author} {\bibinfo {author} {\bibfnamefont {P.}~\bibnamefont {{Meszaros}}},\ }\href@noop {} {\bibfield  {journal} {\bibinfo  {journal} {Astron. \& Astrophys.}\ }\textbf {\bibinfo {volume} {38}},\ \bibinfo {pages} {5} (\bibinfo {year} {1975})}\BibitemShut {NoStop}%
\bibitem [{\citenamefont {{Carr}}(1977)}]{Car77}%
  \BibitemOpen
  \bibfield  {author} {\bibinfo {author} {\bibfnamefont {B.~J.}\ \bibnamefont {{Carr}}},\ }\href@noop {} {\bibfield  {journal} {\bibinfo  {journal} {Astron. \& Astrophys.}\ }\textbf {\bibinfo {volume} {56}},\ \bibinfo {pages} {377} (\bibinfo {year} {1977})}\BibitemShut {NoStop}%
\bibitem [{\citenamefont {Afshordi}\ \emph {et~al.}(2003)\citenamefont {Afshordi}, \citenamefont {McDonald},\ and\ \citenamefont {Spergel}}]{Afshordi:2003zb}%
  \BibitemOpen
  \bibfield  {author} {\bibinfo {author} {\bibfnamefont {N.}~\bibnamefont {Afshordi}}, \bibinfo {author} {\bibfnamefont {P.}~\bibnamefont {McDonald}}, \ and\ \bibinfo {author} {\bibfnamefont {D.~N.}\ \bibnamefont {Spergel}},\ }\href {\doibase 10.1086/378763} {\bibfield  {journal} {\bibinfo  {journal} {Astrophys. J. Lett.}\ }\textbf {\bibinfo {volume} {594}},\ \bibinfo {pages} {L71} (\bibinfo {year} {2003})},\ \Eprint {http://arxiv.org/abs/astro-ph/0302035} {arXiv:astro-ph/0302035} \BibitemShut {NoStop}%
\bibitem [{\citenamefont {Murgia}\ \emph {et~al.}(2019)\citenamefont {Murgia}, \citenamefont {Scelfo}, \citenamefont {Viel},\ and\ \citenamefont {Raccanelli}}]{Murgia:2019duy}%
  \BibitemOpen
  \bibfield  {author} {\bibinfo {author} {\bibfnamefont {R.}~\bibnamefont {Murgia}}, \bibinfo {author} {\bibfnamefont {G.}~\bibnamefont {Scelfo}}, \bibinfo {author} {\bibfnamefont {M.}~\bibnamefont {Viel}}, \ and\ \bibinfo {author} {\bibfnamefont {A.}~\bibnamefont {Raccanelli}},\ }\href {\doibase 10.1103/PhysRevLett.123.071102} {\bibfield  {journal} {\bibinfo  {journal} {Phys. Rev. Lett.}\ }\textbf {\bibinfo {volume} {123}},\ \bibinfo {pages} {071102} (\bibinfo {year} {2019})},\ \Eprint {http://arxiv.org/abs/1903.10509} {arXiv:1903.10509 [astro-ph.CO]} \BibitemShut {NoStop}%
\bibitem [{\citenamefont {{Vasiliev}}\ and\ \citenamefont {{Shchekinov}}(2012)}]{VasShc12}%
  \BibitemOpen
  \bibfield  {author} {\bibinfo {author} {\bibfnamefont {E.~O.}\ \bibnamefont {{Vasiliev}}}\ and\ \bibinfo {author} {\bibfnamefont {Y.~A.}\ \bibnamefont {{Shchekinov}}},\ }\href {\doibase 10.1134/S1063772912020096} {\bibfield  {journal} {\bibinfo  {journal} {Astronomy Reports}\ }\textbf {\bibinfo {volume} {56}},\ \bibinfo {pages} {77} (\bibinfo {year} {2012})},\ \Eprint {http://arxiv.org/abs/1207.5109} {arXiv:1207.5109 [astro-ph.CO]} \BibitemShut {NoStop}%
\bibitem [{\citenamefont {{Dubrovich}}\ and\ \citenamefont {{Grachev}}(2020)}]{DubrovichGra19}%
  \BibitemOpen
  \bibfield  {author} {\bibinfo {author} {\bibfnamefont {V.~K.}\ \bibnamefont {{Dubrovich}}}\ and\ \bibinfo {author} {\bibfnamefont {S.~I.}\ \bibnamefont {{Grachev}}},\ }\href {\doibase 10.1134/S1063773719110021} {\bibfield  {journal} {\bibinfo  {journal} {Astronomy Letters}\ }\textbf {\bibinfo {volume} {45}},\ \bibinfo {pages} {701} (\bibinfo {year} {2020})}\BibitemShut {NoStop}%
\bibitem [{\citenamefont {Dubrovich}\ \emph {et~al.}(2021)\citenamefont {Dubrovich}, \citenamefont {Eroshenko},\ and\ \citenamefont {Grachev}}]{DubEroGra21}%
  \BibitemOpen
  \bibfield  {author} {\bibinfo {author} {\bibfnamefont {V.~K.}\ \bibnamefont {Dubrovich}}, \bibinfo {author} {\bibfnamefont {Y.~N.}\ \bibnamefont {Eroshenko}}, \ and\ \bibinfo {author} {\bibfnamefont {S.~I.}\ \bibnamefont {Grachev}},\ }\href {\doibase 10.1093/mnras/stab689} {\bibfield  {journal} {\bibinfo  {journal} {Mon. Not. Roy. Astron. Soc.}\ }\textbf {\bibinfo {volume} {503}},\ \bibinfo {pages} {3081} (\bibinfo {year} {2021})},\ \Eprint {http://arxiv.org/abs/2007.10896} {arXiv:2007.10896 [astro-ph.CO]} \BibitemShut {NoStop}%
\bibitem [{\citenamefont {{Tashiro}}\ and\ \citenamefont {{Sugiyama}}(2013)}]{2013MNRAS.435.3001T}%
  \BibitemOpen
  \bibfield  {author} {\bibinfo {author} {\bibfnamefont {H.}~\bibnamefont {{Tashiro}}}\ and\ \bibinfo {author} {\bibfnamefont {N.}~\bibnamefont {{Sugiyama}}},\ }\href {\doibase 10.1093/mnras/stt1493} {\bibfield  {journal} {\bibinfo  {journal} {Mon. Not. R. Astron. Soc.}\ }\textbf {\bibinfo {volume} {435}},\ \bibinfo {pages} {3001} (\bibinfo {year} {2013})},\ \Eprint {http://arxiv.org/abs/1207.6405} {arXiv:1207.6405 [astro-ph.CO]} \BibitemShut {NoStop}%
\bibitem [{\citenamefont {{Taylor}}\ \emph {et~al.}(2025)\citenamefont {{Taylor}} \emph {et~al.}}]{2025ApJ...989L...7T}%
  \BibitemOpen
  \bibfield  {author} {\bibinfo {author} {\bibfnamefont {A.~J.}\ \bibnamefont {{Taylor}}} \emph {et~al.},\ }\href {\doibase 10.3847/2041-8213/ade789} {\bibfield  {journal} {\bibinfo  {journal} {Astrophys. J. Lett.}\ }\textbf {\bibinfo {volume} {989}},\ \bibinfo {eid} {L7} (\bibinfo {year} {2025})},\ \Eprint {http://arxiv.org/abs/2505.04609} {arXiv:2505.04609 [astro-ph.GA]} \BibitemShut {NoStop}%
\bibitem [{\citenamefont {{Delvecchio}}\ \emph {et~al.}(2025)\citenamefont {{Delvecchio}}, \citenamefont {{Daddi}}, \citenamefont {{Magnelli}}, \citenamefont {{Elbaz}}, \citenamefont {{Giavalisco}}, \citenamefont {{Traina}}, \citenamefont {{Lanzuisi}}, \citenamefont {{Akins}}, \citenamefont {{Belli}}, \citenamefont {{Casey}}, \citenamefont {{Gentile}}, \citenamefont {{Gruppioni}}, \citenamefont {{Pozzi}},\ and\ \citenamefont {{Zamorani}}}]{2025A&A...704A.313D}%
  \BibitemOpen
  \bibfield  {author} {\bibinfo {author} {\bibfnamefont {I.}~\bibnamefont {{Delvecchio}}}, \bibinfo {author} {\bibfnamefont {E.}~\bibnamefont {{Daddi}}}, \bibinfo {author} {\bibfnamefont {B.}~\bibnamefont {{Magnelli}}}, \bibinfo {author} {\bibfnamefont {D.}~\bibnamefont {{Elbaz}}}, \bibinfo {author} {\bibfnamefont {M.}~\bibnamefont {{Giavalisco}}}, \bibinfo {author} {\bibfnamefont {A.}~\bibnamefont {{Traina}}}, \bibinfo {author} {\bibfnamefont {G.}~\bibnamefont {{Lanzuisi}}}, \bibinfo {author} {\bibfnamefont {H.~B.}\ \bibnamefont {{Akins}}}, \bibinfo {author} {\bibfnamefont {S.}~\bibnamefont {{Belli}}}, \bibinfo {author} {\bibfnamefont {C.~M.}\ \bibnamefont {{Casey}}}, \bibinfo {author} {\bibfnamefont {F.}~\bibnamefont {{Gentile}}}, \bibinfo {author} {\bibfnamefont {C.}~\bibnamefont {{Gruppioni}}}, \bibinfo {author} {\bibfnamefont {F.}~\bibnamefont {{Pozzi}}}, \ and\ \bibinfo {author} {\bibfnamefont {G.}~\bibnamefont {{Zamorani}}},\ }\href {\doibase 10.1051/0004-6361/202557164} {\bibfield  {journal} {\bibinfo
   {journal} {Astron. \& Astrophys.}\ }\textbf {\bibinfo {volume} {704}},\ \bibinfo {eid} {A313} (\bibinfo {year} {2025})},\ \Eprint {http://arxiv.org/abs/2509.07100} {arXiv:2509.07100 [astro-ph.GA]} \BibitemShut {NoStop}%
\bibitem [{\citenamefont {Zhang}\ \emph {et~al.}(2026)\citenamefont {Zhang}, \citenamefont {Liu}, \citenamefont {Bromm},\ and\ \citenamefont {K{\"u}hnel}}]{Zhang:2025oyl}%
  \BibitemOpen
  \bibfield  {author} {\bibinfo {author} {\bibfnamefont {S.}~\bibnamefont {Zhang}}, \bibinfo {author} {\bibfnamefont {B.}~\bibnamefont {Liu}}, \bibinfo {author} {\bibfnamefont {V.}~\bibnamefont {Bromm}}, \ and\ \bibinfo {author} {\bibfnamefont {F.}~\bibnamefont {K{\"u}hnel}},\ }\href {\doibase 10.3847/2041-8213/ae4bd0} {\bibfield  {journal} {\bibinfo  {journal} {Astrophys. J. Lett.}\ }\textbf {\bibinfo {volume} {1000}},\ \bibinfo {pages} {L19} (\bibinfo {year} {2026})},\ \Eprint {http://arxiv.org/abs/2512.14066} {arXiv:2512.14066 [astro-ph.GA]} \BibitemShut {NoStop}%
\bibitem [{\citenamefont {Carr}\ and\ \citenamefont {Silk}(2018)}]{Carr:2018rid}%
  \BibitemOpen
  \bibfield  {author} {\bibinfo {author} {\bibfnamefont {B.}~\bibnamefont {Carr}}\ and\ \bibinfo {author} {\bibfnamefont {J.}~\bibnamefont {Silk}},\ }\href {\doibase 10.1093/mnras/sty1204} {\bibfield  {journal} {\bibinfo  {journal} {Mon. Not. Roy. Astron. Soc.}\ }\textbf {\bibinfo {volume} {478}},\ \bibinfo {pages} {3756} (\bibinfo {year} {2018})},\ \Eprint {http://arxiv.org/abs/1801.00672} {arXiv:1801.00672 [astro-ph.CO]} \BibitemShut {NoStop}%
\bibitem [{\citenamefont {Inman}\ and\ \citenamefont {Ali-Ha{\"\i}moud}(2019)}]{Inman:2019wvr}%
  \BibitemOpen
  \bibfield  {author} {\bibinfo {author} {\bibfnamefont {D.}~\bibnamefont {Inman}}\ and\ \bibinfo {author} {\bibfnamefont {Y.}~\bibnamefont {Ali-Ha{\"\i}moud}},\ }\href {\doibase 10.1103/PhysRevD.100.083528} {\bibfield  {journal} {\bibinfo  {journal} {Phys. Rev. D}\ }\textbf {\bibinfo {volume} {100}},\ \bibinfo {pages} {083528} (\bibinfo {year} {2019})},\ \Eprint {http://arxiv.org/abs/1907.08129} {arXiv:1907.08129 [astro-ph.CO]} \BibitemShut {NoStop}%
\bibitem [{\citenamefont {Delos}\ \emph {et~al.}(2024)\citenamefont {Delos}, \citenamefont {Rantala}, \citenamefont {Young},\ and\ \citenamefont {Schmidt}}]{Delos:2024poq}%
  \BibitemOpen
  \bibfield  {author} {\bibinfo {author} {\bibfnamefont {M.~S.}\ \bibnamefont {Delos}}, \bibinfo {author} {\bibfnamefont {A.}~\bibnamefont {Rantala}}, \bibinfo {author} {\bibfnamefont {S.}~\bibnamefont {Young}}, \ and\ \bibinfo {author} {\bibfnamefont {F.}~\bibnamefont {Schmidt}},\ }\href {\doibase 10.1088/1475-7516/2024/12/005} {\bibfield  {journal} {\bibinfo  {journal} {J. Cosmol. Astropart. Phys.}\ }\textbf {\bibinfo {volume} {12}},\ \bibinfo {pages} {005} (\bibinfo {year} {2024})},\ \Eprint {http://arxiv.org/abs/2410.01876} {arXiv:2410.01876 [astro-ph.CO]} \BibitemShut {NoStop}%
\bibitem [{\citenamefont {Press}\ and\ \citenamefont {Schechter}(1974)}]{Press:1973iz}%
  \BibitemOpen
  \bibfield  {author} {\bibinfo {author} {\bibfnamefont {W.~H.}\ \bibnamefont {Press}}\ and\ \bibinfo {author} {\bibfnamefont {P.}~\bibnamefont {Schechter}},\ }\href {\doibase 10.1086/152650} {\bibfield  {journal} {\bibinfo  {journal} {Astrophys. J.}\ }\textbf {\bibinfo {volume} {187}},\ \bibinfo {pages} {425} (\bibinfo {year} {1974})}\BibitemShut {NoStop}%
\bibitem [{\citenamefont {Bardeen}\ \emph {et~al.}(1986)\citenamefont {Bardeen}, \citenamefont {Bond}, \citenamefont {Kaiser},\ and\ \citenamefont {Szalay}}]{Bardeen:1985tr}%
  \BibitemOpen
  \bibfield  {author} {\bibinfo {author} {\bibfnamefont {J.~M.}\ \bibnamefont {Bardeen}}, \bibinfo {author} {\bibfnamefont {J.~R.}\ \bibnamefont {Bond}}, \bibinfo {author} {\bibfnamefont {N.}~\bibnamefont {Kaiser}}, \ and\ \bibinfo {author} {\bibfnamefont {A.~S.}\ \bibnamefont {Szalay}},\ }\href {\doibase 10.1086/164143} {\bibfield  {journal} {\bibinfo  {journal} {Astrophys. J.}\ }\textbf {\bibinfo {volume} {304}},\ \bibinfo {pages} {15} (\bibinfo {year} {1986})}\BibitemShut {NoStop}%
\bibitem [{\citenamefont {Mo}\ and\ \citenamefont {White}(2002)}]{Mo:2002ft}%
  \BibitemOpen
  \bibfield  {author} {\bibinfo {author} {\bibfnamefont {H.~J.}\ \bibnamefont {Mo}}\ and\ \bibinfo {author} {\bibfnamefont {S.~D.~M.}\ \bibnamefont {White}},\ }\href {\doibase 10.1046/j.1365-8711.2002.05723.x} {\bibfield  {journal} {\bibinfo  {journal} {Mon. Not. Roy. Astron. Soc.}\ }\textbf {\bibinfo {volume} {336}},\ \bibinfo {pages} {112} (\bibinfo {year} {2002})},\ \Eprint {http://arxiv.org/abs/astro-ph/0202393} {arXiv:astro-ph/0202393} \BibitemShut {NoStop}%
\bibitem [{\citenamefont {Merritt}(2013)}]{Merritt2013}%
  \BibitemOpen
  \bibfield  {author} {\bibinfo {author} {\bibfnamefont {D.}~\bibnamefont {Merritt}},\ }\href@noop {} {\emph {\bibinfo {title} {Dynamics and Evolution of Galactic Nuclei}}}\ (\bibinfo  {publisher} {Princeton University Press},\ \bibinfo {address} {Princeton, New Jersey},\ \bibinfo {year} {2013})\BibitemShut {NoStop}%
\bibitem [{\citenamefont {Stasenko}\ and\ \citenamefont {Belotsky}(2023)}]{Stasenko:2023zmf}%
  \BibitemOpen
  \bibfield  {author} {\bibinfo {author} {\bibfnamefont {V.}~\bibnamefont {Stasenko}}\ and\ \bibinfo {author} {\bibfnamefont {K.}~\bibnamefont {Belotsky}},\ }\href {\doibase 10.1093/mnras/stad3032} {\bibfield  {journal} {\bibinfo  {journal} {Mon. Not. Roy. Astron. Soc.}\ }\textbf {\bibinfo {volume} {526}},\ \bibinfo {pages} {4308} (\bibinfo {year} {2023})},\ \Eprint {http://arxiv.org/abs/2307.12924} {arXiv:2307.12924 [astro-ph.CO]} \BibitemShut {NoStop}%
\bibitem [{\citenamefont {Stasenko}(2024)}]{Stasenko:2024pzd}%
  \BibitemOpen
  \bibfield  {author} {\bibinfo {author} {\bibfnamefont {V.}~\bibnamefont {Stasenko}},\ }\href {\doibase 10.1103/PhysRevD.109.123546} {\bibfield  {journal} {\bibinfo  {journal} {Phys. Rev. D}\ }\textbf {\bibinfo {volume} {109}},\ \bibinfo {pages} {123546} (\bibinfo {year} {2024})},\ \Eprint {http://arxiv.org/abs/2403.11325} {arXiv:2403.11325 [astro-ph.CO]} \BibitemShut {NoStop}%
\bibitem [{\citenamefont {Binney}\ and\ \citenamefont {Tremaine}(2008)}]{BinTre2008}%
  \BibitemOpen
  \bibfield  {author} {\bibinfo {author} {\bibfnamefont {J.}~\bibnamefont {Binney}}\ and\ \bibinfo {author} {\bibfnamefont {S.}~\bibnamefont {Tremaine}},\ }\href {https://books.google.ru/books?id=6mF4CKxlbLsC} {\emph {\bibinfo {title} {Galactic Dynamics: Second Edition}}}\ (\bibinfo  {publisher} {Princeton University Press},\ \bibinfo {address} {Princeton, New Jersey},\ \bibinfo {year} {2008})\BibitemShut {NoStop}%
\bibitem [{\citenamefont {{Lynden-Bell}}\ and\ \citenamefont {{Wood}}(1968)}]{1968MNRAS.138..495L}%
  \BibitemOpen
  \bibfield  {author} {\bibinfo {author} {\bibfnamefont {D.}~\bibnamefont {{Lynden-Bell}}}\ and\ \bibinfo {author} {\bibfnamefont {R.}~\bibnamefont {{Wood}}},\ }\href {\doibase 10.1093/mnras/138.4.495} {\bibfield  {journal} {\bibinfo  {journal} {Mon. Not. R. Astron. Soc.}\ }\textbf {\bibinfo {volume} {138}},\ \bibinfo {pages} {495} (\bibinfo {year} {1968})}\BibitemShut {NoStop}%
\bibitem [{\citenamefont {{Cohn}}(1980)}]{1980ApJ...242..765C}%
  \BibitemOpen
  \bibfield  {author} {\bibinfo {author} {\bibfnamefont {H.}~\bibnamefont {{Cohn}}},\ }\href {\doibase 10.1086/158511} {\bibfield  {journal} {\bibinfo  {journal} {Astrophys. J.}\ }\textbf {\bibinfo {volume} {242}},\ \bibinfo {pages} {765} (\bibinfo {year} {1980})}\BibitemShut {NoStop}%
\bibitem [{\citenamefont {Heggie}\ and\ \citenamefont {Hut}(2003)}]{HegHut06}%
  \BibitemOpen
  \bibfield  {author} {\bibinfo {author} {\bibfnamefont {D.}~\bibnamefont {Heggie}}\ and\ \bibinfo {author} {\bibfnamefont {P.}~\bibnamefont {Hut}},\ }\href@noop {} {\emph {\bibinfo {title} {The Gravitational Million-Body Problem: A Multidisciplinary Approach to Star Cluster Dynamics}}}\ (\bibinfo  {publisher} {Cambridge University Press},\ \bibinfo {address} {Cambridge, UK},\ \bibinfo {year} {2003})\BibitemShut {NoStop}%
\bibitem [{\citenamefont {Quinlan}(1996)}]{Quinlan:1996bw}%
  \BibitemOpen
  \bibfield  {author} {\bibinfo {author} {\bibfnamefont {G.~D.}\ \bibnamefont {Quinlan}},\ }\href {\doibase 10.1016/S1384-1076(96)00018-8} {\bibfield  {journal} {\bibinfo  {journal} {New Astron.}\ }\textbf {\bibinfo {volume} {1}},\ \bibinfo {pages} {255} (\bibinfo {year} {1996})},\ \Eprint {http://arxiv.org/abs/astro-ph/9606182} {arXiv:astro-ph/9606182} \BibitemShut {NoStop}%
\bibitem [{\citenamefont {{Khlopov}}\ \emph {et~al.}(2005)\citenamefont {{Khlopov}}, \citenamefont {{Rubin}},\ and\ \citenamefont {{Sakharov}}}]{2005APh....23..265K}%
  \BibitemOpen
  \bibfield  {author} {\bibinfo {author} {\bibfnamefont {M.~Y.}\ \bibnamefont {{Khlopov}}}, \bibinfo {author} {\bibfnamefont {S.~G.}\ \bibnamefont {{Rubin}}}, \ and\ \bibinfo {author} {\bibfnamefont {A.~S.}\ \bibnamefont {{Sakharov}}},\ }\href {\doibase 10.1016/j.astropartphys.2004.12.002} {\bibfield  {journal} {\bibinfo  {journal} {Astropart. Phys.}\ }\textbf {\bibinfo {volume} {23}},\ \bibinfo {pages} {265} (\bibinfo {year} {2005})},\ \Eprint {http://arxiv.org/abs/astro-ph/0401532} {arXiv:astro-ph/0401532} \BibitemShut {NoStop}%
\bibitem [{\citenamefont {Belotsky}\ \emph {et~al.}(2019)\citenamefont {Belotsky}, \citenamefont {Dokuchaev}, \citenamefont {Eroshenko}, \citenamefont {Esipova}, \citenamefont {Khlopov}, \citenamefont {Khromykh}, \citenamefont {Kirillov}, \citenamefont {Nikulin}, \citenamefont {Rubin},\ and\ \citenamefont {Svadkovsky}}]{Belotsky:2018wph}%
  \BibitemOpen
  \bibfield  {author} {\bibinfo {author} {\bibfnamefont {K.~M.}\ \bibnamefont {Belotsky}}, \bibinfo {author} {\bibfnamefont {V.~I.}\ \bibnamefont {Dokuchaev}}, \bibinfo {author} {\bibfnamefont {Y.~N.}\ \bibnamefont {Eroshenko}}, \bibinfo {author} {\bibfnamefont {E.~A.}\ \bibnamefont {Esipova}}, \bibinfo {author} {\bibfnamefont {M.~Y.}\ \bibnamefont {Khlopov}}, \bibinfo {author} {\bibfnamefont {L.~A.}\ \bibnamefont {Khromykh}}, \bibinfo {author} {\bibfnamefont {A.~A.}\ \bibnamefont {Kirillov}}, \bibinfo {author} {\bibfnamefont {V.~V.}\ \bibnamefont {Nikulin}}, \bibinfo {author} {\bibfnamefont {S.~G.}\ \bibnamefont {Rubin}}, \ and\ \bibinfo {author} {\bibfnamefont {I.~V.}\ \bibnamefont {Svadkovsky}},\ }\href {\doibase 10.1140/epjc/s10052-019-6741-4} {\bibfield  {journal} {\bibinfo  {journal} {Eur. Phys. J. C}\ }\textbf {\bibinfo {volume} {79}},\ \bibinfo {pages} {246} (\bibinfo {year} {2019})},\ \Eprint {http://arxiv.org/abs/1807.06590} {arXiv:1807.06590 [astro-ph.CO]} \BibitemShut {NoStop}%
\bibitem [{\citenamefont {Dokuchaev}\ \emph {et~al.}(2005)\citenamefont {Dokuchaev}, \citenamefont {Eroshenko},\ and\ \citenamefont {Rubin}}]{Dokuchaev:2004kr}%
  \BibitemOpen
  \bibfield  {author} {\bibinfo {author} {\bibfnamefont {V.}~\bibnamefont {Dokuchaev}}, \bibinfo {author} {\bibfnamefont {Y.}~\bibnamefont {Eroshenko}}, \ and\ \bibinfo {author} {\bibfnamefont {S.}~\bibnamefont {Rubin}},\ }\href@noop {} {\bibfield  {journal} {\bibinfo  {journal} {Grav. Cosmol.}\ }\textbf {\bibinfo {volume} {11}},\ \bibinfo {pages} {99} (\bibinfo {year} {2005})},\ \Eprint {http://arxiv.org/abs/astro-ph/0412418} {arXiv:astro-ph/0412418} \BibitemShut {NoStop}%
\bibitem [{\citenamefont {Dokuchaev}\ \emph {et~al.}(2008)\citenamefont {Dokuchaev}, \citenamefont {Eroshenko},\ and\ \citenamefont {Rubin}}]{Dokuchaev:2008hz}%
  \BibitemOpen
  \bibfield  {author} {\bibinfo {author} {\bibfnamefont {V.~I.}\ \bibnamefont {Dokuchaev}}, \bibinfo {author} {\bibfnamefont {Y.~N.}\ \bibnamefont {Eroshenko}}, \ and\ \bibinfo {author} {\bibfnamefont {S.~G.}\ \bibnamefont {Rubin}},\ }\href {\doibase 10.1134/S1063772908100016} {\bibfield  {journal} {\bibinfo  {journal} {Astron. Rep.}\ }\textbf {\bibinfo {volume} {52}},\ \bibinfo {pages} {779} (\bibinfo {year} {2008})},\ \Eprint {http://arxiv.org/abs/0801.0885} {arXiv:0801.0885 [astro-ph]} \BibitemShut {NoStop}%
\bibitem [{\citenamefont {{Dokuchaev}}\ \emph {et~al.}(2009)\citenamefont {{Dokuchaev}}, \citenamefont {{Eroshenko}},\ and\ \citenamefont {{Rubin}}}]{2009AstL...35..143D}%
  \BibitemOpen
  \bibfield  {author} {\bibinfo {author} {\bibfnamefont {V.~I.}\ \bibnamefont {{Dokuchaev}}}, \bibinfo {author} {\bibfnamefont {Y.~N.}\ \bibnamefont {{Eroshenko}}}, \ and\ \bibinfo {author} {\bibfnamefont {S.~G.}\ \bibnamefont {{Rubin}}},\ }\href {\doibase 10.1134/S1063773709030013} {\bibfield  {journal} {\bibinfo  {journal} {Astronomy Letters}\ }\textbf {\bibinfo {volume} {35}},\ \bibinfo {pages} {143} (\bibinfo {year} {2009})},\ \Eprint {http://arxiv.org/abs/1003.4158} {arXiv:1003.4158 [astro-ph.CO]} \BibitemShut {NoStop}%
\bibitem [{\citenamefont {{Dokuchaev}}\ \emph {et~al.}(2010)\citenamefont {{Dokuchaev}}, \citenamefont {{Eroshenko}}, \citenamefont {{Rubin}},\ and\ \citenamefont {{Samarchenko}}}]{2010AstL...36..773D}%
  \BibitemOpen
  \bibfield  {author} {\bibinfo {author} {\bibfnamefont {V.~I.}\ \bibnamefont {{Dokuchaev}}}, \bibinfo {author} {\bibfnamefont {Y.~N.}\ \bibnamefont {{Eroshenko}}}, \bibinfo {author} {\bibfnamefont {S.~G.}\ \bibnamefont {{Rubin}}}, \ and\ \bibinfo {author} {\bibfnamefont {D.~A.}\ \bibnamefont {{Samarchenko}}},\ }\href {\doibase 10.1134/S1063773710110022} {\bibfield  {journal} {\bibinfo  {journal} {Astronomy Letters}\ }\textbf {\bibinfo {volume} {36}},\ \bibinfo {pages} {773} (\bibinfo {year} {2010})},\ \Eprint {http://arxiv.org/abs/1010.5325} {arXiv:1010.5325 [astro-ph.CO]} \BibitemShut {NoStop}%
\bibitem [{\citenamefont {Kolb}\ and\ \citenamefont {Tkachev}(1994)}]{Kolb:1994fi}%
  \BibitemOpen
  \bibfield  {author} {\bibinfo {author} {\bibfnamefont {E.~W.}\ \bibnamefont {Kolb}}\ and\ \bibinfo {author} {\bibfnamefont {I.~I.}\ \bibnamefont {Tkachev}},\ }\href {\doibase 10.1103/PhysRevD.50.769} {\bibfield  {journal} {\bibinfo  {journal} {Phys. Rev. D}\ }\textbf {\bibinfo {volume} {50}},\ \bibinfo {pages} {769} (\bibinfo {year} {1994})},\ \Eprint {http://arxiv.org/abs/astro-ph/9403011} {arXiv:astro-ph/9403011} \BibitemShut {NoStop}%
\bibitem [{\citenamefont {Chisholm}(2006)}]{Chi06}%
  \BibitemOpen
  \bibfield  {author} {\bibinfo {author} {\bibfnamefont {J.~R.}\ \bibnamefont {Chisholm}},\ }\href {\doibase 10.1103/PhysRevD.73.083504} {\bibfield  {journal} {\bibinfo  {journal} {Phys. Rev. D}\ }\textbf {\bibinfo {volume} {73}},\ \bibinfo {pages} {083504} (\bibinfo {year} {2006})},\ \Eprint {http://arxiv.org/abs/astro-ph/0509141} {arXiv:astro-ph/0509141} \BibitemShut {NoStop}%
\bibitem [{\citenamefont {Desjacques}\ and\ \citenamefont {Riotto}(2018)}]{DesRio18}%
  \BibitemOpen
  \bibfield  {author} {\bibinfo {author} {\bibfnamefont {V.}~\bibnamefont {Desjacques}}\ and\ \bibinfo {author} {\bibfnamefont {A.}~\bibnamefont {Riotto}},\ }\href {\doibase 10.1103/PhysRevD.98.123533} {\bibfield  {journal} {\bibinfo  {journal} {Phys. Rev. D}\ }\textbf {\bibinfo {volume} {98}},\ \bibinfo {pages} {123533} (\bibinfo {year} {2018})},\ \Eprint {http://arxiv.org/abs/1806.10414} {arXiv:1806.10414 [astro-ph.CO]} \BibitemShut {NoStop}%
\bibitem [{\citenamefont {Young}\ \emph {et~al.}(2014)\citenamefont {Young}, \citenamefont {Byrnes},\ and\ \citenamefont {Sasaki}}]{YouByrSas14}%
  \BibitemOpen
  \bibfield  {author} {\bibinfo {author} {\bibfnamefont {S.}~\bibnamefont {Young}}, \bibinfo {author} {\bibfnamefont {C.~T.}\ \bibnamefont {Byrnes}}, \ and\ \bibinfo {author} {\bibfnamefont {M.}~\bibnamefont {Sasaki}},\ }\href {\doibase 10.1088/1475-7516/2014/07/045} {\bibfield  {journal} {\bibinfo  {journal} {J. Cosmol. Astropart. Phys.}\ }\textbf {\bibinfo {volume} {07}},\ \bibinfo {pages} {045} (\bibinfo {year} {2014})},\ \Eprint {http://arxiv.org/abs/1405.7023} {arXiv:1405.7023 [gr-qc]} \BibitemShut {NoStop}%
\bibitem [{\citenamefont {{Young}}\ and\ \citenamefont {{Byrnes}}(2015)}]{2015PhRvD..91h3521Y}%
  \BibitemOpen
  \bibfield  {author} {\bibinfo {author} {\bibfnamefont {S.}~\bibnamefont {{Young}}}\ and\ \bibinfo {author} {\bibfnamefont {C.~T.}\ \bibnamefont {{Byrnes}}},\ }\href {\doibase 10.1103/PhysRevD.91.083521} {\bibfield  {journal} {\bibinfo  {journal} {Phys. Rev. D}\ }\textbf {\bibinfo {volume} {91}},\ \bibinfo {eid} {083521} (\bibinfo {year} {2015})},\ \Eprint {http://arxiv.org/abs/1411.4620} {arXiv:1411.4620 [astro-ph.CO]} \BibitemShut {NoStop}%
\bibitem [{\citenamefont {{Tada}}\ and\ \citenamefont {{Yokoyama}}(2015)}]{2015PhRvD..91l3534T}%
  \BibitemOpen
  \bibfield  {author} {\bibinfo {author} {\bibfnamefont {Y.}~\bibnamefont {{Tada}}}\ and\ \bibinfo {author} {\bibfnamefont {S.}~\bibnamefont {{Yokoyama}}},\ }\href {\doibase 10.1103/PhysRevD.91.123534} {\bibfield  {journal} {\bibinfo  {journal} {Phys. Rev. D}\ }\textbf {\bibinfo {volume} {91}},\ \bibinfo {eid} {123534} (\bibinfo {year} {2015})},\ \Eprint {http://arxiv.org/abs/1502.01124} {arXiv:1502.01124 [astro-ph.CO]} \BibitemShut {NoStop}%
\bibitem [{\citenamefont {Suyama}\ and\ \citenamefont {Yokoyama}(2019)}]{SuyYok19}%
  \BibitemOpen
  \bibfield  {author} {\bibinfo {author} {\bibfnamefont {T.}~\bibnamefont {Suyama}}\ and\ \bibinfo {author} {\bibfnamefont {S.}~\bibnamefont {Yokoyama}},\ }\href {\doibase 10.1093/ptep/ptz105} {\bibfield  {journal} {\bibinfo  {journal} {PTEP}\ }\textbf {\bibinfo {volume} {2019}},\ \bibinfo {pages} {103E02} (\bibinfo {year} {2019})},\ \Eprint {http://arxiv.org/abs/1906.04958} {arXiv:1906.04958 [astro-ph.CO]} \BibitemShut {NoStop}%
\bibitem [{\citenamefont {Starobinsky}\ and\ \citenamefont {Yokoyama}(1994)}]{Starobinsky:1994mh}%
  \BibitemOpen
  \bibfield  {author} {\bibinfo {author} {\bibfnamefont {A.~A.}\ \bibnamefont {Starobinsky}}\ and\ \bibinfo {author} {\bibfnamefont {J.}~\bibnamefont {Yokoyama}},\ }in\ \href@noop {} {\emph {\bibinfo {booktitle} {{4th Workshop on General Relativity and Gravitation}}}}\ (\bibinfo {year} {1994})\ p.\ \bibinfo {pages} {381},\ \Eprint {http://arxiv.org/abs/gr-qc/9502002} {arXiv:gr-qc/9502002} \BibitemShut {NoStop}%
\bibitem [{\citenamefont {Abbott}\ \emph {et~al.}(2016)\citenamefont {Abbott} \emph {et~al.}}]{LIGOScientific:2016aoc}%
  \BibitemOpen
  \bibfield  {author} {\bibinfo {author} {\bibfnamefont {B.~P.}\ \bibnamefont {Abbott}} \emph {et~al.} (\bibinfo {collaboration} {LIGO Scientific, Virgo}),\ }\href {\doibase 10.1103/PhysRevLett.116.061102} {\bibfield  {journal} {\bibinfo  {journal} {Phys. Rev. Lett.}\ }\textbf {\bibinfo {volume} {116}},\ \bibinfo {pages} {061102} (\bibinfo {year} {2016})},\ \Eprint {http://arxiv.org/abs/1602.03837} {arXiv:1602.03837 [gr-qc]} \BibitemShut {NoStop}%
\bibitem [{\citenamefont {Bird}\ \emph {et~al.}(2016)\citenamefont {Bird}, \citenamefont {Cholis}, \citenamefont {Mu{\~n}oz}, \citenamefont {Ali-Ha{\"\i}moud}, \citenamefont {Kamionkowski}, \citenamefont {Kovetz}, \citenamefont {Raccanelli},\ and\ \citenamefont {Riess}}]{Bird:2016dcv}%
  \BibitemOpen
  \bibfield  {author} {\bibinfo {author} {\bibfnamefont {S.}~\bibnamefont {Bird}}, \bibinfo {author} {\bibfnamefont {I.}~\bibnamefont {Cholis}}, \bibinfo {author} {\bibfnamefont {J.~B.}\ \bibnamefont {Mu{\~n}oz}}, \bibinfo {author} {\bibfnamefont {Y.}~\bibnamefont {Ali-Ha{\"\i}moud}}, \bibinfo {author} {\bibfnamefont {M.}~\bibnamefont {Kamionkowski}}, \bibinfo {author} {\bibfnamefont {E.~D.}\ \bibnamefont {Kovetz}}, \bibinfo {author} {\bibfnamefont {A.}~\bibnamefont {Raccanelli}}, \ and\ \bibinfo {author} {\bibfnamefont {A.~G.}\ \bibnamefont {Riess}},\ }\href {\doibase 10.1103/PhysRevLett.116.201301} {\bibfield  {journal} {\bibinfo  {journal} {Phys. Rev. Lett.}\ }\textbf {\bibinfo {volume} {116}},\ \bibinfo {pages} {201301} (\bibinfo {year} {2016})},\ \Eprint {http://arxiv.org/abs/1603.00464} {arXiv:1603.00464 [astro-ph.CO]} \BibitemShut {NoStop}%
\bibitem [{\citenamefont {Clesse}\ and\ \citenamefont {Garc{\'\i}a-Bellido}(2017)}]{Clesse:2016vqa}%
  \BibitemOpen
  \bibfield  {author} {\bibinfo {author} {\bibfnamefont {S.}~\bibnamefont {Clesse}}\ and\ \bibinfo {author} {\bibfnamefont {J.}~\bibnamefont {Garc{\'\i}a-Bellido}},\ }\href {\doibase 10.1016/j.dark.2016.10.002} {\bibfield  {journal} {\bibinfo  {journal} {Phys. Dark Univ.}\ }\textbf {\bibinfo {volume} {15}},\ \bibinfo {pages} {142} (\bibinfo {year} {2017})},\ \Eprint {http://arxiv.org/abs/1603.05234} {arXiv:1603.05234 [astro-ph.CO]} \BibitemShut {NoStop}%
\bibitem [{\citenamefont {Ali-Ha{\"\i}moud}\ \emph {et~al.}(2017)\citenamefont {Ali-Ha{\"\i}moud}, \citenamefont {Kovetz},\ and\ \citenamefont {Kamionkowski}}]{Ali-Haimoud:2017rtz}%
  \BibitemOpen
  \bibfield  {author} {\bibinfo {author} {\bibfnamefont {Y.}~\bibnamefont {Ali-Ha{\"\i}moud}}, \bibinfo {author} {\bibfnamefont {E.~D.}\ \bibnamefont {Kovetz}}, \ and\ \bibinfo {author} {\bibfnamefont {M.}~\bibnamefont {Kamionkowski}},\ }\href {\doibase 10.1103/PhysRevD.96.123523} {\bibfield  {journal} {\bibinfo  {journal} {Phys. Rev. D}\ }\textbf {\bibinfo {volume} {96}},\ \bibinfo {pages} {123523} (\bibinfo {year} {2017})},\ \Eprint {http://arxiv.org/abs/1709.06576} {arXiv:1709.06576 [astro-ph.CO]} \BibitemShut {NoStop}%
\bibitem [{\citenamefont {Raidal}\ \emph {et~al.}(2019)\citenamefont {Raidal}, \citenamefont {Spethmann}, \citenamefont {Vaskonen},\ and\ \citenamefont {Veerm{\"a}e}}]{Raidal:2018bbj}%
  \BibitemOpen
  \bibfield  {author} {\bibinfo {author} {\bibfnamefont {M.}~\bibnamefont {Raidal}}, \bibinfo {author} {\bibfnamefont {C.}~\bibnamefont {Spethmann}}, \bibinfo {author} {\bibfnamefont {V.}~\bibnamefont {Vaskonen}}, \ and\ \bibinfo {author} {\bibfnamefont {H.}~\bibnamefont {Veerm{\"a}e}},\ }\href {\doibase 10.1088/1475-7516/2019/02/018} {\bibfield  {journal} {\bibinfo  {journal} {J. Cosmol. Astropart. Phys.}\ }\textbf {\bibinfo {volume} {02}},\ \bibinfo {pages} {018} (\bibinfo {year} {2019})},\ \Eprint {http://arxiv.org/abs/1812.01930} {arXiv:1812.01930 [astro-ph.CO]} \BibitemShut {NoStop}%
\bibitem [{\citenamefont {Franciolini}\ \emph {et~al.}(2022)\citenamefont {Franciolini}, \citenamefont {Kritos}, \citenamefont {Berti},\ and\ \citenamefont {Silk}}]{Franciolini:2022ewd}%
  \BibitemOpen
  \bibfield  {author} {\bibinfo {author} {\bibfnamefont {G.}~\bibnamefont {Franciolini}}, \bibinfo {author} {\bibfnamefont {K.}~\bibnamefont {Kritos}}, \bibinfo {author} {\bibfnamefont {E.}~\bibnamefont {Berti}}, \ and\ \bibinfo {author} {\bibfnamefont {J.}~\bibnamefont {Silk}},\ }\href {\doibase 10.1103/PhysRevD.106.083529} {\bibfield  {journal} {\bibinfo  {journal} {Phys. Rev. D}\ }\textbf {\bibinfo {volume} {106}},\ \bibinfo {pages} {083529} (\bibinfo {year} {2022})},\ \Eprint {http://arxiv.org/abs/2205.15340} {arXiv:2205.15340 [astro-ph.CO]} \BibitemShut {NoStop}%
\bibitem [{\citenamefont {Peters}(1964)}]{Peters:1964zz}%
  \BibitemOpen
  \bibfield  {author} {\bibinfo {author} {\bibfnamefont {P.~C.}\ \bibnamefont {Peters}},\ }\href {\doibase 10.1103/PhysRev.136.B1224} {\bibfield  {journal} {\bibinfo  {journal} {Phys. Rev.}\ }\textbf {\bibinfo {volume} {136}},\ \bibinfo {pages} {B1224} (\bibinfo {year} {1964})}\BibitemShut {NoStop}%
\bibitem [{\citenamefont {Eroshenko}(2018)}]{Eroshenko:2016hmn}%
  \BibitemOpen
  \bibfield  {author} {\bibinfo {author} {\bibfnamefont {Y.~N.}\ \bibnamefont {Eroshenko}},\ }\href {\doibase 10.1088/1742-6596/1051/1/012010} {\bibfield  {journal} {\bibinfo  {journal} {J. Phys. Conf. Ser.}\ }\textbf {\bibinfo {volume} {1051}},\ \bibinfo {pages} {012010} (\bibinfo {year} {2018})},\ \Eprint {http://arxiv.org/abs/1604.04932} {arXiv:1604.04932 [astro-ph.CO]} \BibitemShut {NoStop}%
\bibitem [{\citenamefont {Hayasaki}\ \emph {et~al.}(2016)\citenamefont {Hayasaki}, \citenamefont {Takahashi}, \citenamefont {Sendouda},\ and\ \citenamefont {Nagataki}}]{Hayasaki:2009ug}%
  \BibitemOpen
  \bibfield  {author} {\bibinfo {author} {\bibfnamefont {K.}~\bibnamefont {Hayasaki}}, \bibinfo {author} {\bibfnamefont {K.}~\bibnamefont {Takahashi}}, \bibinfo {author} {\bibfnamefont {Y.}~\bibnamefont {Sendouda}}, \ and\ \bibinfo {author} {\bibfnamefont {S.}~\bibnamefont {Nagataki}},\ }\href {\doibase 10.1093/pasj/psw065} {\bibfield  {journal} {\bibinfo  {journal} {Publ. Astron. Soc. Jap.}\ }\textbf {\bibinfo {volume} {68}},\ \bibinfo {pages} {66} (\bibinfo {year} {2016})},\ \Eprint {http://arxiv.org/abs/0909.1738} {arXiv:0909.1738 [astro-ph.CO]} \BibitemShut {NoStop}%
\bibitem [{\citenamefont {Kavanagh}\ \emph {et~al.}(2018)\citenamefont {Kavanagh}, \citenamefont {Gaggero},\ and\ \citenamefont {Bertone}}]{Kavanagh:2018ggo}%
  \BibitemOpen
  \bibfield  {author} {\bibinfo {author} {\bibfnamefont {B.~J.}\ \bibnamefont {Kavanagh}}, \bibinfo {author} {\bibfnamefont {D.}~\bibnamefont {Gaggero}}, \ and\ \bibinfo {author} {\bibfnamefont {G.}~\bibnamefont {Bertone}},\ }\href {\doibase 10.1103/PhysRevD.98.023536} {\bibfield  {journal} {\bibinfo  {journal} {Phys. Rev. D}\ }\textbf {\bibinfo {volume} {98}},\ \bibinfo {pages} {023536} (\bibinfo {year} {2018})},\ \Eprint {http://arxiv.org/abs/1805.09034} {arXiv:1805.09034 [astro-ph.CO]} \BibitemShut {NoStop}%
\bibitem [{\citenamefont {Pilipenko}\ \emph {et~al.}(2022)\citenamefont {Pilipenko}, \citenamefont {Tkachev},\ and\ \citenamefont {Ivanov}}]{Pilipenko:2022emp}%
  \BibitemOpen
  \bibfield  {author} {\bibinfo {author} {\bibfnamefont {S.}~\bibnamefont {Pilipenko}}, \bibinfo {author} {\bibfnamefont {M.}~\bibnamefont {Tkachev}}, \ and\ \bibinfo {author} {\bibfnamefont {P.}~\bibnamefont {Ivanov}},\ }\href {\doibase 10.1103/PhysRevD.105.123504} {\bibfield  {journal} {\bibinfo  {journal} {Phys. Rev. D}\ }\textbf {\bibinfo {volume} {105}},\ \bibinfo {pages} {123504} (\bibinfo {year} {2022})},\ \Eprint {http://arxiv.org/abs/2205.10792} {arXiv:2205.10792 [astro-ph.CO]} \BibitemShut {NoStop}%
\bibitem [{\citenamefont {Jangra}\ \emph {et~al.}(2023)\citenamefont {Jangra}, \citenamefont {Kavanagh},\ and\ \citenamefont {Diego}}]{Jangra:2023mqp}%
  \BibitemOpen
  \bibfield  {author} {\bibinfo {author} {\bibfnamefont {P.}~\bibnamefont {Jangra}}, \bibinfo {author} {\bibfnamefont {B.~J.}\ \bibnamefont {Kavanagh}}, \ and\ \bibinfo {author} {\bibfnamefont {J.~M.}\ \bibnamefont {Diego}},\ }\href {\doibase 10.1088/1475-7516/2023/11/069} {\bibfield  {journal} {\bibinfo  {journal} {J. Cosmol. Astropart. Phys.}\ }\textbf {\bibinfo {volume} {11}},\ \bibinfo {pages} {069} (\bibinfo {year} {2023})},\ \Eprint {http://arxiv.org/abs/2304.05892} {arXiv:2304.05892 [astro-ph.CO]} \BibitemShut {NoStop}%
\bibitem [{\citenamefont {Stasenko}\ and\ \citenamefont {Eroshenko}(2024)}]{Stasenko:2024dui}%
  \BibitemOpen
  \bibfield  {author} {\bibinfo {author} {\bibfnamefont {V.~D.}\ \bibnamefont {Stasenko}}\ and\ \bibinfo {author} {\bibfnamefont {Y.~N.}\ \bibnamefont {Eroshenko}},\ }\href {\doibase 10.1134/S1063773724700348} {\bibfield  {journal} {\bibinfo  {journal} {Astron. Lett.}\ }\textbf {\bibinfo {volume} {50}},\ \bibinfo {pages} {431} (\bibinfo {year} {2024})},\ \Eprint {http://arxiv.org/abs/2412.20859} {arXiv:2412.20859 [astro-ph.CO]} \BibitemShut {NoStop}%
\bibitem [{\citenamefont {Fernandez}\ and\ \citenamefont {Profumo}(2019)}]{FerPro19}%
  \BibitemOpen
  \bibfield  {author} {\bibinfo {author} {\bibfnamefont {N.}~\bibnamefont {Fernandez}}\ and\ \bibinfo {author} {\bibfnamefont {S.}~\bibnamefont {Profumo}},\ }\href {\doibase 10.1088/1475-7516/2019/08/022} {\bibfield  {journal} {\bibinfo  {journal} {J. Cosmol. Astropart. Phys.}\ }\textbf {\bibinfo {volume} {08}},\ \bibinfo {pages} {022} (\bibinfo {year} {2019})},\ \Eprint {http://arxiv.org/abs/1905.13019} {arXiv:1905.13019 [astro-ph.HE]} \BibitemShut {NoStop}%
\bibitem [{\citenamefont {{Tutukov}}\ and\ \citenamefont {{Yungelson}}(1993)}]{TutYun93}%
  \BibitemOpen
  \bibfield  {author} {\bibinfo {author} {\bibfnamefont {A.~V.}\ \bibnamefont {{Tutukov}}}\ and\ \bibinfo {author} {\bibfnamefont {L.~R.}\ \bibnamefont {{Yungelson}}},\ }\href {\doibase 10.1093/mnras/260.3.675} {\bibfield  {journal} {\bibinfo  {journal} {Mon. Not. R. Astron. Soc.}\ }\textbf {\bibinfo {volume} {260}},\ \bibinfo {pages} {675} (\bibinfo {year} {1993})}\BibitemShut {NoStop}%
\bibitem [{\citenamefont {{Lipunov}}\ \emph {et~al.}(1997)\citenamefont {{Lipunov}}, \citenamefont {{Postnov}},\ and\ \citenamefont {{Prokhorov}}}]{Lipetal97}%
  \BibitemOpen
  \bibfield  {author} {\bibinfo {author} {\bibfnamefont {V.~M.}\ \bibnamefont {{Lipunov}}}, \bibinfo {author} {\bibfnamefont {K.~A.}\ \bibnamefont {{Postnov}}}, \ and\ \bibinfo {author} {\bibfnamefont {M.~E.}\ \bibnamefont {{Prokhorov}}},\ }\href@noop {} {\bibfield  {journal} {\bibinfo  {journal} {Astronomy Letters}\ }\textbf {\bibinfo {volume} {23}},\ \bibinfo {pages} {492} (\bibinfo {year} {1997})}\BibitemShut {NoStop}%
\bibitem [{\citenamefont {Grishchuk}\ \emph {et~al.}(2001)\citenamefont {Grishchuk}, \citenamefont {Lipunov}, \citenamefont {Postnov}, \citenamefont {Prokhorov},\ and\ \citenamefont {Sathyaprakash}}]{Grietal01}%
  \BibitemOpen
  \bibfield  {author} {\bibinfo {author} {\bibfnamefont {L.~P.}\ \bibnamefont {Grishchuk}}, \bibinfo {author} {\bibfnamefont {V.~M.}\ \bibnamefont {Lipunov}}, \bibinfo {author} {\bibfnamefont {K.~A.}\ \bibnamefont {Postnov}}, \bibinfo {author} {\bibfnamefont {M.~E.}\ \bibnamefont {Prokhorov}}, \ and\ \bibinfo {author} {\bibfnamefont {B.~S.}\ \bibnamefont {Sathyaprakash}},\ }\href {\doibase 10.1070/PU2001v044n01ABEH000873} {\bibfield  {journal} {\bibinfo  {journal} {Phys. Usp.}\ }\textbf {\bibinfo {volume} {44}},\ \bibinfo {pages} {1} (\bibinfo {year} {2001})}\BibitemShut {NoStop}%
\bibitem [{\citenamefont {{Belczynski}}\ \emph {et~al.}(2010)\citenamefont {{Belczynski}}, \citenamefont {{Dominik}}, \citenamefont {{Bulik}}, \citenamefont {{O'Shaughnessy}}, \citenamefont {{Fryer}},\ and\ \citenamefont {{Holz}}}]{chris1}%
  \BibitemOpen
  \bibfield  {author} {\bibinfo {author} {\bibfnamefont {K.}~\bibnamefont {{Belczynski}}}, \bibinfo {author} {\bibfnamefont {M.}~\bibnamefont {{Dominik}}}, \bibinfo {author} {\bibfnamefont {T.}~\bibnamefont {{Bulik}}}, \bibinfo {author} {\bibfnamefont {R.}~\bibnamefont {{O'Shaughnessy}}}, \bibinfo {author} {\bibfnamefont {C.}~\bibnamefont {{Fryer}}}, \ and\ \bibinfo {author} {\bibfnamefont {D.~E.}\ \bibnamefont {{Holz}}},\ }\href {\doibase 10.1088/2041-8205/715/2/L138} {\bibfield  {journal} {\bibinfo  {journal} {Astrophys. J. Lett.}\ }\textbf {\bibinfo {volume} {715}},\ \bibinfo {pages} {L138} (\bibinfo {year} {2010})},\ \Eprint {http://arxiv.org/abs/1004.0386} {arXiv:1004.0386 [astro-ph.HE]} \BibitemShut {NoStop}%
\bibitem [{\citenamefont {Abac}\ \emph {et~al.}(2025{\natexlab{a}})\citenamefont {Abac} \emph {et~al.}}]{LIGOScientific:2025rsn}%
  \BibitemOpen
  \bibfield  {author} {\bibinfo {author} {\bibfnamefont {A.~G.}\ \bibnamefont {Abac}} \emph {et~al.} (\bibinfo {collaboration} {LIGO Scientific, VIRGO, KAGRA}),\ }\href {\doibase 10.3847/2041-8213/ae0c9c} {\bibfield  {journal} {\bibinfo  {journal} {Astrophys. J. Lett.}\ }\textbf {\bibinfo {volume} {993}},\ \bibinfo {pages} {L25} (\bibinfo {year} {2025}{\natexlab{a}})},\ \Eprint {http://arxiv.org/abs/2507.08219} {arXiv:2507.08219 [astro-ph.HE]} \BibitemShut {NoStop}%
\bibitem [{\citenamefont {{Woosley}}\ \emph {et~al.}(2007)\citenamefont {{Woosley}}, \citenamefont {{Blinnikov}},\ and\ \citenamefont {{Heger}}}]{Wooetal07}%
  \BibitemOpen
  \bibfield  {author} {\bibinfo {author} {\bibfnamefont {S.~E.}\ \bibnamefont {{Woosley}}}, \bibinfo {author} {\bibfnamefont {S.}~\bibnamefont {{Blinnikov}}}, \ and\ \bibinfo {author} {\bibfnamefont {A.}~\bibnamefont {{Heger}}},\ }\href {\doibase 10.1038/nature06333} {\bibfield  {journal} {\bibinfo  {journal} {Nature}\ }\textbf {\bibinfo {volume} {450}},\ \bibinfo {pages} {390} (\bibinfo {year} {2007})},\ \Eprint {http://arxiv.org/abs/0710.3314} {arXiv:0710.3314 [astro-ph]} \BibitemShut {NoStop}%
\bibitem [{\citenamefont {{Woosley}}\ and\ \citenamefont {{Heger}}(2021)}]{WooHeg21}%
  \BibitemOpen
  \bibfield  {author} {\bibinfo {author} {\bibfnamefont {S.~E.}\ \bibnamefont {{Woosley}}}\ and\ \bibinfo {author} {\bibfnamefont {A.}~\bibnamefont {{Heger}}},\ }\href {\doibase 10.3847/2041-8213/abf2c4} {\bibfield  {journal} {\bibinfo  {journal} {Astrophys. J. Lett.}\ }\textbf {\bibinfo {volume} {912}},\ \bibinfo {eid} {L31} (\bibinfo {year} {2021})},\ \Eprint {http://arxiv.org/abs/2103.07933} {arXiv:2103.07933 [astro-ph.SR]} \BibitemShut {NoStop}%
\bibitem [{\citenamefont {De~Luca}\ \emph {et~al.}(2026)\citenamefont {De~Luca}, \citenamefont {Franciolini},\ and\ \citenamefont {Riotto}}]{DeLuca:2025fln}%
  \BibitemOpen
  \bibfield  {author} {\bibinfo {author} {\bibfnamefont {V.}~\bibnamefont {De~Luca}}, \bibinfo {author} {\bibfnamefont {G.}~\bibnamefont {Franciolini}}, \ and\ \bibinfo {author} {\bibfnamefont {A.}~\bibnamefont {Riotto}},\ }\href {\doibase 10.1103/y4x3-v6j2} {\bibfield  {journal} {\bibinfo  {journal} {Phys. Rev. Lett.}\ }\textbf {\bibinfo {volume} {136}},\ \bibinfo {pages} {201401} (\bibinfo {year} {2026})},\ \Eprint {http://arxiv.org/abs/2508.09965} {arXiv:2508.09965 [astro-ph.CO]} \BibitemShut {NoStop}%
\bibitem [{\citenamefont {Dolgov}\ \emph {et~al.}(2020)\citenamefont {Dolgov}, \citenamefont {Kuranov}, \citenamefont {Mitichkin}, \citenamefont {Porey}, \citenamefont {Postnov}, \citenamefont {Sazhina},\ and\ \citenamefont {Simkin}}]{Doletal20}%
  \BibitemOpen
  \bibfield  {author} {\bibinfo {author} {\bibfnamefont {A.~D.}\ \bibnamefont {Dolgov}}, \bibinfo {author} {\bibfnamefont {A.~G.}\ \bibnamefont {Kuranov}}, \bibinfo {author} {\bibfnamefont {N.~A.}\ \bibnamefont {Mitichkin}}, \bibinfo {author} {\bibfnamefont {S.}~\bibnamefont {Porey}}, \bibinfo {author} {\bibfnamefont {K.~A.}\ \bibnamefont {Postnov}}, \bibinfo {author} {\bibfnamefont {O.~S.}\ \bibnamefont {Sazhina}}, \ and\ \bibinfo {author} {\bibfnamefont {I.~V.}\ \bibnamefont {Simkin}},\ }\href {\doibase 10.1088/1475-7516/2020/12/017} {\bibfield  {journal} {\bibinfo  {journal} {J. Cosmol. Astropart. Phys.}\ }\textbf {\bibinfo {volume} {12}},\ \bibinfo {pages} {017} (\bibinfo {year} {2020})},\ \Eprint {http://arxiv.org/abs/2005.00892} {arXiv:2005.00892 [astro-ph.CO]} \BibitemShut {NoStop}%
\bibitem [{\citenamefont {Boybeyi}\ \emph {et~al.}(2026)\citenamefont {Boybeyi}, \citenamefont {Veske},\ and\ \citenamefont {Maibach}}]{Boybeyi:2026lnq}%
  \BibitemOpen
  \bibfield  {author} {\bibinfo {author} {\bibfnamefont {T.}~\bibnamefont {Boybeyi}}, \bibinfo {author} {\bibfnamefont {D.}~\bibnamefont {Veske}}, \ and\ \bibinfo {author} {\bibfnamefont {D.}~\bibnamefont {Maibach}},\ }\href@noop {} {\  (\bibinfo {year} {2026})},\ \Eprint {http://arxiv.org/abs/2606.06478} {arXiv:2606.06478 [gr-qc]} \BibitemShut {NoStop}%
\bibitem [{\citenamefont {Garc{\'\i}a-Bellido}\ \emph {et~al.}(2021)\citenamefont {Garc{\'\i}a-Bellido}, \citenamefont {Nu{\~n}o~Siles},\ and\ \citenamefont {Ruiz~Morales}}]{GarSilMor21}%
  \BibitemOpen
  \bibfield  {author} {\bibinfo {author} {\bibfnamefont {J.}~\bibnamefont {Garc{\'\i}a-Bellido}}, \bibinfo {author} {\bibfnamefont {J.~F.}\ \bibnamefont {Nu{\~n}o~Siles}}, \ and\ \bibinfo {author} {\bibfnamefont {E.}~\bibnamefont {Ruiz~Morales}},\ }\href {\doibase 10.1016/j.dark.2021.100791} {\bibfield  {journal} {\bibinfo  {journal} {Phys. Dark Univ.}\ }\textbf {\bibinfo {volume} {31}},\ \bibinfo {pages} {100791} (\bibinfo {year} {2021})},\ \Eprint {http://arxiv.org/abs/2010.13811} {arXiv:2010.13811 [astro-ph.CO]} \BibitemShut {NoStop}%
\bibitem [{\citenamefont {Belczynski}\ \emph {et~al.}(2020)\citenamefont {Belczynski} \emph {et~al.}}]{Beletal20}%
  \BibitemOpen
  \bibfield  {author} {\bibinfo {author} {\bibfnamefont {K.}~\bibnamefont {Belczynski}} \emph {et~al.},\ }\href {\doibase 10.1051/0004-6361/201936528} {\bibfield  {journal} {\bibinfo  {journal} {Astron. Astrophys.}\ }\textbf {\bibinfo {volume} {636}},\ \bibinfo {pages} {A104} (\bibinfo {year} {2020})},\ \Eprint {http://arxiv.org/abs/1706.07053} {arXiv:1706.07053 [astro-ph.HE]} \BibitemShut {NoStop}%
\bibitem [{\citenamefont {Abac}\ \emph {et~al.}(2026{\natexlab{a}})\citenamefont {Abac} \emph {et~al.}}]{GWTC5}%
  \BibitemOpen
  \bibfield  {author} {\bibinfo {author} {\bibfnamefont {A.~G.}\ \bibnamefont {Abac}} \emph {et~al.} (\bibinfo {collaboration} {LIGO Scientific, VIRGO, KAGRA}),\ }\href@noop {} {\  (\bibinfo {year} {2026}{\natexlab{a}})},\ \Eprint {http://arxiv.org/abs/2605.27226} {arXiv:2605.27226 [astro-ph.HE]} \BibitemShut {NoStop}%
\bibitem [{\citenamefont {{Postnov}}\ and\ \citenamefont {{Kuranov}}(2019)}]{PosKur19}%
  \BibitemOpen
  \bibfield  {author} {\bibinfo {author} {\bibfnamefont {K.~A.}\ \bibnamefont {{Postnov}}}\ and\ \bibinfo {author} {\bibfnamefont {A.~G.}\ \bibnamefont {{Kuranov}}},\ }\href {\doibase 10.1093/mnras/sty3313} {\bibfield  {journal} {\bibinfo  {journal} {Mon. Not. R. Astron. Soc.}\ }\textbf {\bibinfo {volume} {483}},\ \bibinfo {pages} {3288} (\bibinfo {year} {2019})},\ \Eprint {http://arxiv.org/abs/1706.00369} {arXiv:1706.00369 [astro-ph.HE]} \BibitemShut {NoStop}%
\bibitem [{\citenamefont {{Postnov}}\ \emph {et~al.}(2025)\citenamefont {{Postnov}}, \citenamefont {{Shaposhnikov}},\ and\ \citenamefont {{Cherepashchuk}}}]{Posetal26}%
  \BibitemOpen
  \bibfield  {author} {\bibinfo {author} {\bibfnamefont {K.~A.}\ \bibnamefont {{Postnov}}}, \bibinfo {author} {\bibfnamefont {I.~A.}\ \bibnamefont {{Shaposhnikov}}}, \ and\ \bibinfo {author} {\bibfnamefont {A.~M.}\ \bibnamefont {{Cherepashchuk}}},\ }\href {\doibase 10.1093/mnrasl/slaf103} {\bibfield  {journal} {\bibinfo  {journal} {Mon. Not. R. Astron. Soc.}\ }\textbf {\bibinfo {volume} {544}},\ \bibinfo {pages} {L83} (\bibinfo {year} {2025})},\ \Eprint {http://arxiv.org/abs/2509.19981} {arXiv:2509.19981 [astro-ph.HE]} \BibitemShut {NoStop}%
\bibitem [{\citenamefont {Haque}\ \emph {et~al.}(2026)\citenamefont {Haque}, \citenamefont {Iocco},\ and\ \citenamefont {Visinelli}}]{HaqIocVis26}%
  \BibitemOpen
  \bibfield  {author} {\bibinfo {author} {\bibfnamefont {M.~R.}\ \bibnamefont {Haque}}, \bibinfo {author} {\bibfnamefont {F.}~\bibnamefont {Iocco}}, \ and\ \bibinfo {author} {\bibfnamefont {L.}~\bibnamefont {Visinelli}},\ }\href@noop {} {\  (\bibinfo {year} {2026})},\ \Eprint {http://arxiv.org/abs/2603.25795} {arXiv:2603.25795 [astro-ph.CO]} \BibitemShut {NoStop}%
\bibitem [{\citenamefont {Magaraggia}\ and\ \citenamefont {Cappelluti}(2026)}]{MagCap26}%
  \BibitemOpen
  \bibfield  {author} {\bibinfo {author} {\bibfnamefont {A.}~\bibnamefont {Magaraggia}}\ and\ \bibinfo {author} {\bibfnamefont {N.}~\bibnamefont {Cappelluti}},\ }\href {\doibase 10.3847/1538-4357/ae48f9} {\bibfield  {journal} {\bibinfo  {journal} {Astrophys. J.}\ }\textbf {\bibinfo {volume} {1000}},\ \bibinfo {pages} {262} (\bibinfo {year} {2026})},\ \Eprint {http://arxiv.org/abs/2602.21295} {arXiv:2602.21295 [astro-ph.CO]} \BibitemShut {NoStop}%
\bibitem [{\citenamefont {Berezhiani}\ \emph {et~al.}(1996)\citenamefont {Berezhiani}, \citenamefont {Dolgov},\ and\ \citenamefont {Mohapatra}}]{BerDolMoh96}%
  \BibitemOpen
  \bibfield  {author} {\bibinfo {author} {\bibfnamefont {Z.~G.}\ \bibnamefont {Berezhiani}}, \bibinfo {author} {\bibfnamefont {A.~D.}\ \bibnamefont {Dolgov}}, \ and\ \bibinfo {author} {\bibfnamefont {R.~N.}\ \bibnamefont {Mohapatra}},\ }\href {\doibase 10.1016/0370-2693(96)00219-5} {\bibfield  {journal} {\bibinfo  {journal} {Phys. Lett. B}\ }\textbf {\bibinfo {volume} {375}},\ \bibinfo {pages} {26} (\bibinfo {year} {1996})},\ \Eprint {http://arxiv.org/abs/hep-ph/9511221} {arXiv:hep-ph/9511221} \BibitemShut {NoStop}%
\bibitem [{\citenamefont {Okun}(2007)}]{Oku07}%
  \BibitemOpen
  \bibfield  {author} {\bibinfo {author} {\bibfnamefont {L.~B.}\ \bibnamefont {Okun}},\ }\href {\doibase 10.1070/PU2007v050n04ABEH006227} {\bibfield  {journal} {\bibinfo  {journal} {Phys. Usp.}\ }\textbf {\bibinfo {volume} {50}},\ \bibinfo {pages} {380} (\bibinfo {year} {2007})}\BibitemShut {NoStop}%
\bibitem [{\citenamefont {Blinnikov}(2014)}]{Bli14}%
  \BibitemOpen
  \bibfield  {author} {\bibinfo {author} {\bibfnamefont {S.~I.}\ \bibnamefont {Blinnikov}},\ }\href {\doibase 10.3367/UFNe.0184.201402h.0194} {\bibfield  {journal} {\bibinfo  {journal} {Phys. Usp.}\ }\textbf {\bibinfo {volume} {57}},\ \bibinfo {pages} {183} (\bibinfo {year} {2014})}\BibitemShut {NoStop}%
\bibitem [{\citenamefont {{Martel}}\ and\ \citenamefont {{Poisson}}(1999)}]{1999PhRvD..60l4008M}%
  \BibitemOpen
  \bibfield  {author} {\bibinfo {author} {\bibfnamefont {K.}~\bibnamefont {{Martel}}}\ and\ \bibinfo {author} {\bibfnamefont {E.}~\bibnamefont {{Poisson}}},\ }\href {\doibase 10.1103/PhysRevD.60.124008} {\bibfield  {journal} {\bibinfo  {journal} {Phys. Rev. D}\ }\textbf {\bibinfo {volume} {60}},\ \bibinfo {eid} {124008} (\bibinfo {year} {1999})},\ \Eprint {http://arxiv.org/abs/gr-qc/9907006} {arXiv:gr-qc/9907006 [gr-qc]} \BibitemShut {NoStop}%
\bibitem [{\citenamefont {{Sasaki}}\ \emph {et~al.}(2018)\citenamefont {{Sasaki}}, \citenamefont {{Suyama}}, \citenamefont {{Tanaka}},\ and\ \citenamefont {{Yokoyama}}}]{2018CQGra..35f3001S}%
  \BibitemOpen
  \bibfield  {author} {\bibinfo {author} {\bibfnamefont {M.}~\bibnamefont {{Sasaki}}}, \bibinfo {author} {\bibfnamefont {T.}~\bibnamefont {{Suyama}}}, \bibinfo {author} {\bibfnamefont {T.}~\bibnamefont {{Tanaka}}}, \ and\ \bibinfo {author} {\bibfnamefont {S.}~\bibnamefont {{Yokoyama}}},\ }\href {\doibase 10.1088/1361-6382/aaa7b4} {\bibfield  {journal} {\bibinfo  {journal} {Classical and Quantum Gravity}\ }\textbf {\bibinfo {volume} {35}},\ \bibinfo {eid} {063001} (\bibinfo {year} {2018})},\ \Eprint {http://arxiv.org/abs/1801.05235} {arXiv:1801.05235 [astro-ph.CO]} \BibitemShut {NoStop}%
\bibitem [{\citenamefont {Andr{\'e}s-Carcasona}\ \emph {et~al.}(2026)\citenamefont {Andr{\'e}s-Carcasona}, \citenamefont {Iovino}, \citenamefont {Vallejo-Pag{\`e}s}, \citenamefont {Vaskonen}, \citenamefont {Veerm{\"a}e}, \citenamefont {Mart{\'\i}nez},\ and\ \citenamefont {Mir}}]{2026arXiv260515749A}%
  \BibitemOpen
  \bibfield  {author} {\bibinfo {author} {\bibfnamefont {M.}~\bibnamefont {Andr{\'e}s-Carcasona}}, \bibinfo {author} {\bibfnamefont {A.~J.}\ \bibnamefont {Iovino}}, \bibinfo {author} {\bibfnamefont {E.}~\bibnamefont {Vallejo-Pag{\`e}s}}, \bibinfo {author} {\bibfnamefont {V.}~\bibnamefont {Vaskonen}}, \bibinfo {author} {\bibfnamefont {H.}~\bibnamefont {Veerm{\"a}e}}, \bibinfo {author} {\bibfnamefont {M.}~\bibnamefont {Mart{\'\i}nez}}, \ and\ \bibinfo {author} {\bibfnamefont {L.~M.}\ \bibnamefont {Mir}},\ }\href@noop {} {\  (\bibinfo {year} {2026})},\ \Eprint {http://arxiv.org/abs/2605.15749} {arXiv:2605.15749 [astro-ph.CO]} \BibitemShut {NoStop}%
\bibitem [{\citenamefont {Jedamzik}(2020)}]{Jedamzik:2020ypm}%
  \BibitemOpen
  \bibfield  {author} {\bibinfo {author} {\bibfnamefont {K.}~\bibnamefont {Jedamzik}},\ }\href {\doibase 10.1088/1475-7516/2020/09/022} {\bibfield  {journal} {\bibinfo  {journal} {J. Cosmol. Astropart. Phys.}\ }\textbf {\bibinfo {volume} {09}},\ \bibinfo {pages} {022} (\bibinfo {year} {2020})},\ \Eprint {http://arxiv.org/abs/2006.11172} {arXiv:2006.11172 [astro-ph.CO]} \BibitemShut {NoStop}%
\bibitem [{\citenamefont {Tkachev}\ \emph {et~al.}(2020)\citenamefont {Tkachev}, \citenamefont {Pilipenko},\ and\ \citenamefont {Yepes}}]{Tkachev:2020uin}%
  \BibitemOpen
  \bibfield  {author} {\bibinfo {author} {\bibfnamefont {M.}~\bibnamefont {Tkachev}}, \bibinfo {author} {\bibfnamefont {S.}~\bibnamefont {Pilipenko}}, \ and\ \bibinfo {author} {\bibfnamefont {G.}~\bibnamefont {Yepes}},\ }\href {\doibase 10.1093/mnras/staa3103} {\bibfield  {journal} {\bibinfo  {journal} {Mon. Not. Roy. Astron. Soc.}\ }\textbf {\bibinfo {volume} {499}},\ \bibinfo {pages} {4854} (\bibinfo {year} {2020})},\ \Eprint {http://arxiv.org/abs/2009.07813} {arXiv:2009.07813 [astro-ph.CO]} \BibitemShut {NoStop}%
\bibitem [{\citenamefont {Vaskonen}\ and\ \citenamefont {Veerm{\"a}e}(2020)}]{Vaskonen:2019jpv}%
  \BibitemOpen
  \bibfield  {author} {\bibinfo {author} {\bibfnamefont {V.}~\bibnamefont {Vaskonen}}\ and\ \bibinfo {author} {\bibfnamefont {H.}~\bibnamefont {Veerm{\"a}e}},\ }\href {\doibase 10.1103/PhysRevD.101.043015} {\bibfield  {journal} {\bibinfo  {journal} {Phys. Rev. D}\ }\textbf {\bibinfo {volume} {101}},\ \bibinfo {pages} {043015} (\bibinfo {year} {2020})},\ \Eprint {http://arxiv.org/abs/1908.09752} {arXiv:1908.09752 [astro-ph.CO]} \BibitemShut {NoStop}%
\bibitem [{\citenamefont {{Quinlan}}\ and\ \citenamefont {{Shapiro}}(1989)}]{QuinShap89}%
  \BibitemOpen
  \bibfield  {author} {\bibinfo {author} {\bibfnamefont {G.~D.}\ \bibnamefont {{Quinlan}}}\ and\ \bibinfo {author} {\bibfnamefont {S.~L.}\ \bibnamefont {{Shapiro}}},\ }\href {\doibase 10.1086/167745} {\bibfield  {journal} {\bibinfo  {journal} {Astrophys. J.}\ }\textbf {\bibinfo {volume} {343}},\ \bibinfo {pages} {725} (\bibinfo {year} {1989})}\BibitemShut {NoStop}%
\bibitem [{\citenamefont {Stasenko}\ and\ \citenamefont {Kirillov}(2021)}]{Stasenko:2021vmm}%
  \BibitemOpen
  \bibfield  {author} {\bibinfo {author} {\bibfnamefont {V.~D.}\ \bibnamefont {Stasenko}}\ and\ \bibinfo {author} {\bibfnamefont {A.~A.}\ \bibnamefont {Kirillov}},\ }\href {\doibase 10.3390/physics3020026} {\bibfield  {journal} {\bibinfo  {journal} {MDPI Physics}\ }\textbf {\bibinfo {volume} {3}},\ \bibinfo {pages} {372} (\bibinfo {year} {2021})},\ \Eprint {http://arxiv.org/abs/2105.14523} {arXiv:2105.14523 [astro-ph.CO]} \BibitemShut {NoStop}%
\bibitem [{\citenamefont {Abbott}\ \emph {et~al.}(2023)\citenamefont {Abbott} \emph {et~al.}}]{KAGRA:2021duu}%
  \BibitemOpen
  \bibfield  {author} {\bibinfo {author} {\bibfnamefont {R.}~\bibnamefont {Abbott}} \emph {et~al.} (\bibinfo {collaboration} {KAGRA, VIRGO, LIGO Scientific}),\ }\href {\doibase 10.1103/PhysRevX.13.011048} {\bibfield  {journal} {\bibinfo  {journal} {Phys. Rev. X}\ }\textbf {\bibinfo {volume} {13}},\ \bibinfo {pages} {011048} (\bibinfo {year} {2023})},\ \Eprint {http://arxiv.org/abs/2111.03634} {arXiv:2111.03634 [astro-ph.HE]} \BibitemShut {NoStop}%
\bibitem [{\citenamefont {{Hills}}(1975)}]{1975AJ.....80..809H}%
  \BibitemOpen
  \bibfield  {author} {\bibinfo {author} {\bibfnamefont {J.~G.}\ \bibnamefont {{Hills}}},\ }\href {\doibase 10.1086/111815} {\bibfield  {journal} {\bibinfo  {journal} {Astron. J.}\ }\textbf {\bibinfo {volume} {80}},\ \bibinfo {pages} {809} (\bibinfo {year} {1975})}\BibitemShut {NoStop}%
\bibitem [{\citenamefont {{Heggie}}(1975)}]{1975MNRAS.173..729H}%
  \BibitemOpen
  \bibfield  {author} {\bibinfo {author} {\bibfnamefont {D.~C.}\ \bibnamefont {{Heggie}}},\ }\href {\doibase 10.1093/mnras/173.3.729} {\bibfield  {journal} {\bibinfo  {journal} {Mon. Not. R. Astron. Soc.}\ }\textbf {\bibinfo {volume} {173}},\ \bibinfo {pages} {729} (\bibinfo {year} {1975})}\BibitemShut {NoStop}%
\bibitem [{\citenamefont {Mr{\'o}z}\ \emph {et~al.}(2024)\citenamefont {Mr{\'o}z} \emph {et~al.}}]{Mroz:2024mse}%
  \BibitemOpen
  \bibfield  {author} {\bibinfo {author} {\bibfnamefont {P.}~\bibnamefont {Mr{\'o}z}} \emph {et~al.},\ }\href {\doibase 10.1038/s41586-024-07704-6} {\bibfield  {journal} {\bibinfo  {journal} {Nature}\ }\textbf {\bibinfo {volume} {632}},\ \bibinfo {pages} {749} (\bibinfo {year} {2024})},\ \Eprint {http://arxiv.org/abs/2403.02386} {arXiv:2403.02386 [astro-ph.GA]} \BibitemShut {NoStop}%
\bibitem [{\citenamefont {Stasenko}(2025)}]{Stasenko:2025vqz}%
  \BibitemOpen
  \bibfield  {author} {\bibinfo {author} {\bibfnamefont {V.}~\bibnamefont {Stasenko}},\ }\href {\doibase 10.1016/j.dark.2025.102138} {\bibfield  {journal} {\bibinfo  {journal} {Phys. Dark Univ.}\ }\textbf {\bibinfo {volume} {50}},\ \bibinfo {pages} {102138} (\bibinfo {year} {2025})},\ \Eprint {http://arxiv.org/abs/2506.11290} {arXiv:2506.11290 [astro-ph.CO]} \BibitemShut {NoStop}%
\bibitem [{\citenamefont {Koushiappas}\ and\ \citenamefont {Loeb}(2017)}]{Koushiappas:2017kqm}%
  \BibitemOpen
  \bibfield  {author} {\bibinfo {author} {\bibfnamefont {S.~M.}\ \bibnamefont {Koushiappas}}\ and\ \bibinfo {author} {\bibfnamefont {A.}~\bibnamefont {Loeb}},\ }\href {\doibase 10.1103/PhysRevLett.119.221104} {\bibfield  {journal} {\bibinfo  {journal} {Phys. Rev. Lett.}\ }\textbf {\bibinfo {volume} {119}},\ \bibinfo {pages} {221104} (\bibinfo {year} {2017})},\ \Eprint {http://arxiv.org/abs/1708.07380} {arXiv:1708.07380 [astro-ph.CO]} \BibitemShut {NoStop}%
\bibitem [{\citenamefont {Madau}\ and\ \citenamefont {Dickinson}(2014)}]{Madau:2014bja}%
  \BibitemOpen
  \bibfield  {author} {\bibinfo {author} {\bibfnamefont {P.}~\bibnamefont {Madau}}\ and\ \bibinfo {author} {\bibfnamefont {M.}~\bibnamefont {Dickinson}},\ }\href {\doibase 10.1146/annurev-astro-081811-125615} {\bibfield  {journal} {\bibinfo  {journal} {Ann. Rev. Astron. Astrophys.}\ }\textbf {\bibinfo {volume} {52}},\ \bibinfo {pages} {415} (\bibinfo {year} {2014})},\ \Eprint {http://arxiv.org/abs/1403.0007} {arXiv:1403.0007 [astro-ph.CO]} \BibitemShut {NoStop}%
\bibitem [{\citenamefont {Belczynski}\ \emph {et~al.}(2016)\citenamefont {Belczynski}, \citenamefont {Holz}, \citenamefont {Bulik},\ and\ \citenamefont {O'Shaughnessy}}]{Belczynski:2016obo}%
  \BibitemOpen
  \bibfield  {author} {\bibinfo {author} {\bibfnamefont {K.}~\bibnamefont {Belczynski}}, \bibinfo {author} {\bibfnamefont {D.~E.}\ \bibnamefont {Holz}}, \bibinfo {author} {\bibfnamefont {T.}~\bibnamefont {Bulik}}, \ and\ \bibinfo {author} {\bibfnamefont {R.}~\bibnamefont {O'Shaughnessy}},\ }\href {\doibase 10.1038/nature18322} {\bibfield  {journal} {\bibinfo  {journal} {Nature}\ }\textbf {\bibinfo {volume} {534}},\ \bibinfo {pages} {512} (\bibinfo {year} {2016})},\ \Eprint {http://arxiv.org/abs/1602.04531} {arXiv:1602.04531 [astro-ph.HE]} \BibitemShut {NoStop}%
\bibitem [{\citenamefont {Abac}\ \emph {et~al.}(2026{\natexlab{b}})\citenamefont {Abac} \emph {et~al.}}]{ET:2025xjr}%
  \BibitemOpen
  \bibfield  {author} {\bibinfo {author} {\bibfnamefont {A.}~\bibnamefont {Abac}} \emph {et~al.} (\bibinfo {collaboration} {ET}),\ }\href {\doibase 10.1088/1475-7516/2026/03/081} {\bibfield  {journal} {\bibinfo  {journal} {J. Cosmol. Astropart. Phys.}\ }\textbf {\bibinfo {volume} {03}},\ \bibinfo {pages} {081} (\bibinfo {year} {2026}{\natexlab{b}})},\ \Eprint {http://arxiv.org/abs/2503.12263} {arXiv:2503.12263 [gr-qc]} \BibitemShut {NoStop}%
\bibitem [{\citenamefont {Evans}\ \emph {et~al.}(2021)\citenamefont {Evans} \emph {et~al.}}]{CE:2021}%
  \BibitemOpen
  \bibfield  {author} {\bibinfo {author} {\bibfnamefont {M.}~\bibnamefont {Evans}} \emph {et~al.},\ }\href@noop {} {\  (\bibinfo {year} {2021})},\ \Eprint {http://arxiv.org/abs/2109.09882} {arXiv:2109.09882 [astro-ph.IM]} \BibitemShut {NoStop}%
\bibitem [{\citenamefont {Abbott}\ \emph {et~al.}(2017)\citenamefont {Abbott} \emph {et~al.}}]{LIGOScientific:2016wof}%
  \BibitemOpen
  \bibfield  {author} {\bibinfo {author} {\bibfnamefont {B.~P.}\ \bibnamefont {Abbott}} \emph {et~al.} (\bibinfo {collaboration} {LIGO Scientific}),\ }\href {\doibase 10.1088/1361-6382/aa51f4} {\bibfield  {journal} {\bibinfo  {journal} {Class. Quant. Grav.}\ }\textbf {\bibinfo {volume} {34}},\ \bibinfo {pages} {044001} (\bibinfo {year} {2017})},\ \Eprint {http://arxiv.org/abs/1607.08697} {arXiv:1607.08697 [astro-ph.IM]} \BibitemShut {NoStop}%
\bibitem [{\citenamefont {Phinney}(2001)}]{Phinney:2001di}%
  \BibitemOpen
  \bibfield  {author} {\bibinfo {author} {\bibfnamefont {E.~S.}\ \bibnamefont {Phinney}},\ }\href@noop {} {\  (\bibinfo {year} {2001})},\ \Eprint {http://arxiv.org/abs/astro-ph/0108028} {arXiv:astro-ph/0108028} \BibitemShut {NoStop}%
\bibitem [{\citenamefont {Maggiore}(2007)}]{Magg2007}%
  \BibitemOpen
  \bibfield  {author} {\bibinfo {author} {\bibfnamefont {M.}~\bibnamefont {Maggiore}},\ }\href@noop {} {\emph {\bibinfo {title} {Gravitational Waves. Vol. 1: Theory and Experiments}}}\ (\bibinfo  {publisher} {Oxford University Press},\ \bibinfo {address} {New York City},\ \bibinfo {year} {2007})\BibitemShut {NoStop}%
\bibitem [{\citenamefont {Abac}\ \emph {et~al.}(2025{\natexlab{b}})\citenamefont {Abac} \emph {et~al.}}]{LIGOScientific:2025kry}%
  \BibitemOpen
  \bibfield  {author} {\bibinfo {author} {\bibfnamefont {A.~G.}\ \bibnamefont {Abac}} \emph {et~al.} (\bibinfo {collaboration} {LIGO Scientific, VIRGO, KAGRA}),\ }\href@noop {} {\  (\bibinfo {year} {2025}{\natexlab{b}})},\ \Eprint {http://arxiv.org/abs/2510.26848} {arXiv:2510.26848 [gr-qc]} \BibitemShut {NoStop}%
\bibitem [{\citenamefont {Matarrese}\ \emph {et~al.}(1994)\citenamefont {Matarrese}, \citenamefont {Pantano},\ and\ \citenamefont {Saez}}]{MatPanSae94}%
  \BibitemOpen
  \bibfield  {author} {\bibinfo {author} {\bibfnamefont {S.}~\bibnamefont {Matarrese}}, \bibinfo {author} {\bibfnamefont {O.}~\bibnamefont {Pantano}}, \ and\ \bibinfo {author} {\bibfnamefont {D.}~\bibnamefont {Saez}},\ }\href {\doibase 10.1103/PhysRevLett.72.320} {\bibfield  {journal} {\bibinfo  {journal} {Phys. Rev. Lett.}\ }\textbf {\bibinfo {volume} {72}},\ \bibinfo {pages} {320} (\bibinfo {year} {1994})},\ \Eprint {http://arxiv.org/abs/astro-ph/9310036} {arXiv:astro-ph/9310036} \BibitemShut {NoStop}%
\bibitem [{\citenamefont {Bugaev}\ and\ \citenamefont {Klimai}(2010)}]{BugKli10}%
  \BibitemOpen
  \bibfield  {author} {\bibinfo {author} {\bibfnamefont {E.~V.}\ \bibnamefont {Bugaev}}\ and\ \bibinfo {author} {\bibfnamefont {P.~A.}\ \bibnamefont {Klimai}},\ }\href@noop {} {\bibfield  {journal} {\bibinfo  {journal} {JETP Letters}\ }\textbf {\bibinfo {volume} {91}},\ \bibinfo {pages} {1} (\bibinfo {year} {2010})}\BibitemShut {NoStop}%
\bibitem [{\citenamefont {Kohri}\ and\ \citenamefont {Terada}(2018)}]{KohTer18}%
  \BibitemOpen
  \bibfield  {author} {\bibinfo {author} {\bibfnamefont {K.}~\bibnamefont {Kohri}}\ and\ \bibinfo {author} {\bibfnamefont {T.}~\bibnamefont {Terada}},\ }\href {\doibase 10.1103/PhysRevD.97.123532} {\bibfield  {journal} {\bibinfo  {journal} {Phys. Rev. D}\ }\textbf {\bibinfo {volume} {97}},\ \bibinfo {pages} {123532} (\bibinfo {year} {2018})},\ \Eprint {http://arxiv.org/abs/1804.08577} {arXiv:1804.08577 [gr-qc]} \BibitemShut {NoStop}%
\bibitem [{\citenamefont {Pi}\ and\ \citenamefont {Sasaki}(2020)}]{PiSas20}%
  \BibitemOpen
  \bibfield  {author} {\bibinfo {author} {\bibfnamefont {S.}~\bibnamefont {Pi}}\ and\ \bibinfo {author} {\bibfnamefont {M.}~\bibnamefont {Sasaki}},\ }\href {\doibase 10.1088/1475-7516/2020/09/037} {\bibfield  {journal} {\bibinfo  {journal} {J. Cosmol. Astropart. Phys.}\ }\textbf {\bibinfo {volume} {09}},\ \bibinfo {pages} {037} (\bibinfo {year} {2020})},\ \Eprint {http://arxiv.org/abs/2005.12306} {arXiv:2005.12306 [gr-qc]} \BibitemShut {NoStop}%
\bibitem [{\citenamefont {Inomata}\ and\ \citenamefont {Terada}(2020)}]{InoTer20}%
  \BibitemOpen
  \bibfield  {author} {\bibinfo {author} {\bibfnamefont {K.}~\bibnamefont {Inomata}}\ and\ \bibinfo {author} {\bibfnamefont {T.}~\bibnamefont {Terada}},\ }\href {\doibase 10.1103/PhysRevD.101.023523} {\bibfield  {journal} {\bibinfo  {journal} {Phys. Rev. D}\ }\textbf {\bibinfo {volume} {101}},\ \bibinfo {pages} {023523} (\bibinfo {year} {2020})},\ \Eprint {http://arxiv.org/abs/1912.00785} {arXiv:1912.00785 [gr-qc]} \BibitemShut {NoStop}%
\bibitem [{\citenamefont {Dom{\`e}nech}(2021)}]{Dom21}%
  \BibitemOpen
  \bibfield  {author} {\bibinfo {author} {\bibfnamefont {G.}~\bibnamefont {Dom{\`e}nech}},\ }\href {\doibase 10.3390/universe7110398} {\bibfield  {journal} {\bibinfo  {journal} {Universe}\ }\textbf {\bibinfo {volume} {7}},\ \bibinfo {pages} {398} (\bibinfo {year} {2021})},\ \Eprint {http://arxiv.org/abs/2109.01398} {arXiv:2109.01398 [gr-qc]} \BibitemShut {NoStop}%
\bibitem [{\citenamefont {Vaskonen}\ and\ \citenamefont {Veerm{\"a}e}(2021)}]{VasVee21}%
  \BibitemOpen
  \bibfield  {author} {\bibinfo {author} {\bibfnamefont {V.}~\bibnamefont {Vaskonen}}\ and\ \bibinfo {author} {\bibfnamefont {H.}~\bibnamefont {Veerm{\"a}e}},\ }\href {\doibase 10.1103/PhysRevLett.126.051303} {\bibfield  {journal} {\bibinfo  {journal} {Phys. Rev. Lett.}\ }\textbf {\bibinfo {volume} {126}},\ \bibinfo {pages} {051303} (\bibinfo {year} {2021})},\ \Eprint {http://arxiv.org/abs/2009.07832} {arXiv:2009.07832 [astro-ph.CO]} \BibitemShut {NoStop}%
\bibitem [{\citenamefont {{Saito}}\ and\ \citenamefont {{Yokoyama}}(2009)}]{2009PhRvL.102p1101S}%
  \BibitemOpen
  \bibfield  {author} {\bibinfo {author} {\bibfnamefont {R.}~\bibnamefont {{Saito}}}\ and\ \bibinfo {author} {\bibfnamefont {J.}~\bibnamefont {{Yokoyama}}},\ }\href {\doibase 10.1103/PhysRevLett.102.161101} {\bibfield  {journal} {\bibinfo  {journal} {Phys. Rev. Lett.}\ }\textbf {\bibinfo {volume} {102}},\ \bibinfo {eid} {161101} (\bibinfo {year} {2009})},\ \Eprint {http://arxiv.org/abs/0812.4339} {arXiv:0812.4339 [astro-ph]} \BibitemShut {NoStop}%
\bibitem [{\citenamefont {Postnov}\ \emph {et~al.}(2025)\citenamefont {Postnov}, \citenamefont {Porayko},\ and\ \citenamefont {Pshirkov}}]{PosPorPsh25}%
  \BibitemOpen
  \bibfield  {author} {\bibinfo {author} {\bibfnamefont {K.~A.}\ \bibnamefont {Postnov}}, \bibinfo {author} {\bibfnamefont {N.~K.}\ \bibnamefont {Porayko}}, \ and\ \bibinfo {author} {\bibfnamefont {M.~S.}\ \bibnamefont {Pshirkov}},\ }\href {\doibase 10.3367/UFNe.2024.11.039812} {\bibfield  {journal} {\bibinfo  {journal} {Phys. Usp.}\ }\textbf {\bibinfo {volume} {68}},\ \bibinfo {pages} {146} (\bibinfo {year} {2025})},\ \Eprint {http://arxiv.org/abs/2502.00080} {arXiv:2502.00080 [astro-ph.IM]} \BibitemShut {NoStop}%
\bibitem [{\citenamefont {{Nakama}}\ and\ \citenamefont {{Suyama}}(2015)}]{2015PhRvD..92l1304N}%
  \BibitemOpen
  \bibfield  {author} {\bibinfo {author} {\bibfnamefont {T.}~\bibnamefont {{Nakama}}}\ and\ \bibinfo {author} {\bibfnamefont {T.}~\bibnamefont {{Suyama}}},\ }\href {\doibase 10.1103/PhysRevD.92.121304} {\bibfield  {journal} {\bibinfo  {journal} {Phys. Rev. D}\ }\textbf {\bibinfo {volume} {92}},\ \bibinfo {eid} {121304} (\bibinfo {year} {2015})},\ \Eprint {http://arxiv.org/abs/1506.05228} {arXiv:1506.05228 [gr-qc]} \BibitemShut {NoStop}%
\bibitem [{\citenamefont {{Nakama}}\ and\ \citenamefont {{Suyama}}(2016)}]{2016PhRvD..94d3507N}%
  \BibitemOpen
  \bibfield  {author} {\bibinfo {author} {\bibfnamefont {T.}~\bibnamefont {{Nakama}}}\ and\ \bibinfo {author} {\bibfnamefont {T.}~\bibnamefont {{Suyama}}},\ }\href {\doibase 10.1103/PhysRevD.94.043507} {\bibfield  {journal} {\bibinfo  {journal} {Phys. Rev. D}\ }\textbf {\bibinfo {volume} {94}},\ \bibinfo {eid} {043507} (\bibinfo {year} {2016})},\ \Eprint {http://arxiv.org/abs/1605.04482} {arXiv:1605.04482 [gr-qc]} \BibitemShut {NoStop}%
\bibitem [{\citenamefont {Kumar}(2025)}]{Kumar:2025jfi}%
  \BibitemOpen
  \bibfield  {author} {\bibinfo {author} {\bibfnamefont {U.}~\bibnamefont {Kumar}},\ }\href {\doibase 10.1103/ymhr-9711} {\bibfield  {journal} {\bibinfo  {journal} {Phys. Rev. D}\ }\textbf {\bibinfo {volume} {112}},\ \bibinfo {pages} {084027} (\bibinfo {year} {2025})},\ \Eprint {http://arxiv.org/abs/2507.10033} {arXiv:2507.10033 [gr-qc]} \BibitemShut {NoStop}%
\bibitem [{\citenamefont {Sakharov}\ \emph {et~al.}(2021)\citenamefont {Sakharov}, \citenamefont {Eroshenko},\ and\ \citenamefont {Rubin}}]{SakEroRub21}%
  \BibitemOpen
  \bibfield  {author} {\bibinfo {author} {\bibfnamefont {A.~S.}\ \bibnamefont {Sakharov}}, \bibinfo {author} {\bibfnamefont {Y.~N.}\ \bibnamefont {Eroshenko}}, \ and\ \bibinfo {author} {\bibfnamefont {S.~G.}\ \bibnamefont {Rubin}},\ }\href {\doibase 10.1103/PhysRevD.104.043005} {\bibfield  {journal} {\bibinfo  {journal} {Phys. Rev. D}\ }\textbf {\bibinfo {volume} {104}},\ \bibinfo {pages} {043005} (\bibinfo {year} {2021})},\ \Eprint {http://arxiv.org/abs/2104.08750} {arXiv:2104.08750 [hep-ph]} \BibitemShut {NoStop}%
\bibitem [{\citenamefont {Zel’dovich}\ and\ \citenamefont {Novikov}(1966)}]{ZelNov65}%
  \BibitemOpen
  \bibfield  {author} {\bibinfo {author} {\bibfnamefont {Y.~B.}\ \bibnamefont {Zel’dovich}}\ and\ \bibinfo {author} {\bibfnamefont {I.~D.}\ \bibnamefont {Novikov}},\ }\href {\doibase 10.1070/PU1966v008n04ABEH002990} {\bibfield  {journal} {\bibinfo  {journal} {Sov. Phys. Usp.}\ }\textbf {\bibinfo {volume} {8}},\ \bibinfo {pages} {522} (\bibinfo {year} {1966})}\BibitemShut {NoStop}%
\bibitem [{\citenamefont {Carr}\ \emph {et~al.}(2010{\natexlab{a}})\citenamefont {Carr}, \citenamefont {Harada},\ and\ \citenamefont {Maeda}}]{CarHarMae10}%
  \BibitemOpen
  \bibfield  {author} {\bibinfo {author} {\bibfnamefont {B.~J.}\ \bibnamefont {Carr}}, \bibinfo {author} {\bibfnamefont {T.}~\bibnamefont {Harada}}, \ and\ \bibinfo {author} {\bibfnamefont {H.}~\bibnamefont {Maeda}},\ }\href {\doibase 10.1088/0264-9381/27/18/183101} {\bibfield  {journal} {\bibinfo  {journal} {Class. Quant. Grav.}\ }\textbf {\bibinfo {volume} {27}},\ \bibinfo {pages} {183101} (\bibinfo {year} {2010}{\natexlab{a}})},\ \Eprint {http://arxiv.org/abs/1003.3324} {arXiv:1003.3324 [gr-qc]} \BibitemShut {NoStop}%
\bibitem [{\citenamefont {Gorkavyi}\ and\ \citenamefont {Tyul'bashev}(2021)}]{GorTul21}%
  \BibitemOpen
  \bibfield  {author} {\bibinfo {author} {\bibfnamefont {N.~N.}\ \bibnamefont {Gorkavyi}}\ and\ \bibinfo {author} {\bibfnamefont {S.}~\bibnamefont {Tyul'bashev}},\ }\href {\doibase 10.1134/S199034132103007X} {\bibfield  {journal} {\bibinfo  {journal} {Astrophys. Bull.}\ }\textbf {\bibinfo {volume} {76}},\ \bibinfo {pages} {229} (\bibinfo {year} {2021})},\ \Eprint {http://arxiv.org/abs/2110.11897} {arXiv:2110.11897 [astro-ph.CO]} \BibitemShut {NoStop}%
\bibitem [{\citenamefont {{Gazta{\~n}aga}}(2026)}]{Gaz26}%
  \BibitemOpen
  \bibfield  {author} {\bibinfo {author} {\bibfnamefont {E.}~\bibnamefont {{Gazta{\~n}aga}}},\ }\href {\doibase 10.1103/pr4p-6m49} {\bibfield  {journal} {\bibinfo  {journal} {Phys. Rev.~D}\ }\textbf {\bibinfo {volume} {113}},\ \bibinfo {eid} {043544} (\bibinfo {year} {2026})},\ \Eprint {http://arxiv.org/abs/2602.17702} {arXiv:2602.17702 [physics.gen-ph]} \BibitemShut {NoStop}%
\bibitem [{\citenamefont {{Carr}}(1979)}]{Car79}%
  \BibitemOpen
  \bibfield  {author} {\bibinfo {author} {\bibfnamefont {B.~J.}\ \bibnamefont {{Carr}}},\ }\href {\doibase 10.1093/mnras/189.1.123} {\bibfield  {journal} {\bibinfo  {journal} {Mon. Not. R. Astron. Soc.}\ }\textbf {\bibinfo {volume} {189}},\ \bibinfo {pages} {123} (\bibinfo {year} {1979})}\BibitemShut {NoStop}%
\bibitem [{\citenamefont {{Carr}}(1981)}]{Car81}%
  \BibitemOpen
  \bibfield  {author} {\bibinfo {author} {\bibfnamefont {B.~J.}\ \bibnamefont {{Carr}}},\ }\href {\doibase 10.1093/mnras/194.3.639} {\bibfield  {journal} {\bibinfo  {journal} {Mon. Not. R. Astron. Soc.}\ }\textbf {\bibinfo {volume} {194}},\ \bibinfo {pages} {639} (\bibinfo {year} {1981})}\BibitemShut {NoStop}%
\bibitem [{\citenamefont {Poulin}\ \emph {et~al.}(2017)\citenamefont {Poulin}, \citenamefont {Serpico}, \citenamefont {Calore}, \citenamefont {Clesse},\ and\ \citenamefont {Kohri}}]{Pouetal17}%
  \BibitemOpen
  \bibfield  {author} {\bibinfo {author} {\bibfnamefont {V.}~\bibnamefont {Poulin}}, \bibinfo {author} {\bibfnamefont {P.~D.}\ \bibnamefont {Serpico}}, \bibinfo {author} {\bibfnamefont {F.}~\bibnamefont {Calore}}, \bibinfo {author} {\bibfnamefont {S.}~\bibnamefont {Clesse}}, \ and\ \bibinfo {author} {\bibfnamefont {K.}~\bibnamefont {Kohri}},\ }\href {\doibase 10.1103/PhysRevD.96.083524} {\bibfield  {journal} {\bibinfo  {journal} {Phys. Rev. D}\ }\textbf {\bibinfo {volume} {96}},\ \bibinfo {pages} {083524} (\bibinfo {year} {2017})},\ \Eprint {http://arxiv.org/abs/1707.04206} {arXiv:1707.04206 [astro-ph.CO]} \BibitemShut {NoStop}%
\bibitem [{\citenamefont {Ali-Ha{\"\i}moud}\ and\ \citenamefont {Kamionkowski}(2017)}]{AliKam17}%
  \BibitemOpen
  \bibfield  {author} {\bibinfo {author} {\bibfnamefont {Y.}~\bibnamefont {Ali-Ha{\"\i}moud}}\ and\ \bibinfo {author} {\bibfnamefont {M.}~\bibnamefont {Kamionkowski}},\ }\href {\doibase 10.1103/PhysRevD.95.043534} {\bibfield  {journal} {\bibinfo  {journal} {Phys. Rev. D}\ }\textbf {\bibinfo {volume} {95}},\ \bibinfo {pages} {043534} (\bibinfo {year} {2017})},\ \Eprint {http://arxiv.org/abs/1612.05644} {arXiv:1612.05644 [astro-ph.CO]} \BibitemShut {NoStop}%
\bibitem [{\citenamefont {Serpico}(2025)}]{Ser24}%
  \BibitemOpen
  \bibfield  {author} {\bibinfo {author} {\bibfnamefont {P.~D.}\ \bibnamefont {Serpico}},\ }\enquote {\bibinfo {title} {{Cosmic Microwave Background and~Accretion}},}\ in\ \href {\doibase 10.1007/978-981-97-8887-3_24} {\emph {\bibinfo {booktitle} {{Primordial Black Holes}}}},\ \bibinfo {editor} {edited by\ \bibinfo {editor} {\bibfnamefont {C.}~\bibnamefont {Byrnes}}, \bibinfo {editor} {\bibfnamefont {G.}~\bibnamefont {Franciolini}}, \bibinfo {editor} {\bibfnamefont {T.}~\bibnamefont {Harada}}, \bibinfo {editor} {\bibfnamefont {P.}~\bibnamefont {Pani}}, \ and\ \bibinfo {editor} {\bibfnamefont {M.}~\bibnamefont {Sasaki}}}\ (\bibinfo {year} {2025})\ pp.\ \bibinfo {pages} {595--616},\ \Eprint {http://arxiv.org/abs/2406.12489} {arXiv:2406.12489 [astro-ph.CO]} \BibitemShut {NoStop}%
\bibitem [{\citenamefont {Casanueva-Villarreal}\ \emph {et~al.}(2024)\citenamefont {Casanueva-Villarreal}, \citenamefont {Tissera}, \citenamefont {Padilla}, \citenamefont {Liu}, \citenamefont {Bromm}, \citenamefont {Pedrosa}, \citenamefont {Bignone},\ and\ \citenamefont {Dominguez-Tenreiro}}]{Casetal24}%
  \BibitemOpen
  \bibfield  {author} {\bibinfo {author} {\bibfnamefont {C.}~\bibnamefont {Casanueva-Villarreal}}, \bibinfo {author} {\bibfnamefont {P.~B.}\ \bibnamefont {Tissera}}, \bibinfo {author} {\bibfnamefont {N.}~\bibnamefont {Padilla}}, \bibinfo {author} {\bibfnamefont {B.}~\bibnamefont {Liu}}, \bibinfo {author} {\bibfnamefont {V.}~\bibnamefont {Bromm}}, \bibinfo {author} {\bibfnamefont {S.}~\bibnamefont {Pedrosa}}, \bibinfo {author} {\bibfnamefont {L.}~\bibnamefont {Bignone}}, \ and\ \bibinfo {author} {\bibfnamefont {R.}~\bibnamefont {Dominguez-Tenreiro}},\ }\href {\doibase 10.1051/0004-6361/202449650} {\bibfield  {journal} {\bibinfo  {journal} {Astron. Astrophys.}\ }\textbf {\bibinfo {volume} {688}},\ \bibinfo {pages} {A183} (\bibinfo {year} {2024})},\ \Eprint {http://arxiv.org/abs/2405.02206} {arXiv:2405.02206 [astro-ph.GA]} \BibitemShut {NoStop}%
\bibitem [{\citenamefont {Takhistov}\ \emph {et~al.}(2022)\citenamefont {Takhistov}, \citenamefont {Lu}, \citenamefont {Gelmini}, \citenamefont {Hayashi}, \citenamefont {Inoue},\ and\ \citenamefont {Kusenko}}]{Taketal22}%
  \BibitemOpen
  \bibfield  {author} {\bibinfo {author} {\bibfnamefont {V.}~\bibnamefont {Takhistov}}, \bibinfo {author} {\bibfnamefont {P.}~\bibnamefont {Lu}}, \bibinfo {author} {\bibfnamefont {G.~B.}\ \bibnamefont {Gelmini}}, \bibinfo {author} {\bibfnamefont {K.}~\bibnamefont {Hayashi}}, \bibinfo {author} {\bibfnamefont {Y.}~\bibnamefont {Inoue}}, \ and\ \bibinfo {author} {\bibfnamefont {A.}~\bibnamefont {Kusenko}},\ }\href {\doibase 10.1088/1475-7516/2022/03/017} {\bibfield  {journal} {\bibinfo  {journal} {J. Cosmol. Astropart. Phys.}\ }\textbf {\bibinfo {volume} {03}},\ \bibinfo {pages} {017} (\bibinfo {year} {2022})},\ \Eprint {http://arxiv.org/abs/2105.06099} {arXiv:2105.06099 [astro-ph.GA]} \BibitemShut {NoStop}%
\bibitem [{\citenamefont {{Gaggero}}\ \emph {et~al.}(2017)\citenamefont {{Gaggero}}, \citenamefont {{Bertone}}, \citenamefont {{Calore}}, \citenamefont {{Connors}}, \citenamefont {{Lovell}}, \citenamefont {{Markoff}},\ and\ \citenamefont {{Storm}}}]{2017PhRvL.118x1101G}%
  \BibitemOpen
  \bibfield  {author} {\bibinfo {author} {\bibfnamefont {D.}~\bibnamefont {{Gaggero}}}, \bibinfo {author} {\bibfnamefont {G.}~\bibnamefont {{Bertone}}}, \bibinfo {author} {\bibfnamefont {F.}~\bibnamefont {{Calore}}}, \bibinfo {author} {\bibfnamefont {R.~M.~T.}\ \bibnamefont {{Connors}}}, \bibinfo {author} {\bibfnamefont {M.}~\bibnamefont {{Lovell}}}, \bibinfo {author} {\bibfnamefont {S.}~\bibnamefont {{Markoff}}}, \ and\ \bibinfo {author} {\bibfnamefont {E.}~\bibnamefont {{Storm}}},\ }\href {\doibase 10.1103/PhysRevLett.118.241101} {\bibfield  {journal} {\bibinfo  {journal} {Phys. Rev. Lett.}\ }\textbf {\bibinfo {volume} {118}},\ \bibinfo {eid} {241101} (\bibinfo {year} {2017})},\ \Eprint {http://arxiv.org/abs/1612.00457} {arXiv:1612.00457 [astro-ph.HE]} \BibitemShut {NoStop}%
\bibitem [{\citenamefont {Giddings}\ and\ \citenamefont {Mangano}(2008)}]{GidMan08}%
  \BibitemOpen
  \bibfield  {author} {\bibinfo {author} {\bibfnamefont {S.~B.}\ \bibnamefont {Giddings}}\ and\ \bibinfo {author} {\bibfnamefont {M.~L.}\ \bibnamefont {Mangano}},\ }\href {\doibase 10.1103/PhysRevD.78.035009} {\bibfield  {journal} {\bibinfo  {journal} {Phys. Rev. D}\ }\textbf {\bibinfo {volume} {78}},\ \bibinfo {pages} {035009} (\bibinfo {year} {2008})},\ \Eprint {http://arxiv.org/abs/0806.3381} {arXiv:0806.3381 [hep-ph]} \BibitemShut {NoStop}%
\bibitem [{\citenamefont {Montero-Camacho}\ \emph {et~al.}(2019)\citenamefont {Montero-Camacho}, \citenamefont {Fang}, \citenamefont {Vasquez}, \citenamefont {Silva},\ and\ \citenamefont {Hirata}}]{Monetal19}%
  \BibitemOpen
  \bibfield  {author} {\bibinfo {author} {\bibfnamefont {P.}~\bibnamefont {Montero-Camacho}}, \bibinfo {author} {\bibfnamefont {X.}~\bibnamefont {Fang}}, \bibinfo {author} {\bibfnamefont {G.}~\bibnamefont {Vasquez}}, \bibinfo {author} {\bibfnamefont {M.}~\bibnamefont {Silva}}, \ and\ \bibinfo {author} {\bibfnamefont {C.~M.}\ \bibnamefont {Hirata}},\ }\href {\doibase 10.1088/1475-7516/2019/08/031} {\bibfield  {journal} {\bibinfo  {journal} {J. Cosmol. Astropart. Phys.}\ }\textbf {\bibinfo {volume} {08}},\ \bibinfo {pages} {031} (\bibinfo {year} {2019})},\ \Eprint {http://arxiv.org/abs/1906.05950} {arXiv:1906.05950 [astro-ph.CO]} \BibitemShut {NoStop}%
\bibitem [{\citenamefont {Frolov}(1976)}]{Fro76}%
  \BibitemOpen
  \bibfield  {author} {\bibinfo {author} {\bibfnamefont {V.~P.}\ \bibnamefont {Frolov}},\ }\href {\doibase 10.1070/PU1976v019n03ABEH005141} {\bibfield  {journal} {\bibinfo  {journal} {Phys. Usp.}\ }\textbf {\bibinfo {volume} {19}},\ \bibinfo {pages} {244} (\bibinfo {year} {1976})}\BibitemShut {NoStop}%
\bibitem [{\citenamefont {Bugaev}\ and\ \citenamefont {Konishchev}(2002)}]{BugKon02}%
  \BibitemOpen
  \bibfield  {author} {\bibinfo {author} {\bibfnamefont {E.~V.}\ \bibnamefont {Bugaev}}\ and\ \bibinfo {author} {\bibfnamefont {K.~V.}\ \bibnamefont {Konishchev}},\ }\href {\doibase 10.1103/PhysRevD.65.123005} {\bibfield  {journal} {\bibinfo  {journal} {Phys. Rev. D}\ }\textbf {\bibinfo {volume} {65}},\ \bibinfo {pages} {123005} (\bibinfo {year} {2002})},\ \Eprint {http://arxiv.org/abs/astro-ph/0005295} {arXiv:astro-ph/0005295} \BibitemShut {NoStop}%
\bibitem [{\citenamefont {Lu}\ \emph {et~al.}(2024)\citenamefont {Lu}, \citenamefont {Picker},\ and\ \citenamefont {Kusenko}}]{ZacPicKus24}%
  \BibitemOpen
  \bibfield  {author} {\bibinfo {author} {\bibfnamefont {Y.}~\bibnamefont {Lu}}, \bibinfo {author} {\bibfnamefont {Z.~S.~C.}\ \bibnamefont {Picker}}, \ and\ \bibinfo {author} {\bibfnamefont {A.}~\bibnamefont {Kusenko}},\ }\href {\doibase 10.1103/PhysRevD.109.123016} {\bibfield  {journal} {\bibinfo  {journal} {Phys. Rev. D}\ }\textbf {\bibinfo {volume} {109}},\ \bibinfo {pages} {123016} (\bibinfo {year} {2024})},\ \Eprint {http://arxiv.org/abs/2312.15062} {arXiv:2312.15062 [astro-ph.GA]} \BibitemShut {NoStop}%
\bibitem [{\citenamefont {{Zel'dovich}}\ \emph {et~al.}(1977)\citenamefont {{Zel'dovich}}, \citenamefont {{Starobinskii}}, \citenamefont {{Khlopov}},\ and\ \citenamefont {{Chechetkin}}}]{Zeletal77}%
  \BibitemOpen
  \bibfield  {author} {\bibinfo {author} {\bibfnamefont {Y.~B.}\ \bibnamefont {{Zel'dovich}}}, \bibinfo {author} {\bibfnamefont {A.~A.}\ \bibnamefont {{Starobinskii}}}, \bibinfo {author} {\bibfnamefont {M.~Y.}\ \bibnamefont {{Khlopov}}}, \ and\ \bibinfo {author} {\bibfnamefont {V.~M.}\ \bibnamefont {{Chechetkin}}},\ }\href@noop {} {\bibfield  {journal} {\bibinfo  {journal} {Soviet Astronomy Letters}\ }\textbf {\bibinfo {volume} {3}},\ \bibinfo {pages} {110} (\bibinfo {year} {1977})}\BibitemShut {NoStop}%
\bibitem [{\citenamefont {Carr}\ \emph {et~al.}(2010{\natexlab{b}})\citenamefont {Carr}, \citenamefont {Kohri}, \citenamefont {Sendouda},\ and\ \citenamefont {Yokoyama}}]{Caretal10}%
  \BibitemOpen
  \bibfield  {author} {\bibinfo {author} {\bibfnamefont {B.~J.}\ \bibnamefont {Carr}}, \bibinfo {author} {\bibfnamefont {K.}~\bibnamefont {Kohri}}, \bibinfo {author} {\bibfnamefont {Y.}~\bibnamefont {Sendouda}}, \ and\ \bibinfo {author} {\bibfnamefont {J.}~\bibnamefont {Yokoyama}},\ }\href {\doibase 10.1103/PhysRevD.81.104019} {\bibfield  {journal} {\bibinfo  {journal} {Phys. Rev. D}\ }\textbf {\bibinfo {volume} {81}},\ \bibinfo {pages} {104019} (\bibinfo {year} {2010}{\natexlab{b}})},\ \Eprint {http://arxiv.org/abs/0912.5297} {arXiv:0912.5297 [astro-ph.CO]} \BibitemShut {NoStop}%
\bibitem [{\citenamefont {Zel'dovich}\ and\ \citenamefont {Starobinsky}(1976)}]{ZelSta76}%
  \BibitemOpen
  \bibfield  {author} {\bibinfo {author} {\bibfnamefont {Y.~B.}\ \bibnamefont {Zel'dovich}}\ and\ \bibinfo {author} {\bibfnamefont {A.~A.}\ \bibnamefont {Starobinsky}},\ }\href@noop {} {\bibfield  {journal} {\bibinfo  {journal} {JETP Letters}\ }\textbf {\bibinfo {volume} {24}},\ \bibinfo {pages} {571} (\bibinfo {year} {1976})}\BibitemShut {NoStop}%
\bibitem [{\citenamefont {Khlopov}\ \emph {et~al.}(2006)\citenamefont {Khlopov}, \citenamefont {Barrau},\ and\ \citenamefont {Grain}}]{KhlBarGra06}%
  \BibitemOpen
  \bibfield  {author} {\bibinfo {author} {\bibfnamefont {M.~Y.}\ \bibnamefont {Khlopov}}, \bibinfo {author} {\bibfnamefont {A.}~\bibnamefont {Barrau}}, \ and\ \bibinfo {author} {\bibfnamefont {J.}~\bibnamefont {Grain}},\ }\href {\doibase 10.1088/0264-9381/23/6/004} {\bibfield  {journal} {\bibinfo  {journal} {Class. Quant. Grav.}\ }\textbf {\bibinfo {volume} {23}},\ \bibinfo {pages} {1875} (\bibinfo {year} {2006})},\ \Eprint {http://arxiv.org/abs/astro-ph/0406621} {arXiv:astro-ph/0406621} \BibitemShut {NoStop}%
\bibitem [{\citenamefont {Chattopadhyay}\ \emph {et~al.}(2025)\citenamefont {Chattopadhyay}, \citenamefont {Chaudhuri},\ and\ \citenamefont {Khlopov}}]{ChaChaKhl25}%
  \BibitemOpen
  \bibfield  {author} {\bibinfo {author} {\bibfnamefont {P.}~\bibnamefont {Chattopadhyay}}, \bibinfo {author} {\bibfnamefont {A.}~\bibnamefont {Chaudhuri}}, \ and\ \bibinfo {author} {\bibfnamefont {M.~Y.}\ \bibnamefont {Khlopov}},\ }\href {\doibase 10.1142/S0219887824502517} {\bibfield  {journal} {\bibinfo  {journal} {Int. J. Geom. Meth. Mod. Phys.}\ }\textbf {\bibinfo {volume} {22}},\ \bibinfo {pages} {2450251} (\bibinfo {year} {2025})},\ \Eprint {http://arxiv.org/abs/2209.11288} {arXiv:2209.11288 [hep-ph]} \BibitemShut {NoStop}%
\bibitem [{\citenamefont {Hawking}(1975)}]{Haw75}%
  \BibitemOpen
  \bibfield  {author} {\bibinfo {author} {\bibfnamefont {S.~W.}\ \bibnamefont {Hawking}},\ }\href {\doibase 10.1007/BF02345020} {\bibfield  {journal} {\bibinfo  {journal} {Commun. Math. Phys.}\ }\textbf {\bibinfo {volume} {43}},\ \bibinfo {pages} {199} (\bibinfo {year} {1975})},\ \bibinfo {note} {[Erratum: Commun.Math.Phys. 46, 206 (1976)]}\BibitemShut {NoStop}%
\bibitem [{\citenamefont {Zel'dovich}(1976)}]{Zel76}%
  \BibitemOpen
  \bibfield  {author} {\bibinfo {author} {\bibfnamefont {Y.~B.}\ \bibnamefont {Zel'dovich}},\ }\href@noop {} {\bibfield  {journal} {\bibinfo  {journal} {JETP Letters}\ }\textbf {\bibinfo {volume} {24}},\ \bibinfo {pages} {25} (\bibinfo {year} {1976})}\BibitemShut {NoStop}%
\bibitem [{\citenamefont {{Turner}}\ and\ \citenamefont {{Schramm}}(1979)}]{TurSch79}%
  \BibitemOpen
  \bibfield  {author} {\bibinfo {author} {\bibfnamefont {M.~S.}\ \bibnamefont {{Turner}}}\ and\ \bibinfo {author} {\bibfnamefont {D.~N.}\ \bibnamefont {{Schramm}}},\ }\href {\doibase 10.1038/279303a0} {\bibfield  {journal} {\bibinfo  {journal} {Nature}\ }\textbf {\bibinfo {volume} {279}},\ \bibinfo {pages} {303} (\bibinfo {year} {1979})}\BibitemShut {NoStop}%
\bibitem [{\citenamefont {Dolgov}(1981)}]{Dol81}%
  \BibitemOpen
  \bibfield  {author} {\bibinfo {author} {\bibfnamefont {A.~D.}\ \bibnamefont {Dolgov}},\ }\href {\doibase 10.1103/PhysRevD.24.1042} {\bibfield  {journal} {\bibinfo  {journal} {Phys. Rev. D}\ }\textbf {\bibinfo {volume} {24}},\ \bibinfo {pages} {1042} (\bibinfo {year} {1981})}\BibitemShut {NoStop}%
\bibitem [{\citenamefont {Dolgov}(1992)}]{Dol92}%
  \BibitemOpen
  \bibfield  {author} {\bibinfo {author} {\bibfnamefont {A.~D.}\ \bibnamefont {Dolgov}},\ }\href {\doibase 10.1016/0370-1573(92)90107-B} {\bibfield  {journal} {\bibinfo  {journal} {Phys. Rep.}\ }\textbf {\bibinfo {volume} {222}},\ \bibinfo {pages} {309} (\bibinfo {year} {1992})}\BibitemShut {NoStop}%
\bibitem [{\citenamefont {Nagatani}(1999)}]{Nag99}%
  \BibitemOpen
  \bibfield  {author} {\bibinfo {author} {\bibfnamefont {Y.}~\bibnamefont {Nagatani}},\ }\href {\doibase 10.1103/PhysRevD.59.041301} {\bibfield  {journal} {\bibinfo  {journal} {Phys. Rev. D}\ }\textbf {\bibinfo {volume} {59}},\ \bibinfo {pages} {041301} (\bibinfo {year} {1999})},\ \Eprint {http://arxiv.org/abs/hep-ph/9811485} {arXiv:hep-ph/9811485} \BibitemShut {NoStop}%
\bibitem [{\citenamefont {Fujita}\ \emph {et~al.}(2014)\citenamefont {Fujita}, \citenamefont {Kawasaki}, \citenamefont {Harigaya},\ and\ \citenamefont {Matsuda}}]{Fujetal14}%
  \BibitemOpen
  \bibfield  {author} {\bibinfo {author} {\bibfnamefont {T.}~\bibnamefont {Fujita}}, \bibinfo {author} {\bibfnamefont {M.}~\bibnamefont {Kawasaki}}, \bibinfo {author} {\bibfnamefont {K.}~\bibnamefont {Harigaya}}, \ and\ \bibinfo {author} {\bibfnamefont {R.}~\bibnamefont {Matsuda}},\ }\href {\doibase 10.1103/PhysRevD.89.103501} {\bibfield  {journal} {\bibinfo  {journal} {Phys. Rev. D}\ }\textbf {\bibinfo {volume} {89}},\ \bibinfo {pages} {103501} (\bibinfo {year} {2014})},\ \Eprint {http://arxiv.org/abs/1401.1909} {arXiv:1401.1909 [astro-ph.CO]} \BibitemShut {NoStop}%
\bibitem [{\citenamefont {{Wu}}\ \emph {et~al.}(2022)\citenamefont {{Wu}}, \citenamefont {{Pinetti}},\ and\ \citenamefont {{Silk}}}]{2022PhRvL.128c1102W}%
  \BibitemOpen
  \bibfield  {author} {\bibinfo {author} {\bibfnamefont {Y.-P.}\ \bibnamefont {{Wu}}}, \bibinfo {author} {\bibfnamefont {E.}~\bibnamefont {{Pinetti}}}, \ and\ \bibinfo {author} {\bibfnamefont {J.}~\bibnamefont {{Silk}}},\ }\href {\doibase 10.1103/PhysRevLett.128.031102} {\bibfield  {journal} {\bibinfo  {journal} {Phys. Rev. Lett.}\ }\textbf {\bibinfo {volume} {128}},\ \bibinfo {eid} {031102} (\bibinfo {year} {2022})},\ \Eprint {http://arxiv.org/abs/2109.09875} {arXiv:2109.09875 [astro-ph.CO]} \BibitemShut {NoStop}%
\bibitem [{\citenamefont {Guedens}\ \emph {et~al.}(2002)\citenamefont {Guedens}, \citenamefont {Clancy},\ and\ \citenamefont {Liddle}}]{GueClaLid02}%
  \BibitemOpen
  \bibfield  {author} {\bibinfo {author} {\bibfnamefont {R.}~\bibnamefont {Guedens}}, \bibinfo {author} {\bibfnamefont {D.}~\bibnamefont {Clancy}}, \ and\ \bibinfo {author} {\bibfnamefont {A.~R.}\ \bibnamefont {Liddle}},\ }\href {\doibase 10.1103/PhysRevD.66.043513} {\bibfield  {journal} {\bibinfo  {journal} {Phys. Rev. D}\ }\textbf {\bibinfo {volume} {66}},\ \bibinfo {pages} {043513} (\bibinfo {year} {2002})},\ \Eprint {http://arxiv.org/abs/astro-ph/0205149} {arXiv:astro-ph/0205149} \BibitemShut {NoStop}%
\bibitem [{\citenamefont {Chapline}(1975)}]{Cha75}%
  \BibitemOpen
  \bibfield  {author} {\bibinfo {author} {\bibfnamefont {G.~F.}\ \bibnamefont {Chapline}},\ }\href {\doibase 10.1038/253251a0} {\bibfield  {journal} {\bibinfo  {journal} {Nature}\ }\textbf {\bibinfo {volume} {253}},\ \bibinfo {pages} {251} (\bibinfo {year} {1975})}\BibitemShut {NoStop}%
\bibitem [{\citenamefont {{Carr}}\ and\ \citenamefont {{K{\"u}hnel}}(2020)}]{2020ARNPS..70..355C}%
  \BibitemOpen
  \bibfield  {author} {\bibinfo {author} {\bibfnamefont {B.}~\bibnamefont {{Carr}}}\ and\ \bibinfo {author} {\bibfnamefont {F.}~\bibnamefont {{K{\"u}hnel}}},\ }\href {\doibase 10.1146/annurev-nucl-050520-125911} {\bibfield  {journal} {\bibinfo  {journal} {Annual Review of Nuclear and Particle Science}\ }\textbf {\bibinfo {volume} {70}},\ \bibinfo {pages} {355} (\bibinfo {year} {2020})},\ \Eprint {http://arxiv.org/abs/2006.02838} {arXiv:2006.02838 [astro-ph.CO]} \BibitemShut {NoStop}%
\bibitem [{\citenamefont {Green}\ and\ \citenamefont {Kavanagh}(2021)}]{2021JPhG...48d3001G}%
  \BibitemOpen
  \bibfield  {author} {\bibinfo {author} {\bibfnamefont {A.~M.}\ \bibnamefont {Green}}\ and\ \bibinfo {author} {\bibfnamefont {B.~J.}\ \bibnamefont {Kavanagh}},\ }\href {\doibase 10.1088/1361-6471/abc534} {\bibfield  {journal} {\bibinfo  {journal} {J. Phys. G}\ }\textbf {\bibinfo {volume} {48}},\ \bibinfo {pages} {043001} (\bibinfo {year} {2021})},\ \Eprint {http://arxiv.org/abs/2007.10722} {arXiv:2007.10722 [astro-ph.CO]} \BibitemShut {NoStop}%
\bibitem [{\citenamefont {{Graham}}\ \emph {et~al.}(2015)\citenamefont {{Graham}}, \citenamefont {{Rajendran}},\ and\ \citenamefont {{Varela}}}]{2015PhRvD..92f3007G}%
  \BibitemOpen
  \bibfield  {author} {\bibinfo {author} {\bibfnamefont {P.~W.}\ \bibnamefont {{Graham}}}, \bibinfo {author} {\bibfnamefont {S.}~\bibnamefont {{Rajendran}}}, \ and\ \bibinfo {author} {\bibfnamefont {J.}~\bibnamefont {{Varela}}},\ }\href {\doibase 10.1103/PhysRevD.92.063007} {\bibfield  {journal} {\bibinfo  {journal} {Phys. Rev. D}\ }\textbf {\bibinfo {volume} {92}},\ \bibinfo {eid} {063007} (\bibinfo {year} {2015})},\ \Eprint {http://arxiv.org/abs/1505.04444} {arXiv:1505.04444 [hep-ph]} \BibitemShut {NoStop}%
\bibitem [{\citenamefont {Capela}\ \emph {et~al.}(2013{\natexlab{a}})\citenamefont {Capela}, \citenamefont {Pshirkov},\ and\ \citenamefont {Tinyakov}}]{CapPshTin13-2}%
  \BibitemOpen
  \bibfield  {author} {\bibinfo {author} {\bibfnamefont {F.}~\bibnamefont {Capela}}, \bibinfo {author} {\bibfnamefont {M.}~\bibnamefont {Pshirkov}}, \ and\ \bibinfo {author} {\bibfnamefont {P.}~\bibnamefont {Tinyakov}},\ }\href {\doibase 10.1103/PhysRevD.87.123524} {\bibfield  {journal} {\bibinfo  {journal} {Phys. Rev. D}\ }\textbf {\bibinfo {volume} {87}},\ \bibinfo {pages} {123524} (\bibinfo {year} {2013}{\natexlab{a}})},\ \Eprint {http://arxiv.org/abs/1301.4984} {arXiv:1301.4984 [astro-ph.CO]} \BibitemShut {NoStop}%
\bibitem [{\citenamefont {{Kouvaris}}\ and\ \citenamefont {{Tinyakov}}(2014)}]{2014PhRvD..90d3512K}%
  \BibitemOpen
  \bibfield  {author} {\bibinfo {author} {\bibfnamefont {C.}~\bibnamefont {{Kouvaris}}}\ and\ \bibinfo {author} {\bibfnamefont {P.}~\bibnamefont {{Tinyakov}}},\ }\href {\doibase 10.1103/PhysRevD.90.043512} {\bibfield  {journal} {\bibinfo  {journal} {Phys. Rev. D}\ }\textbf {\bibinfo {volume} {90}},\ \bibinfo {eid} {043512} (\bibinfo {year} {2014})},\ \Eprint {http://arxiv.org/abs/1312.3764} {arXiv:1312.3764 [astro-ph.SR]} \BibitemShut {NoStop}%
\bibitem [{\citenamefont {Garcia-Bellido}\ and\ \citenamefont {Hawkins}(2024)}]{GarHaw24}%
  \BibitemOpen
  \bibfield  {author} {\bibinfo {author} {\bibfnamefont {J.}~\bibnamefont {Garcia-Bellido}}\ and\ \bibinfo {author} {\bibfnamefont {M.}~\bibnamefont {Hawkins}},\ }\href {\doibase 10.3390/universe10120449} {\bibfield  {journal} {\bibinfo  {journal} {Universe}\ }\textbf {\bibinfo {volume} {10}},\ \bibinfo {pages} {449} (\bibinfo {year} {2024})},\ \Eprint {http://arxiv.org/abs/2402.00212} {arXiv:2402.00212 [astro-ph.GA]} \BibitemShut {NoStop}%
\bibitem [{\citenamefont {Hawkins}\ and\ \citenamefont {Garc{\'\i}a-Bellido}(2025)}]{HawGar25}%
  \BibitemOpen
  \bibfield  {author} {\bibinfo {author} {\bibfnamefont {M.~R.~S.}\ \bibnamefont {Hawkins}}\ and\ \bibinfo {author} {\bibfnamefont {J.}~\bibnamefont {Garc{\'\i}a-Bellido}},\ }\href {\doibase 10.1093/mnras/staf1826} {\bibfield  {journal} {\bibinfo  {journal} {Mon. Not. Roy. Astron. Soc.}\ }\textbf {\bibinfo {volume} {544}},\ \bibinfo {pages} {1950} (\bibinfo {year} {2025})},\ \Eprint {http://arxiv.org/abs/2509.05400} {arXiv:2509.05400 [astro-ph.GA]} \BibitemShut {NoStop}%
\bibitem [{\citenamefont {Mr{\'o}z}\ \emph {et~al.}(2025)\citenamefont {Mr{\'o}z} \emph {et~al.}}]{Mroetal25}%
  \BibitemOpen
  \bibfield  {author} {\bibinfo {author} {\bibfnamefont {P.}~\bibnamefont {Mr{\'o}z}} \emph {et~al.},\ }\href {\doibase 10.32023/0001-5237/75.3.1} {\bibfield  {journal} {\bibinfo  {journal} {Acta Astron.}\ }\textbf {\bibinfo {volume} {75}},\ \bibinfo {pages} {173} (\bibinfo {year} {2025})},\ \Eprint {http://arxiv.org/abs/2510.20005} {arXiv:2510.20005 [astro-ph.CO]} \BibitemShut {NoStop}%
\bibitem [{\citenamefont {Gurevich}\ \emph {et~al.}(1997)\citenamefont {Gurevich}, \citenamefont {Zybin},\ and\ \citenamefont {Sirota}}]{GurZybSir97}%
  \BibitemOpen
  \bibfield  {author} {\bibinfo {author} {\bibfnamefont {A.~V.}\ \bibnamefont {Gurevich}}, \bibinfo {author} {\bibfnamefont {K.~P.}\ \bibnamefont {Zybin}}, \ and\ \bibinfo {author} {\bibfnamefont {V.~A.}\ \bibnamefont {Sirota}},\ }\href {\doibase 10.1070/PU1997v040n09ABEH000281} {\bibfield  {journal} {\bibinfo  {journal} {Phys. Usp.}\ }\textbf {\bibinfo {volume} {40}},\ \bibinfo {pages} {869} (\bibinfo {year} {1997})}\BibitemShut {NoStop}%
\bibitem [{\citenamefont {Yokoyama}(1998{\natexlab{b}})}]{Yok98}%
  \BibitemOpen
  \bibfield  {author} {\bibinfo {author} {\bibfnamefont {J.}~\bibnamefont {Yokoyama}},\ }\href {\doibase 10.1103/PhysRevD.58.107502} {\bibfield  {journal} {\bibinfo  {journal} {Phys. Rev. D}\ }\textbf {\bibinfo {volume} {58}},\ \bibinfo {pages} {107502} (\bibinfo {year} {1998}{\natexlab{b}})},\ \Eprint {http://arxiv.org/abs/gr-qc/9804041} {arXiv:gr-qc/9804041} \BibitemShut {NoStop}%
\bibitem [{\citenamefont {Carr}\ \emph {et~al.}(2017)\citenamefont {Carr}, \citenamefont {Raidal}, \citenamefont {Tenkanen}, \citenamefont {Vaskonen},\ and\ \citenamefont {Veerm{\"a}e}}]{Caretal17}%
  \BibitemOpen
  \bibfield  {author} {\bibinfo {author} {\bibfnamefont {B.}~\bibnamefont {Carr}}, \bibinfo {author} {\bibfnamefont {M.}~\bibnamefont {Raidal}}, \bibinfo {author} {\bibfnamefont {T.}~\bibnamefont {Tenkanen}}, \bibinfo {author} {\bibfnamefont {V.}~\bibnamefont {Vaskonen}}, \ and\ \bibinfo {author} {\bibfnamefont {H.}~\bibnamefont {Veerm{\"a}e}},\ }\href {\doibase 10.1103/PhysRevD.96.023514} {\bibfield  {journal} {\bibinfo  {journal} {Phys. Rev. D}\ }\textbf {\bibinfo {volume} {96}},\ \bibinfo {pages} {023514} (\bibinfo {year} {2017})},\ \Eprint {http://arxiv.org/abs/1705.05567} {arXiv:1705.05567 [astro-ph.CO]} \BibitemShut {NoStop}%
\bibitem [{\citenamefont {{Niikura}}\ \emph {et~al.}(2019)\citenamefont {{Niikura}}, \citenamefont {{Takada}}, \citenamefont {{Yasuda}}, \citenamefont {{Lupton}}, \citenamefont {{Sumi}}, \citenamefont {{More}}, \citenamefont {{Kurita}}, \citenamefont {{Sugiyama}}, \citenamefont {{More}}, \citenamefont {{Oguri}},\ and\ \citenamefont {{Chiba}}}]{2019NatAs...3..524N}%
  \BibitemOpen
  \bibfield  {author} {\bibinfo {author} {\bibfnamefont {H.}~\bibnamefont {{Niikura}}}, \bibinfo {author} {\bibfnamefont {M.}~\bibnamefont {{Takada}}}, \bibinfo {author} {\bibfnamefont {N.}~\bibnamefont {{Yasuda}}}, \bibinfo {author} {\bibfnamefont {R.~H.}\ \bibnamefont {{Lupton}}}, \bibinfo {author} {\bibfnamefont {T.}~\bibnamefont {{Sumi}}}, \bibinfo {author} {\bibfnamefont {S.}~\bibnamefont {{More}}}, \bibinfo {author} {\bibfnamefont {T.}~\bibnamefont {{Kurita}}}, \bibinfo {author} {\bibfnamefont {S.}~\bibnamefont {{Sugiyama}}}, \bibinfo {author} {\bibfnamefont {A.}~\bibnamefont {{More}}}, \bibinfo {author} {\bibfnamefont {M.}~\bibnamefont {{Oguri}}}, \ and\ \bibinfo {author} {\bibfnamefont {M.}~\bibnamefont {{Chiba}}},\ }\href {\doibase 10.1038/s41550-019-0723-1} {\bibfield  {journal} {\bibinfo  {journal} {Nature Astronomy}\ }\textbf {\bibinfo {volume} {3}},\ \bibinfo {pages} {524} (\bibinfo {year} {2019})},\ \Eprint {http://arxiv.org/abs/1701.02151} {arXiv:1701.02151 [astro-ph.CO]} \BibitemShut {NoStop}%
\bibitem [{\citenamefont {{Mu{\~n}oz}}\ \emph {et~al.}(2016)\citenamefont {{Mu{\~n}oz}}, \citenamefont {{Kovetz}}, \citenamefont {{Dai}},\ and\ \citenamefont {{Kamionkowski}}}]{2016PhRvL.117i1301M}%
  \BibitemOpen
  \bibfield  {author} {\bibinfo {author} {\bibfnamefont {J.~B.}\ \bibnamefont {{Mu{\~n}oz}}}, \bibinfo {author} {\bibfnamefont {E.~D.}\ \bibnamefont {{Kovetz}}}, \bibinfo {author} {\bibfnamefont {L.}~\bibnamefont {{Dai}}}, \ and\ \bibinfo {author} {\bibfnamefont {M.}~\bibnamefont {{Kamionkowski}}},\ }\href {\doibase 10.1103/PhysRevLett.117.091301} {\bibfield  {journal} {\bibinfo  {journal} {Phys. Rev. Lett.}\ }\textbf {\bibinfo {volume} {117}},\ \bibinfo {eid} {091301} (\bibinfo {year} {2016})},\ \Eprint {http://arxiv.org/abs/1605.00008} {arXiv:1605.00008 [astro-ph.CO]} \BibitemShut {NoStop}%
\bibitem [{\citenamefont {Popov}\ \emph {et~al.}(2018)\citenamefont {Popov}, \citenamefont {Postnov},\ and\ \citenamefont {Pshirkov}}]{PopPosPsh18}%
  \BibitemOpen
  \bibfield  {author} {\bibinfo {author} {\bibfnamefont {S.~B.}\ \bibnamefont {Popov}}, \bibinfo {author} {\bibfnamefont {K.~A.}\ \bibnamefont {Postnov}}, \ and\ \bibinfo {author} {\bibfnamefont {M.~S.}\ \bibnamefont {Pshirkov}},\ }\href {\doibase 10.3367/UFNe.2018.03.038313} {\bibfield  {journal} {\bibinfo  {journal} {Phys. Usp.}\ }\textbf {\bibinfo {volume} {61}},\ \bibinfo {pages} {965} (\bibinfo {year} {2018})}\BibitemShut {NoStop}%
\bibitem [{\citenamefont {{Popov}}\ and\ \citenamefont {{Postnov}}(2010)}]{2010vaoa.conf..129P}%
  \BibitemOpen
  \bibfield  {author} {\bibinfo {author} {\bibfnamefont {S.~B.}\ \bibnamefont {{Popov}}}\ and\ \bibinfo {author} {\bibfnamefont {K.~A.}\ \bibnamefont {{Postnov}}},\ }in\ \href {\doibase 10.48550/arXiv.0710.2006} {\emph {\bibinfo {booktitle} {Evolution of Cosmic Objects through their Physical Activity}}},\ \bibinfo {editor} {edited by\ \bibinfo {editor} {\bibfnamefont {H.~A.}\ \bibnamefont {{Harutyunian}}}, \bibinfo {editor} {\bibfnamefont {A.~M.}\ \bibnamefont {{Mickaelian}}}, \ and\ \bibinfo {editor} {\bibfnamefont {Y.}~\bibnamefont {{Terzian}}}}\ (\bibinfo {year} {2010})\ pp.\ \bibinfo {pages} {129--132},\ \Eprint {http://arxiv.org/abs/0710.2006} {arXiv:0710.2006 [astro-ph]} \BibitemShut {NoStop}%
\bibitem [{\citenamefont {Popov}\ and\ \citenamefont {Postnov}(2013)}]{2013arXiv1307.4924P}%
  \BibitemOpen
  \bibfield  {author} {\bibinfo {author} {\bibfnamefont {S.~B.}\ \bibnamefont {Popov}}\ and\ \bibinfo {author} {\bibfnamefont {K.~A.}\ \bibnamefont {Postnov}},\ }\href@noop {} {\  (\bibinfo {year} {2013})},\ \Eprint {http://arxiv.org/abs/1307.4924} {arXiv:1307.4924 [astro-ph.HE]} \BibitemShut {NoStop}%
\bibitem [{\citenamefont {{Gould}}(1992)}]{Gou92}%
  \BibitemOpen
  \bibfield  {author} {\bibinfo {author} {\bibfnamefont {A.}~\bibnamefont {{Gould}}},\ }\href {\doibase 10.1086/186279} {\bibfield  {journal} {\bibinfo  {journal} {Astrophys. J. Lett.}\ }\textbf {\bibinfo {volume} {386}},\ \bibinfo {pages} {L5} (\bibinfo {year} {1992})}\BibitemShut {NoStop}%
\bibitem [{\citenamefont {Gorton}\ and\ \citenamefont {Green}(2022)}]{GorGre22}%
  \BibitemOpen
  \bibfield  {author} {\bibinfo {author} {\bibfnamefont {M.}~\bibnamefont {Gorton}}\ and\ \bibinfo {author} {\bibfnamefont {A.~M.}\ \bibnamefont {Green}},\ }\href {\doibase 10.1088/1475-7516/2022/08/035} {\bibfield  {journal} {\bibinfo  {journal} {J. Cosmol. Astropart. Phys.}\ }\textbf {\bibinfo {volume} {08}},\ \bibinfo {pages} {035} (\bibinfo {year} {2022})},\ \Eprint {http://arxiv.org/abs/2203.04209} {arXiv:2203.04209 [astro-ph.CO]} \BibitemShut {NoStop}%
\bibitem [{\citenamefont {Toshchenko}\ \emph {et~al.}(2025)\citenamefont {Toshchenko}, \citenamefont {Baklanov}, \citenamefont {Belotsky},\ and\ \citenamefont {Blinnikov}}]{Tosetal26}%
  \BibitemOpen
  \bibfield  {author} {\bibinfo {author} {\bibfnamefont {K.~A.}\ \bibnamefont {Toshchenko}}, \bibinfo {author} {\bibfnamefont {P.~V.}\ \bibnamefont {Baklanov}}, \bibinfo {author} {\bibfnamefont {K.~M.}\ \bibnamefont {Belotsky}}, \ and\ \bibinfo {author} {\bibfnamefont {S.~I.}\ \bibnamefont {Blinnikov}},\ }\href {\doibase 10.1134/S1063773725700562} {\bibfield  {journal} {\bibinfo  {journal} {Astron. Lett.}\ }\textbf {\bibinfo {volume} {51}},\ \bibinfo {pages} {461} (\bibinfo {year} {2025})},\ \Eprint {http://arxiv.org/abs/2603.23412} {arXiv:2603.23412 [astro-ph.CO]} \BibitemShut {NoStop}%
\bibitem [{\citenamefont {Clesse}\ and\ \citenamefont {Garc{\'\i}a-Bellido}(2018)}]{CleGar17}%
  \BibitemOpen
  \bibfield  {author} {\bibinfo {author} {\bibfnamefont {S.}~\bibnamefont {Clesse}}\ and\ \bibinfo {author} {\bibfnamefont {J.}~\bibnamefont {Garc{\'\i}a-Bellido}},\ }\href {\doibase 10.1016/j.dark.2018.08.004} {\bibfield  {journal} {\bibinfo  {journal} {Phys. Dark Univ.}\ }\textbf {\bibinfo {volume} {22}},\ \bibinfo {pages} {137} (\bibinfo {year} {2018})},\ \Eprint {http://arxiv.org/abs/1711.10458} {arXiv:1711.10458 [astro-ph.CO]} \BibitemShut {NoStop}%
\bibitem [{\citenamefont {Capela}\ \emph {et~al.}(2013{\natexlab{b}})\citenamefont {Capela}, \citenamefont {Pshirkov},\ and\ \citenamefont {Tinyakov}}]{CapPshTin13-1}%
  \BibitemOpen
  \bibfield  {author} {\bibinfo {author} {\bibfnamefont {F.}~\bibnamefont {Capela}}, \bibinfo {author} {\bibfnamefont {M.}~\bibnamefont {Pshirkov}}, \ and\ \bibinfo {author} {\bibfnamefont {P.}~\bibnamefont {Tinyakov}},\ }\href {\doibase 10.1103/PhysRevD.87.023507} {\bibfield  {journal} {\bibinfo  {journal} {Phys. Rev. D}\ }\textbf {\bibinfo {volume} {87}},\ \bibinfo {pages} {023507} (\bibinfo {year} {2013}{\natexlab{b}})},\ \Eprint {http://arxiv.org/abs/1209.6021} {arXiv:1209.6021 [astro-ph.CO]} \BibitemShut {NoStop}%
\bibitem [{\citenamefont {Pani}\ and\ \citenamefont {Loeb}(2014)}]{PanLoe14}%
  \BibitemOpen
  \bibfield  {author} {\bibinfo {author} {\bibfnamefont {P.}~\bibnamefont {Pani}}\ and\ \bibinfo {author} {\bibfnamefont {A.}~\bibnamefont {Loeb}},\ }\href {\doibase 10.1088/1475-7516/2014/06/026} {\bibfield  {journal} {\bibinfo  {journal} {J. Cosmol. Astropart. Phys.}\ }\textbf {\bibinfo {volume} {06}},\ \bibinfo {pages} {026} (\bibinfo {year} {2014})},\ \Eprint {http://arxiv.org/abs/1401.3025} {arXiv:1401.3025 [astro-ph.CO]} \BibitemShut {NoStop}%
\bibitem [{\citenamefont {Capela}\ \emph {et~al.}(2014)\citenamefont {Capela}, \citenamefont {Pshirkov},\ and\ \citenamefont {Tinyakov}}]{CapPshTin14}%
  \BibitemOpen
  \bibfield  {author} {\bibinfo {author} {\bibfnamefont {F.}~\bibnamefont {Capela}}, \bibinfo {author} {\bibfnamefont {M.}~\bibnamefont {Pshirkov}}, \ and\ \bibinfo {author} {\bibfnamefont {P.}~\bibnamefont {Tinyakov}},\ }\href@noop {} {\  (\bibinfo {year} {2014})},\ \Eprint {http://arxiv.org/abs/1402.4671} {arXiv:1402.4671 [astro-ph.CO]} \BibitemShut {NoStop}%
\bibitem [{\citenamefont {Carr}\ \emph {et~al.}(1994)\citenamefont {Carr}, \citenamefont {Gilbert},\ and\ \citenamefont {Lidsey}}]{CarGilLid94}%
  \BibitemOpen
  \bibfield  {author} {\bibinfo {author} {\bibfnamefont {B.~J.}\ \bibnamefont {Carr}}, \bibinfo {author} {\bibfnamefont {J.~H.}\ \bibnamefont {Gilbert}}, \ and\ \bibinfo {author} {\bibfnamefont {J.~E.}\ \bibnamefont {Lidsey}},\ }\href {\doibase 10.1103/PhysRevD.50.4853} {\bibfield  {journal} {\bibinfo  {journal} {Phys. Rev. D}\ }\textbf {\bibinfo {volume} {50}},\ \bibinfo {pages} {4853} (\bibinfo {year} {1994})},\ \Eprint {http://arxiv.org/abs/astro-ph/9405027} {arXiv:astro-ph/9405027} \BibitemShut {NoStop}%
\bibitem [{\citenamefont {Chen}\ and\ \citenamefont {Adler}(2003)}]{CheAdl03}%
  \BibitemOpen
  \bibfield  {author} {\bibinfo {author} {\bibfnamefont {P.}~\bibnamefont {Chen}}\ and\ \bibinfo {author} {\bibfnamefont {R.~J.}\ \bibnamefont {Adler}},\ }\href {\doibase 10.1016/S0920-5632(03)02088-7} {\bibfield  {journal} {\bibinfo  {journal} {Nucl. Phys. B Proc. Suppl.}\ }\textbf {\bibinfo {volume} {124}},\ \bibinfo {pages} {103} (\bibinfo {year} {2003})},\ \Eprint {http://arxiv.org/abs/gr-qc/0205106} {arXiv:gr-qc/0205106} \BibitemShut {NoStop}%
\bibitem [{\citenamefont {{MacGibbon}}(1987)}]{MacGib87}%
  \BibitemOpen
  \bibfield  {author} {\bibinfo {author} {\bibfnamefont {J.~H.}\ \bibnamefont {{MacGibbon}}},\ }\href {\doibase 10.1038/329308a0} {\bibfield  {journal} {\bibinfo  {journal} {Nature}\ }\textbf {\bibinfo {volume} {329}},\ \bibinfo {pages} {308} (\bibinfo {year} {1987})}\BibitemShut {NoStop}%
\bibitem [{\citenamefont {Barrow}\ \emph {et~al.}(1992)\citenamefont {Barrow}, \citenamefont {Copeland},\ and\ \citenamefont {Liddle}}]{Barrow:1992hq}%
  \BibitemOpen
  \bibfield  {author} {\bibinfo {author} {\bibfnamefont {J.~D.}\ \bibnamefont {Barrow}}, \bibinfo {author} {\bibfnamefont {E.~J.}\ \bibnamefont {Copeland}}, \ and\ \bibinfo {author} {\bibfnamefont {A.~R.}\ \bibnamefont {Liddle}},\ }\href {\doibase 10.1103/PhysRevD.46.645} {\bibfield  {journal} {\bibinfo  {journal} {Phys. Rev. D}\ }\textbf {\bibinfo {volume} {46}},\ \bibinfo {pages} {645} (\bibinfo {year} {1992})}\BibitemShut {NoStop}%
\bibitem [{\citenamefont {Volkova}\ and\ \citenamefont {Dokuchaev}(1994)}]{DokVol}%
  \BibitemOpen
  \bibfield  {author} {\bibinfo {author} {\bibfnamefont {L.~V.}\ \bibnamefont {Volkova}}\ and\ \bibinfo {author} {\bibfnamefont {V.~I.}\ \bibnamefont {Dokuchaev}},\ }\href {http://jetpletters.ru/ps/0/article_19982.shtml} {\bibfield  {journal} {\bibinfo  {journal} {JETP Letters}\ }\textbf {\bibinfo {volume} {60}},\ \bibinfo {pages} {76} (\bibinfo {year} {1994})}\BibitemShut {NoStop}%
\bibitem [{\citenamefont {Luo}\ \emph {et~al.}(2012)\citenamefont {Luo}, \citenamefont {Hanasoge}, \citenamefont {Tromp},\ and\ \citenamefont {Pretorius}}]{Luoetal12}%
  \BibitemOpen
  \bibfield  {author} {\bibinfo {author} {\bibfnamefont {Y.}~\bibnamefont {Luo}}, \bibinfo {author} {\bibfnamefont {S.}~\bibnamefont {Hanasoge}}, \bibinfo {author} {\bibfnamefont {J.}~\bibnamefont {Tromp}}, \ and\ \bibinfo {author} {\bibfnamefont {F.}~\bibnamefont {Pretorius}},\ }\href {\doibase 10.1088/0004-637X/751/1/16} {\bibfield  {journal} {\bibinfo  {journal} {Astrophys. J.}\ }\textbf {\bibinfo {volume} {751}},\ \bibinfo {pages} {16} (\bibinfo {year} {2012})},\ \Eprint {http://arxiv.org/abs/1203.3806} {arXiv:1203.3806 [astro-ph.CO]} \BibitemShut {NoStop}%
\bibitem [{\citenamefont {{Batygin}}\ \emph {et~al.}(2019)\citenamefont {{Batygin}}, \citenamefont {{Adams}}, \citenamefont {{Brown}},\ and\ \citenamefont {{Becker}}}]{Batetal19}%
  \BibitemOpen
  \bibfield  {author} {\bibinfo {author} {\bibfnamefont {K.}~\bibnamefont {{Batygin}}}, \bibinfo {author} {\bibfnamefont {F.~C.}\ \bibnamefont {{Adams}}}, \bibinfo {author} {\bibfnamefont {M.~E.}\ \bibnamefont {{Brown}}}, \ and\ \bibinfo {author} {\bibfnamefont {J.~C.}\ \bibnamefont {{Becker}}},\ }\href {\doibase 10.1016/j.physrep.2019.01.009} {\bibfield  {journal} {\bibinfo  {journal} {Phys. Rep.}\ }\textbf {\bibinfo {volume} {805}},\ \bibinfo {pages} {1} (\bibinfo {year} {2019})},\ \Eprint {http://arxiv.org/abs/1902.10103} {arXiv:1902.10103 [astro-ph.EP]} \BibitemShut {NoStop}%
\bibitem [{\citenamefont {{Lykawka}}\ and\ \citenamefont {{Ito}}(2023)}]{LykIto23}%
  \BibitemOpen
  \bibfield  {author} {\bibinfo {author} {\bibfnamefont {P.~S.}\ \bibnamefont {{Lykawka}}}\ and\ \bibinfo {author} {\bibfnamefont {T.}~\bibnamefont {{Ito}}},\ }\href {\doibase 10.3847/1538-3881/aceaf0} {\bibfield  {journal} {\bibinfo  {journal} {Astron. J.}\ }\textbf {\bibinfo {volume} {166}},\ \bibinfo {eid} {118} (\bibinfo {year} {2023})},\ \Eprint {http://arxiv.org/abs/2308.13765} {arXiv:2308.13765 [astro-ph.EP]} \BibitemShut {NoStop}%
\bibitem [{\citenamefont {Scholtz}\ and\ \citenamefont {Unwin}(2020)}]{Scholtz:2019csj}%
  \BibitemOpen
  \bibfield  {author} {\bibinfo {author} {\bibfnamefont {J.}~\bibnamefont {Scholtz}}\ and\ \bibinfo {author} {\bibfnamefont {J.}~\bibnamefont {Unwin}},\ }\href {\doibase 10.1103/PhysRevLett.125.051103} {\bibfield  {journal} {\bibinfo  {journal} {Phys. Rev. Lett.}\ }\textbf {\bibinfo {volume} {125}},\ \bibinfo {pages} {051103} (\bibinfo {year} {2020})},\ \Eprint {http://arxiv.org/abs/1909.11090} {arXiv:1909.11090 [hep-ph]} \BibitemShut {NoStop}%
\bibitem [{\citenamefont {Siraj}\ and\ \citenamefont {Loeb}(2020)}]{SirLoe20}%
  \BibitemOpen
  \bibfield  {author} {\bibinfo {author} {\bibfnamefont {A.}~\bibnamefont {Siraj}}\ and\ \bibinfo {author} {\bibfnamefont {A.}~\bibnamefont {Loeb}},\ }\href {\doibase 10.3847/2041-8213/aba119} {\bibfield  {journal} {\bibinfo  {journal} {Astrophys. J. Lett.}\ }\textbf {\bibinfo {volume} {898}},\ \bibinfo {pages} {L4} (\bibinfo {year} {2020})},\ \Eprint {http://arxiv.org/abs/2005.12280} {arXiv:2005.12280 [astro-ph.HE]} \BibitemShut {NoStop}%
\bibitem [{\citenamefont {Arbey}\ and\ \citenamefont {Auffinger}(2020)}]{Arbey:2020urq}%
  \BibitemOpen
  \bibfield  {author} {\bibinfo {author} {\bibfnamefont {A.}~\bibnamefont {Arbey}}\ and\ \bibinfo {author} {\bibfnamefont {J.}~\bibnamefont {Auffinger}},\ }\href@noop {} {\  (\bibinfo {year} {2020})},\ \Eprint {http://arxiv.org/abs/2006.02944} {arXiv:2006.02944 [gr-qc]} \BibitemShut {NoStop}%
\bibitem [{\citenamefont {{Eroshenko}}\ and\ \citenamefont {{Popova}}(2023)}]{EroPop23}%
  \BibitemOpen
  \bibfield  {author} {\bibinfo {author} {\bibfnamefont {Y.~N.}\ \bibnamefont {{Eroshenko}}}\ and\ \bibinfo {author} {\bibfnamefont {E.~A.}\ \bibnamefont {{Popova}}},\ }\href {\doibase 10.1134/S1063772923110057} {\bibfield  {journal} {\bibinfo  {journal} {Astronomy Reports}\ }\textbf {\bibinfo {volume} {67}},\ \bibinfo {pages} {1193} (\bibinfo {year} {2023})},\ \Eprint {http://arxiv.org/abs/2312.05902} {arXiv:2312.05902 [astro-ph.EP]} \BibitemShut {NoStop}%
\bibitem [{\citenamefont {Rojas}\ \emph {et~al.}(2021)\citenamefont {Rojas}, \citenamefont {Duprat}, \citenamefont {Engrand}, \citenamefont {Dartois}, \citenamefont {Delauche}, \citenamefont {Godard}, \citenamefont {Gounelle}, \citenamefont {Carrillo-Sánchez}, \citenamefont {Pokorný},\ and\ \citenamefont {Plane}}]{Rojetal21}%
  \BibitemOpen
  \bibfield  {author} {\bibinfo {author} {\bibfnamefont {J.}~\bibnamefont {Rojas}}, \bibinfo {author} {\bibfnamefont {J.}~\bibnamefont {Duprat}}, \bibinfo {author} {\bibfnamefont {C.}~\bibnamefont {Engrand}}, \bibinfo {author} {\bibfnamefont {E.}~\bibnamefont {Dartois}}, \bibinfo {author} {\bibfnamefont {L.}~\bibnamefont {Delauche}}, \bibinfo {author} {\bibfnamefont {M.}~\bibnamefont {Godard}}, \bibinfo {author} {\bibfnamefont {M.}~\bibnamefont {Gounelle}}, \bibinfo {author} {\bibfnamefont {J.}~\bibnamefont {Carrillo-Sánchez}}, \bibinfo {author} {\bibfnamefont {P.}~\bibnamefont {Pokorný}}, \ and\ \bibinfo {author} {\bibfnamefont {J.}~\bibnamefont {Plane}},\ }\href {\doibase https://doi.org/10.1016/j.epsl.2021.116794} {\bibfield  {journal} {\bibinfo  {journal} {Earth and Planetary Science Letters}\ }\textbf {\bibinfo {volume} {560}},\ \bibinfo {pages} {116794} (\bibinfo {year} {2021})}\BibitemShut {NoStop}%
\bibitem [{\citenamefont {Siraj}\ and\ \citenamefont {Loeb}(2021)}]{SirLoe21}%
  \BibitemOpen
  \bibfield  {author} {\bibinfo {author} {\bibfnamefont {A.}~\bibnamefont {Siraj}}\ and\ \bibinfo {author} {\bibfnamefont {A.}~\bibnamefont {Loeb}},\ }\href@noop {} {\  (\bibinfo {year} {2021})},\ \Eprint {http://arxiv.org/abs/2103.04995} {arXiv:2103.04995 [astro-ph.CO]} \BibitemShut {NoStop}%
\bibitem [{\citenamefont {{Siraj}}\ and\ \citenamefont {{Loeb}}(2021)}]{SirLoe21-2}%
  \BibitemOpen
  \bibfield  {author} {\bibinfo {author} {\bibfnamefont {A.}~\bibnamefont {{Siraj}}}\ and\ \bibinfo {author} {\bibfnamefont {A.}~\bibnamefont {{Loeb}}},\ }\href {\doibase 10.3847/2515-5172/ac0c1a} {\bibfield  {journal} {\bibinfo  {journal} {Research Notes of the American Astronomical Society}\ }\textbf {\bibinfo {volume} {5}},\ \bibinfo {eid} {145} (\bibinfo {year} {2021})},\ \Eprint {http://arxiv.org/abs/2104.07672} {arXiv:2104.07672 [astro-ph.EP]} \BibitemShut {NoStop}%
\bibitem [{\citenamefont {{Escriv{\`a}}}\ \emph {et~al.}(2024)\citenamefont {{Escriv{\`a}}}, \citenamefont {{K{\"u}hnel}},\ and\ \citenamefont {{Tada}}}]{2024bheg.book..261E}%
  \BibitemOpen
  \bibfield  {author} {\bibinfo {author} {\bibfnamefont {A.}~\bibnamefont {{Escriv{\`a}}}}, \bibinfo {author} {\bibfnamefont {F.}~\bibnamefont {{K{\"u}hnel}}}, \ and\ \bibinfo {author} {\bibfnamefont {Y.}~\bibnamefont {{Tada}}},\ }in\ \href {\doibase 10.1016/B978-0-32-395636-9.00012-8} {\emph {\bibinfo {booktitle} {Black Holes in the Era of Gravitational-Wave Astronomy}}},\ \bibinfo {editor} {edited by\ \bibinfo {editor} {\bibfnamefont {M.}~\bibnamefont {{Arca Sedda}}}, \bibinfo {editor} {\bibfnamefont {E.}~\bibnamefont {{Bortolas}}}, \ and\ \bibinfo {editor} {\bibfnamefont {M.}~\bibnamefont {{Spera}}}}\ (\bibinfo {year} {2024})\ pp.\ \bibinfo {pages} {261--377}\BibitemShut {NoStop}%
\bibitem [{\citenamefont {{Harada}}(2024)}]{2024Univ...10..444H}%
  \BibitemOpen
  \bibfield  {author} {\bibinfo {author} {\bibfnamefont {T.}~\bibnamefont {{Harada}}},\ }\href {\doibase 10.3390/universe10120444} {\bibfield  {journal} {\bibinfo  {journal} {Universe}\ }\textbf {\bibinfo {volume} {10}},\ \bibinfo {eid} {444} (\bibinfo {year} {2024})},\ \Eprint {http://arxiv.org/abs/2409.01934} {arXiv:2409.01934 [gr-qc]} \BibitemShut {NoStop}%
\bibitem [{\citenamefont {Bagui}\ \emph {et~al.}(2025)\citenamefont {Bagui} \emph {et~al.}}]{2025LRR....28....1B}%
  \BibitemOpen
  \bibfield  {author} {\bibinfo {author} {\bibfnamefont {E.}~\bibnamefont {Bagui}} \emph {et~al.} (\bibinfo {collaboration} {LISA Cosmology Working Group}),\ }\href {\doibase 10.1007/s41114-024-00053-w} {\bibfield  {journal} {\bibinfo  {journal} {Living Rev. Rel.}\ }\textbf {\bibinfo {volume} {28}},\ \bibinfo {pages} {1} (\bibinfo {year} {2025})},\ \Eprint {http://arxiv.org/abs/2310.19857} {arXiv:2310.19857 [astro-ph.CO]} \BibitemShut {NoStop}%
\bibitem [{\citenamefont {Carr}\ and\ \citenamefont {Green}(2025)}]{Green2025}%
  \BibitemOpen
  \bibfield  {author} {\bibinfo {author} {\bibfnamefont {B.~J.}\ \bibnamefont {Carr}}\ and\ \bibinfo {author} {\bibfnamefont {A.~M.}\ \bibnamefont {Green}},\ }\enquote {\bibinfo {title} {{The History of Primordial Black Holes}},}\ \ (\bibinfo {year} {2025})\ \Eprint {http://arxiv.org/abs/2406.05736} {arXiv:2406.05736 [astro-ph.CO]} \BibitemShut {NoStop}%
\bibitem [{\citenamefont {Carr}\ \emph {et~al.}(2026)\citenamefont {Carr}, \citenamefont {Iovino}, \citenamefont {Perna}, \citenamefont {Vaskonen},\ and\ \citenamefont {Veerm{\"a}e}}]{2026NCimR..49..225C}%
  \BibitemOpen
  \bibfield  {author} {\bibinfo {author} {\bibfnamefont {B.}~\bibnamefont {Carr}}, \bibinfo {author} {\bibfnamefont {A.~J.}\ \bibnamefont {Iovino}}, \bibinfo {author} {\bibfnamefont {G.}~\bibnamefont {Perna}}, \bibinfo {author} {\bibfnamefont {V.}~\bibnamefont {Vaskonen}}, \ and\ \bibinfo {author} {\bibfnamefont {H.}~\bibnamefont {Veerm{\"a}e}},\ }\href {\doibase 10.1007/s40766-026-00080-z} {\bibfield  {journal} {\bibinfo  {journal} {Riv. Nuovo Cim.}\ }\textbf {\bibinfo {volume} {49}},\ \bibinfo {pages} {225} (\bibinfo {year} {2026})},\ \Eprint {http://arxiv.org/abs/2601.06024} {arXiv:2601.06024 [astro-ph.CO]} \BibitemShut {NoStop}%
\bibitem [{\citenamefont {Byrnes}\ \emph {et~al.}(2025)\citenamefont {Byrnes}, \citenamefont {Franciolini}, \citenamefont {Harada}, \citenamefont {Pani},\ and\ \citenamefont {Sasaki}}]{PBHs}%
  \BibitemOpen
  \bibinfo {editor} {\bibfnamefont {C.}~\bibnamefont {Byrnes}}, \bibinfo {editor} {\bibfnamefont {G.}~\bibnamefont {Franciolini}}, \bibinfo {editor} {\bibfnamefont {T.}~\bibnamefont {Harada}}, \bibinfo {editor} {\bibfnamefont {P.}~\bibnamefont {Pani}}, \ and\ \bibinfo {editor} {\bibfnamefont {M.}~\bibnamefont {Sasaki}},\ eds.,\ \href {\doibase 10.1007/978-981-97-8887-3} {\emph {\bibinfo {title} {{Primordial Black Holes}}}},\ Springer Series in Astrophysics and Cosmology\ (\bibinfo  {publisher} {Springer},\ \bibinfo {year} {2025})\BibitemShut {NoStop}%
\bibitem [{\citenamefont {{Green A M}}(2025)}]{Gre25}%
  \BibitemOpen
  \bibfield  {author} {\bibinfo {author} {\bibnamefont {{Green A M}}},\ }\href@noop {} {\enquote {\bibinfo {title} {{\it Observational probes of Primordial Black Hole dark matter}},}\ }\bibinfo {howpublished} {``Zooming in on primordial black holes'' online seminar hosted by Sam Young, October 6} (\bibinfo {year} {2025})\BibitemShut {NoStop}%
\bibitem [{\citenamefont {Rey}(1987)}]{Rey:1986zk}%
  \BibitemOpen
  \bibfield  {author} {\bibinfo {author} {\bibfnamefont {S.-J.}\ \bibnamefont {Rey}},\ }\href {\doibase 10.1016/0550-3213(87)90058-7} {\bibfield  {journal} {\bibinfo  {journal} {Nucl. Phys. B}\ }\textbf {\bibinfo {volume} {284}},\ \bibinfo {pages} {706} (\bibinfo {year} {1987})}\BibitemShut {NoStop}%
\bibitem [{\citenamefont {Feynman}\ and\ \citenamefont {Hibbs}(2010)}]{FeynHibb}%
  \BibitemOpen
  \bibfield  {author} {\bibinfo {author} {\bibfnamefont {R.~P.}\ \bibnamefont {Feynman}}\ and\ \bibinfo {author} {\bibfnamefont {A.~R.}\ \bibnamefont {Hibbs}},\ }\href@noop {} {\emph {\bibinfo {title} {Quantum Mechanics and Path Integrals}}},\ \bibinfo {edition} {emended}\ ed.,\ edited by\ \bibinfo {editor} {\bibfnamefont {D.~F.}\ \bibnamefont {Styer}}\ (\bibinfo  {publisher} {Dover Publications},\ \bibinfo {address} {Mineola, NY},\ \bibinfo {year} {2010})\BibitemShut {NoStop}%
\end{thebibliography}%

\newpage

\section*{Appendix II}
\setcounter{footnote}{0} 
\interfootnotelinepenalty=10000
{
  \renewcommand{\thefootnote}{*}
  
{\large \bf THE HYPOTHESIS OF CORES RETARDED DURING EXPANSION AND THE HOT COSMOLOGICAL MODEL\footnotemark[1]}

\footnotetext[1]{The article is reproduced with the kind permission of I.D. Novikov and the ``Astronomicheskii Zhurnal''.}
}

\medskip

{\large \bf Ya. B. Zel'dovich and I. D. Novikov}

\medskip

Sov. Astron. -- AJ, Vol.~10, No.~4, pp.~602--603, January--February, 1967; Translated from Astronomicheskii Zhurnal, Vol.~43, No.~4, pp.~758--760, July--August, 1966

\medskip

Original article submitted March 14, 1966

\bigskip

The existence of bodies with dimensions less than \(R_g = 2GM/c^2\) at the early stages of expansion of the cosmological model leads to a strong accretion of radiation by these bodies. If further calculations confirm that accretion is catastrophically high, the hypothesis on cores retarded during expansion [3, 4] will conflict with observational data.

\bigskip

Ambartsumyan has long held the view that stars and galaxies formed from hypothetical superdense D-objects, whose material never passed through a low-density state before conversion into ordinary celestial
bodies [1, 2]. In 1964 one of the authors of this note advanced the hypothesis [3] that the D-objects might be individual pieces of matter, or cores, in an expanding Friedmann universe, which for reasons that we shall not review here have been retarded in their expansion for an external observer, are located inside their Schwarzschild spheres of radius \(R = R_g = 2GM/c^2\), and emerge from these spheres only after a long period has passed since the beginning of the general expansion. Subsequently, and independently, a similar hypothesis was advanced by Ne'eman [4].

This hypothesis was treated within the scope of a cosmological model in which the density of all forms of radiation was considered negligibly small.

The discovery of general cosmic radiation with $T \approx 3^\circ$~K would appear to argue for a hot cosmological model. This circumstance radically changes the situation regarding the retarded-cores hypothesis. If these bodies had existed at an early epoch in the expansion, when the mean radiation density \(\rho_{\mathrm{r}}\) was large and much greater than the baryon density \(\rho_{\mathrm{b}}\), then accretion (gravitational capture) of radiation by the retarded cores might have been catastrophically strong.\footnote{In a radiation-free model there is a solution in which a retarded core is surrounded by a region of lower density; the mean density does not change, and there is no accretion. Obviously, the reason for the difference in the hot-model case is the motion of quanta, equivalent to a quantum-gas pressure $p=\varepsilon/3$, unlike the dust for which $p=0$. Accretion has been considered in [5-7].} Let us estimate this effect.

Stationary accretion of relativistic gas by a gravitating center of mass \(M\) may be regarded either as the motion of noninteracting particles (if the mean free path is greater than \(R_g\)) or as the motion of a continuous medium. Both approaches yield the same expression, in order of magnitude, for the mass variation of the body. The formula for noninteracting particles is
\begin{equation}
\frac{dM}{dt} = \frac{27}{4}\pi R_g^2 c \rho_{\mathrm{r}}.
\tag{1}
\end{equation}
Substituting
\(\rho_{\mathrm{r}} = 3/(32\pi G t^2)\), as is the case for the Friedmann model, and integrating over the interval from \(t=t_0\) to time \(t=t_1\), when
\(\rho_{\mathrm{r}} = \rho_{\mathrm{b}} \simeq
10^{-17}\,\mathrm{g/cm^3}\), with
\(t_1 \approx 3\cdot 10^{11}\,\mathrm{sec}\) (at subsequent times
\(\rho_{\mathrm{r}}\) will decline more rapidly, like \(t^{-8/3}\), and
its contribution will be small), we obtain
\begin{equation}
M_1 = M_0
\left(
1-\frac{81}{32}\frac{GM_0}{c^3}
\left(\frac{1}{t_0}-\frac{1}{t_1}\right)
\right)^{-1}.
\tag{2}
\end{equation}
Equation (2) may yield infinite values of \(M_1\), for
\begin{equation}
t_0 \leq
\left(\frac{81}{32}\frac{GM_0}{c^3}\right)
\left(
1+\frac{81}{32}\frac{GM_0}{c^3 t_1}
\right)^{-1}.
\tag{3}
\end{equation}
To a certain extent this singularity is a formal one since Eq.~(2) is applicable only for a stationary flux, and it is clear that the mass of a core within the Schwarzschild sphere cannot increase because of accretion by more than the value of the mass that a perturbation wave can penetrate up to time \(t_1\) at the velocity of light (approximately equal to the sound velocity), beginning at the time \(t_0\) of collapse of the core, or at \(t_0=0\) if the core has existed from the start of the expansion.

In the first place, then, \(M_1\) should definitely be less than the mass of the radiations in the metagalaxy out to the optical horizon at time \(t_1\);\footnote{Or at the present time. The important thing is that the limit on $M_1$ is very high.} that is, \(M_1 \leq\sim 10^{16} M_\odot\). This limit is very high; captured radiation will not participate in the cosmological expansion.\footnote{Such a body is similar to a geon [8]: its mass is much greater than the rest mass of the baryons comprising it. The distinction from specific Wheeler models is that a nonstationary collapsing formation is considered essential here.}

Secondly, if the core has existed since the start of the expansion, then until the time when the optical horizon encompasses a mass of order \(M_0\) of surrounding material the mass of the core will not grow appreciably; it is this time that should be used in Eq.~(2) in place of \(t_0\).

This epoch is determined by the relation
\begin{equation}
t_c = \frac{GM_0}{c^3},
\tag{4}
\end{equation}
and corresponds to the mean density of the cosmological model, being equal to the characteristic density of gravitational self-closure for a body of mass \(M_0\), namely \(\rho_{\mathrm{c}} = M_0/(\frac{4}{3}\pi R_g^3)\).

Of course these considerations give only the order of magnitude of \(t_0\). From formulas (3) and (4) it follows \((t_1 \gg t_0)\), that \(t_c\) is of just the order of the critical value \(t_0\), giving a formal singularity, or more explicitly implying an accretion of radiation mass out to the current optical horizon.

Thus the answer depends essentially on the numerical coefficient in the exact nonstationary expression for \(dM/dt\). If up to the time when the mean density equals the gravitational self-closure density \(\rho_{\mathrm{c}} = M_0/(\frac{4}{3}\pi R_g^3)\) the expression for \(dM/dt\) differs from Eq.~(1) by a factor significantly less than unity (more accurately, \(<32/81\)), then the increase in the mass of the retarded cores due to accretion of radiation will be of the same order as the mass of the cores themselves;\footnote{In this event it would be interesting to study the observable effects associated with collisions between expanding cores and relativistic gas falling in from outside.} otherwise the captured mass would tend to be catastrophically large.

The value of the factor should be determined by more detailed calculations, and it may depend on the particular assumptions made. One might show that the hypothesis of nuclei retarded in expansion and existing from the early stages of expansion is contrary to observation in certain respects.

Finally, we note that if gravitational collapse of any body occurs at a time when the mean radiation density is less than the gravitational self-closure density of the body, then accretion of radiation will already be unimportant.

Apart from the question of the accretion of radiation, the fraction \(\alpha\) of all the material that suffered collapse at an early stage (when \(\rho_{\mathrm{r}} \gg \rho_{\mathrm{b}}\)) should be small. In fact, \(\rho_{\mathrm{r}}/\rho_{\mathrm{b}}\) will decline during the expansion, while the radiation in the cores will not participate in the cosmological expansion but will have \(\rho_{\mathrm{c}}/\rho_{\mathrm{b}} = \mathrm{const}\), where \(\rho_{\mathrm{c}}\) is the density of the collapsed masses averaged over all space. The observations now imply \(\rho_{\mathrm{c}}/\rho_{\mathrm{b}} < 160\). One can thereby show that $\alpha < 10^{-6}(M/M_\odot)^{1/2}$, where \(M\) is the mass of the collapsed bodies. Similar considerations provide bounds on the initial spectrum of inhomogeneity fluctuations.

\bigskip

LITERATURE CITED

\bigskip

1. V. A. Ambartsumyan, Scientific Works [in Russian] (Erevan, 1960), Vol. 2.

2. V. A. Ambartsumyan, Rapport 13, Conseil de physique Solvay, Brussels (1964).

3. I. D. Novikov, Astron. Zh. 41, 1075 (1964) [Soviet Astronomy – AJ, Vol. 8, p. 85].

4. Y. Ne’eman, Astrophys. J. 141, 1303 (1965).

5. E. E. Salpeter, Astrophys. J. 140, 796 (1964).

6. Ya. B. Zel’dovich, DAN SSSR 155, 67 (1964) [Soviet Physics – Doklady, Vol. 9, p. 195].

7. Ya. B. Zel’dovich and I. D. Novikov, Uspekhi Fiz. Nauk
86, 433 (1965).

8. J. A. Wheeler, Phys. Rev. 97, 511 (1955).

\end{document}